\documentclass[11pt]{article}
\usepackage{epsfig,amsfonts,amsthm}
\usepackage{amsmath,amssymb}
\usepackage{float}
\usepackage{cite}
\usepackage{xcolor}
\usepackage{braket}
\usepackage{booktabs}
\usepackage{makecell}
\usepackage{multirow}
\usepackage{makecell}
\usepackage{slashed}
\usepackage{mdframed}
\newtheorem{theorem}{Theorem}
\usepackage{setspace}

\usepackage{geometry}
\usepackage{titlesec}
\usepackage{hyperref}
\titleclass{\subsubsubsection}{straight}[\subsection]
\newcounter{subsubsubsection}[subsubsection]
\renewcommand\thesubsubsubsection{\thesubsubsection.\arabic{subsubsubsection}}
 
\titleformat{\subsubsubsection}
  {\normalfont\normalsize\bfseries}{\thesubsubsubsection}{1em}{}
\titlespacing*{\subsubsubsection}
{0pt}{3.25ex plus 1ex minus .2ex}{1.5ex plus .2ex}
\makeatletter
\renewcommand\paragraph{\@startsection{paragraph}{5}{\z@}%
  {3.25ex \@plus1ex \@minus.2ex}%
  {-1em}%
  {\normalfont\normalsize\bfseries}}
\renewcommand\subparagraph{\@startsection{subparagraph}{6}{\parindent}%
  {3.25ex \@plus1ex \@minus .2ex}%
  {-1em}%
  {\normalfont\normalsize\bfseries}}
\def\toclevel@subsubsubsection{4}
\def\toclevel@paragraph{5}
\def\toclevel@subparagraph{6}
\def\l@subsubsubsection{\@dottedtocline{4}{7em}{4em}}
\def\l@paragraph{\@dottedtocline{5}{10em}{5em}}
\def\l@subparagraph{\@dottedtocline{6}{14em}{6em}}
\makeatother
\newcommand{\be}{\begin{equation}}
\newcommand{\ee}{\end{equation}}
\newcommand{\bea}{\begin{eqnarray}}
\newcommand{\eea}{\end{eqnarray}}

\renewcommand{\Im}{\mathrm{Im }}

\newcommand{\lr}[1]{ \langle #1 \rangle}
\newcommand{\Tr}{\mathrm{Tr}}
\newcommand{\Z}{\mathbb{Z}}
\newcommand{\mmatrix}[4]{ \left(\! \begin{array}{ccc}#1 & #2 \\ #3 & #4 \end{array}\!\right) }

\providecommand{\id}{{\boldsymbol{1}}}
\newcommand{\diag}{\mathrm{diag}}
\newcommand{\Aut}{\mathrm{Aut}}
\newcommand{\Ker}{\mathrm{Ker}}
\newcommand{\fdf}[2]{(\phi^\dagger_{#1}\phi_{#2})}
\newcommand{\fdfn}[2]{\phi^\dagger_{#1}\phi_{#2}}

\definecolor{darkred}{rgb}{0.7,0.0,0.0}

\newcommand{\gray}{\color{gray}}
\definecolor{darkgreen}{rgb}{0.0,0.5,0.0}

\def\lsim{\mathrel{\rlap{\lower4pt\hbox{\hskip1pt$\sim$}}
\raise1pt\hbox{$<$}}}         
\def\gsim{\mathrel{\rlap{\lower4pt\hbox{\hskip1pt$\sim$}}
\raise1pt\hbox{$>$}}}         

\begin{document}

\begin{titlepage}
    \centering
    
    {\large \textsc{Undergraduate Thesis}}\\[0.6cm]

    \includegraphics[width=0.215\textwidth]{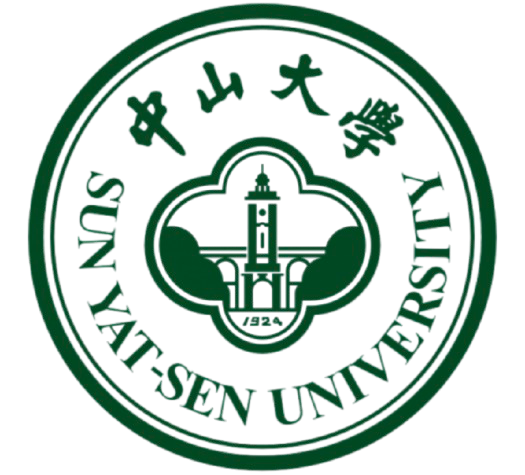} \quad
    \includegraphics[width=0.5\textwidth]{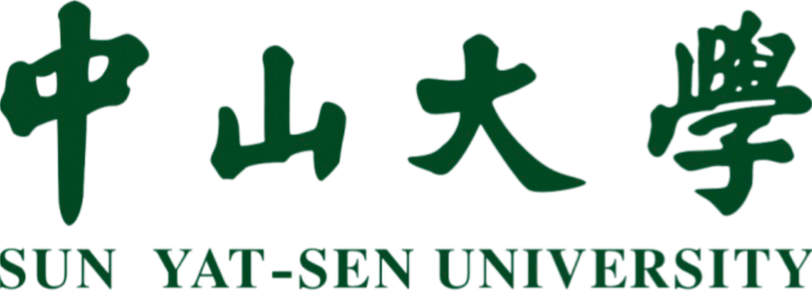}\\
    \;\\[1.0cm]

    {\Huge \textbf{Classification of the Discrete} \\[-9pt] 
    \textbf{Symmetries for the} \\[10pt] 
    \textbf{Four-Higgs-Doublet Model}}\\[2.0cm]
    
    \today\\[3cm]

    {\Large Jiazhen Shao \\ {\small\tt Email: shaojzh5@mail2.sysu.edu.cn\,,\;shaojzh5@gmail.com}}\\[1.5cm]

    \textit{School of Physics and Astronomy, Sun Yat-Sen University,\\ 519082 Zhuhai, Guangdong, P.R. China}\\[1.5cm]
    
\end{titlepage}

\newpage

\centerline{\Large ACKNOWLEDGEMENTS}
\;\\[-10pt]

Over the years, the achievements I have made as an undergraduate physics student still feel unbelievable to me, especially considering that just a few years ago, I was grappling with whether to change my major—continuing to pursue my dream in physics and astronomy while also studying engineering.

First, I would like to say a few words to my family. Without their support, I would not have had the chance to ignore all practical considerations and continue to chase my childhood dream of exploring the beautiful laws of nature and the universe. Thank you for understanding and supporting my somewhat naïve and carefree career plans.

During my undergraduate studies, I am especially grateful to Professor Pengming Zhang. We had several conversations, and his deep insights into various fields of physics guided me in choosing quantum field theory and elementary particle physics as my primary research focus as an undergraduate. Most importantly, he introduced me to Professor Igor Ivanov, who provided me with the greatest academic support. As my primary research advisor, he not only taught me quantum field theory and particle physics but also gave me a comprehensive experience of scientific research, including publishing papers and presenting at academic conferences.

\newpage

\centerline{\Large ABSTRACT}
\;\\[-10pt]

The multi-Higgs model (NHDM) is a class of new physics models that go beyond the Standard Model with fewer assumptions, namely by postulating the existence of multiple Higgs doublets, while yielding a rich phenomenology. Among them, the Four-Higgs-Doublet model (4HDM) has attracted increasing attention in recent years, with nearly a hundred papers dedicated to its study. Scientists have attempted to use it to explain many phenomena that the Standard Model cannot account for, such as dark matter, fermion mass hierarchies, neutrino masses, cosmological phase transitions, and cosmic strings, among others. In these efforts, imposing finite symmetries on the 4HDM is a standard procedure, because finite symmetries are central to explaining many physical phenomena, such as dark matter, and can also constrain excessively large parameter spaces technically. 

Therefore, given the recent attention to 4HDM and the importance of finite symmetries in model building, it is crucial from a mathematical perspective to classify all possible finite symmetries, as the classification results can guide model construction and phenomenological studies. So far, two scenarios of discrete symmetries of 4HDM have been classified, but there are still cases that haven't been studied. During our study, we found out situations that have never appeared in 2HDM or 3HDM, and we developed many new techniques to help us handle the difficulties. Therefore, before any further progresses are made towards a full classification, it is timely to revisit the current progress of 4HDM discrete symmetry classification in a beginner-friendly manner, and to provide a look towards future research. 

It's worth noting that this is an extended version of the original undergraduate thesis, written in Chinese, submitted by the author in Spring 2024 to the School of Physics and Astronomy, Sun Yat-sen University (Zhuhai, China). We added the latest calculation results and provide an English version, making this work more publicized. 

\newpage
\tableofcontents
\bigskip

\hrule

\newpage

\section{The Standard Model and Beyond}

From Born, Heisenberg, and Jordan's paper \cite{Born:1926uzf} on using matrix mechanics to describe radiation fields in 1926 to the discovery of the Higgs boson in 2012 \cite{CMS:2012qbp,ATLAS:2012yve}, our understanding of the elementary particle physics has been developing for about 100 years, and the model we have now, namely the Standard Model of particle physics (SM) is successful in accounting for a huge range of experimental data ranging from ultra-cold atoms to high energy colliders. 

\begin{figure}[h]
    \centering
    \includegraphics[scale=0.3]{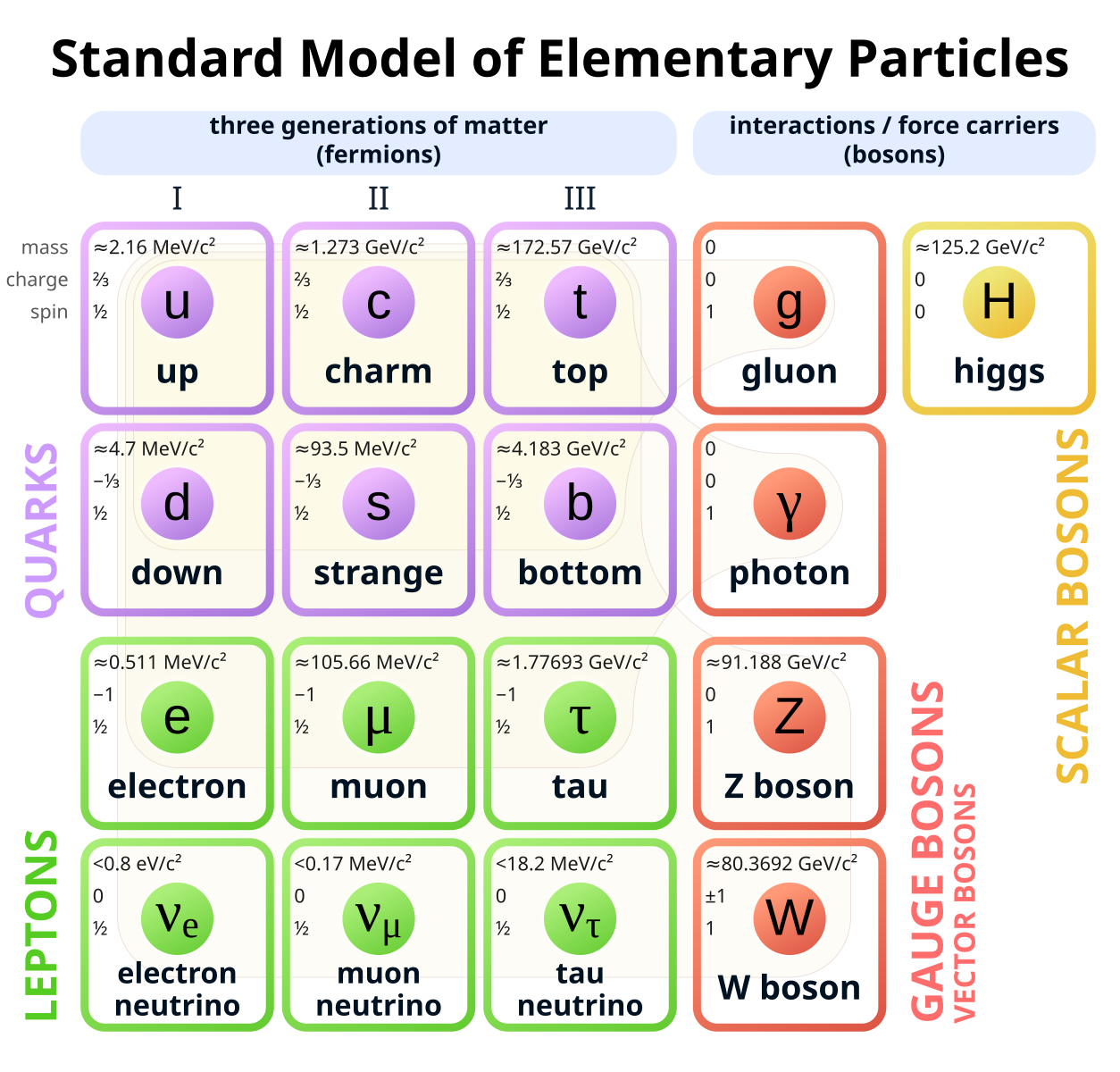}
    \caption{Collection of elementary particles/fields and interactions in the Standard Model of Particle Physics. }
    \label{fig:SM}
\end{figure}

The SM uses the mathematical framework of quantum field theory to describe what is our universe is composed of: 12 fermionic fields and their anti-matter fields with spin 1/2, and what are the interactions among them: 4 gauge bosonic fields with spin 1 coming from gauge symmetry breaking and one scalar boson causing the symmetry breaking. Details in Figure \ref{fig:SM}. 

In this section, we provide a review of the construction of the Standard Model, including its mathematical formulation and how this formulation describes matter and interactions. We will delve into a more detailed explanation of the summary Eq.~\eqref{sketch-SM}. The Standard Model is known to be the theoretical framework that describes the most elementary particles—the elementary particles that constitute the natural world—and the interactions between these particles. Therefore, it encompasses the part that describes interactions, including the description of the elementary particles (gauge bosons) that mediate electromagnetic, strong, and weak interactions (the first line of Eq.~\eqref{sketch-SM}). We will also see that the gauge bosons responsible for mediating these interactions originate from gauge symmetries. The model further includes the description of the elementary particles that constitute matter (the second line of Eq.~\eqref{sketch-SM}): quarks and leptons. Most importantly, the terms describing the masses of both the matter particles and the gauge bosons are not gauge-invariant. Hence, we need to introduce a mechanism to make these particles acquire mass, in accordance with experimental observations (the third and fourth lines of Eq.~\eqref{sketch-SM}). This mechanism is the electroweak symmetry breaking mechanism, also known as the Brout-Englert-Higgs mechanism. Concepts such as the Higgs doublet and the Higgs boson are central to this mechanism.
\begin{equation}
\begin{aligned}
\mathcal{L}_{SM} \overset{\mbox{\tiny summary}}{\simeq}  
& -\frac{1}{4}F^{\mu\nu}F_{\mu\nu} + {\gray \frac{1}{2}m_b^2A^\mu A_\mu} & \mbox{ Gauge Boson sector}\\
& + i\overline{\psi}\slashed{D}\psi - {\gray m_f\overline{\psi}\psi} & \mbox{Matter Field sector} \\
& + Y\overline{Q}_L H \psi_R + h.c. & \mbox{ Yukawa sector} \\
& + (D^\mu H)^\dagger D_\mu H - V(H) & \mbox{ Higgs sector}
\end{aligned}\label{sketch-SM}
\end{equation}

First, let's focus on how the gauge theory within the Standard Model uses symmetries to describe interactions.

\subsection{The Standard Model's Description of Interactions—Gauge Theory}\label{SM-Gauge}

In the Standard Model, the theory describing gauge interactions is known as gauge theory. This theory requires the model to possess a symmetry known as gauge symmetry. To preserve this symmetry, a vector boson—referred to as a gauge boson—must be introduced to mediate the interaction. In fact, the core of gauge theory and the essence of gauge symmetry arise from the ``redundant information'' inherent in using mathematical equations to describe physical processes.

From the earliest exposure to physics in middle school, when learning about uniform linear motion and the equation relating distance to speed and time. We notice that as we are considering more and more complicated physical processes, the equations seem to become increasingly complex. Initially, only one or two physical quantities are connected by algebraic formulas, but gradually, terms involving squares, trigonometric functions, differentials, integrals, and special functions appear. The physical quantities themselves also become more complex: from positive numbers to the entire set of real numbers, then vectors, infinitesimals, tensors, and spinors. 

As we strive to describe a broader range of natural phenomena with greater precision, the models we use become more intricate. Consider three of the most successful theories in the history of physics:

\begin{itemize}
    \item Newtonian classical mechanics: Despite being an approximation for macroscopic, low-speed, and weak gravitational systems, it remains widely used in engineering. However, even this ``simple'' theory is unsolvable for systems involving more than three bodies.
    \item General relativity: A masterpiece of mathematics, science, and art, it extends human understanding to the scientific study of the entire universe. Nevertheless, exact solutions to the field equations are extremely rare, and even a single-body system with charge and rotation is challenging, while the two-body problem is even more difficult to solve analytically.
    \item Quantum electrodynamics (QED): Renowned for its predictive accuracy, it is considered one of the most successful theories in human history. However, the theory is rife with infinities, requiring renormalization, and even the vacuum exhibits fluctuations, rendering it unsolvable in a closed form. 
\end{itemize}

As we go down the list provided above, the physical processes we study transition from simple classical mechanical systems to extreme macroscopic or microscopic systems, the physical models we use become increasingly complex. Yet, one thing is certain: within these complex physical quantities, some complexities are redundant, meaning they retain excessive mathematical details that are unnecessary for describing the world. We will now examine this mathematical redundancy, known as gauge freedom, through two examples.

Consider first the classical theory describing electromagnetic phenomena. When studying electrodynamics, we know that the real physical quantities around us are the electric field and magnetic field. However, solving differential vector field equations (such as Maxwell's equations) is often difficult, so we use mathematical techniques to construct potentials $(\phi, \mathbf{A})$ that help solve these problems:
\begin{equation}
    \mathbf{E} = -\nabla \phi - \frac{\partial \mathbf{A}}{\partial t}
\end{equation}
\begin{equation}
    \mathbf{B} = \nabla \times \mathbf{A}
\end{equation}
In this construction, solving for two vector fields $\mathbf{E}, \mathbf{B}$ is transformed into solving for a vector field $\mathbf{A}$ and a scalar field $\phi$, greatly simplifying the task. However, performing the following transformation on the potentials does not affect the real physical quantities: the electric and magnetic fields.
\begin{equation}
\left\{
    \begin{aligned}
    & \phi\mapsto\phi^\prime = \phi - \frac{\partial \Lambda}{\partial t} \\
    & \mathbf{A}\mapsto \mathbf{A}^\prime = \mathbf{A} + \nabla \Lambda \\
    \end{aligned}
\right.\label{EaM-potential-transformation}
\end{equation}
Thus, the potentials $(\phi, \mathbf{A})$ contain redundant mathematical details. This is similar to choosing the zero point of gravitational potential energy in mechanics: the choice of zero point does not affect the actual physics, and the choice of zero point for the scalar field $\Lambda$ does not affect the physical discussion in electrodynamics.

Next, consider quantum mechanics. In quantum mechanics, the quantum state is represented by a state vector, or a ``ket'', $\ket{a}$. According to the probabilistic interpretation, the inner product of the ket to itself represents a probability density, and the measured value of an observable $A$ is defined as:
\begin{equation}
    \langle A \rangle = \frac{\braket{a|A|a}}{\braket{a|a}}
\end{equation}
From this definition, it is evident that $c\ket{a}$, where $c$ is any non-zero complex number, or $c-$number, represents the same physical state as $\ket{a}$. Even if we impose the condition $\braket{a|a} = 1$, ensuring that probability is normalized, $e^{i\theta}\ket{a}$ and $\ket{a}$ remain indistinguishable from an experimental perspective. Therefore, multiplying the ket by any complex number with a modulus of 1 does not change the real quantum state. Thus, the ket itself, as a vector in a complex vector space, inherently contains redundant mathematical details.

These redundant mathematical details can be viewed as a form of symmetry. If we consider Eq.~\eqref{EaM-potential-transformation} and $\ket{\psi}\mapsto e^{i\theta}\ket{\psi}$ as transformations, and the physical laws (such as Maxwell's equations and the Schrödinger equation) as the entities that remain invariant under these transformations, we can interpret these mathematical redundancies as symmetries of the physical laws, known as gauge symmetries.

Having introduced the concept of gauge symmetry, we note that gauge symmetry is not merely a mathematical redundancy that complicates our research. In fact, if we impose stronger conditions on gauge symmetry, different physical theories can be related! This stronger condition is local gauge invariance. Here, we require that the state vector undergoes a phase transformation not by a uniform phase factor $e^{i\theta}$ but by a phase factor $e^{i\theta(x)}$ that varies with position. Since the phase $e^{i\theta}$ belongs to the group $U(1)$, we refer to the gauge transformation $\psi\mapsto e^{i\theta}\psi$ as a $U(1)$ gauge transformation.

Now consider both electromagnetic theory and quantum mechanics. The equation of motion describing the non-relativistic quantum theory of electromagnetic interactions is given by Eq.~\eqref{QED-classical}, obtained by applying the minimal substitution procedure to the Schrödinger equation $\displaystyle i\hbar\partial_t\ket{\psi} = H\ket{\psi}$, which replaces the Hamiltonian $ H $ in the Schrödinger equation with the non-relativistic Hamiltonian for a charged particle $ \displaystyle H = \frac{(\mathbf{p} - q\mathbf{A})^2}{2m} + q\phi $ under the substitution $ \mathbf{p} \mapsto -i\hbar\nabla $.
\begin{equation}
    i\hbar\frac{\partial \psi}{\partial t} = \left[\frac{1}{2m}\left(\frac{\hbar}{i}\nabla - q\mathbf{A}\right)^2 + q\phi \right]\psi\label{QED-classical}
\end{equation}
Apparently, the physical law described by Eq.~\eqref{QED-classical} remains invariant under the global $U(1)$ gauge transformation $\ket{\psi}\mapsto e^{i\theta}\ket{\psi}$, meaning that the quantum theory of electromagnetic interactions under non-relativistic conditions respects global $U(1)$ gauge symmetry. However, if we upgrade the gauge transformation and require that the $U(1)$ gauge transformation is local, i.e., $\ket{\psi}\mapsto e^{i\theta(x)}\ket{\psi}$, the original equation will change due to the presence of spatial derivatives on the right side of Eq.~\eqref{QED-classical}, which will also involve derivatives of $\theta(x)$. Thus, the local $U(1)$ gauge transformation cannot serve as a symmetry of Eq.~\eqref{QED-classical}. But what if we simultaneously apply a gauge transformation to the electromagnetic potentials $(\phi, \mathbf{A})$?

We set $\displaystyle \theta(x) = \frac{q\Lambda(x)}{\hbar}$, and perform a local $U(1)$ gauge transformation on the state vector while simultaneously applying the gauge transformation to the electromagnetic potentials as shown in Eq.~\eqref{QED-classical-transformation}.
\begin{equation}
\left\{
    \begin{aligned}
    & \phi\to\phi^\prime = \phi - \frac{\partial \Lambda}{\partial t} \\
    & \mathbf{A}\to \mathbf{A}^\prime = \mathbf{A} + \nabla \Lambda \\
    & \psi \to \psi^\prime = e^{\frac{iq\Lambda}{\hbar}}\psi
    \end{aligned}
\right.\label{QED-classical-transformation}
\end{equation}
It can be shown that the quantum theory of electromagnetic interactions under non-relativistic conditions, described by Eq.~\eqref{QED-classical}, remains invariant under the transformations in Eq.~\eqref{QED-classical-transformation}! Thus, the local gauge transformation connects the two gauge freedom. We can rewrite Eq.~\eqref{QED-classical}, defining the covariant derivative as in Eq.~\eqref{QED-classical-cov-def}. This allows us to introduce $(\phi, \mathbf{A})$ without using the minimal substitution procedure, which would replace the Hamilton

Thus, we demonstrated that the quantum theory of electromagnetic interactions, as represented by Eq.~\eqref{QED-classical}, remains invariant under the transformation outlined in Eq.~\eqref{QED-classical-transformation}, and this invariance establishes a connection between the two gauge freedoms through local gauge transformations. By modifying Eq.~\eqref{QED-classical} and defining the covariant derivative as in Eq.~\eqref{QED-classical-cov-def}, we can bypass the minimal substitution procedure. Consequently, it is no longer necessary to assume that the $ A_\mu $ field exclusively represents the electromagnetic field.

In fact, $ A_\mu $ is intrinsically linked to the $ U(1) $ local gauge symmetry and is thus referred to as the $ U(1) $ gauge field. The electromagnetic field is a specific example of a $ U(1) $ gauge field; there may exist other physical fields that are also $ U(1) $ gauge fields but are not electromagnetic fields. For example, in the Standard Model, the hypercharge gauge field $ U(1)_Y $ is not the electromagnetic gauge field; only after electroweak symmetry breaking during which it combines with the third component of the $ SU(2)_L $ gauge field, the physical $ U(1)_{em} $ gauge field is formed, which corresponds to the electromagnetic field.
\begin{equation}
\left\{
\begin{aligned}
    & \frac{\hbar}{i}\nabla \mapsto \frac{\hbar}{i}\mathbf{D} =  \frac{\hbar}{i}\nabla - q\mathbf{A} \\
    & i\hbar\partial_t \mapsto i\hbar D_t =  i\hbar\partial_t - q\phi
\end{aligned}
\right.\label{QED-classical-cov-def}
\end{equation}

The original Eq.~\eqref{QED-classical} can then be rewritten in a simplified form Eq.~\eqref{QED-classical-cov}, which closely resembles the free particle Schrödinger Eq.~\eqref{QED-classical-free}:
\begin{equation}
i\hbar \partial_t \psi = -\frac{\hbar^2}{2m}\mathbf{\nabla}^2 \psi\label{QED-classical-free}
\end{equation} 
\begin{equation}
i\hbar D_t \psi = -\frac{\hbar^2}{2m}\mathbf{D}^2 \psi\label{QED-classical-cov}
\end{equation}

Therefore, we conclude that if we upgrade the global gauge symmetry of the Lagrangian or equation of motion, which describes the dynamics of the matter field (whether it be a wave function or a fermion field in particle physics), to consider local gauge transformations, the original Lagrangian or equation of motion will no longer possess this local gauge symmetry. To restore local gauge symmetry to the Lagrangian or equation of motion, it is necessary to introduce an additional physical quantity $ A_\mu $, known as the gauge field, which mediates the interaction. The introduction of the gauge field is achieved by replacing the ordinary derivative with the covariant derivative.

We have outlined the core concepts of gauge theory, and the next step is to apply these ideas to particle physics. In this context, particles are no longer described by state vectors or wave functions but by fields. The Lagrangian for a free fermion field is given by Eq.~\eqref{fermion-free}:
\begin{equation}
\mathcal{L} = \overline{\psi}(i\slashed{\partial} - m)\psi\label{fermion-free}
\end{equation}

Clearly, this Lagrangian possesses global $ U(1) $ gauge invariance, under the transformation $\psi \mapsto e^{-ie\Lambda}\psi$. If we require the $ U(1) $ gauge transformation to be local, i.e., $\Lambda$ becomes a function of spacetime coordinates $\Lambda(x^\mu)$ (denoted as $\Lambda(x)$), then we must introduce an auxiliary field $ A_\mu $, modify the ordinary derivative to the covariant derivative Eq.~\eqref{cov-di}, and impose the transformation $ A_\mu \mapsto A_\mu - \partial_\mu \Lambda $ under the gauge transformation.
\begin{equation}
\partial_\mu \mapsto D_\mu = \partial_\mu - ieA_\mu\label{cov-di}
\end{equation}

It can be verified that the modified Lagrangian $\overline{\psi}(i\slashed{D} - m)\psi$ remains invariant under the local gauge transformation $\psi \mapsto e^{-ie\Lambda(x)}\psi,\,A_\mu \mapsto A_\mu - \partial_\mu \Lambda$. Since the mass term $m\overline{\psi}\psi$ is evidently gauge invariant, we need only verify the gauge invariance of the kinetic term $i\overline{\psi}\slashed{D}\psi$ and the $U(1)$ gauge transformation invariance of $ -\frac{1}{4}F^{\mu\nu}F_{\mu\nu} $:
\begin{equation}
\begin{aligned}
i\overline{\psi}\slashed{D}\psi 
& \mapsto i\overline{\psi}e^{ie\Lambda(x)}\gamma^\mu\left[\partial_\mu - ie(A_\mu - \partial_\mu\Lambda(x) )\right]e^{-ie\Lambda(x)}\psi \\
& = i\overline{\psi}e^{ie\Lambda(x)}\gamma^\mu\left[\psi \partial_\mu e^{-ie\Lambda} + e^{-ie\Lambda}\partial_\mu \psi - ieA_\mu \psi + ie\cdot e^{-ie\Lambda}\psi \partial_\mu\Lambda\right] \\
& = i\overline{\psi}e^{ie\Lambda(x)}\gamma^\mu\left[-ie\cdot e^{-ie\Lambda}\psi\partial_\mu \Lambda + e^{-ie\Lambda}\partial_\mu \psi - ieA_\mu \psi + ie\cdot e^{-ie\Lambda}\psi \partial_\mu\Lambda\right]\\
& = i\overline{\psi}e^{ie\Lambda(x)}\gamma^\mu e^{-ie\Lambda}(\partial_\mu - ieA_\mu)\psi\\
& = i\overline{\psi}\slashed{D}\psi
\end{aligned}
\end{equation}
\begin{equation}
\begin{aligned}
F_{\mu\nu} & = \partial_\mu A_\nu - \partial_\nu A_\mu \\
& \mapsto \partial_\mu (A_\nu - \partial_\nu\Lambda) - \partial_\nu(A_\mu - \partial_\mu\Lambda) \\
& = \partial_\mu A_\nu - \partial_\nu A_\mu - \partial_\mu\partial_\nu\Lambda + \partial_\nu\partial_\mu\Lambda \\
& = F_{\mu\nu}
\end{aligned}
\end{equation}

It is important to note that in Quantum Electrodynamics (QED), the Dirac fermion mass term is gauge invariant. However, since weak interactions exhibit $ CP $ violation, the Standard Model before symmetry breaking does not use Dirac fermions but rather Weyl fermions (or chiral fermions), whose mass terms are not gauge invariant. This point will be discussed in detail in the next section. Similarly, the gauge boson mass term $ \displaystyle \frac{1}{2}m^2A^\mu A_\mu $ is not gauge invariant. Therefore, in the original Lagrangian, which possesses local gauge invariance, all fields: gauge boson fields and Weyl fermion fields, are mass-less. However, experimental observations reveal that all fermions, neutrinos included, and the weak interaction bosons $ W^{\pm} $ and $ Z $ have masses, see Figure \ref{fig:SM}. Thus, gauge symmetry is spontaneously broken through the electroweak symmetry breaking mechanism, giving mass to the matter fields and $ W^{\pm} $ and $ Z $ bosons. However, how neutrinos get their mess is still a mystery. Although there are models describing neutrino mass, further investigations, experiments, and observations are still needed to select one model from the sea of neutrino mass models. 

Consequently, we derive a theory with local $ U(1) $ gauge invariance, as expressed by Eq.~\eqref{QED-L}, which is the QED Lagrangian. Here, $ \displaystyle -\frac{1}{4}F^{\mu\nu}F_{\mu\nu} $ represents the dynamics of the $ U(1) $ gauge field $ A_\mu $, whose variation leads to the Maxwell equations. 
\begin{equation}
\mathcal{L}_{QED} = -\frac{1}{4}F^{\mu\nu}F_{\mu\nu} + \overline{\psi}(i\slashed{D} - m)\psi = -\frac{1}{4}F^{\mu\nu}F_{\mu\nu} + i\overline{\psi}\slashed{\partial}\psi - m\overline{\psi}\psi + e\,\overline{\psi}\gamma^\mu \psi A_\mu
\label{QED-L}
\end{equation}

Using the same principles, we can derive the Lagrangians for other gauge theories. For example, gluons, which mediate the strong interaction, correspond to the $ SU(3)_c $ gauge field, where the subscript $ c $ stands for color, referring to the color charge in Quantum Chromodynamics (QCD). The Standard Model describes the three fundamental interactions: strong, weak, and electromagnetic. Therefore, we naturally expect three corresponding gauge fields. Besides the $ U(1)_{em} $ electromagnetic field, the other gauge fields arise from non-abelian gauge transformations and exhibit local non-abelian gauge invariance. Before introducing the specific interaction gauge fields, such as the $ SU(3)_c $ gauge field for the strong interaction, we will first discuss non-abelian gauge theory in general terms. However, before proceeding, we will introduce several useful equations that are convenient for extending to the non-abelian gauge theory context.

First, under gauge transformations, we can rewrite the transformation of the covariant derivative as follows: 
\begin{equation}
D_\mu \mapsto e^{-ie\Lambda}D_\mu e^{ie\Lambda}\label{cov-dir-transformation}
\end{equation}
This makes the gauge invariance of $i\overline{\psi}\slashed{D}\psi$ quite evident. It is important to note that the definition in Eq.~\eqref{cov-dir-transformation} is established at the operator level, and its proof requires acting on an auxiliary quantity. Similarly, we can rewrite $F_{\mu\nu}$ as:
\begin{equation}
F_{\mu\nu} = \frac{i}{e}[D_\mu, D_\nu]
\end{equation}
Again, the proof requires introducing an auxiliary state vector $\Psi$ and showing that $F_{\mu\nu}\Psi = \frac{i}{e}[D_\mu, D_\nu]\Psi$.

In the context of non-abelian gauge transformations, the fermion field cannot consist of just a single component; therefore, we represent the fermion field as $\psi_i$, where $i$ is an index in the gauge transformation space, distinct from the Dirac spinor indices and spacetime coordinates. Under a non-abelian gauge transformation, the fermion field transforms as $\psi_i \mapsto U_{ij}\psi_j$. The free Lagrangian for the fermion field, summed over the index $k$, is expressed as:
\begin{equation}
\mathcal{L} = \overline{\psi}_k(i\delta_{kl}\slashed{\partial} - m\delta_{kl})\psi_l = \overline{\psi}_k(i\slashed{\partial} - m)\psi_k
\end{equation}
It is evident that $\overline{\psi}_k(i\slashed{\partial} - m)\psi_k$ remains invariant under the global transformation $\psi_i\mapsto U_{ij}\psi_j$. However, if we demand that the gauge transformation $U_{ij}$ be local, then the spacetime derivatives in the kinetic term will cause the transformed Lagrangian to differ from the original. To ensure that the Lagrangian retains local gauge invariance, we must again replace the ordinary derivative with the covariant derivative. It is crucial to note that the introduced auxiliary field $A_\mu$ is a matrix rather than a scalar, because the covariant derivative must act on $\psi_k$. Next, we study a bit further of the matrix $A_\mu$.

In the $U(1)$ gauge theory, the local $U(1)$ transformation is represented by $e^{-i\Lambda(x)}$. After replacing the derivative with the covariant derivative, the $\partial_\mu\Lambda$ term coming from locality of the gauge transformation is eliminated. Hence, $A_\mu$ and $\partial_\mu \Lambda$ reside in the same space: the Lie algebra of the $U(1)$ group. Extending this to a non-abelian gauge theory, we consider a non-abelian transformation forming a Lie group $G$. The local $G$ transformation is represented by the exponential map $U = e^{-ig\theta^a(x)T^a}$, where $\theta^a(x)T^a$ is a vector in the Lie algebra of the group $G$, and the integral curve obtained after exponentiation is the set of group elements parameterized by continuous coordinates within the Lie group. By analogy with the abelian case, the matrix $A_\mu$ also belongs to the Lie algebra, so $A_\mu = A_\mu^aT^a$, and the covariant derivative is given by:
\begin{equation}
\partial_\mu \equiv \mathbf{1}\partial_\mu \mapsto D_\mu = \partial_\mu - igA_\mu = \partial_\mu - igA^a_\mu T^a \equiv \delta_{ij}\partial_\mu - igA^a_\mu(T^a)_{ij} = D_{\mu,ij}
\end{equation}
The gauge transformation of the gauge field $A_\mu$ becomes more complex. The method to determine the transformation rule for $A_\mu$ is to ensure that $D_\mu\psi \equiv D_{\mu,ij}\psi_j\mapsto U^\dagger _{ik}D_{\mu,kl}\psi_l\equiv U^{\dagger}D_\mu\psi$. We find that when the transformation rule is as in Eq.~\eqref{non-abelian-A-trans}, the invariance of the kinetic term for the fermion field is preserved:
\begin{equation}
A_\mu\mapsto UA_\mu U^\dagger + \frac{i}{g}U\partial_\mu U^\dagger\label{non-abelian-A-trans}
\end{equation}
Similarly, we can compactly express the transformation rule for the gauge field as $D_\mu\mapsto UD_\mu U^\dagger$. For the kinetic term of the gauge field, we have:
\begin{equation}
F_{\mu\nu} = \frac{i}{g}[D_\mu, D_\nu] = \partial_\mu A_\nu - \partial_\nu A_\mu - ig[A_\mu, A_\nu]
\end{equation}
We observe that $F_{\mu\nu}$ includes an additional term $-ig[A_\mu, A_\nu]$, which arises because the gauge field $A_\mu$ is a non-commutative matrix. These features result from the gauge group being non-abelian.

Next, we consider the kinetic term for the gauge field. Clearly, $ -\frac{1}{4}F^{\mu\nu}F_{\mu\nu} $ is no longer feasible because $F_{\mu\nu}$ is a matrix, while the Lagrangian must be a real scalar. To resolve this issue, we first examine how $F_{\mu\nu}$ transforms under gauge transformations. Since two matrices undergoing the same similarity transformation will have their commutator transformed in the same way, based on $D_\mu\mapsto UD_\mu U^\dagger$, we anticipate that $F_{\mu\nu}\mapsto UF_{\mu\nu}U^\dagger$. The proof is as follows:
\begin{equation}
\begin{aligned}
\frac{g}{i}F_{\mu\nu} & = [D_\mu, D_\nu] \mapsto [UD_\mu U^\dagger, UD_\nu U^\dagger] \\
& = U[D_\mu U^\dagger, UD_\nu U^\dagger] + [U, UD_\nu U^\dagger]D_\mu U^\dagger \\
& = UD_\mu [U^\dagger, UD_\nu U^\dagger] + U[D_\mu, UD_\nu U^\dagger]U^\dagger + [U, UD_\nu U^\dagger]D_\mu U^\dagger \\
& = UD_\mu UD_\nu[U^\dagger, U^\dagger] + UD_\mu U[U^\dagger, D_\nu]U^\dagger + UD_\mu [U^\dagger, U]D_\nu U^\dagger \\
& + U UD_\nu [D_\mu, U^\dagger] U^\dagger + UU [D_\mu, D_\nu]U^\dagger U^\dagger + U [D_\mu, U]D_\nu U^\dagger U^\dagger \\
& + [U, U]D_\nu U^\dagger D_\mu U^\dagger + U[U, D_\nu]U^\dagger D_\mu U^\dagger + UD_\nu[U, U^\dagger]D_\mu U^\dagger \\
& = UD_\mu U[U^\dagger, D_\nu]U^\dagger + U UD_\nu [D_\mu, U^\dagger] U^\dagger + UU [D_\mu, D_\nu]U^\dagger U^\dagger \\
& + U [D_\mu, U]D_\nu U^\dagger U^\dagger + U[U, D_\nu]U^\dagger D_\mu U^\dagger \\
& = UD_\mu UU^\dagger D_\nu U^\dagger - UD_\mu U D_\nu U^\dagger U^\dagger + UUD_\nu D_\mu U^\dagger U^\dagger - UUD_\nu U^\dagger D_\mu U^\dagger \\
& + UUD_\mu D_\nu U^\dagger U^\dagger - UU D_\nu D_\mu U^\dagger U^\dagger + UD_\mu U D_\nu U^\dagger U^\dagger - UUD_\mu D_\nu U^\dagger U^\dagger \\
& + UUD_\nu U^\dagger D_\mu U^\dagger - U D_\nu U U^\dagger D_\mu U^\dagger\\
& = UD_\mu D_\nu U^\dagger - U D_\nu D_\mu U^\dagger = U[D_\mu, D_\nu]U^\dagger = \frac{g}{i}U F_{\mu\nu} U^\dagger
\end{aligned}
\end{equation}
Thus, we have $ F_{\mu\nu} \mapsto UF_{\mu\nu}U^\dagger $.

Next, based on this transformation rule, we need to construct a gauge-invariant real scalar quantity using $ F_{\mu\nu} $. According to the cyclic property of the trace operation:
\begin{equation}
\Tr(A_1A_2A_3\dots A_n) = \Tr(A_2A_3\dots A_n A_1)
\end{equation}

we find that $ \Tr(F^{\mu\nu}F_{\mu\nu}) $ is gauge-invariant. Therefore, the Lagrangian for the gauge field can be chosen as $ \Tr(F^{\mu\nu}F_{\mu\nu}) $. Typically, a factor of $-1/2$ is added at the beginning, which is a conventional notation that ensures the coefficient returns to $-1/4$ in the abelian case. Thus, the full Lagrangian for non-abelian gauge theory is given by:
\begin{equation}
\mathcal{L} = - \frac{1}{2}\Tr(F^{\mu\nu}F_{\mu\nu}) + \overline{\psi}_k(i\slashed{D}_{kl} - m)\psi_l
\end{equation}

When there are multiple flavors of fermions, we must include each flavor of fermions. In practice, quantum chromodynamics (QCD) involves 3 generations of quarks, totaling 6 quarks, each with a different mass. The gauge group is $ SU(3) $, leading to the QCD Lagrangian:
\begin{equation}
\mathcal{L}_{QCD} = -\frac{1}{2}\Tr(G^{\mu\nu}G_{\mu\nu}) + \sum_{I = 1}^6\overline{\psi}^I_k(i\slashed{D}_{kl} - m^I)\psi^I_l
\end{equation}

Here, the index $ I $ labels the quark types, corresponding to the $ u, d, s, c, b, t $ quarks, and $ G_{\mu\nu} $ is the strength of the $ SU(3)_c $ gauge bosons, commonly referred to as gluons.

Earlier, we introduced the method of representing an element of a Lie algebra using a basis of the Lie algebra. For the Lie algebra $\mathfrak{su}(3)$, the basis elements satisfy the relations $\Tr(T^aT^b) = \frac{1}{2}\delta^{ab}$ and $[T^a, T^b] = if^{abc}T^c$. We can further obtain the following expressions using these relations:
\begin{equation}
\begin{aligned}
F_{\mu\nu} 
& = F^a_{\mu\nu}T^a = \partial_\mu A_\nu^aT^a - \partial_\nu A_\mu^a T^a - ig A^b_\mu A^c_\nu [T^b, T^c] \\
& = T^a(\partial_\mu A_\nu^a - \partial_\nu A^a_\mu + gf^{abc}A_\mu^bA_\nu^c)
\end{aligned}
\end{equation}

This leads to:
\begin{equation}
\begin{aligned}
-\frac{1}{2}\Tr(F^{\mu\nu} F_{\mu\nu}) = -\frac{1}{2}F^{a\mu\nu}F^b_{\mu\nu}\Tr(T^aT^b) = -\frac{1}{4}F^{a\mu\nu}F^a_{\mu\nu}
\end{aligned}
\end{equation}
From the above relation, we can easily see the convention of having a $-1/2$ coefficient in front of $\Tr (F^{\mu\nu})F_{\mu\nu}$ indeed gives us the coefficient $-1/4$ when we bypass the gauge group from a non-abelian one to an abelian one: the sum over index $a$ in abelian gauge symmetry is trivial since $a\equiv 1$

Thus, the QCD Lagrangian can also be written as Eq.~\eqref{QCD-L}, with the gray-shaded term to be discussed later:
\begin{equation}
\mathcal{L}_{QCD} = -\frac{1}{4}G^{a\mu\nu}G^a_{\mu\nu} + \sum_{I = 1}^6\overline{\psi}^I_k(i\slashed{D}_{kl} - m^I)\psi^I_l + \gray{\frac{\theta}{32\pi^2}g^2G^{a\mu\nu}\tilde{G}^a_{\mu\nu}} \label{QCD-L}
\end{equation}

The Standard Model also includes the weak interaction, whose gauge group is $SU(2)$. However, the gauge bosons that mediate the weak interaction are massive, and the mass terms for gauge bosons are not gauge-invariant. Therefore, the $SU(2)$ gauge symmetry associated with the weak interaction is spontaneously broken. To describe the weak interaction, we need to first introduce the concept of electroweak symmetry breaking and the Higgs mechanism. Before doing so, it is necessary to explain how the Standard Model describes fermions.

In the QED Lagrangian Eq.~\eqref{QED-L} and the QCD Lagrangian Eq.~\eqref{QCD-L}, the fermion fields are Dirac fermions, whose mass terms are gauge-invariant and do not violate CP symmetry. However, experimental observations indicate that CP symmetry is not conserved, suggesting the need to introduce fermions that can potentially violate CP symmetry.

In fact, gauge bosons can also violate CP symmetry. The gray-shaded term in Eq.~\eqref{QCD-L}, $G^{a\mu\nu}\tilde{G}^a_{\mu\nu} = \frac{1}{2}\epsilon^{\mu\nu\rho\sigma}G_{a\mu\nu}G^a_{\rho\sigma}$, is gauge-invariant and CP-violating. However, experimental observations suggest that this term is extremely small, with the coefficient $\theta \leq 10^{-10}$, indicating that CP violation from the fermion sector needs to be considered.

\subsection{Matter Fields in the Standard Model}\label{SM-fermion}

In the previous section, we introduced the particles that constitute matter—fermions—using Dirac spinor fields $\psi$, which are mathematically represented as a $4 \times 1$ matrix. Historically, the Dirac spinor was first introduced to address the problem of negative energy in relativistically covariant quantum mechanics, with the core idea being to take the square root of the relativistic particle Hamiltonian $H^2 = m^2 + p^2$ to obtain a linear theory.

The fermion fields encountered in QED and QCD are all Dirac spinors that obey anti-commutative canonical quantization relations. Here, the consequence of anti-commutative canonical quantization relations is that, as fermions, matter particles must obey the Pauli exclusion principle, which requires that the state vector changes sign when two fermions are exchanged. In the language of field theory, a two-fermion state in Fock space, $\ket{x_1,s; x_2,t}$, denotes a fermion with spin $s$ at spacetime coordinate $x_1$ and another fermion with spin $t$ at coordinate $x_2$. Exchanging these positions is equivalent to swapping their creation operators:
\begin{equation}
\ket{x_1,s; x_2,t} = \psi^\dagger(x_1,s)\psi^\dagger(x_2,t)\ket{0} = -\ket{x_2,t; x_1,s} = -\psi^\dagger(x_2,t)\psi^\dagger(x_1,s)\ket{0}
\end{equation}

Thus, in the language of field theory, the anti-symmetry of fermions implies that fermion fields satisfy $\psi_1 \psi_2 = -\psi_2\psi_1$. Moreover, since fermions are spin-1/2 particles, the fermion field should also be a spinor. Hence, fermion fields possess two identities: anti-symmetric and spinor, which is why we refer to fermion fields as anti-symmetric spinor fields. In general, a spinor is represented by a column vector, which is not inherently anti-symmetric, and the exchange of two column vectors is not well-defined. Therefore, a fermion field, such as a Dirac spinor field, is typically expanded in momentum space as follows:
\begin{equation}
\psi(x) = \sum_{s = \pm}\int \frac{dp^3}{(2\pi)^3 2\omega}\left[a_p^su_p^s e^{-ip\cdot x} + b_p^{\dagger s}v_p^se^{ip\cdot x}\right]
\end{equation}

Here, the mode expansion separates the momentum-space Dirac spinors $u, v$ from the anti-symmetric momentum-space creation operators $a_p, b^\dagger_p$.

However, not all fermions are represented as Dirac spinor fields. In field theory, the equation of motion for a field characterizes a type of particle, and the equation of motion must be Lorentz covariant. Therefore, the number of possible particle types is, in principle, determined by the representations of the Lorentz group. The trivial representation corresponds to scalar particles, while the defining 4-D representation corresponds to vector particles, etc. Thus, the study of the representations of the Lorentz group is, to some extent, the study of the classification of particle types.

It can be shown that the Lorentz algebra is the direct sum of two $\mathfrak{su}(2)$ algebras, which we distinguish as the left-handed $\mathfrak{su}(2)_L$ and the right-handed $\mathfrak{su}(2)_R$. For reference, see Part II of Srednicki's \textit{Quantum Field Theory} \cite{Srednicki:2007qs}. Therefore, representations of the Lorentz group can be constructed using representations of the $SU(2)$ group, denoted as $(a,b)$, where $a$ represents the dimension of the $SU(2)_L$ representation and $b$ represents the dimension of the $SU(2)_R$ representation. Each different pair of values for $a$ and $b$ corresponds to a different type of particle: the scalar $(1,1)$ is the direct sum of two trivial $SU(2)$ representations; the Dirac spinor is the direct sum of the fundamental representations of $SU(2)_L$ and $SU(2)_R$, i.e., $(2,1)\oplus(1,2)$; and the vector is the direct product of the fundamental representations of $SU(2)_L$ and $SU(2)_R$, i.e., $(2,2)\simeq (2,1)\otimes(1,2)$.

Therefore, the Dirac spinor is not an irreducible representation of the Lorentz group. More fundamental fermionic building blocks are the left-handed chiral fermion $\xi_a$ in the $(2,1)$ space and the right-handed chiral fermion $\chi^{\dagger \dot{a}}$ in the $(1,2)$ space, each of which is a 2-dimensional representation. The Dirac spinor is their direct sum, so a Dirac spinor field can be decomposed as:
\begin{equation}
\psi = 
\begin{pmatrix}
\xi \\
\chi
\end{pmatrix} 
= 
\begin{pmatrix}
\xi \\
0
\end{pmatrix} +
\begin{pmatrix}
0 \\
\chi
\end{pmatrix} 
= \psi_L + \psi_R
\end{equation}
There are a set of mathematical formalism and conventions for manipulating the $(2,1)$ and $(1,2)$ spinors and their indexes, which is somewhat subtle and beyond the scoop of this thesis. For simplicity, we write the two-component spinors as four-component spinors $\psi_L$ and $\psi_R$ by padding them with two zeros.

The significance of chiral fermions lies in their potential to cause $CP$ violation. Hence, in constructing the Standard Model, we use chiral fermions to represent all matter particles; they only combine into Dirac fermions after electroweak symmetry breaking, resulting in the QED Lagrangian and QCD Lagrangian. However, unlike Dirac fermions, when chiral fermions are used as the matter fields before symmetry breaking, the mass term $m\overline{\psi}\psi = m(\overline{\psi}_L\psi_R + \overline{\psi}_R\psi_L)$ no longer exhibits gauge invariance. This is because the left-handed and right-handed chiral spinors in principle transform differently under gauge transformations: after transformation, the mass term becomes $m(\overline{\psi}_L\psi_Re^{i(\theta_L-\theta_R)} + \overline{\psi}_R\psi_Le^{i(\theta_R - \theta_L)})\neq m(\overline{\psi}_L\psi_R + \overline{\psi}_R\psi_L)$. Therefore, fermions do not have mass before symmetry breaking. To give fermions mass after symmetry breaking, we introduce the Yukawa sector involving the Higgs field $\phi$ of the form $\overline{\psi}_L\phi \psi_R + h.c.$. By using the gauge transformation of the Higgs doublet $\phi \mapsto e^{-i(\theta_L - \theta_R)}\phi$, the difference in gauge transformations of the left- and right-handed chiral spinors, the factor $e^{i(\theta_L - \theta_R)}$ cancels out, making the Yukawa sector gauge-invariant before symmetry breaking. The role of the Yukawa sector is to give fermions mass after symmetry breaking: the Higgs field, by spontaneously acquiring a non-zero vacuum expectation value (VEV), breaks the local gauge symmetry, and the $\phi$ field in the Yukawa term transforms into the sum of a non-zero VEV $v$ and the physical Higgs field $h$, with the $v$ part combining with the left- and right-handed chiral fermions to form a Dirac fermion mass term. This will be discussed in more detail in the next section.

\subsection{Electroweak Symmetry Breaking and the Higgs Mechanism}\label{SM-Higgs}

After introducing how the Standard Model describes matter particles and interactions, we proceed to a detailed discussion of the Standard Model, including its construction, electroweak symmetry breaking (the Higgs mechanism), and the post-symmetry breaking Standard Model. The notations in this section follow the TASI 2013 lectures \cite{Logan:2014jla}.

\subsubsection{Standard Model Before Electroweak Symmetry Breaking}

\begin{table}[htbp]
\renewcommand\arraystretch{1.2}
\centering
\setlength{\tabcolsep}{0.5mm}
\begin{tabular}{c|c|c|c}
\toprule
\multicolumn{2}{c|}{Fermion Fields} & \multicolumn{2}{c}{Boson Fields} \\
\midrule
Field & Representation & Field & Representation  \\
\midrule
\makecell{Left-handed Quark\\ Doublet $q^I_{L} = \left( \begin{array}{c} u^I_L \\ d^I_L\end{array}\right )$ }& $(3, 2, \frac{1}{6})$ & $SU(3)_c$ Gauge Field $G^a_{\mu}$ & $(8_{adj}, 1, 0)$  \\
\midrule
\makecell{Right-handed \\ Up Quark $u^I_{R}$} & $(3, 1, \frac{2}{3})$ & $SU(2)_L$ Gauge Field $W^i_\mu$ & $(1, 3_{adj}, 0)$  \\
\midrule
\makecell{Right-handed\\ Down Quark $d^I_{R}$} & $(3, 1, -\frac{1}{3})$ & $U(1)_Y$ Gauge Field $B_\mu$ & $(1, 1, 1_{adj})$ \\
\midrule
\makecell{Left-handed Lepton\\ Doublet $l^I_{L} = \left( \begin{array}{c} \nu^I_L \\ e^I_L\end{array}\right )$} & $(1, 2, -\frac{1}{2})$ & Higgs Doublet $H = \left( \begin{array}{c} H^+ \\ H^0\end{array}\right )$ & $(1, 2, \frac{1}{2})$ \\
\midrule
\makecell{Right-handed \\Charged Lepton $e^I_{R}$} & $(1, 1, -1)$ & & \\
\bottomrule
\end{tabular}
\caption{Quantum fields in the Standard Model and their representations under the gauge group $SU(3)_c \times SU(2)_L \times U(1)_Y$ before symmetry breaking. The subscripts $L, R$ on the fermion fields denote whether the fermion is a left-handed or right-handed chiral fermion, while the superscript $I = 1,2,3$ indicates the fermion generation. In the gauge field representations, $8_{adj}$, $3_{adj}$, and $1_{adj}$ indicate that these are in the adjoint representation of the respective gauge group. Additionally, the doublets refer to the fundamental representation of $SU(2)_L$, and they are written out explicitly because of their importance in the symmetry breaking mechanism. For fields/particles after the electroweak symmetry breaking, see Figure \ref{fig:SM}}
\label{SM-fields-before-ssb}
\end{table}

\begin{equation}
\mathcal{L}_{SM} = \mathcal{L}_{gauge} + \mathcal{L}_{f} + \mathcal{L}_{Yuk} + \mathcal{L}_H \label{SM-L-before-SSB}
\end{equation}

Before electroweak symmetry breaking, the Lagrangian of the Standard Model is given by Eq. Eq.~\eqref{SM-L-before-SSB}, and the quantum fields and their representations under the gauge group are shown in Table \ref{SM-fields-before-ssb}. The Lagrangian of the Standard Model is locally gauge-invariant under the gauge group $SU(3)_c \times SU(2)_L \times U(1)_Y$. Here, $SU(2)_L$ should be distinguished from the left-handed $SU(2)_L$ component of the Lorentz group: the $SU(2)_L$ in the Standard Model is a gauge group, where the gauge transformations are $SU(2)$ transformations. The subscript $L$ indicates that it only acts on left-handed chiral fermions, and all right-handed chiral fermions don't feel the $SU(2)_L$ gauge transformations. Since the gauge group is a direct product of three Lie groups, we introduce three gauge fields: the $SU(3)_c$ gauge field strength $G^a_{\mu\nu}$, the $SU(2)_L$ gauge field strength $W^i_{\mu\nu}$, and the $U(1)_Y$ gauge field strength $B_{\mu\nu}$. Therefore, the $\mathcal{L}_{gauge}$ part of Eq. Eq.~\eqref{SM-L-before-SSB} is:
\begin{equation}
\mathcal{L}_{gauge} = -\frac{1}{4}G^{a\mu\nu}G^a_{\mu\nu} - \frac{1}{4}W^{i\mu\nu}W^i_{\mu\nu} - \frac{1}{4}B^{\mu\nu}B_{\mu\nu}\label{SM-L-gauge}
\end{equation}

Next is the fermion part $\mathcal{L}_f$. According to Table \ref{SM-fields-before-ssb}, the covariant derivatives of different chiral fermions differ depending on whether they interact with the gauge transformations. For example, the right-handed chiral fermions are invariant under $SU(2)_L$ gauge transformations, so their covariant derivatives do not include the $-ig_1\frac{\sigma^i}{2} W^i_\mu \equiv -ig_1\tau^iW^i_\mu$ term. It is also important to note that the last number in the parentheses in Table \ref{SM-fields-before-ssb} represents the hypercharge $Y$ of the corresponding field, and their covariant derivatives have a $U(1)_Y$ covariant term $-ig_2YB_\mu$. As an example, we write out the covariant derivative for the left-handed quark doublet:
\begin{equation}
D_\mu q^I_{L} = \left(\partial_\mu - ig_sG^a_\mu T^a - ig_1W^i_\mu \tau^i - ig_2YB_\mu \right)q^I_{L}
\end{equation}

It is worth noting that we seem to be adding matrices of different dimensions together, but these matrices represent transformations in different spaces, and they should be understood as the tensor product of a non-trivial transformation in one space with a trivial transformation in another space. For instance, $T^a$ should be understood as $T^a \otimes 1_{SU(2)_L} \otimes 1_{U(1)_Y}$.

Therefore, the $\mathcal{L}_f$ part of Eq. Eq.~\eqref{SM-L-before-SSB} is as follows:
\begin{equation}
\mathcal{L}_{f}  = \sum_{I = 1}^3\left( i\,\overline{q}^I_L\slashed{D}q^I_L + i\,\overline{l}^I_L\slashed{D}l^I_L + i\,\overline{u}^I_R\slashed{D}u^I_R + i\, \overline{d}^I_R\slashed{D}d^I_R + i\, \overline{e}^I_R\slashed{D}e^I_R \right)\label{SM-L-fermion}
\end{equation}

$\mathcal{L}_f$ defines the kinetic terms of all chiral fermion fields in Table \ref{SM-fields-before-ssb} and their coupling terms with the gauge bosons. As an example, we again take the left-handed quark doublet $q^I_L$ and expand $i\,\overline{q}^I_L\slashed{D}q^I_L$ explicitly in Eq.~\eqref{SM-Leftquark-expand}, where we write the identity of the $SU(2)_L$ gauge group as a matrix and the identity of the hyper charge gauge group  $U(1)_Y$ as $1_{U(1)_Y}$ and the identity of the $SU(3)_c$ gauge group as $\delta_{kl}$.
\begin{equation}
\begin{aligned}
i\,\overline{q}^I_L\slashed{D}q^I_L & = i\sum_{I = 1}^3\sum_{k,l = 1}^3 (\overline{u}^I_{L,k}, \overline{d}^I_{L,k})\gamma^\mu \left[\delta_{kl}\otimes \left(\begin{array}{cc} 1& 0\\ 0 & 1\end{array}\right) \otimes 1_{U(1)_Y}\partial_\mu \right.\\
& - ig_sG^a_\mu T^a_{kl}\otimes \left(\begin{array}{cc} 1& 0\\ 0 & 1\end{array}\right)\otimes 1_{U(1)_Y} - ig_1 \delta_{kl}\otimes \left(\begin{array}{cc} W^3_\mu & W^1_\mu - iW^2_\mu \\ W^1_\mu + iW^2_\mu & -W^3_\mu \end{array}\right) \otimes 1_{U(1)_Y} \\
& - ig_2 \delta_{kl}\otimes \left. \left(\begin{array}{cc} 1& 0\\ 0 & 1\end{array}\right)\otimes YB_\mu \right]\label{SM-Leftquark-expand}
\begin{pmatrix}
u^I_{L,l} \\
d^I_{L,l}
\end{pmatrix}
\end{aligned}
\end{equation}
Note that the bar over the fermion field indicates that the field undergoes Hermitian conjugation in the four-component spinor space, followed by multiplication with the matrix $ \beta = \left(\begin{array}{cc} 0 & I \\ I & 0 \end{array}\right)$, while in other spaces, only the Hermitian conjugation operation is performed. In the above expansion, the lowercase Latin letters $k$ and $l$ denote the color indices of the quarks. 

Next, we consider the Yukawa sector. As discussed at the end of the previous section, to be consistent with the experimentally observed $CP$ violation, we must use chiral fermions in constructing the model. However, this introduces the problem that chiral fermions transform differently under gauge transformations. For example, in a $U(1)$ gauge theory, $\psi_L \mapsto e^{-i\theta_L}\psi_L$, $\psi_R \mapsto e^{-i\theta_R}\psi_R$, so the Dirac fermion mass term $m\overline{\psi}\psi = m(\overline{\psi}_L\psi_R + \overline{\psi}_R\psi_L)\mapsto m(\overline{\psi}_L\psi_R e^{i(\theta_L - \theta_R)} + \overline{\psi}_R\psi_L e^{i(\theta_R - \theta_L)})$, which no longer possesses gauge invariance. We introduce the Yukawa sector and a Higgs field, which transform under the gauge transformation as $\phi \mapsto \phi e^{-i(\theta_R - \theta_L)}$, canceling the phase difference between left- and right-handed chiral fermions. So the Yukawa sector respects the gauge symmetry. After symmetry breaking, the Higgs field acquires a VEV, giving mass to fermions. 

However, in practice, the situation differs from the abelian $U(1)$ gauge theory used to illustrate the essence of Yukawa sector above. This is because the realistic gauge transformations are non-abelian. In the actual construction, we need to take account of the mathematical subtly of the non-abelian gauge symmetry and obtain the following Yukawa sector:
\begin{equation}
\mathcal{L}_{Yuk} = -\sum_{I,J = 1}^3\left[ \Gamma^u_{IJ}\,\overline{q}^I_{L}\tilde{H}u^J_R + \Gamma^d_{IJ}\overline{q}^I_L H d^J_{R} + \Gamma^e_{IJ}\overline{l}^I_L H e^J_R + h.c. \right]\label{SM-L-Yukawa}
\end{equation}

where $\tilde{H} \equiv i\sigma^2 H$, which ensures that the up quarks also acquire mass after symmetry breaking. We will elaborate on this in detail during the discussion of the symmetry-breaking mechanism later on. Additionally, the Yukawa parameter is upgraded to a $\Gamma$ matrices from the Yukawa scalar in the simple $U(1)$ gauge toy theory. Similarly, we can extract a term from $\mathcal{L}_{Yuk}$, such as the lepton term, and expand it as follows:
\begin{equation}
\Gamma^e_{IJ}\,\overline{l}^I_L H e^J_R + h.c. 
= \Gamma^e_{IJ}\,(\overline{\nu}^I_L, \overline{e}^I_L)
\begin{pmatrix}
H^+ \\
H^0
\end{pmatrix}
e^J_R + 
\Gamma^{e*}_{JI}\, \overline{e}^J_R (H^-, H^{0*})
\begin{pmatrix}
\nu^I_L \\
e^I_L
\end{pmatrix}\label{Yukawa-lepton}
\end{equation}

It can be verified that under gauge transformations, the Yukawa term remains invariant: the difference in transformations for the chiral fermions under the electroweak gauge group $SU(2)_L \times U(1)_Y$ is canceled out by the transformation of the Higgs doublet $H$, ensuring that the entire Yukawa term remains invariant under $SU(3)_c \times SU(2)_L \times U(1)_Y$ transformations.

Finally, let us consider the Higgs sector $\mathcal{L}_H$. According to Table \ref{SM-fields-before-ssb}, the Higgs field is an $SU(2)_L$ doublet with hypercharge $1/2$, so its covariant derivative is given by:
\begin{equation}
D_\mu H = \left(\partial_\mu - ig_1W^i_\mu\tau^i - \frac{ig_2}{2}B_\mu \right)H
\end{equation}

We know that the kinetic term for a scalar field is $(D^\mu H)^\dagger D_\mu H$. Additionally, the Higgs sector $\mathcal{L}_H$ includes a crucial Higgs potential term $V(H)$:
\begin{equation}
V(H) = \mu^2 H^\dagger H + \lambda (H^\dagger H)^2
\end{equation}

Thus:
\begin{equation}
\mathcal{L}_H = (D^\mu H)^\dagger D_\mu H - V(H)\label{SM-L-Higgs}
\end{equation}

Electroweak symmetry breaking is realized through the spontaneous evolution of the parameters in $V(H)$: when $\mu^2 > 0$, the minimum of $V(H)$ is located at the origin. When the parameter $\mu^2$ spontaneously evolves to a value less than zero, the minimum of $V(H)$ is no longer at the origin, and we need to expand around the non-zero minimum according to perturbation theory. This is the core idea behind the Higgs mechanism.

\subsubsection{Electroweak Symmetry Breaking}

As previously discussed, to align with experimental observations of $CP$ violation, we constructed the model using Chiral fermions. This approach results in all fermions being mass-less. Additionally, the gauge boson mass terms don't respect gauge symmetry by construction. However, the experimentally observed $W$ and $Z$ bosons do possess non-zero mass. Therefore, we need a mechanism to break gauge symmetry, allowing fermion mass terms to appear in the Lagrangian and giving mass to the $W$ and $Z$ bosons. This mechanism is known as the Higgs mechanism.

To explain the core idea of the Higgs mechanism, we first consider a simplified abelian Higgs model, rather than the full Standard Model with the $SU(2)_L \times U(1)_Y$ gauge group. In this model, the Higgs field is a complex scalar field $\Phi$, and the gauge group is $U(1)$. We write down the Lagrangian for this simplified model, where the superscript $t$ denotes that the abelian Higgs model is a ``toy'' model.
\begin{equation}
\mathcal{L}^t = \mathcal{L}^t_{gauge} + \mathcal{L}^t_{f} + \mathcal{L}^t_{Yuk} + \mathcal{L}^t_{\Phi}
\end{equation}
\begin{equation}
\mathcal{L}^t_{gauge} = -\frac{1}{4}F^{\mu\nu}F_{\mu\nu}
\end{equation}
\begin{equation}
\mathcal{L}^t_{f} = i\,\overline{\psi}_L\slashed{D}\psi_L + i\,\overline{\psi}_R\slashed{D}\psi_R
\end{equation}
\begin{equation}
\mathcal{L}^t_{Yuk} = -\Gamma\, \overline{\psi}_L\psi_R \Phi - \Gamma^*\,\overline{\psi}_R\psi_L \Phi^*
\end{equation}
\begin{equation}
\mathcal{L}^t_{\Phi} = (D^\mu \Phi)^\dagger D_\mu \Phi - V(\Phi) = (D^\mu \Phi)^\dagger D_\mu \Phi - \mu^2\Phi^\dagger\Phi - \lambda(\Phi^\dagger\Phi)^2
\end{equation}

The covariant derivative is $D_\mu = \partial_\mu - igqA_\mu$, where $q$ represents the $U(1)$ charge of different fields. Under $U(1)$ gauge transformations, the transformations for the left-handed and right-handed Chiral fermion fields, gauge boson field, and Higgs field are:
\begin{equation}
\left\{
\begin{aligned}
\psi_L & \mapsto e^{-iq_L\theta(x)}\psi_L\\
\psi_R & \mapsto e^{-iq_R\theta(x)}\psi_R\\
A_\mu & \mapsto A_\mu - \frac{1}{g}\partial_\mu\theta \\
\Phi & \mapsto e^{-iq_\Phi\theta(x)}\Phi\,,\;(q_\Phi = q_L - q_R)
\end{aligned}
\right.
\end{equation}

It is straightforward to verify that $\mathcal{L}^t$ is invariant under these gauge transformations.

Next, we consider the vacuum of the $\Phi$ field defined by $V(\Phi)$.
In perturbation theory, the vacuum is the stationary point of the classical field before quantization. To carry out perturbation theory, we expand the Lagrangian around the vacuum and quantize the field. The quadratic terms represent the mass terms of the excitation particles and higher-order terms represent interaction, which can be further analyzed using time-dependent perturbation theory or path integrals.

Examining $V(\Phi)$, we find that $\lambda$ must be positive; otherwise, there would be no minimum energy: energy could decrease indefinitely. Such a theory is pathological, where any perturbation strong enough for large penetration of the potential wall would cause the vacuum state to collapse continuously, physically resulting in a large amount of energy and particle excitations appearing from thin air. Next, we examine the value of $\mu^2$. If $\mu^2 > 0$, $V(\Phi)$ has a unique minimum (the minimum value) at $\Phi = 0$. Perturbation expansion around $\Phi = 0$ shows that both fermions and gauge bosons remain mass-less, and the $U(1)$ gauge symmetry is preserved, which is not the desired outcome. 

Considering $\mu^2 < 0$, we find that $V(\Phi) = -|\mu^2|\Phi^\dagger\Phi + \lambda(\Phi^\dagger\Phi)^2$ has a minimum at $|\Phi|^2 = \frac{|\mu^2|}{2\lambda}$, which we denote as $|\phi| = v = \sqrt{\frac{|\mu^2|}{\lambda}}$. At this point, the VEV is:
\begin{equation}
\langle \Phi\rangle = \frac{v}{\sqrt{2}}e^{i\alpha}
\end{equation}

Without loss of generality, we choose a specific $\alpha$, such as $\alpha = 0$, as the vacuum and expand the field $\Phi$ around it:
\begin{equation}
\Phi(x) = \frac{1}{\sqrt{2}}(v+h(x))e^{ia(x)}\label{toy-H}
\end{equation}

Note that although the vacuum state is set at $\frac{v}{\sqrt{2}}$, there are still two degrees of freedom in the field $\Phi$ near the vacuum point. Hence, there are two real scalar fields, which we choose as the radial field $h(x)$ and the tangential field $a(x)$.

Next, we substitute the formula Eq.~\eqref{toy-H} into $\mathcal{L}^t_\Phi$. To compute this more complex quantity, we first substitute it into the Higgs potential and obtain:
\begin{equation}
V(\Phi) = -\frac{|\mu^2|}{2}(v+h)^2 + \frac{\lambda}{4}(v+h)^4
\end{equation}

We find that only the $h(x)$ field has a mass term (quadratic term) $\left(-\frac{|\mu^2|}{2} + \frac{3\lambda v^2}{2}\right)h^2 = \frac{|\mu^2|}{4}h^2$, hence $m_h = \frac{|\mu|}{\sqrt{2}}$. The $a(x)$ field does not have a mass term, so the $a$ particle is a mass-less gauge boson. It is noteworthy that if we choose a different perturbation method from Eq.~\eqref{toy-H}, such as $(v+h+ia)$, the $a$ field still remains mass-less. We refer to $a$ as the Goldstone boson. Since we have not observed interactions mediated by mass-less scalar gauge bosons experimentally, Goldstone bosons are generally considered nonphysical, and thus we omit the $a(x)$ field in the following discussion.

Next, we consider $D_\mu\Phi$. When computing this quantity, we do not include the $h(x)$ and $a(x)$ fields from Eq.~\eqref{toy-H}, as this would only yield their kinetic terms $\frac{1}{2}\partial^\mu h\partial_\mu h$ and $\frac{1}{2}\partial^\mu a\partial_\mu a$, along with coupling with gauge bosons. While these coupling terms are crucial for scattering processes, they are not particularly useful for illustrating how the Higgs mechanism gives mass to the gauge bosons. Therefore, we substitute $\Phi = \frac{v}{\sqrt{2}}$ into the kinetic terms, and obtain:
\begin{equation}
\begin{aligned}
D_\mu\Phi & = (\partial_\mu - igq_\phi A_\mu)\frac{v}{\sqrt{2}} = -\frac{igq_\Phi v}{\sqrt{2}} A_\mu v
\end{aligned}
\end{equation}
\begin{equation}
\text{Thus:} \qquad (D^\mu\Phi)^\dagger D_\mu\Phi = \frac{g^2q_\Phi^2 v^2}{2}A^\mu A_\mu
\end{equation}

Therefore, the gauge boson $A_\mu$ acquires a mass $m_A = gq_\Phi v$.

Next, we consider the Yukawa sector, which shows how the Higgs mechanism gives mass to fermions. Substituting $\Phi = \frac{v+h}{\sqrt{2}}$ into $\mathcal{L}^t_{Yuk}$, we get:
\begin{equation}
\begin{aligned}
\mathcal{L}^t_{Yuk} & = -\Gamma\, \overline{\psi}_L\psi_R\frac{v+h}{\sqrt{2}} - \Gamma^*\, \overline{\psi}_R\psi_L\frac{v+h}{\sqrt{2}} \\
& = -\frac{|\Gamma|(v+h)}{\sqrt{2}}(\overline{\psi}_L\psi_R + \overline{\psi}_R\psi_L) = -\frac{|\Gamma|(v+h)}{\sqrt{2}}\overline{\psi}\psi
\end{aligned}
\end{equation}

In the second step, we absorbed the phase factor $\Gamma = |\Gamma|e^{i\theta}$ into the definition of the fermion field. Thus, we have composed Chiral fermions into Dirac fermions, giving the Dirac fermions a mass $m_\psi = \frac{v|\Gamma|}{\sqrt{2}}$. The other term represents the coupling between the fermions and the physical Higgs particle $h$. Since the VEV $v$ is constant, we also observe that the larger the fermion mass, the larger the Yukawa parameter $\Gamma$, and thus the stronger the coupling of that fermion with the physical Higgs. This is a very important phenomena of Higgs-fermion coupling.

After discussing the essence of the Higgs mechanism in the abelian Higgs model, we extend the abelian gauge group $U(1)$ to the real life case: the electroweak gauge group $SU(2)_L \times U(1)_Y$. The idea remains unchanged; we only need to address the additional mathematical complexities.

Similar to the abelian Higgs model, we first consider the Higgs potential. To achieve spontaneous symmetry breaking of $SU(2)_L \times U(1)_Y$, we must choose $\lambda > 0$ and have a negative coefficient for the quadratic term $H^\dagger H$, i.e., $V(H) = -\mu^2 H^\dagger H + \lambda (H^\dagger H)^2$. Here, we take $\mu^2 > 0$. The Higgs potential reaches its minimum at $|H|^2 = \frac{\mu^2}{2\lambda}$, and we denote this as $v = \frac{\mu}{\sqrt{\lambda}}$. However, since $H$ is now an $SU(2)_L$ doublet, we choose the vacuum state as:
\begin{equation}
\langle H\rangle = \frac{1}{\sqrt{2}}e^{-iY\theta - i\xi^i\tau^i}
\begin{pmatrix}
0 \\
v
\end{pmatrix}
\end{equation}

Here, the factor $e^{-i\alpha}$ in the $U(1)$ case becomes $e^{-iY\theta - i\xi^i\tau^i}$ because the gauge group is now $SU(2)_L \times U(1)_Y$. It can be verified that substituting $\langle H\rangle$ into $V(H)$ yields the minimum value. Similarly, perturbation around $\langle H\rangle$ yields four real degrees of freedom: one physical Higgs degree of freedom $h(x)$, and three Goldstone bosons $G^{\pm}(x) = G^1(x) \pm iG^2(x)$, $G^0(x)$.
\begin{equation}
H = 
\begin{pmatrix}
G^{+}(x)\\
\frac{v + h(x) + iG^0(x)}{\sqrt{2}}\label{real-H}
\end{pmatrix}
\end{equation}

Substituting Eq.~\eqref{real-H} into the Higgs potential $V(H)$, we get:
\begin{equation}
\begin{aligned}
V = & -\frac{\mu^2}{2}\left(\sqrt{2}G^+G^- + (v+h)^2 + (G^0)^2\right) + \frac{\lambda}{4} \left(\sqrt{2}G^+G^- + (v+h)^2 + (G^0)^2\right)^2 \\
& = -\frac{\lambda v^2}{\sqrt{2}}G^+G^- -\frac{\lambda v^2}{2}(v+h)^2 -\frac{\lambda v^2}{2}(G^0)^2 + \frac{\lambda}{\sqrt{2}}(G^+G^-)(v+h)^2 + \frac{\lambda}{2}(G^0)^2(v+h)^2\\ 
& + \frac{\lambda}{2\sqrt{2}}(G^+G^-)^2 + \frac{\lambda}{4}(v+h)^4 + \frac{\lambda}{4}(G^0)^4 + \frac{\lambda}{\sqrt{2}}(G^+G^-)(G^0)^2 
\end{aligned}
\end{equation}

The last line of the above formula contains quartic terms representing interactions, while the third line contains quadratic terms, which correspond to the mass terms of these particles. We find that the mass terms for $G^\pm$ and $G^0$ are zero, and the mass of $h$ is $m_h^2 = 2\lambda v^2$. Therefore, in subsequent calculations, we will ignore the nonphysical Goldstone bosons. To reiterate, Goldstone bosons are mass-less, and the interactions they mediate are long-ranged, which have not been observed experimentally.

Next, we calculate the mass of the gauge bosons by evaluating $(D^\mu H)^\dagger D_\mu H$. Similarly, we omit substituting in $h$ here because our goal is to determine the gauge boson masses. The terms involving $h$ correspond to its kinetic terms and interactions with the gauge bosons, which are not relevant to our current calculation.
\begin{equation}
\begin{aligned}
D_\mu H & = \frac{1}{\sqrt{2}}\left(\partial_\mu - \frac{ig_1}{2}W^i_\mu\sigma^i - \frac{ig_2}{2}B_\mu\right)
\begin{pmatrix}
0 \\
v
\end{pmatrix} \\
& = -\frac{i}{2\sqrt{2}}
\begin{pmatrix}
g_2B_\mu + g_1W^3_\mu & \sqrt{2}g_1W^+_\mu \\
\sqrt{2}g_1 W^-_\mu & g_2B_\mu - g_1W^3_\mu
\end{pmatrix}
\begin{pmatrix}
0 \\
v
\end{pmatrix} \\
& = -\frac{i}{2\sqrt{2}}
\begin{pmatrix}
\sqrt{2}g_1vW^+ \\ 
\sqrt{g_1^2 + g_2^2}Z_\mu v
\end{pmatrix}
\end{aligned}
\end{equation}

We define $\sqrt{2}W^\pm_\mu = W^1_\mu \mp iW^2_\mu$ and $g_2B_\mu - g_1W^3_\mu = \sqrt{g_1^2 + g_2^2}Z_\mu$. Thus, we have:
\begin{equation}
\begin{aligned}
(D^\mu H)^\dagger D_\mu H & = \frac{1}{8}\left(\sqrt{2}g_1vW^{-\mu}, v\sqrt{g_1^2 + g_2^2}\,Z^\mu\right)
\begin{pmatrix}
\sqrt{2}g_1v W^+_\mu \\
v\sqrt{g_1^2 + g_2^2}\,Z_\mu
\end{pmatrix} \\
& = \frac{g_1^2v^2}{4}W^{+\mu}W_\mu^- + \frac{v^2(g_1^2 + g_2^2)}{8}Z^\mu Z_\mu = m_W^2W^{+\mu}W_\mu^- + \frac{1}{2}m^2_ZZ^\mu Z_\mu
\end{aligned}
\end{equation}

Here, $W^\pm_\mu$ bosons are vector fields and their mass term lacks a 1/2 coefficient. Therefore, the masses of the gauge bosons mediating weak interactions are: $m_W = \frac{g_1 v}{2}$ and $m_Z = \frac{v\sqrt{g_1^2 + g_2^2}}{2}$.

Next, we consider how fermions acquire mass. Starting with the lepton Yukawa term Eq.~\eqref{Yukawa-lepton}, we similarly ignore the Goldstone bosons and substitute Eq.~\eqref{real-H} for the physical Higgs field $h$:
\begin{equation}
\begin{aligned}
\frac{1}{\sqrt{2}}\Gamma^e_{IJ}\,\overline{l}^I_L H e^J_R + \text{h.c.} 
& = \frac{1}{\sqrt{2}}\Gamma^e_{IJ}\,(\overline{\nu}^I_L, \overline{e}^I_L)
\begin{pmatrix}
0 \\
v
\end{pmatrix}
e^J_R + \frac{1}{\sqrt{2}}
\Gamma^{e*}_{JI}\, \overline{e}^J_R (0, v)
\begin{pmatrix}
\nu^I_L \\
e^I_L
\end{pmatrix} \\
& = \frac{v}{\sqrt{2}}\left(\Gamma^e_{IJ}\,\overline{e}^I_L e^J_R + \Gamma^{e*}_{JI}\overline{e}^J_R e^I_L \right)
\end{aligned}
\end{equation}

We observe that, due to our choice of the Higgs VEV being in the lower component, neutrinos $\nu$ remain mass-less. Next, we need to select a special basis in the lepton flavor space to diagonalize the Yukawa matrix $\Gamma^e$. In principle, the choice of the $\Gamma^e$ matrix is quite arbitrary during model construction, but experimentally, since the three generations of charged leptons have unequal non-zero masses, the diagonal elements obtained after diagonalizing $\Gamma^e$ should be three unequal real numbers. Thus:
\begin{equation}
\frac{1}{\sqrt{2}}\Gamma^e_{IJ}\,\overline{l}^I_L H e^J_R + \text{h.c.} 
= \frac{vM^e_I\delta_{IJ}}{\sqrt{2}}\left(\,\overline{e}^I_L e^J_R + \overline{e}^J_R e^I_L \right) = \frac{vM^e_I}{\sqrt{2}}\overline{e}^I e^I
\end{equation}

Thus, the masses of the charged leptons are: $m_{e,\mu, \tau} = \frac{v M^e_{e, \mu, \tau}}{\sqrt{2}}$.

The situation with quarks is quite similar. However, it is important to note that, due to the choice of the Higgs VEV being in the lower component, up-type quarks cannot acquire mass if we only consider $\Gamma^d_{IJ}\, \overline{q}^I_L H d^J_R + h.c.$ To address this, we introduce $\tilde{H} = i\sigma^2 H$, leading to:
\begin{equation}
\langle \tilde{H}\rangle = \frac{i}{\sqrt{2}}\begin{pmatrix}0 & -i \\ i & 0\end{pmatrix}\begin{pmatrix}0 \\ v\end{pmatrix} = \frac{1}{\sqrt{2}}\begin{pmatrix}v \\ 0\end{pmatrix}
\end{equation}

We then include the Yukawa term $\Gamma^u_{IJ}\, \overline{q}^I_L \tilde{H} u^J_R + \text{h.c.}$ to allow up-type quarks to also acquire mass.

From this, we can infer that if we wish neutrinos to also acquire mass, a feasible method is to add the following term to the lepton Yukawa Lagrangian:
\begin{equation}
\mathcal{L}^\nu_{Yuk} = \Gamma^\nu_{IJ}\,\overline{l}^I_L \tilde{H} \nu^J_R + h.c.
\end{equation}

Indeed, the observation of neutrino oscillations indicates that neutrinos have non-zero masses, making $\mathcal{L}^\nu_{Yuk}$ an extension beyond the Standard Model. However, the neutrino mass models in real life is more complicated than the naive $\mathcal{L}^\nu_{Yuk}$ term we added above. First above all, the nature of neutrino is still unclear--whether it's Dirac fermion as described by SM or Majorana fermion, which is a type of spin-1/2 particle whose anti-particle is itself. For introduction to neutrino models, see Chapter 9 of the Book \cite{Langacker:2010zza} by Paul Langacker.

After briefly motivating studying new physics beyond SM by mentioning neutrino masses, in the next section, we will delve deeper into motivating beyond SM physics research and introduce the main focus of this thesis: multi-Higgs doublet (NHDM) models and specifically, 4 Higgs doublet model (4HDM).

\subsection{Necessity for New Physics Beyond The SM} \label{intro-NHDM}

The discovery of the Higgs boson in 2012 \cite{CMS:2012qbp, ATLAS:2012yve} marked the completion of the Standard Model, which describes the fundamental laws governing particle physics. However, it became evident that the Standard Model is not the ultimate theory of elementary particles, as there remain numerous phenomena it cannot explain.

Current experimental observations show the existence of three types of neutrinos—$\nu_e,\;\nu_\mu,\;\nu_\tau$—which are elementary particles predicted by the Standard Model to have zero mass and to interact only via weak interactions. However, recent observations have demonstrated that neutrinos undergo ``oscillations'' during their propagation. For instance, an initially produced electron neutrino $\nu_e$ may transform into a muon neutrino $\nu_\mu$ after traveling for some time, contradicting the Standard Model's predictions: neutrinos don't change flavor during propagation. The widely accepted explanation is that neutrinos possess non-zero mass, but each flavored neutrino's mass is not a definite value; rather, it is a quantum superposition. The three neutrinos, $\{\ket{\nu_e}, \ket{\nu_\mu}, \ket{\nu_\tau}\}$, are the basis vectors in the neutrino flavor space, but these vectors are not eigenstates of the mass operator, so they are not eigenstates of the time evolution operator. As a result, neutrino oscillations occur during propagation, i.e., the flavor of the neutrino changes. However, in the Standard Model, neutrinos are strictly mass-less, and thus it cannot account for this oscillation phenomenon. Though to many people, SM with simple neutrino mass term is also considered SM, reasons such as too many additional Yukawa parameters still motivate people to study neutrino models beyond SM. 

\begin{figure}[h]
    \centering
    \includegraphics[scale=0.4]{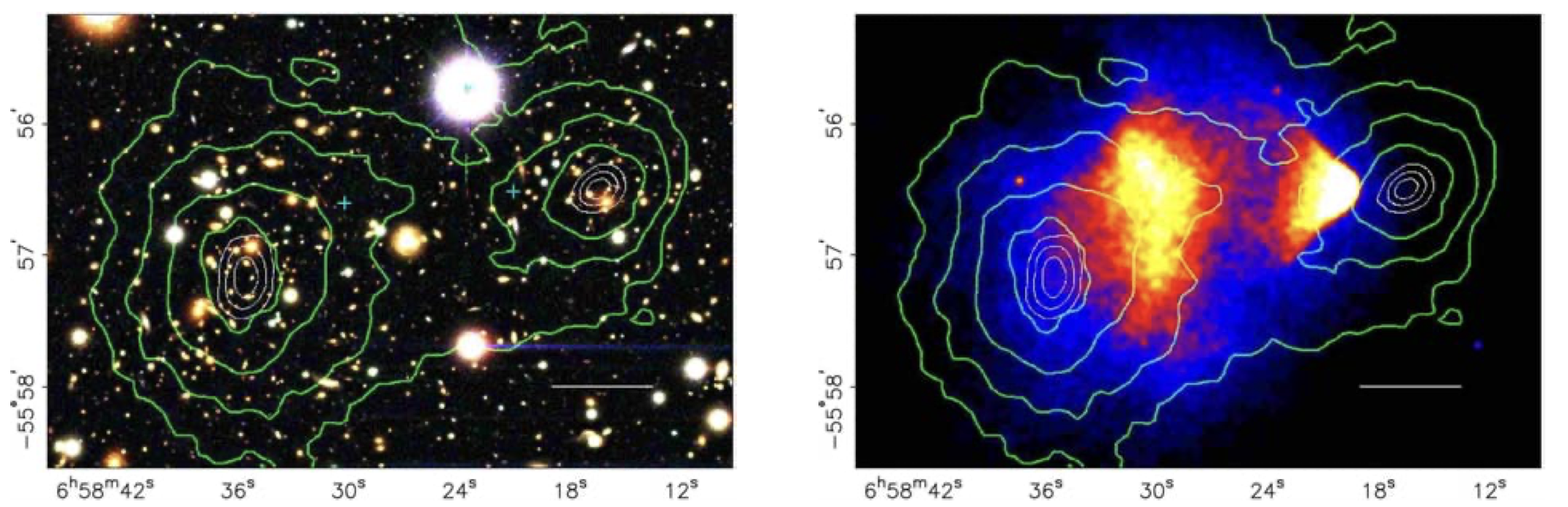}
    \caption{Observation of the merging cluster 1E0657-558. \cite{Clowe:2006eq}. The white bar marks 200 kpc distance in the figure; The green contour indicates the mass distribution measured by gravitational lensing; The color shaded reign in right panel is X-ray signals radiated by hot baryon gas.}
    \label{fig:MergingCluster}
\end{figure}

Looking into the universe, astronomers can measure the mass distribution of galaxy clusters through Einstein's gravitational lensing effect, and they can observe the distribution of luminous celestial bodies using telescopes across various wavelengths. These observations reveal that luminous matter (including celestial bodies and hot baryonic gases) only accounts for about a quarter of the total mass involved in gravitational interactions. When observing the rotation speeds of spiral galaxy arms, astronomers found that the rotational velocity of luminous objects around the galaxy center is almost independent of their distance from the center, contradicting the gravitational theory prediction $v = \sqrt{\frac{GM}{R}}$, which suggests $v \sim R^{-1/2}$. Additionally, when observing galaxy cluster collisions, such as the Bullet Cluster (1E0657-558), the mass distribution measured via gravitational lensing is significantly offset from the distribution of luminous matter observed in X-ray telescopes \cite{Clowe:2006eq}. The mass centers of the two galaxy clusters have already moved apart, but the luminous hot gas, which constitutes nearly 90\% of the baryonic mass, lags behind. See Figure \ref{fig:MergingCluster}.  It's worthy pointing out that optically visible galaxies, which are made of stars, comets, and so on, are formally described in cosmology as ``dust'' which have no statistical properties like temperature and pressure. So when they collide, most probably the celestial bodies just fly by. Therefore, it's the hot intergalactic gas observed by X-ray telescopes in the galaxy clusters that collide and is offset from gravitational lensing. The optically observed galaxies actually match the gravitational lensing observations quite well. All these observations suggest the presence of a large amount of non-luminous, massive matter in galaxies that interacts only weakly with itself and baryonic matter. The identity of this matter remains unknown, and it has been named ``dark matter''. The mainstream view is that dark matter consists of a new particle not included in the Standard Model, and many beyond-the-Standard-Model candidate theories have been proposed to solve this mystery.

\begin{figure}[h]
    \centering
    \includegraphics[scale=0.45]{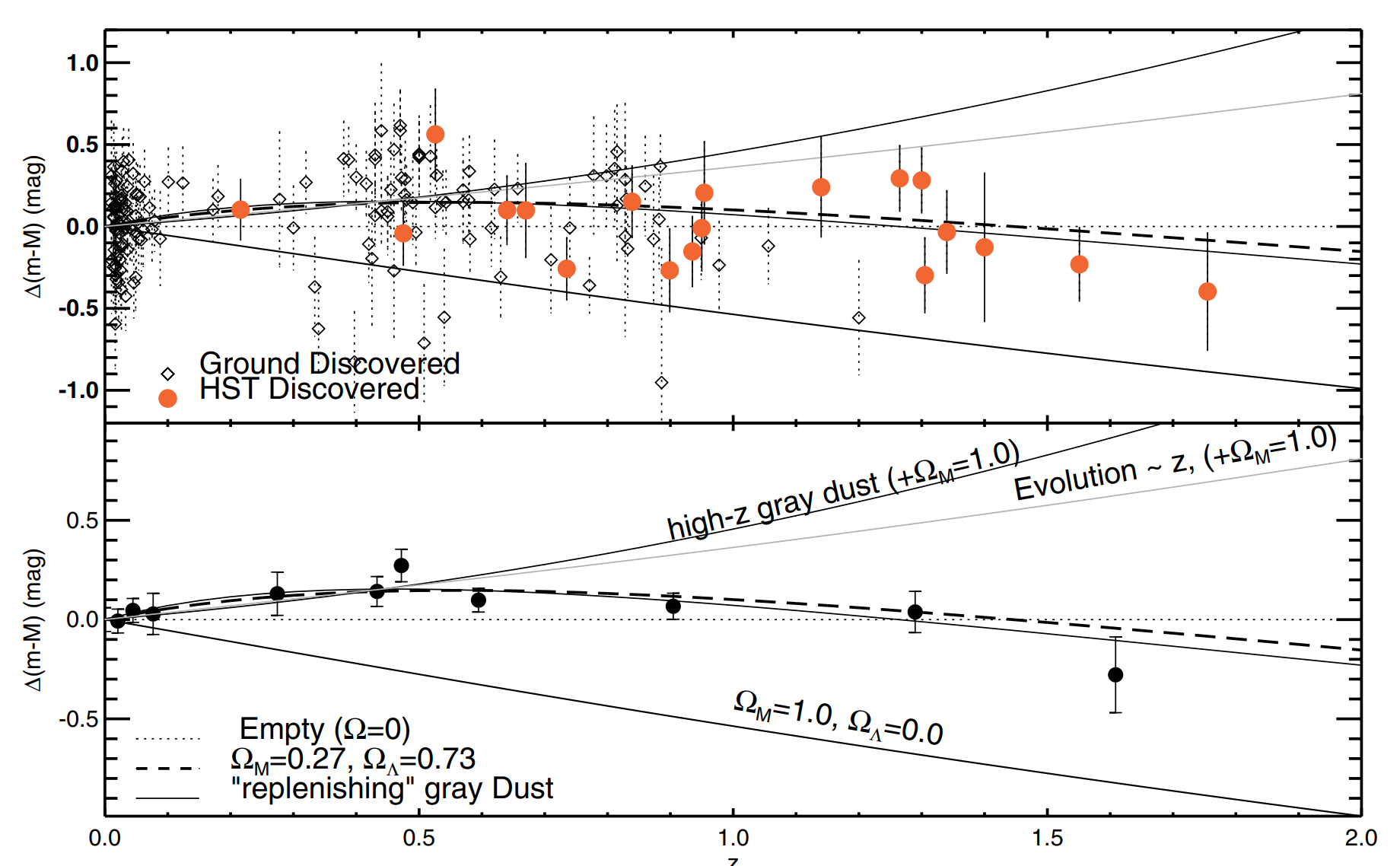}
    \caption{Observed SNe Ia data plotted in brightness-distance diagram with predictions from $\Lambda$CDM model. \cite{Rubakov:2017xzr}. We use the red-shift $z$ to indicate distance and the $m-M = 5\log_{10}(r_{ph}/\mbox{Mpc}) + 25$ to show the brightness of SNE Ia. We see the best fit is under $\Omega_M = 0.27$, $\Omega_\Lambda = 0.73$, meaning the dark energy density -- non-relativistic matter (dark matter and usual matter together) density ratio is around 3 to 1. Data source \cite{SupernovaSearchTeam:2004lze}. }
    \label{fig:SNe Ia}
\end{figure}

In addition, there are another great mystery in the cosmos apart from the dark matter. From the \text{$\Lambda$}CDM model of cosmology, one can calculate the brightness-distance relation for a kind of standard candle in the universe: SNe Ia supernova. See Figure \ref{fig:SNe Ia}. In $\Lambda$CDM model, the standard model of cosmology based on general relativity and SM, the density of non-relativistic matter $\rho_M$, relativistic matter (radiation) $\rho_{rad}$, cosmology constant term $\rho_\Lambda$, and the effective curvature density $\rho_{curv}$ are all input parameters. We introduce relative density $\Omega_i = \rho_i/\rho_c$ where $\rho_c = \frac{3}{8\pi}H_0^2M^2_{Pl}\simeq 0.52\times 10^{-5}\mbox{GeV/cm}^3$ is the total density or critical density. From the 2.7K background radiation, we can estimate the $\rho_{rad}$, and obtain $\Omega_{rad}\simeq 10^{-4}$, enabling us to ignore the contribution from radiation. The effective curvature density, which is actually a geometric term in the Friedmann's equation, is relatively small too, $|\Omega_{curv}|\leq 0.02$, so we our spacetime is almost flat, and can omit $\Omega_{curv}$ too. So the input parameters for cosmological evolution are $\Omega_M$ and $\Omega_\Lambda$. As can be seen from Figure \ref{fig:SNe Ia}, when $\Omega_M = 0.27, \Omega_\Lambda = 0.73$, the prediction from $\Lambda$CDM model best matches the observed SNe Ia data. 

The $\Omega_\Lambda$ term comes from the cosmological constant term in Einstein's equation, whose equation of state is peculiar: $\rho = -p$, meaning this matter has negative pressure, ``stretching'' the cosmos to expand. Because of this property, people name it ``dark energy'', which is also not explained my SM and General Relativity. 

What's more, there are other puzzles of vacuum energy, or the cosmological constant. Observations indicate that the vacuum energy density of the universe is $\rho_c \simeq 10^{-46}\mbox{GeV}^4$ in natural unit. However, the predicted vacuum energy densities from the Standard Model's strong and electroweak interactions, as well as from some unified theories' gravitational interactions, are:
\begin{equation}
\rho_{strong} \simeq 1\mbox{GeV}^4\;,\quad \rho_{ew} \simeq 10^8\mbox{GeV}^4\;,\quad \rho_{G} \simeq 10^{76}\mbox{GeV}^4
\end{equation}

These predicted values are vastly larger than the observed value. The phenomena discussed above represent some of the most famous cases that cannot be explained by the current Standard Model of particle physics or by Einstein's gravitational theory. In addition to these, there are other anomalies such as the muon anomalous magnetic moment, a $\sim 3\sigma$ deviation in $B$-meson decay from the Standard Model prediction, new scalar particle hints at 152 GeV ($\sim 3.9\sigma$) and 95 GeV ($\sim 3.8\sigma$), and more \cite{2309.03870}. Each of these indicates the limitations of the Standard Model, making it necessary to explore new physics beyond it.

\subsection{Multi-Higgs Doublet Model as New Physics Candidate}\label{NHDM-4HDM}

Among the many beyond-the-Standard-Model theories, multi-Higgs doublet models (NHDM) have garnered significant attention. These models hypothesize the existence of $N$ Higgs doublets, while keeping other aspects of the Standard Model unchanged. { In other words, NHDM can be understood as a humble extension of the idea of generations of particles: there are three generations of fermions, so maybe it's possible that nature has multiple generations of Higgs-like particles.} After spontaneous symmetry breaking, multi-Higgs models predict a rich phenomenology, such as multiple scalar particles, which aligns with the new scalar particle hints at 95 GeV and 152 GeV. They also predict a more complex vacuum phase transition scenario and the existence of topological structures like cosmic strings. While most research within this broad category of multi-Higgs models has focused on the $N = 2$ and $N = 3$ cases, the $N = 4$ case also has a research history spanning nearly 40 years, with close to a hundred papers discussing it. This model has been employed to address various experimental observations that the Standard Model cannot explain.

Let's first review the historical development of NHDM with 4HDM in particular. In 1973, T. D. Lee was the first to consider a { spontaneously $T$ symmetry-breaking model} with two scalar doublets \cite{Lee:1973iz}, marking the first paper in the field of two-Higgs-doublet models (2HDM). The idea { is based on the fact that} in some range of the parameter space (which happens to be quite a large range) of the potential $V(\phi_2, \phi_2)$, the two scalar fields have VEV's that have a relative phase difference $e^{i\theta}$,  ($\theta\neq 0 \mbox{ or } \pi$). 
\begin{equation}
\langle \phi_1\rangle_{VEV} = \rho_1\;,\quad \langle\phi_2\rangle_{VEV} = \rho_2e^{i\theta}\;,\quad\mbox{where }\rho_1, \rho_2, \theta\in \mathbb{R}\,,\;\mbox{and } \theta\neq 0\mbox{ or }\pi
\end{equation}
Due to the anti-unitary nature of the time reversal operator: $iT = -Ti$, when we apply perturbation theory around the VEV, the Lagrangian is no longer $T$ invariant, leading to spontaneous $T$ violation. So 2HDM is a somewhat natural candidate for spontaneous $T$ violation, which provided enough freedom in parameter space, avoiding fine-tuning of parameters. 

Later in 1976, S. Weinberg explored a new possibility for $CP$ violation \cite{Weinberg:1976hu} through a model with three Higgs doublets. The idea of this paper is what follows. From T. D. Lee's paper we know that 2HDM naturally suggests $CP$ violation, avoiding fine-tuning of or extremely small parameters. However, when considering a model with more than one Higgs doublet, flavor-changing-neutral-current (FCNC) usually arises in tree level diagrams, which is nonphysical since FCNC is strongly suppressed by experiments. Therefore we need mechanisms such as discrete symmetries so that FCNC is not allowed in tree level when constructing multi-Higgs models. It so happens that in 2HDM, such discrete symmetry to suppress FCNC can't co-exist with the desired $CP$ violation through Higgs sector. This problem comes from the little algebraic freedom in 2HDM, which can be solved with another Higgs doublet introduced. This paper became the first study in the domain of three-Higgs-doublet models (3HDM). 

The idea of a four-Higgs-doublet model (4HDM) was proposed just one year after Weinberg’s 3HDM. In 1977, Bjorken and Weinberg published the first study discussing a four-Higgs model \cite{Bjorken:1977vt}, where they addressed the violation of muon number through scalar field mixing. The Yukawa sector they used is:
\begin{equation}
\begin{aligned}
\mathcal{L}_{Yuk} = & -y_1\begin{pmatrix}\overline{\nu}^e_L & \overline{e}_L\end{pmatrix}\begin{pmatrix}\phi_1^+\\ \phi_1^0\end{pmatrix}e_R 
-y_2\begin{pmatrix}\overline{\nu}^e_L & \overline{e}_L\end{pmatrix}\begin{pmatrix}\phi_2^+\\ \phi_2^0\end{pmatrix}\mu_R \\
& -y_3\begin{pmatrix}\overline{\nu}^\mu_L & \overline{\mu}_L\end{pmatrix}\begin{pmatrix}\phi_3^+\\ \phi_3^0\end{pmatrix}e_R
-y_4\begin{pmatrix}\overline{\nu}^\mu_L & \overline{\mu}_L\end{pmatrix}\begin{pmatrix}\phi_4^+\\ \phi_4^0\end{pmatrix}\mu_R \label{Yuk-early-days}
\end{aligned}
\end{equation}
in which the 4 Higgs doublets are correlated with the 4 Yukawa parameters which form a 2-by-2 Yukawa matrix. This is different from NHDM Yukawa sector we discuss today, where for each Higgs doublet, there is a Yukawa matrix, see Eq.~\eqref{4HDM-L-Yukawa} in later sections. From Eq.~\eqref{Yuk-early-days}, we easily find out that through $\phi_2$ and $\phi_3$, the two leptons $e, \mu$ can be mixed. The strength of the mixing is determined by the VEV of $\phi_2$ and $\phi_3$ and the Yukawa matrix element $y_2$ and $y_3$. 

In multi-Higgs doublet model studies, most people construct the multi-Higgs model with finite symmetries. Wyler was one of the earliest to consider a 3HDM with $S_4$ symmetry in his paper \cite{Wyler:1979fe}. 
The reason for considering three Higgs doublets and the group $S_4$ is that $S_4$ is { one of very few }groups among all other well known finite groups that can make the three fermions and Higgs transform as triplets in the finite group space. This $S_4$ model addresses Cabibbo universality and { accounts for experimentally observed} lifetime of $b$ quark and the branching ratio $\Gamma(b\to c + \mbox{leptons})/\Gamma(b\to u + \mbox{leptons})$. 
However, due to the algebraic structure of the $S_4$ symmetric 3HDM, the symmetry could not be fully broken: $S_4\to S_3$ if the VEV is real and nonzero, and $S_4\to \Z_3$ if the VEV is complex and nonzero. This results in the presence of mass-less quarks, or zero Cabibbo angle, which contradicts experimental observations. It's worthy pointing out that such difficulties arise in many multi-Higgs models.  The problem could be circumvented if we introduce another Higgs doublet $\phi_0$, which is a $S_4$ singlet, and transforms under $S_4$ as sign flip, breaking $S_4$ completely, resulting in a (3+1)HDM. Another approach addressing this problem is to be highly selective towards the discrete groups we use to construct the model: although some discrete symmetries leads to pathology such as degenerate quark mass, there are other, relatively few, discrete groups having special structure that don't lead to problems.  
This issue is further addressed in \cite{Leurer:1992wg, GonzalezFelipe:2014mcf}. 
A similar method was used in \cite{Cree:2011uy}, where the authors considered a soft-breaking $\Z_2\times\Z_2\times\Z_2$ symmetry and divided the fermions into three categories: leptons, quarks with charge $2/3$, and quarks with charge $-1/3$. Each class was coupled with its own Higgs doublet (N = 3). As the author pointed out, this model has somewhat strict constraints resulting from the slim algebraic freedom of 3HDM: coupling parameters of charged Higgs and leptons must be relatively close to unity with stringent relations among them. If one additional Higgs is introduced, which doesn't couple to fermions, such strict constraints could be relaxed. Thus, we see from here that 4HDM usually arise to explore ideas that are not allowed in 2HDM and 3HDM because 4HDM provides more algebraic freedom. 

Another source of considering multi-Higgs doublet model is supersymmetry (SUSY). In the minimal supersymmetric Standard Model (MSSM), two Higgs doublets are required to give mass to up-quarks and down-quarks respectively. 
This is because the conjugate $\tilde{\phi} = i\sigma_2\phi$ and it's Yukawa couplings we use in the standard model are not allowed in supersymmetry since the superpotential is holomorphic thus containing no conjugated fields. 
The supersymmetric extension of SM with four Higgs doublets (SS-4HDM) is motivated because it's the next-to-leading-minimal supersymmetric SM, since we know that Higgs doublets in SUSY must come in pairs. With one additional Higgs doublet pair, we have more freedom to circumvent experimental constraints where MSSM is vulnerable, and to realize ideas that are not allowed in MSSM due to the slim algebraic freedom.  In \cite{Dutta:2018yos}, the authors studied this SS-4HDM to address the Yukawa unification problem and the little hierarchy problem. In \cite{Arroyo-Urena:2019lzv}, the authors also studied SS-4HDM with a more constrained parameter space. This is achieved by having two Higgs doublets $H_1 = H_{u1}, H_4 = H_{u4}$ coupling to the up-quarks and one $H_2 = H_d$ to down-quarks and one $H_3 = H_l$ to leptons. Then the absence of observed FCNC placed strong constraints on the $\langle H_1, H_4\rangle$ parameter space. More literature studying SS-4HDM or supersymmetrized 2HDM: \cite{Drees:1988fc,Griest:1989ew,Griest:1990vh,Nelson:1993vc,Krasnikov:1993qd,Masip:1995sm,Aranda:2000zf,Marshall:2010qi,Kawase:2011az,Clark:2011cv,Yagyu:2012qp,Grossman:2014xia,Dutta:2018yos}, which offer more degrees of freedom than the MSSM. 

Moreover, 4HDM has its own research motivations beyond solving issues that 2HDM or 3HDM cannot. It can provide a potential solution to the fermion mass hierarchy problem. The known mass of the lightest fermion, the electron, is $m_e = 0.511$ MeV, while the heaviest fermion, the top quark, has a mass of $m_t = 173.1$ GeV—a difference of several orders of magnitude. Other fermions are scattered across this wide range of masses. Why do fermion masses span such a large range? This is known as the fermion mass hierarchy problem. 4HDM offers a viable solution: in \cite{Rodejohann:2019izm}, the authors categorized fermions into four groups based on their mass proximity: $\{t\}$, $\{b, c, \tau\}$, $\{\mu, s\}$, and $\{d, u, e\}$. Each group was coupled to its own Higgs doublet, resulting in a model with four Higgs doublets. In this way, the mass of fermions in each group is determined by the VEV of the associated Higgs doublet, thus explaining the fermion mass hierarchy through the distribution of 4HDM vacuum states. The mixing parameters between fermions are explained by the discrete symmetries of the 4HDM potential. A similar study in \cite{Goncalves:2023ydf} proposed a model where all quarks share a single Higgs doublet, while the three generations of leptons each have their own. Both of these studies are variations of the private Higgs model \cite{Porto:2007ed,Porto:2008hb}, in which each fermion is coupled to its own private Higgs doublet. Recent works such as \cite{CarcamoHernandez:2021osw, Vien:2020aif} also explore using 4HDM with discrete symmetries to address fermion-related issues, as the spontaneous breaking of discrete symmetry has rich phenomenological implications.

Discrete symmetries can also prevent certain physical processes from occurring, thereby predicting the existence of stable particles that do not decay. These stable, weakly interacting particles are suitable candidates for dark matter. The earliest consideration of this mechanism was the $\Z_2$ symmetric 2HDM, introduced by Deshpande and Ma in 1977 \cite{Deshpande:1977rw}, now known as the inert Higgs doublet model. The main idea is that the lightest $\Z_2$ odd particle is protected from decaying to any other particles, which could be taken as a dark matter candidate. 

Of course one can consider a model with discrete symmetry which is broken, leaving a residual symmetry, protecting certain particles from decay. Several studies have been conducted on 4HDM-based models: for example, those with $D_4\times\Z_n$ symmetries \cite{Meloni:2011cc,Lavoura:2011ry}, $A_4$ symmetries \cite{Meloni:2010sk,Boucenna:2011tj,deAdelhartToorop:2011ad,Bonilla:2023pna}, or cyclic symmetries \cite{Ivanov:2012hc}. Importantly, this mechanism cannot extend to the Yukawa sector because, as mentioned earlier, the algebraic structure of the model would lead to nonphysical zero masses for some quarks \cite{GonzalezFelipe:2013yhh,GonzalezFelipe:2013xok}. Therefore, when constructing models with partially broken discrete symmetries, only certain Higgs doublets within the Higgs potential can be assigned discrete symmetry, and these Higgs doublets cannot couple to fermions. Such Higgs doublets are referred to as inert Higgs doublets.

4HDM has also been explored as a candidate model for addressing the neutrino mass problem. For example, \cite{Ma:2001dn} employed $A_4$ symmetry to study neutrino masses and mixing coefficients, and similar studies can be found in \cite{He:2006dk,Grimus:2008tt,Grimus:2008nf,Grimus:2008vg,Grimus:2009sq,Grimus:2009pg,Ferreira:2011hw,Park:2011zt,BenTov:2012tg,Grossman:2014xia}.

\subsection{The Goal of This Thesis and 4HDM Lagrangian}

From previous discussion we see that the NHDM family is a widely studied class of beyond-standard-model physics. Within this family, there are a lot of, approaching hundreds, literature studying 4HDM because 4HDM not only enables the realization of mechanisms that 2HDM and 3HDM cannot due to algebraic constraints but also, when endowed with discrete symmetries, serves as a promising theoretical candidate for addressing significant issues such as neutrino masses and dark matter. We also see the significance of discrete symmetries in NHDM model building. However, a common difficulty in this research area is the lack of physical intuition or prior knowledge to guide researchers in selecting the appropriate discrete symmetry group for 4HDM: which finite group should be chosen from among the many options? In addition, the uniqueness of the relationship between the symmetry group and the 4HDM model remains unclear: can the same finite symmetry group be used for two distinct 4HDM models with different physical predictions? Moreover, once a symmetry group is chosen, how can we ensure that this symmetry is the maximal symmetry of the 4HDM model, meaning that the model does not inadvertently possess additional symmetries, such as continuous symmetries? These questions clearly present challenges to scholars working in the field of 4HDM. Therefore, with the 4HDM papers approaching a hundred and the difficulties in 4HDM model building given above, we find it essential to systematically study and classify the finite symmetries that can be applied to 4HDM models, and this is the primary focus of this work. 

Next, we examine what does the 4HDM Lagrangian looks like. The 4HDM assumes the existence of four Higgs doublets, while the $SU(3)_c$, $SU(2)_L$, and $U(1)_Y$ gauge bosons, and fermions remain as consistent with the Standard Model as possible. Therefore, its gauge Lagrangian $\mathcal{L}_{gauge}$ is identical to the Standard Model gauge field Lagrangian given in Eq.~\eqref{SM-L-gauge}. Next, we consider the fermion dynamics and gauge interaction terms. Since the fermion sector in the 4HDM is exactly the same as that of the Standard Model too, its fermion Lagrangian $\mathcal{L}_{f}$ is identical to the one in Eq.~\eqref{SM-L-fermion}.

The primary differences from the Standard Model lie in the Yukawa and the Higgs sector. First, consider the Yukawa sector: in the most general case, each Higgs doublet interacts with the fermions. Therefore, the Yukawa Lagrangian is given by:
\begin{equation}
\mathcal{L}_{Yuk} = -\sum_{k = 1}^4\sum_{I,J = 1}^3\left[ \Gamma^{u,k}_{IJ}\,\overline{q}^I_{L}\tilde{\phi}_k u^J_R + \Gamma^{d,k}_{IJ}\overline{q}^I_L \phi_k d^J_R + \Gamma^{e,k}_{IJ}\overline{l}^I_L \phi_k e^J_R + h.c. \right]\label{4HDM-L-Yukawa}
\end{equation}
Here, the index $k$ represents the Higgs flavor, and each flavor has a Yukawa interaction similar to the Standard Model, resulting in four sets of Yukawa matrices. In practice, we can choose to set some of these terms to zero.

Next, consider the Higgs Lagrangian $\mathcal{L}_H$:
\begin{equation}
\mathcal{L}_H = \sum_{I = 1}^4 (D^\mu \phi_i)^\dagger D_\mu \phi_i - V(\phi)
\end{equation}
where the Higgs potential $V(\phi)$ still only includes quadratic and quartic terms, which ensures the renormalizability of the theory:
\begin{equation}
V = m^2_{ij}\fdf{i}{j} + \Lambda_{ijkl}\fdf{i}{j}\fdf{k}{l}
\end{equation}

Thus, in the 4HDM, only the Yukawa and the Higgs sector differ from the Standard Model. Any additional discrete symmetries, or Higgs flavor symmetries, are applied to $\mathcal{L}_{Yuk}$ and $\mathcal{L}_H$. Moreover, the kinetic term in $\mathcal{L}_{H}$ contains the gauge bosons, and we do not wish to impose additional global discrete symmetries on the gauge bosons, as this could lead to nonphysical couplings between interactions. While discrete symmetries can be imposed on the Yukawa terms, the variations in fermion sectors across different models make a unified analysis of such symmetries difficult. Therefore, a systematic analysis that can provide guidance for all 4HDM studies is the classification of discrete symmetries in the 4HDM potential. Thus, by ``we will be studying the discrete symmetries of 4HDM'', we mean that we will study the discrete symmetries of 4HDM potential $V(\phi_i)$ in particular. Overall, this thesis is based on our previous research papers \cite{Shao:2023oxt,Shao:2024ibu}.

\newpage

\section{Basics of Finite Group Theory}\label{Finite_group_intro}

Before discussing the research methodology and performing detailed calculations, it is necessary to introduce the basics of finite group theory, as this is a topic unfamiliar to many physicists.

\subsection{Basic Definitions and Terminology}

Let us begin with the definition of finite groups. We assume that the readers are already familiar with the basic axioms of groups. A finite group is a group with a finite number of elements. It is important to note that in discussions of finite groups, the term “order” has two distinct meanings depending on whether it is applied to the group itself or to an element of the group. The order of a finite group $G$, denoted by $|G|$, refers to the number of elements in $G$. The order of an element $g \in G$ is the smallest positive integer $n$ such that $g^n = e$, where $e$ is the identity element of the group. As examples, the following are all finite groups:

\textbf{Additive Group of Integers Modulo $m$:} The elements of this group are integers, and the group operation is defined as addition modulo $m$. Under this definition, two integers that are congruent modulo $m$ are considered the same group element, while integers that are not congruent modulo $m$ are different elements. There are exactly $m$ distinct elements in this group, corresponding to the residue $0, 1, 2, \ldots, m-1$ under division by $m$. Therefore, this group contains the elements ${\overline{0}, \overline{1}, \overline{2}, \ldots, \overline{m-1}}$, where $\overline{k}$ denotes the set of all integers congruent to $k$ modulo $m$: ${k - 2m, k - m, k, k + m, k + 2m, \ldots}$. This additive group is denoted by $\Z_m$.

Sometimes we write group multiplication additively or vice versa: $a + a = 2a \simeq a^2$. If we want to write the group $\Z_m$ multiplicatively, we write 1 as the identity instead of 0, and define an element $a$ as $\overline{1}$, then $\Z_m$ can then be defined as ${1, a, a^2, \ldots, a^{m-1}}$, with $a^m = 1$. This group is called the cyclic group of order $m$. Here, because every elements in $\Z_m$ can be written as some powers of $a$, we call $a$ the generator of the group. The additive group of integers and the cyclic group share the same structure. 
Just like dimension to vector spaces, the order $m$ of a cyclic group is the characteristics of a cyclic group: two cyclic groups are ``identical'' or formally, isomorphic, if they have the same order. 
Once the order is known, the structure of the cyclic group is uniquely determined. The cyclic group $\Z_m$ can be presented as $\langle a \,|\, a^m = e \rangle$. In the following discussion, both $e$ and $1$ will be treated as the identity element. 

The cyclic group can also be defined as $a = e^{\frac{2\pi i}{m}}$, which similarly satisfies $a^m = 1$. A key property of cyclic groups is that the group operation, defined by the multiplication law, satisfies the commutative property: $a^k a^l = a^{k+l} = a^{l+k} = a^l a^k$. This means that any two elements in a cyclic group commute with each other, and such groups are called abelian groups. All cyclic groups are abelian.

\textbf{Symmetry Groups of Polygon:} Consider the group formed by all transformations that preserve the symmetry of a regular square in the plane. The set of these transformations, under the operation of composition, forms a group, denoted by $D_4$. The transformations that preserve the symmetry of a square include $a$: a counterclockwise rotation by $90^\circ$ around the center, and $b$: a reflection across a diagonal. All other symmetry-preserving transformations can be expressed as combinations of $a$ and $b$. Therefore, every element in $D_4$ can be written as a product of $a$ and $b$, meaning $D_4$ has two generators. If a group has only one generator, knowing its order is enough to determine the group structure (it is a cyclic group). However, when a group has multiple generators, knowing their individual orders is not sufficient, because in general elements in the group doesn't necessarily commute; one must also know the relations between the generators to determine the structure of the group. In this case, knowing $a^4 = b^2 = e$, where $e$ is the identity map, isn't enough. We can assume $a$ and $b$ commute, then this relation doesn't satisfy the geometric of a square: because the net effect of counterclockwise rotation of $90^\circ$ then reflection is different from reflection then counterclockwise rotation of $90^\circ$. The geometric properties of the square tell us that reflecting and then rotating counterclockwise by $90^\circ$, followed by another reflection, is equivalent to a clockwise rotation by $90^\circ$. Thus, the relation between $a$ and $b$ is given by $bab = a^{-1}$. Therefore, we can describe $D_4$ as $D_4 \simeq \langle a, b \,|\, a^4 = b^2 = e\,, \, bab = a^{-1} \rangle$. 

More generally, if we consider symmetry of a general polygon with $n$ edges instead of a simple square, the symmetry group is denoted by $D_{n}$, which satisfy $D_{n}\simeq \langle a, b\,|\,a^n = b^2 = e\,,\, bab = a^{-1} \rangle$ where $a$ represents the counterclockwise rotation around the center by $360^\circ/n$ degrees while $b$ still represents reflection. Note that in mathematical context the symmetry group of a polygon with $n$ edges is usually denoted as $D_{2n}$, where $2n$ indicates the order of the group. However, in this text, we will be adopting the notation $D_n$, which is more popular among physicists, instead of $D_{2n}$. 

\textbf{Permutation Groups or Transformation Groups on Finite Sets:}
The axioms for group multiplication suggest that the composition of maps is a natural candidate for group multiplications. In fact, the idea of group is abstracted from studying sets of transformations.
The $D_n$ group is one example. Now, consider a more abstract situation where the objects being transformed have no specific geometric shape: consider an arbitrary finite set $\Omega$ and the transformations on $\Omega$. For convenience, we can label the elements of $\Omega$ as $1, 2, 3, \dots, n$, where $n$ is the number of elements in $\Omega$. A transformation on $\Omega$ permutes the elements in some way. For example, when $n=7$, we can define a transformation $a$ as $[1, 2, 3, 4, 5, 6, 7] \mapsto [3, 2, 5, 6, 7, 4, 1]$. A notation for representing such cycles is $(i, j, k, l , \dots, m, n)$, meaning a cyclic permutation: $i \mapsto j \mapsto k \mapsto l \mapsto \dots \mapsto m \mapsto n \mapsto i$. For example, the transformation $1 \mapsto 3 \mapsto 5 \mapsto 7 \mapsto 1, , 4 \leftrightarrow 6$ can be written as $a = (1357)(46)$. Thus,
\begin{equation}
a[1,2,3,4,5,6,7] = (1357)(46)[1,2,3,4,5,6,7] = (1357)[1,2,3,6,5,4,7] = [3,2,5,6,7,4,1].
\end{equation}
The set of all transformations on $\Omega$ forms a group under composition of map, denoted by $Sym(\Omega)$ or $S_{\Omega}$. It is straightforward to verify that $S_{\Omega}$ satisfies the group axioms, meaning that $S_{\Omega}$ is indeed a group. We can also denote $S_\Omega$ by $S_n$ where $n$ is the number of elements in $\Omega$. In fact, $S_n$ is more frequently used. 

After introducing some basic definitions through examples, we now consider the relationships between different groups. To achieve this, we can focus on maps between finite groups. Since finite groups can be viewed as finite sets, we can define maps on them. Consider a map $\sigma: G \to G^\prime$ between two finite groups $G$ and $G^\prime$. Because groups are more than mere sets -- they are equipped with an additional structure of group multiplication; therefore, there are additional topics we can discuss beyond simple maps from set to set. Here are some terminology:
\begin{itemize}
\item \textbf{Kernel of a map, $\Ker\,\sigma$}: This is a subset of $G$ and is defined as $\Ker\,\sigma \equiv \{g \in G \,|\, \sigma(g) = e^\prime \in G^\prime\}$, i.e., the kernel consists of elements in $G$ that are mapped to the identity element of $G^\prime$. If $\Ker\,\sigma = \{e\}$, we call the map an injective map, or an injection, sometimes also known as an one-one map, which does not have multiple elements in $G$ mapped to the same element in $G^\prime$.

\item \textbf{Image of a map}, $\Im\,\sigma$, or $\sigma[G]$: This is the set of elements in $G^\prime$ that are the images of elements from $G$, defined as $\Im\,\sigma = \sigma[G] = \{\sigma(g) \in G^\prime \,|\, \forall g \in G\}$. An equivalent definition is $\Im\,\sigma = \{g^\prime \in G^\prime \,|\, \exists g \in G, \, \text{s.t.} \, g^\prime = \sigma(g)\}$. If the image of a map covers all of $G^\prime$, i.e., $\Im\,\sigma = G^\prime$, the map is called a surjective map, or an onto map. Note that if a map is both injective and surjective, it is called bijective, or a bijection, which establishes a one-to-one correspondence between the elements of the two sets. Naturally, the number of elements in both sets must be the same.
\end{itemize}

After introducing the concept of a map between two finite groups, a natural question arises: do the elements in the image of $G$ under the map respect the group multiplication in $G^\prime$? That is, for $g_1, g_2 \in G$, does the following property hold: $\sigma(g_1 g_2) = \sigma(g_1) \sigma(g_2)$? In general, arbitrarily defined maps do not necessarily have this desirable property. A map that does have this property is called a group homomorphism. Naturally, a homomorphism can also be bijective, in which case it is called an isomorphism. Two groups related by an isomorphism have identical group structures. A homomorphism from $G$ to itself is called an endomorphism, and an isomorphism from $G$ to itself is called an automorphism. Although injective and surjective homomorphisms have their own specific mathematical names, they are too detailed for this discussion and will not be elaborated here.

\subsection{Subgroups -- Small Groups in a Larger Group}\label{group-theory-subgroups}

A finite group, first and foremost, is a finite set, and thus we can define subsets within it. A natural question that follows is whether a subset forms a group under the group multiplication. If the subset is closed under the group multiplication, then it inherits the structure of a group and is called a subgroup. Let us now consider some examples of subgroups.

\textbf{Subgroups of a Cyclic Group:} For a cyclic group $\Z_n$, an important property is that if $s$ divides $n$ (denoted by $s|n$), meaning that $s$ is a divisor of $n$, then $\Z_n$ has a subgroup $\Z_s$. This is because $s|n$, which means there exists an integer $l$ such that $n = sl$. If we think of $a^l$ as a new generator $a^\prime$, the group generated by $a^\prime$ is also a cyclic group, with each element being an element of the original group $\Z_n$, and the order of this group is $s$. This follows from $(a^l)^s = a^{ls} = a^n = e$, thus proving that $\Z_s$ is a subgroup of $\Z_n$. 

We use $\Z_s \leq \Z_n$ to denote that $\Z_s$ is a subgroup of $\Z_n$. If it is a proper subgroup, i.e., the order of the subgroup is less than the order of the original group, we write $\Z_s < \Z_n$.

Now, if we consider the converse and assume that $\Z_s$ is a subgroup of $\Z_n$, can we conclude that $s|n$? The answer is yes. In fact, for any finite group, this conclusion holds: if $H \leq G$, then there exists an integer $k$ such that $|G| = k|H|$. We call the integer $k$ the index of $H$ in $G$, denoted as $[G:H]$, so that $|G| = [G:H] |H|$. This result is known as Lagrange's Theorem.

The proof of this theorem requires the knowledge of an additional structure, known as the cosets of a subgroup. The coset of a subgroup $H$ is defined as: $gH \equiv \{gh \,|\, h \in H\}$. It's worthy pointing out that there is a distinction between left-cosets and right-cosets. All cosets we use throughout the text are left-cosets, so unless pointed out, whenever we mention cosets, we mean left-cosets. A subgroup can have multiple cosets, such as $g_1H, g_2H, \dots$. We claim that all cosets form a disjoint union of the group $G$, i.e. Eq.~\eqref{intro-subgroup-1}. We will show this union is disjoint in subsequent texts. 
\begin{equation}
G = \bigcup_{g\in G} gH\;,\mbox{and}\quad \forall\, g,g^\prime\in G\;, \quad g H \cap g^\prime H = \emptyset\;. \quad  \mbox{ denote this as } G = \bigsqcup_{g\in G} gH \label{intro-subgroup-1}
\end{equation}

An equivalent way to express that a set is a disjoint union is by defining an equivalence relation ``$\sim$'' on the set. Here are the axioms of equivalence relation:
\begin{itemize}
    \item Reflexivity: $a \sim a$
    \item Symmetry: $a \sim b \Rightarrow b \sim a$
    \item Transitivity: $a \sim b$ and $b \sim c \Rightarrow a \sim c$
\end{itemize}

All elements equivalent to $g$ form a set: the equivalence class, which we denote as $\overline{g}$. From the converse-negative proposition of the ``Transitivity'' axiom above, which is $a\not\sim c\Rightarrow a\not\sim b \mbox{ or } a\not\sim c$, we conclude that if $g\not\sim g^\prime$, then $\forall g^{\prime\prime}\in \overline{g}^\prime$, $g\not\sim g^{\prime\prime}$, since $g^\prime \sim g^{\prime\prime}$, so $\forall g\in \overline{g},  g\notin\overline{g}^\prime$. Adopting the same reasoning, we conclude that $\forall g^{\prime\prime}\in \overline{g}^\prime,  g^{\prime\prime}\notin\overline{g}$. Hence, $\overline{g} \cap \overline{g^\prime} = \emptyset$. 
The union of all equivalence classes covers the set, which is obvious. Thus, a disjoint union of a set is equivalent to defining an equivalence relation on the set.

Applying this result to group theory, if we can find a nice way to define some equivalence relation on $G$, and if the relation further satisfies $\overline{g}=gH$, then Eq.~\eqref{intro-subgroup-1} is proven. 
This equivalence relation is easy to find and can be expressed as:
\begin{equation}
g \sim g^\prime \Leftrightarrow g^{-1}g^\prime \in H \label{intro-subgroup-2}
\end{equation}
It is straightforward to verify that this relation satisfies the axioms of an equivalence relation. In addition, $\overline{g} = \{k \in G \,|\, k \sim g\} = \{k \in G \,|\, g^{-1}k \in H\} = \{k \in G \,|\, \exists h \in H \;\text{such that}\; g^{-1}k = h\} = \{k \in G \,|\, \exists h \in H \;\text{such that}\; k = gh\} = \{gh \,|\, h \in H\} = gH$.
Thus $\overline{g} = gH$. Therefore, all cosets of a subgroup $H$ is a disjoint union of $G$. 

Once Eq.~\eqref{intro-subgroup-1} is established, we can easily derive the relationship Eq.~\eqref{intro-subgroup-3}. It is easy to see that if we further prove that the number of elements in each coset equals the order of the subset $H$, then we can prove Lagrange's Theorem.
\begin{equation}
|G| = \sum_k |g_kH| \label{intro-subgroup-3}
\end{equation}
To compare the number of elements in two finite sets, it suffices to check if there is a bijection between them. There is indeed a bijection between the subset $H$ and any coset $gH$, defined as $\sigma: H \to gH\,,\; \sigma(h) = gh$. Using the properties of group multiplication, it can be verified that this map is a bijection. Hence, Lagrange's Theorem is proven.

However, it is important to note that although the converse of Lagrange's Theorem holds for cyclic groups—i.e., if $s$ divides $n$, then $\Z_s$ is a subgroup of $\Z_n$—this is not generally true for arbitrary finite groups. That is, for a finite group $G$ of order $n$, if $s$ divides $n$, it does not necessarily imply that $G$ has a subgroup of order $s$. However, if $s$ is a prime number and divides $|G|$, then by Cauchy's Theorem, such a subgroup must exist and is isomorphic to $\Z_s$. But it is difficult to take a step further if $s$ is not prime. There are in fact Sylow's theorems, upgraded version of the Cauchy's theorem, that tells us more about the subgroups of a group $G$ given the group's order $|G|$. 

Sylow's theorems focuses on Sylow $p$-subgroups, groups of order $p^n$ where $p$ is prime, of the group $G$ under question. In addition, the group $\Z_p$ is not a Sylow $p$-subgroup of $G$ if $p^2$ divides $|G|$: Sylow $p$-subgroup is in some sense ``maximal'': a subgroup is a Sylow $p$-subgroup (of order $p^n$) of $G$ if $|G| = p^n m$ where $m$ and $p$ are co-prime. The theorems claim that all Sylow $p$-subgroups of a group exist, and are conjugate to each other, therefore isomorphic. Here, two subgroups $H,K$ of $G$ are conjugate means that there is $g\in G$ such that $H\simeq g^{-1}Kg$.  What's more, the number $n_s$ of all the mutually conjugate Sylow $p$-subgroups satisfy $n_s \equiv 1 (\mbox{mod } p)$. For a proof of Sylow's theorem, see Chapter 4 of \cite{rotman2012introduction}. 

Therefore, Sylow's theorems tell us about the subgroup structure of a group $G$ in horizontal ``slices'': subgroups of $G$ in order $p_i^{n_i}$ if we prime decompose $|G|$ in $|G| = p_1^{n_1}p_2^{n_2}...$ Another perspective is also important: consider a series of subgroups of $G$, which give us a somewhat perpendicular view of the structure of $G$. We will come to this view shortly afterwards. Before that, let's study cosets a bit further to help us define another very important subgroup: normal subgroups of a group.

We can further consider the set of cosets $\{gH \,|\, g \in G\}$, and on this set, we can define a ``group operation'' as in Eq.~\eqref{intro-subgroup-4}. 
\begin{equation}
(g_1H)(g_2H) \equiv \{g_1h_1g_2h_2 \,|\, h_1, h_2 \in H\} \overset{?}{=} g_1g_2H \label{intro-subgroup-4}
\end{equation}
Under the operation defined in Eq.~\eqref{intro-subgroup-4}, the set $\{gH \,|\, g \in G\}$ may form a group, which is called the quotient group $G/H$. However, it is not guaranteed that the definition in Eq.~\eqref{intro-subgroup-4} will work for every subgroup $H$. 

The reason is that $g_1h_1g_2h_2 = g_1g_2g_2^{-1}h_1g_2h_2 = g_1g_2(g_2^{-1}h_1g_2)h_2 \in g_1g_2(g_2^{-1}h_1g_2)H \neq g_1g_2H$. The primary issue is that $(g_2^{-1}h_1g_2)$ is not necessarily an element of $H$. That's why we put a ``?'' on top of the second equal sign in Eq.~\eqref{intro-subgroup-4}.  

Thus, for the set of cosets to form a group, or in other words, for the quotient group $G/H$ to be well-defined, $H$ must satisfy the condition $g^{-1}Hg = H$ for all $g \in G$. A subgroup $H$ of $G$ that satisfies this condition is called a normal subgroup and is denoted as $H \triangleleft G$. To distinguish it from a general subgroup $H$, we will use the letter $N$ to denote a normal subgroup. The formal definition of a normal subgroup is given in Eq.~\eqref{intro-subgroup-normal}.
\begin{equation}
N \triangleleft G \quad \text{if and only if} \quad \forall g \in G\,,\; g^{-1}Ng = N. \label{intro-subgroup-normal}
\end{equation}

Let us now see some examples of normal subgroups.

\textbf{The kernel of a group homomorphism:} For any group homomorphism $\sigma: G \to \tilde{G}$, the kernel of the map, $\Ker\,\sigma$, is always a normal subgroup of $G$, i.e., $\Ker\,\sigma \triangleleft G$. To prove this, we need to show that $\forall g \in G$, we have $g^{-1}(\Ker\,\sigma)g = \Ker\,\sigma$, which means proving both $g^{-1}(\Ker\,\sigma)g \subseteq \Ker\,\sigma$ and $g^{-1}(\Ker\,\sigma)g \supseteq \Ker\,\sigma$. 
First prove $g^{-1}(\Ker\,\sigma)g \subseteq \Ker\,\sigma$: Consider any element $h \in g^{-1}(\Ker\,\sigma)g$, so by the definition of $g^{-1}(\Ker\,\sigma)g$, there exists $k \in \Ker\,\sigma$ such that $h = g^{-1}kg$. Since $\sigma(h) = \sigma(g^{-1}kg) = \sigma(g^{-1})\sigma(k)\sigma(g) = \sigma(g)^{-1}\tilde{e}\sigma(g) = \tilde{e}$, we have $h \in \Ker\,\sigma$. 
Next, to prove $g^{-1}(\Ker\,\sigma)g \supseteq \Ker\,\sigma$: Consider any $a \in \Ker\,\sigma$. Since $\sigma(gag^{-1}) = \sigma(g)\sigma(a)\sigma(g^{-1}) = \sigma(g)\tilde{e}\sigma(g)^{-1} = \tilde{e}$, it follows that $gag^{-1} \in \Ker\,\sigma$. Therefore, there exists $b \in \Ker\,\sigma$ such that $gag^{-1} = b$, meaning that for any $a \in \Ker\,\sigma$, we have $a = g^{-1}bg$. Thus, $a \in g^{-1}(\Ker\,\sigma)g$. 

\textbf{The center of a group:} The center of a group $G$, denoted $Z(G)$, is the set of elements in $G$ that commute with every elements of $G$. In mathematical notation, $Z(G) \equiv \{c \in G \,|\, cg = gc \,,\; \forall g \in G\}$. It is easy to verify that $Z(G) \triangleleft G$: For any $g \in G$, we first prove that $g^{-1}Z(G)g \subseteq Z(G)$. Consider an arbitrary element of $g^{-1}Z(G)g$, which takes the form $g^{-1}cg$, where $c \in Z(G)$. By the definition of $Z(G)$, we have $g^{-1}cg = g^{-1}gc = c \in Z(G)$. 
Next, we prove that $Z(G) \subseteq g^{-1}Z(G)g$. Take any $c \in Z(G)$, and since $c = g^{-1}gc = g^{-1}cg \in g^{-1}Z(G)g$, we conclude the proof. A less strict version is a centralizer of an element $a$, defined as $C_G(a) \equiv \{g\in G\,|\, ga = ag\}$. One can also define the centralizer of a subgroup $C_G(A)\equiv \{g\in G\,|\,g \mbox{ commutes with all elements in } A\}$. However, although a centralizer is a subgroup, it isn't necessarily normal. 
Another important structure which are not necessarily normal subgroup is the normalizer of a subgroup: For any subgroup $A \leq G$, the normalizer $N_G(A)$ is defined as $N_G(A) \equiv \{g \in G \,|\, g^{-1}Ag = A\}$. 
Note that the normalizer of all diagonal matrices are called monomial matrices, which has only one element on each row and column with all other elements being 0. Monomial matrices will be the main character on the stage of this paper when we later discuss non-abelian transformations of 4HDM. 

\textbf{The commutator subgroup:} The commutator subgroup of a group, also called the derived subgroup, denoted $G' = [G, G]$, is defined as the group generated by all commutators $[g_1, g_2] \equiv g_1g_2g_1^{-1}g_2^{-1}$ for $g_1, g_2 \in G$: $[G, G] = \langle \{[g_1, g_2] \,|\, g_1, g_2 \in G\} \rangle$. Strictly speaking, we should also specify the relations among the generators, but since the structure of $G$ is already determined, the relations between different commutators $[g_1, g_2]$ are implicitly defined by the structure of $G$, leaving no ambiguity for the definition of $[G,G]$.

After defining the commutator subgroup, we can further calculate the commutator subgroup of $G'$, denoted $G'' \simeq G^{(2)}$, and further compute the commutator subgroup of $G^{(2)}$, denoted $G^{(3)}$, and so on. It is important to note that this sequence is not trivial and may not terminate. 

To prove that the commutator subgroup $[G, G] \triangleleft G$, we proceed as follows: For any $g \in G$, take any $z \in [G, G]$. By the definition of the commutator subgroup, we have $[g, z] = gzg^{-1}z^{-1} \in [G, G]$. Therefore, $gzg^{-1} \in [G, G]$, implying that $g[G, G]g^{-1} \subseteq [G, G]$. Note that in the definition of a normal subgroup Eq.~\eqref{intro-subgroup-normal}, we can replace $g^{-1}Ng$ with $gNg^{-1}$, and the resulting definition is isomorphic to the original one. Reversing the steps of proving $g[G, G]g^{-1} \subseteq [G, G]$, we can show that $[G, G] \subseteq g[G, G]g^{-1}$. Therefore, $[G, G] \triangleleft G$.

The commutator subgroup has many interesting properties, such as $G/[G,G]$ being an abelian group. However, the proofs of these properties are not straightforward. Before delving into them, we need to introduce a key result, namely the Isomorphism Theorems. 

Consider a group homomorphism $\sigma: G \to \tilde{G}$. We can then define $\Ker\,\sigma$ and $\Im\,\sigma$. The Isomorphism Theorem-I asserts the existence of the following group isomorphism: $G/\Ker\,\sigma \simeq \Im\,\sigma$. The proof of the theorem is as follows:

\begin{proof}
First, we know that $\Ker\,\sigma \triangleleft G$, so the quotient group $G/\Ker\,\sigma$ is well-defined, and it is composed of cosets $\{g\,\Ker\,\sigma\}$, with group multiplication of cosets given by the formula Eq.~\eqref{intro-subgroup-4}. Next, consider the following map $\psi$:
\begin{equation}
\psi: G/\Ker\,\sigma \to \Im\,\sigma\;,\quad \psi(g\,\Ker\,\sigma) = \sigma(g)\label{intro-subgroup-psi}
\end{equation}
First, we prove that $\psi$ is a group homomorphism: $\psi(g_1\Ker\,\sigma\cdot g_2\Ker\,\sigma) = \psi(g_1g_2\Ker\,\sigma) = \sigma(g_1g_2) = \sigma(g_1)\sigma(g_2) = \psi(g_1\Ker\,\sigma)\cdot\psi(g_1\Ker\,\sigma)$, so $\psi$ is a homomorphism. Now, we prove that $\psi$ is injective: $g_1\Ker\,\sigma = g_2\Ker\,\sigma \Leftrightarrow g_1^{-1}g_2 \in \Ker\,\sigma \Leftrightarrow \sigma(g_1^{-1}g_2) = \tilde{e} \in \tilde{G} \Leftrightarrow \sigma(g_1)^{-1}\sigma(g_2) = \tilde{e} \Leftrightarrow \sigma(g_1) = \sigma(g_2)$, so $\psi$ is injective. The second $\Leftrightarrow$ comes from the equivalence relation Eq.~\eqref{intro-subgroup-2}. Finally, $\psi$ is surjective, since every element $\sigma(g)$ of $\Im\,\sigma$ has a pre-image $g \in G$, which also gives $\psi$ a pre-image $g\Ker\,\sigma \in G/\Ker\,\sigma$. Therefore, we have shown that the map Eq.~\eqref{intro-subgroup-psi} is a bijective homomorphism, hence an isomorphism.
\end{proof}

Now, using the Isomorphism Theorem, we consider the quotient group $G/[G,G]$. First, we aim to construct a group homomorphism $f$ such that $\Ker\,f = [G,G]$, allowing us to study $G/[G,G]$ via the structure of $\Im\,f$. We proceed as follows: if $\Im\,f$ is an abelian group, what would the implications be? For any $f(a), f(b) \in \Im\,f$, we have that $\Im\,f$ is abelian $\Leftrightarrow f(a)f(b) = f(b)f(a) \Leftrightarrow f(ab) = f(ba) \Leftrightarrow f(ab)f(ba)^{-1} = e \Leftrightarrow f(ab)f((ba)^{-1}) = e \Leftrightarrow f(ab)f(a^{-1}b^{-1}) = e \Leftrightarrow f(aba^{-1}b^{-1}) = e \Leftrightarrow f([a,b]) = e \Leftrightarrow [a,b] \in \Ker\,f \Leftrightarrow [G,G] \subseteq \Ker\,f$, where the last $\Leftrightarrow$ follows from the arbitrariness of $a$ and $b$. 

After showing that $\Im\,f$ is abelian $\Leftrightarrow [G,G] \subseteq \Ker\,f$, that $G/[G,G]$ is abelian is easily shown by specifying $f$ as the natural homomorphism, $f: G\to G/[G,G]\,,\;f(g) = g[G,G]$.

Furthermore, when we are considering a sequence of derived subgroups $G^\prime, G^{(2)}, G^{(3)}...$, we have: $G^{(i)}/G^{(i+1)}$ is abelian for all $i$. At this point, we encounter a first example of a sequence of subgroups of a given group $G$ mentioned immediately after the introduction to Sylow's theorems: the sequence of derived subgroups. Now let's study this sequence a bit further to define solvable groups, which is of central importance in this work. 

Before proceeding, we give an intuitive picture of $G/[G,G]$ and the commutator subgroup $[G,G]$. Consider the case where $G$ is an abelian group. In this case, since all elements of an abelian group commute with each other, we have that $\forall\,a,b \in G,\; [a,b] = aba^{-1}b^{-1} = e$, meaning $[G,G] \simeq \{e\}$. Therefore, when $G$ is an abelian group, $G/[G,G] \simeq G/\{e\} \simeq G$. For non-abelian groups, the commutator subgroup $[G,G]$ usually has a non-trivial structure, and its order reflects the degree of non-abelianness of $G$ to some extent: we expect that the larger $[G,G]$, the more non-abelian $G$ is. However, this is not a nice way to quantify the non-abelianness of a group. 

The solvability of a group provides a better quantification. In \cite{Ivanov:2012fp}, the author provides a comprehensive introduction to solvable groups and the basics of finite groups for physics community. Intuitively, the concept of solvable groups provides a quantification of the non-abelianness of a group compared to the order of $[G,G]$. 

Let us first consider an abelian group. Because all elements commute in an abelian group, the group is as simple to study using natural numbers, integers, real numbers, or complex numbers. Abelian groups can be studied using sets of numbers because multiplication among numbers is also commutative (this is not the proof). However, for non-abelian groups, where the group operation is not commutative, the study often relies on matrices, and this approach has developed into a branch of group theory called representation theory. As we proved earlier, the commutator subgroup of an abelian group is trivial, $[G,G] \simeq \{e\}$. If $[G,G]$ is non-trivial, but $G^{(2)} \simeq \{e\}$, we say that $G$ is very close to being abelian. If $G^{(3)}$ is trivial, then $G$ is further from being abelian, but not as far from being abelian as the group $H$ whose $H^{(4)}\simeq \{e\}$, and so on. If this process of repeatedly taking commutator subgroups leads to some commutator subgroup $G^{(n-1)}$ being abelian, and $G^{(n)} \simeq \{e\}$, we define such a group as a solvable group. In this sense, solvable groups can be compared to polynomials, where taking successive derivatives eventually reduces the polynomial to zero. This could help us understand why commutator subgroups are also known as derived subgroups. 

However, not all groups are solvable. Some groups cannot be reduced to the trivial group $\{e\}$ by calculating the series of commutator subgroups. We call such groups insoluble or non-solvable groups. Since all abelian groups are solvable, it follows that all non-solvable groups are non-abelian.

There is another, more general, equivalent definition of a solvable group: a group $G$ is called solvable if there exists a normal series of subgroups as shown in Eq.~\eqref{intro-subgroup-solvable-normal-series}, and each quotient $G_i/G_{i+1}$ is an abelian group.
\begin{equation}
G \simeq G_0 \triangleright G_1 \triangleright G_2 \triangleright \dots \triangleright G_{n-1} \triangleright G_n \simeq \{e\} \label{intro-subgroup-solvable-normal-series}
\end{equation}

The proof of the equivalence between these two definitions proceeds as follows: first, we prove that if $G^{(k)} \simeq \{e\}$ for some $k$, then there exists a normal series as in Eq.~\eqref{intro-subgroup-solvable-normal-series}, and each quotient $G_i/G_{i+1}$ is an abelian group. Since $G^{(i)} \triangleright G^{(i+1)}$, we can simply take $G_i \simeq G^{(i)}$ and let $k = n$, which proves the first part. After setting $G_i \simeq G^{(i)}$, to prove that each quotient $G_i/G_{i+1}$ is an abelian group, it suffices to show that $G^{(i)}/G^{(i+1)}$ is an abelian group, a result we have already established.

Next, we prove that if there exists a normal series as in Eq.~\eqref{intro-subgroup-solvable-normal-series}, with each quotient $G_i/G_{i+1}$ being abelian, then there exists some $k$ such that $G^{(k)} \simeq \{e\}$. Consider the natural homomorphism $\tau_i: G_i \to G_i/G_{i+1}$. Since $\Im\,\tau_i \simeq G_i/G_{i+1}$ is an abelian group, we deduce from our previous results that $\Ker\,\tau_i \supseteq [G_i, G_i]$. However, $\Ker\,\tau_i \simeq G_{i+1}$, so we have $[G_i,G_i] \subseteq G_{i+1}$, i.e., $G^{(i+1)} \subseteq G_{i+1}$. Since $G_n \simeq \{e\}$ for some $n$, we conclude that $G^{(n)} \subseteq G_n \simeq \{e\}$, hence $G^{(n)} \subseteq \{e\}$. Since $G^{(n)}$ is a group, we must have $e \in G^{(n)}$, so $G^{(n)} \simeq \{e\}$. This completes the proof. We note here that since $G^{i}\leq G_i$, the derived series is the ``shortest'' normal series. 

The reason for introducing the second definition of solvable groups is that we can examine the normal series Eq.~\eqref{intro-subgroup-solvable-normal-series} from bottom: the trivial group $\{e\}$ towards the group $G_0$ itself: if each step can be thought of as building a larger group from a smaller one, then the meaning of a solvable group is that it can be constructed step by step from simpler groups, and most importantly, the ``simpler'' groups we use are abelian groups. The next question arise is: what happens if some of the factor groups $G_i/G_{i+1}$ are not abelian? We need to discuss the normal series of a group further. 

It can be proven that any finite group has a unique maximal normal series, and by maximal, we mean for each $G_i$ in the series, it's successive normal subgroup $G_{i+1}$ in the series is the maximal normal subgroup of $G_i$. This is the Jordan-Hödler theorem. We call such series composition series. It's important to note that if $N$ is the maximal normal subgroup of $G$, then the factor group $G/N$ is simple. By simple group, we mean it has no non-trivial normal subgroups. We sketch the proof of $G/N$ being simple as follows: suppose the factor group $G/N$ is not simple, then it has a proper normal subgroup $H\triangleleft G/N$. Then by the correspondence theorem (theorem 2.28 in \cite{rotman2012introduction}), which is a more sophisticated version of the isomorphism theorem, we must have a normal subgroup $\overline{H}$ such that $N\triangleleft \overline{H}$, contradicting that $N$ is maximal. 

Therefore, we can now provide an answer to the question: what happens if $G_i/G_{i+1}$ is not abelian. In general, every finite groups has a normal series, the maximal of which has the following property: the factor groups $G_i/G_{i+1}$ is either simple and abelian, which is $\Z_p$, or simple non-abelian. If each $G_i/G_{i+1}$ is $\Z_p$ for some prime $p$, then the group is solvable. Although the discussion of normal series here seems a bit too detailed and subtle group theoretically, involving using theorems after theorems with fancy names, these discussions are useful when constructing non-abelian groups for 4HDM in later texts and attempt to look for new methods of constructing non-abelian groups to address current difficulties. 

So far, we have discussed the concept of subgroups, which are the smaller groups within a larger group; from the discussion of normal series, when viewing the series from bottom to up, we come to another equally important concept in group theory: how to construct larger groups from smaller ones using operations such as direct products, semidirect products, free products, and so on, to extend the group.

\subsection{Group Multiplication -- Building Larger Groups Using Small Ones}\label{section-split}

Next, we consider how to construct a larger group from smaller groups such that, hopefully, these smaller groups are subgroups of the larger one. Since a group is a set, it naturally has the properties of a set. We know that the Cartesian product of two sets can be defined as: $A \times B \equiv \{(a,b) \mid a \in A, b \in B\}$. Similarly, two groups can be viewed as sets, and we can define their Cartesian product. The follow-up question is whether it's possible to upgrade the product space to a group. We define the group multiplication on two elements from the Cartesian product space as Eq.~\eqref{intro-subgroup-extensions-direct-product-def}.
\begin{equation}
(n_1,h_1)(n_2,h_2) \equiv (n_1n_2, h_1h_2)
\label{intro-subgroup-extensions-direct-product-def}
\end{equation}

which can be shown to satisfy the axioms of group multiplication. Hence, we have defined a group structure on the Cartesian product of two groups, and we refer to this larger group $G$ as the external direct product of the two groups, denoted $G \simeq N \times H$. It can be verified that both $N$ and $H$ are subgroups of $G$ and are normal subgroups. This is because $N \simeq \{(n, e_h) \mid n \in N\} \subseteq G$. It's straight forward to show that $\forall g\in G, g^{-1}Ng = N$.

The external direct product is also referred to as the direct product. We add ``external'' in front because the definition depends on the Cartesian product of sets. In fact, it is possible to define the direct product of groups without relying on the Cartesian product of sets. The essence of direct product constructed via Cartesian product is that elements of $N$ and $H$ in $G$ commute with each other, because
$$
nh \equiv (n, e_h)(e_n, h) = (n, h) = (e_n, h)(n, e_h) \equiv hn,
$$

which leads us to define an internal direct product abstractly, based on the essence that $n, h$ commute:
$$
G \equiv \langle n, h \mid \forall n \in N, h \in H, nh = hn \rangle.
$$

This means that the elements of $G$ consist of all elements from $N$ and $H$ along with their products, with the additional requirement that elements from $N$ and $H$ commute, i.e., $nh = hn$ for all $n \in N$ and $h \in H$. It can be proven that the internal and external direct products are isomorphic, so aside from their definitions, we generally refer to both as direct products.

Thus, we have found a way to construct a larger group $G$ from two smaller groups $N$ and $H$. However, this is hardly enough. From the definition of the internal direct product, we observe that the ``building blocks'' used to construct the larger group $G$ are the elements of the smaller groups $N$ and $H$, while the ``glue'' involves defining the group multiplication structure among these elements. The direct product ``glue'' imposes a strong abelian property, requiring that, apart from the possible non-abelian structure within $N$ and $H$, all other multiplications must be abelian. Therefore, the direct product of abelian groups is necessarily abelian. In some cases, we may wish to define a more complex group from two abelian groups, which necessitates defining a new group multiplication.

Starting from the direct product, we can define a new product by examining the definition of the direct product from another perspective. In the direct product, $nh = hn$. By slightly rewriting this equivalence, we obtain:
$$
h^{-1}nh = n,
$$

which means if we view $h$ as a transformation acting on $N$, we define:
\begin{equation}
h(n) \equiv n^h \equiv h^{-1}nh,
\label{intro-subgroup-extensions-automorphism-def}
\end{equation}

allowing us to reconsider the direct product from a new perspective: $H$ forms a group of transformations on $N$, but each transformation is the identity map, i.e., an identical automorphism. The action of $h$ on $n$ in Eq.~\eqref{intro-subgroup-extensions-automorphism-def} is referred to as a conjugation in mathematics. Since it arises from the rewriting of the commutation relation $nh = hn$, conjugation reflects the commutative relationship to some extent. 

Because we can view elements of $H$ as automorphisms acting on $N$, and since all elements of $H$ in the direct product act as identity automorphisms, we can naturally consider cases where some elements of $H$ act as non-trivial automorphisms on $N$. That is, we consider $H \leq \Aut(N)$, where $h^{-1}nh = n'\neq n$. This allows us to extend the definition of group multiplication. Due to the properties of conjugation and automorphisms, the new definition ensures that $N \triangleleft G$. Additionally, we want the product group $G$ to contain information about $H$, so we require that $G/N \simeq H$. This is, in fact, the definition of a group extension, denoted by $G \simeq N.H$:
$$
G \simeq N.H \Leftrightarrow N \triangleleft G, \quad G/N \simeq H.
$$

We refer to the process of obtaining $G \simeq N.H$ as extending $H$ by $N$. It is worthy pointing out (page 66 of \cite{isaacs2008finite}) that in the definition of group extension above, by $N\triangleleft G$, we mean there is a normal subgroup $N_0$ of $G$ that is isomorphic to $N$. So, it may happen that in $G$, there are several different copies of normal subgroup $N_0$, which may or may not be conjugate to each other, and it may happen that in each different choice of $N_0$, the extension is different. Therefore, the resulting group after extension is not unique, meaning that ``group extension'' is more of a term describing the process and steps of the extension rather than the final result. 

Among all group extensions, one particular case that requires special attention is the semi-direct product, or split extension, denoted by $G\simeq N\rtimes H$. Unlike general extensions, which are referred to as non-split extensions, a split extension not only requires $N\triangleleft G$ and $G/N\simeq H$, but also demands that $G = NH \equiv \{nh \mid n \in N, h \in H\}$ and that $N \cap H = \{e\}$. The definitions can be summarized as follows:
\begin{equation}
G \simeq N.H \quad \Leftrightarrow \quad N \triangleleft G,\;\; G/N \simeq H
\label{intro-non-split-def}
\end{equation}
\begin{equation}
G \simeq N\rtimes H \quad \Leftrightarrow \quad G \simeq N.H\;;\;\; G = NH \equiv \{nh \mid n \in N, h \in H\},\;\; N \cap H = \{e\}
\label{intro-split-def}
\end{equation}

From the definition of a group extension, we can get $|G| = |N| \cdot |H|$. This is because $H \simeq G/N$, and by Lagrange's theorem, $|G| = [G : N] \cdot |N|$. Since $[G : N] = |G/N| = |H|$, we have proven that $|G| = |N| \cdot |H|$. 

As pointed out in page 66 of \cite{isaacs2008finite}, group extension is a term describing a process, which doesn't indicate a unique resulting extended group. A surprising fact is that we can arrive at the same group by split extension (Eq.~\eqref{intro-split-def}) or non-split extension (Eq.~\eqref{intro-non-split-def}). We will see example of this in later texts. 

The key distinction between split extensions and general non-split extensions is that $G \simeq N\rtimes H$ contains a subgroup isomorphic to $H$. We can illustrate this with some examples. Consider $N \simeq \mathbb{Z}_2$ and $H \simeq \mathbb{Z}_2$, with the generators of $N$ and $H$ denoted by $a$ and $b$, respectively. Based on the properties of extending $H$, we know that the group $G$ after the extension will satisfy $|G| = |N| \cdot |H|$, leading to $|G| = 4$. Since there are only two groups of order 4: $\mathbb{Z}_4$ and $\mathbb{Z}_2 \times \mathbb{Z}_2$, we expect that extending $H \simeq \mathbb{Z}_2$ by $N \simeq \mathbb{Z}_2$ will result in one of these two groups.

Let us consider the split extension first. Because $N \triangleleft G$ and $H \leq G$, the extended group $G$ must have two distinct $\mathbb{Z}_2$ subgroups. The only possible choice is $\mathbb{Z}_2 \times \mathbb{Z}_2$, so we have $G \simeq N\rtimes H \simeq \langle a, b \mid a^2 = b^2 = (ab)^2 = e \rangle \simeq \mathbb{Z}_2 \times \mathbb{Z}_2$. For the non-split extension, the only remaining possibility is $G \simeq \mathbb{Z}_4$. Since $\mathbb{Z}_4$ has only one $\mathbb{Z}_2$ subgroup, the other $\mathbb{Z}_2$ cannot be embedded in the extended group $\mathbb{Z}_4$.  We find that in this simple case, the semidirect product yields the same group structure as the direct product.

Now, let's consider more complex examples, such as $N \simeq \mathbb{Z}_4$ with $H \simeq \mathbb{Z}_2$. Starting from the definition of the extension $N.H \simeq \mathbb{Z}_4.\mathbb{Z}_2 \Leftrightarrow  H \simeq G/N$ and $N \triangleleft G$, we treat $H$ as a subgroup of $\Aut(N)$, the automorphism group of $N$. In the case of $\mathbb{Z}_4.\mathbb{Z}_2$, since $\Aut(\mathbb{Z}_4) \simeq \mathbb{Z}_2$, the generator $b$ of $H \simeq \mathbb{Z}_2$ has two choices: non-trivial automorphism $b^{-1}ab = a^3$ and trivial automorphism $b^{-1}ab = a$. Thus, we obtained the relation between $a,b$ in the extended group $G\simeq N.H$. However, the structure of $G$ is not fully determined, as the order of $b$ in the extended group $G$ remains unspecified. Since $H \simeq G/N \Rightarrow b^2 \in N$ in the extended group, the structure of the extension depends on the value of $b^2$ in $N$.

First, consider $b^2 = e$ in $G$. This means $H$ is a subgroup of $G$, resulting in a semidirect product and gives $G \simeq \langle a, b \,|\, a^4 = b^2 = e,\, b^{-1}ab = a^3 \rangle \simeq D_4$, leading to a split extension: $D_4 \simeq \mathbb{Z}_4 \rtimes \mathbb{Z}_2$. Alternatively, $b^2$ can take other elements in $N \simeq \mathbb{Z}_4$. For example, if $b^2 = a$, the group has a single generator $b$, because $a$ can be expressed solely by $b$. Since $a^4 = (b^2)^4 = b^8 = e$, this group is a cyclic group $\mathbb{Z}_8 \simeq \mathbb{Z}_4.\mathbb{Z}_2$. Another possibility is $b^2 = a^3$, where we again obtain $\mathbb{Z}_8$ since both $a$ and $a^3$ can be treated as generators in $\mathbb{Z}_4$. Finally, if $b^2 = a^2$, we obtain a non-abelian group $G \simeq \langle a, b \,|\, a^4 = e,\, b^2 = a^2,\, b^{-1}ab = a^3 \rangle \simeq Q_4$, resulting in a non-split extension $Q_4 \simeq \mathbb{Z}_4.\mathbb{Z}_2$.

Since we are primarily interested in the non-abelian groups resulting from extensions (as we have already identified all the abelian symmetries in the 4HDM), we will not discuss abelian groups resulting from extensions further. It is important to note the definitions of $D_4$ and $Q_4$: $D_4 \simeq \langle a, b \,|\, a^4 = b^2 = e,\, b^{-1}ab = a^3 \rangle$ and $Q_4 \simeq \langle a, b \,|\, a^4 = e,\, b^2 = a^2,\, b^{-1}ab = a^3 \rangle$, both presented using the generators-and-relations method. A more detailed discussion of presenting a group using this method will follow in the next section.

In the examples above, both $N$ and $H$ are cyclic groups with a single generator. From these discussions, we see that since $H \simeq G/N$, the generator $b$ of the cyclic group $H$ satisfies $b^{|b|} \in N \triangleleft G$ in the extended group, where $|b|$ is the original order of $b$ in $H$. Since the abelian symmetries $A$ in 4HDM are not only cyclic groups but also direct products of several cyclic groups, we must consider more complex extensions than those like $\mathbb{Z}_4.\mathbb{Z}_2$. In such cases, $N$ and $H$ will have more than one generator. As an example, consider $N \simeq \mathbb{Z}_2 \times \mathbb{Z}_2$ and $H \leq \Aut(\mathbb{Z}_2 \times \mathbb{Z}_2) \simeq S_3$.

To see why $\Aut(\mathbb{Z}_2 \times \mathbb{Z}_2) \simeq S_3$, although $\mathbb{Z}_2 \times \mathbb{Z}_2$ has two generators, it has three nontrivial elements: $a_1$, $a_2$, and $a_1a_2$. If we treat $a_1a_2$ as a new $a_1$, the structure of $\mathbb{Z}_2 \times \mathbb{Z}_2$ remains unchanged. Further investigation shows that these three elements are interchangeable, and any permutation or cyclic rotation of them leaves the structure of $\mathbb{Z}_2 \times \mathbb{Z}_2$ intact. The group formed by all such permutations is $S_3$, thus proving $\Aut(\mathbb{Z}_2 \times \mathbb{Z}_2) \simeq S_3$. The nontrivial subgroups of $S_3$ are $\mathbb{Z}_2$, $\mathbb{Z}_3$, and $S_3$ itself, allowing us to consider possible extensions: $(\mathbb{Z}_2 \times \mathbb{Z}_2).\mathbb{Z}_2$, $(\mathbb{Z}_2 \times \mathbb{Z}_2).\mathbb{Z}_3$, and $(\mathbb{Z}_2 \times \mathbb{Z}_2).S_3$.

First, let us consider the simpler case of $(\Z_2 \times \Z_2).\Z_2$, where $N \simeq \Z_2 \times \Z_2$ has two generators, denoted by $a_1$ and $a_2$, and $H$ is generated by a single automorphism $b$ of order two. According to our earlier discussion, without loosing generality, we can choose $b$ such that $a_1 \overset{b}{\leftrightarrow} a_2$. This reveals the first distinction from the case where $N$ is a cyclic group: in this case, we need to define the automorphism's action on each generator of $N$, as $N$ now has more than one generator. Thus, we obtain two relations: $b^{-1}a_1b = a_2$ and $b^{-1}a_2b = a_1$, as opposed to a single relation in the case of $N \simeq \Z_4$, where $b^{-1}ab = a^3$. Next, we must determine the order of $b$ in the extended group, which requires us to select one element in $\Z_2 \times \Z_2$ to be the image of $b^2$. All possible choices are: $b^2 = a_1$, $b^2 = a_2$, $b^2 = e$, and $b^2 = a_1a_2$. 

The first two options, $b^2 = a_1$ or $b^2 = a_2$, imply that $a_1$ or $a_2$ can be expressed in terms of $b$, leading to some pathologies. For example, if $b^2 = a_1$, together with $b^{-1}a_1b = a_2$, we obtain $a_2 = b^2$, so that $a_1^2 = b^4 = e$, and $a_1a_2 = b^4 = e$. This definition leads to an extended group isomorphic to $\Z_4$, but clearly, $\Z_2 \times \Z_2$ is not a normal subgroup of $\Z_4$, making this extension invalid. The same reasoning applies to the case $b^2 = a_2$, leading to a similarly problematic extension of $\Z_4$.

The remaining two possibilities are $b^2 = e$ and $b^2 = a_1a_2$. First, consider the case $b^2 = e$. This clearly corresponds to a split extension, as the extended group has the structure of $H \simeq \Z_2$, yielding the following extension: $(\Z_2 \times \Z_2) \rtimes \Z_2 \simeq \langle a_1, a_2, b \,|\, a_1^2 = a_2^2 = b^2 = [a_1, a_2] = e,\, a_1b = ba_2,\, a_2b = ba_1 \rangle \simeq D_4$. The reason that this presentation also corresponds to the dihedral group $D_4$ will be explained in more detail in the next section. 

In the final case, $b^2 = a_1a_2$, we find that $b^4 = (a_1a_2)^2 = e$, and using the relation $a_1b = ba_2$, applying $a_1$ on both sides gives $a_1ba_1 = ba_2a_1 = b^3 = b^{-1}$. Since $a_1^2 = e$, we obtain $a_2 = a_1b^2$. Thus, $a_2$ can be expressed in terms of $a_1$ and $b$, making $a_2$ a redundant generator in the extended group. Therefore, the group structure is given by $\langle a_1, b \,|\, a_1^2 = b^4 = e,\, a_1^{-1}ba_1 = b^{-1} \rangle$, which is the definition of the group $D_4$. Although the choice $b^2 = a_1a_2$ initially appears to yield a non-split extension, the resulting group is isomorphic to the one obtained from the split extension. Upon further examination, we find that $\Z_2 \times \Z_2 \triangleleft D_4$, but $H \simeq \Z_2 \simeq \langle b \rangle$ is not a subgroup of $D_4$. Thus, from a definitional perspective, $b^2 = a_1a_2$ still corresponds to a non-split extension process. 

We then proceed to the case of $(\Z_2 \times \Z_2).\Z_3$, where we have a completely analogous discussion. Here, $c$ represents a cyclic permutation of $a_1$, $a_2$, and $a_1a_2$. Without loss of generality, let $a_1 \overset{c}{\mapsto} a_2$ and $a_2 \overset{c}{\mapsto} a_1a_2$. We next select the value of $c^3$ from $\Z_2 \times \Z_2$. After calculation, we find that the choices $c^3 = a_1$, $c^3 = a_2$, and $c^3 = a_1a_2$ all lead to pathological extensions where the extended group $G$ is isomorphic to $\Z_3$, making non-split extensions impossible. The only viable option is the split extension $c^3 = e$, yielding $\langle a_1, a_2, c \,|\, a_1^2 = a_2^2 = [a_1, a_2] = c^3 = e,\, c^{-1}a_1c = a_2,\, c^{-1}a_2c = a_1a_2 \rangle \simeq A_4$. Hence, we obtain $(\Z_2 \times \Z_2) \rtimes \Z_3 \simeq A_4$. The group $A_4$ is a special case of alternating group $A_n$, which is a subgroup of $S_n$. 

Finally, we consider the most complex case, namely $(\Z_2 \times \Z_2).S_3$. Here, not only does $N$ have more than one generator, but $H$ also has more than one generator. Therefore, when constructing the extension, we need to define how the order-2 generator $b$ of $H \simeq S_3$ acts as an automorphism of $N \simeq \Z_2 \times \Z_2$. Additionally, we must also define how the order-3 generator $c$ of $H \simeq S_3$ acts as an automorphism of $N$. As discussed earlier, $c$ can only form a split extension, where $a_1 \overset{c}{\mapsto} a_2$ and $a_2 \overset{c}{\mapsto} a_1a_2$, and $c^3 = e$. For $b$, both split and non-split extensions are possible, and the resulting groups are isomorphic. Thus, we discuss two cases below, starting with $a_1 \overset{b}{\leftrightarrow} a_2$ and $b^2 = e$.

But before that, it's worthy pointing out that unlike the cases of $(\Z_2 \times \Z_2).\Z_k$ for $k = 2, 3$, even after determining the values of $b^2$ and $c^3$ in $\Z_2 \times \Z_2$, the structure of the extended group is not uniquely determined. This is because the relationship between $b$ and $c$ in the extended group remains undefined. This extra complexity arises when both $N$ and $H$ have more than one generator, compared to the case when $H$ is a cyclic group. However, since $H \simeq G/N$, once the relation between $b$ and $c$ in $H$ is expressed as $f(b, c) = e_h$, we have $f(b, c) \in N \simeq \Z_2 \times \Z_2$ in the extended group. In $S_3$, we have $S_3 \simeq \langle b, c \,|\, b^2 = c^3 = (bc)^2 = e \rangle$, so $f(b, c) = (bc)^2$. Thus, to determine the structure of the extension $(\Z_2 \times \Z_2).S_3$, we need to consider four additional sub-cases under the assumption that the automorphisms are structured as $b^2 = c^3 = e$: the cases where $f(b, c) = (bc)^2 \in N \simeq \Z_2 \times \Z_2$.

First, we examine the split extension case where $(bc)^2 = e \in G$. In this case, we obtain the group $G \simeq \langle a_1, a_2, b, c \,|\, a_1^2 = a_2^2 = [a_1, a_2] = b^2 = c^3 = (bc)^2 = e,\, b^{-1}a_1b = a_2,\, b^{-1}a_2b = a_1,\, c^{-1}a_1c = a_2,\, c^{-1}a_2c = a_1a_2 \rangle \simeq S_4$. The details of how this presentation corresponds to the symmetric group $S_4$ will be explained in the next section. Thus, we obtain the split extension $(\Z_2 \times \Z_2) \rtimes S_3 \simeq S_4$.

Next, we consider the remaining non-split extension sub-cases: $(bc)^2 = a_1$, $(bc)^2 = a_2$, and $(bc)^2 = a_1a_2$, as well as four additional sub-cases under the broader category where $b^2 = a_1a_2$. Many of these sub-cases lead to pathological definitions, but it is necessary to examine each one to determine which lead to pathological extensions and which yield valid extensions satisfying $|G| = |N| \cdot |H|$. Readers are encouraged to carry out analyzing these remaining seven sub-cases which will not be discussed in detail here. 

As we can see, when neither $N$ nor $H$ is a cyclic group, the complexity of the discussion increases significantly. In particular, when both $N$ and $H$ have more than one generator, we need to conduct case-by-case analysis not only to determine how each $h \in H$ acts as an automorphism of $N$ and permutes its generators, but also to consider the value of $h^{|h|} \in N$, which introduces further sub-cases. Moreover, within the sub-cases defined by these two conditions, we must also determine $f(h_i) \in N$ and its specific value, leading to yet more sub-cases that need to be examined. It is easy to imagine that as the number of generators in $N$ and $H$ increases to three or more, the complexity of the analysis becomes immense. This is precisely the challenge we face when discussing the direct products of cyclic groups and their extensions. For instance, as observed in Table \ref{table-abelian}, which contains all abelian symmetry groups of 4HDM we discovered, $N$ can take the form $\Z_2 \times \Z_2 \times \Z_2$, and in this case, $H \leq \Aut(N) \simeq GL(3, 2)$, where $GL(3, 2)$ is a group of order 168 with more than three generators. We even have the case of $N \simeq (\Z_2)^4$, where $H \leq \Aut(N) \simeq A_8$, and $A_8$ is a group of order $\frac{8!}{2} = 20160$. To manage this complexity, one strategy is to use computers to assist with the algebraic calculations involved, which we applied in our study. But before delving into how we use computers, one method needs to be introduced, namely presenting a group using generators-and-relations. This method enables us to study a group in much detail and provides a link between our construction of groups manually and the database of computers.

\subsection{Presenting a Group by Generators-and-relations}

Before delving into the detailed calculations for the classification of the four-Higgs-doublet model (4HDM), we need to introduce some additional group theory concepts, specifically the presentation of a group using generators and relations. This method will be extensively used when describing the abstract structure of groups in the following sections. Additionally, this approach facilitates interfacing with computers, allowing us to use code to avoid tedious manual calculations. 

Before we introduce the method of group presentation by generators and relations, we need to cover one additional method for constructing larger groups from smaller ones, known as the free product. The concept of a free product is a natural progression from the group extension discussed in the previous section to the presentation of groups via generators and relations.

To review, in the previous section, we introduced the direct product, followed by a discussion of its limitations—namely, its strong abelian nature. To address this, we modified the commutation relations and defined $H$ as a subgroup of $\Aut(N)$, resulting in split (semidirect product) or non-split extension, as a method for constructing larger groups. However, this method also has its shortcomings, as it relies on defining $H \leq \Aut(N)$. Sometimes, we want to grant $H$ more freedom, rather than constraining it to be a subgroup of $\Aut(N)$. Moreover, we desire the resulting large group to exhibit less abelian behavior than the direct product. With these considerations, we define the free product.

First, consider a set of elements, which we refer to as the generating set $S$. The elements of the set $S$ are denoted by the letters $a_i$, where the index $i$ can take a discrete or continuous range of values. Thus, we have $S = \{a_i \,|\, i \in I\}$, where $I$ is the index set. We define another set $F$ consisting of all possible words formed by these letters. For example, $a_1a_2a_2 \in F$, and $a_3a_{27}a_{9}a_2a_0a_1a_1 \in F$, and so on. To give a simple example, if $S$ consists of the 26 letters of the English alphabet, then $cat \in F$ and $banana \in F$, etc.

In $F$, we define a multiplication operation: the product of two words is simply their concatenation into a longer word. For instance, $(a_1a_2a_2) \cdot (a_3a_4a_2) = a_1a_2a_2a_3a_4a_2$. It is easy to see that, under this definition, the multiplication is associative: $(w_1w_2)w_3 = w_1(w_2w_3)$. The associativity enables us to unambiguously define powers: for any letter $a \in S$, we denote the repetition of $a$ $n$ times as $\overbrace{aaa \dots a}^{n \text{ occurrences}} = a^n$.

There exists a special element in $F$, denoted as $1$, which represents the empty word. It is not difficult to see that the empty word satisfies the property: $\forall w \in F$, we have $1 \cdot w = w \cdot 1$. Thus, the empty word acts as the identity element.

Next, we define the inverse of a word. First, consider the inverse of a letter. For any letter $a$, its inverse $b$ satisfies $ab = ba = 1$, and we denote this as $b = a^{-1}$. After defining the inverse of a letter, we can extend this to define the inverse of any word, following the recursive rule: $(ab)^{-1} = b^{-1}a^{-1}$ and $(w_1w_2)^{-1} = w_2^{-1}w_1^{-1}$.

From the above, we see that $F$, under the defined multiplication, forms a group. We refer to $F$ as the free group generated by $S$, denoted by $F \simeq \langle S \rangle$. It is important to note that there is only one way to construct a free group from one generating set, and the number of elements in the generating set $S$ is the characteristics of a free group. For more details, see Chapter III of \cite{massey2019basic}. This enables us to define the symbol $F(n)$ unambiguously, where $n$ means the number of letters in $S$, and we call $F(n)$ the free group of rank $n$. Sometimes we list letters in the bracket if the list of letters are not too many, for instance, $F(a,b)$ if $S = \{a,b\}$. 

As the term ``free'' in ``free group'' suggests, it represents the most unrestricted construction of a group while preserving the axiomatic definition of a group. For example, consider a generating set consisting of a single letter. The elements in the free group generated by this set take the form $\overbrace{aaa \dots a}^{n \text{ occurrences}}$, where $n$ is an integer. When $n = 0$, this corresponds to the empty word $1$, and when $n < 0$, it represents a word formed by the inverse of the letter $a$. Hence, we have $F \simeq \langle a \rangle \simeq \{a^n \,|\, n \in \mathbb{Z}\}$. Therefore, $F \simeq \langle a \rangle \simeq \mathbb{Z}$, meaning that a free group generated by a single element is isomorphic to the infinite cyclic group.

Next, we point out an important property: any group with a single generator can be expressed as a quotient group of the free group of rank one, $\mathbb{Z}$. For instance, consider the cyclic group $\mathbb{Z}_t$, whose generator $\tilde{a}$ satisfies $\tilde{a}^t = e$. Define a homomorphism $\mathbb{Z} \overset{\varphi}{\to} \mathbb{Z}_t$ with $\Ker \, \varphi = \langle a^t \rangle$. By the first isomorphism theorem, we then have $\mathbb{Z}_t \simeq \mathbb{Z}/\Ker \, \varphi \simeq \mathbb{Z}/\langle a^t \rangle \equiv \mathbb{Z}/t\mathbb{Z}$. Since any group with a single generator must be a cyclic group of some order, this property is proven. We can represent all cyclic groups as $\mathbb{Z}_t \simeq \mathbb{Z}/t\mathbb{Z} \simeq \langle \tilde{a} \,|\, \tilde{a}^t = e \rangle$, where the list of generators is on the left side of the vertical bar, and the relations equivalent to the generators of $\Ker \, \varphi$ are on the right side. For example, $\tilde{a}^t \in \Ker \, \varphi$, so we list the relation $\tilde{a}^t = e$ on the right side. 

Building on this property, one might ask whether a group with two generators can be presented as a quotient group of the free group of rank two, $F(2)$, via a homomorphism $\varphi$. As an example, consider the dihedral group $D_4$. Since $D_4 \simeq \langle a,b \,|\, a^4 = b^2 = e, \, b^{-1}ab = a^{-1} \rangle$, it is natural to take the free group $F(2) \simeq F(a,b)$. From the relations $a^4 = b^2 = e$ and $b^{-1}ab = a^{-1}$, we define $\varphi: F(a,b)\to G$ which sends $a^4, b^2, b^{-1}aba$ to the identity in $G$. In practice, we just replace $a^4, b^2, b^{-1}aba$ by the empty word $1$ whenever they appear in a word in $F(a,b)$ to get $G$. A subtle point is that since $b^{-1}aba = e\Leftrightarrow aba = b$, when $aba$ appears in a word, it should be identified with the word with all $aba$'s replaced with $b$, therefore, there are certain technical challenges. After addressing all the technical challenges, we can show that $\varphi$ is a homomorphism, and the group $G$ is $D_4$. Thus $D_4\simeq F(a,b)/\Ker\,\varphi$.

Going further, one might wonder whether groups with more generators can all be represented in the form $F(n)/\Ker \, \varphi$. Is this form unique? From the two possible presentation of $D_4$ we discussed while carrying out group extension $(\Z_2\times\Z_2).\Z_2$, we see that this form is not unique, i.e. $\exists m, n$ and homomorphism maps $\sigma, \tau$, s.t. $G \simeq F(m)/\Ker \, \tau \simeq F(n)/\Ker \, \sigma$. It's no easy matter to answer questions such as: given a group $G$, how many different ways can it be presented as $F(n)/\Ker\,\varphi$, and are there some general patterns to present a specific class of $G$, like $D_n$ with various $n$ in the form $F(n)/\Ker\,\varphi$. 

Thus far, we have introduced the use of generators and relations to represent a group, denoted as $G \simeq \langle \text{S}\,|\, \Ker\,\varphi \rangle$. Despite the challenges discussed above, this approach remains widely adopted due to its general applicability and the feasibility of representing group elements using letters, a method enabling us to study a group in much detail practically. As for the challenges, a branch of group theory—combinatorial group theory—is dedicated to addressing such issues. The generator-relation method of representing groups also finds broad applications in algebraic topology, where fundamental groups of topological spaces, derived via the Van Kampen theorem, are typically expressed in this manner.

Although combinatorial group theory systematically studies groups presented through generators and relations, in practical terms, if computer programs are available to assist, they can save us from the need for extensive specialized knowledge of combinatorial group theory. In fact, there exists a computational algebra system known as {\tt GAP} (Groups, Algorithms, and Programming), the use of which greatly reduces the computational workload. The following section provides a brief introduction to this computational tool.

\subsection{{\tt GAP}: Computer Database and Computational Tools for Finite Algebra}\label{GAP-intro}

{\tt GAP} is a free, open-source programming language designed for computing in discrete algebra, with a particular emphasis on group theory. It comes equipped with a large database on finite groups and functional tools, such as those for determining the subgroups of a given group or its automorphism group (see \cite{GAP} for more details). In addition to its powerful finite group computational capabilities, {\tt GAP} also supports operations in other algebraic structures, such as Lie algebra, ring theory, field theory, and module theory. It even includes a database related to crystal symmetry.

In this work, we use the {\tt GAP} database for finite groups, known as {\tt SmallGroup}, and a range of functions, with the aid of presenting a group using generators and relations. Below, we show how we employed this programming language from three perspectives: the {\tt SmallGroup} database, the built-in functions for group theory calculations, and how we use presenting groups via generators-and-relations to link our manual work and {\tt GAP}. 

Firstly, we introduce {\tt SmallGroup}, the database of finite groups.
Although the classification of finite groups, particularly that of finite simple groups, has been completed from a mathematical standpoint (e.g., through solvable group theory and group extensions), the sheer number of groups makes it impractical to store them all. Hence, the {\tt SmallGroup} database in {\tt GAP} only stores a subset of finite groups, including:
\begin{itemize}
    \item All groups of order less than or equal to 2000, except for those of order 1024. total number: 423,164,062;
    \item Groups of order at most 50,000 that are not divisible by a cube (395,703 groups);
    \item Groups of order $p^7$ for $p = 3,5,7,11$ (907,489 groups);
    \item Groups of order $p^n$ for $n \leq 6$ and any prime $p$;
    \item Groups of order $q^n \cdot p$, where $q^n$ divides $2^8$, $3^6$, $5^5$, or $7^4$, and $p$ is any prime distinct from $q$;
    \item Groups whose order contains no square factors;
    \item Groups whose order decomposes into at most three primes.
\end{itemize}

Each group $G$ in the {\tt SmallGroup} database is assigned a unique identifier $(N,k)$, where $N = |G|$ is the order of $G$ and $k$ is the index of $G$ in the list of groups of order $N$. For example, consider the case $N = 4$. There are only two groups of order 4: $\Z_4$ and $\Z_2 \times \Z_2$. In the database, $\Z_4$ is assigned the identifier $(4,1)$, while $\Z_2 \times \Z_2$ is labeled as $(4,2)$. 

Next, {\tt GAP} includes a series of built-in functions that allow users to access information from the {\tt SmallGroup} database and perform related operations, such as direct products of groups, group extensions, and more. We provide a list in Table \ref{GAP-functions}, which enumerates some of the functions that were used in our research.

Finally, we consider how to use the method of presenting a group via generators-and-relations method in {\tt GAP}. 
According to the theory of group presentation using generators and relations, a group can be presented as the quotient of a free group, $F(n)/\Ker\,\varphi$. However, this formal form is not necessarily unique, and investigating its general properties can be quite challenging. Despite this, it remains highly practical for specific problems since group elements are represented as a string of letters, which becomes especially convenient with the aid of the powerful computational abilities of a computer. 

In practical applications, although we have access to the {\tt SmallGroup} database and various built-in functions, we often need to establish a connection between the specific problem we study manually and the database. This involves determining the {\tt GAP id}—the identifier in the {\tt SmallGroup} database—for the group structure obtained in our manual calculations. Only after this can we use {\tt GAP} to explore the detailed properties of the group. In many cases, we can express a group using generators and relations based on the problem at hand. Then, we simply input this information into {\tt GAP} using its syntax. 

For instance, consider how to present the group $D_4 \simeq \langle a,b\,|\,a^4 = b^2 = e,\,b^{-1}ab = a^{-1}\rangle$ in {\tt GAP} using generators and relations. The code for this is as follows:

\begin{table}[H]
\centering
{
    \tt
    \begin{tabular}{ll}
    gap>> F:= FreeGroup(2);;  & \# Create free group $F(2)$\\
    gap>> a:= F.f1;;    &   \# Define generator $a$ \\
    gap>> b:= F.f2;; &  \# Define generator $b$ \\
    gap>> G:= F/[a\^{}4, b\^{}2, b\^{}-1*a*b*a];; &  \# Input the relations\\
     & \# corresponding to $\Ker\,\varphi$ as elements \\
     & \# of the free group in the code\\
    gap>> IsGroup(G); & \# Check if $G$ is a valid group \\
    true & \# Returns true\\
    gap>> IdGroup(G); & \# Retrieve the {\tt GAP} id of $G$ \\
     8, 3  & \# The GAP id of $G$ is (8,3)\\
    \end{tabular}
}
\end{table}

Another important code, which lists the subgroups of a given group up to conjugacy class, is necessary to provide. Take $G\simeq SL(3,2)$ as an example. This code prints all subgroups of $SL(3,2)$ up to conjugacy classes, and explicitly list the generators of each representative subgroup of a conjugacy class. Note that the list {\tt L} in the code already contains all the information, and the follow up code is specifically written to display the generators when the given group $G$ is a linear group such as $SL(3,2)$. When the given group $G$ is, for example, $D_4$, the code needs to be adjusted as is indicated after ``\#''

\begin{table}[H]
\centering
{
    \tt
    \begin{tabular}{ll}
    gap>> G:=SL(3,2);; & \\
    gap>> L:=List(ConjugacyClassesSubgroups(G), g->Representative(g));;  & \\
    gap>> for ele in L do   &    \\
    > Gen:=GeneratorsOfGroup(ele); &   \\
    > N:=Size(Gen); &  \\
    > Print("Name:"); &  \\
    > Print(IdGroup(ele)); & \\
    > Print("$\backslash$n"); &  \\
    > for i in [1..N] do &  \\
    > Display(List(Gen[i], IntVecFFE)); \, \# change to "Print(Gen[i])"  &  \\
    > od;  &  \\
    > Print("$\backslash$n$\backslash$n");  &  \\
    > od;  &  \\
    \end{tabular}
}
\end{table}

Lastly, we provide a table of useful {\tt GAP} built-in functions for group theory calculations, which we will use extensively in the study. 

\begin{table}[H]
\centering
\begin{tabular}{c|c}
\toprule
Function Prototype & Function Description \\
\midrule
{\tt IdGroup(G);} & Returns the {\tt SmallGroup} id of group $G$ \\
{\tt CyclicGroup(N);} & Returns the cyclic group $\Z_N$ \\
{\tt DihedralGroup(2N);} & Returns the dihedral group $D_N$ \\
{\tt FreeGroup(N);} & Returns the free group $F(N)$ \\
{\tt DirectProduct(G1,G2);} & Returns the direct product $G_1 \times G_2$ of $G_1$ and $G_2$ \\
{\tt ConjugacyClassesSubgroups(G);} & Returns all conjugacy classes of subgroups of $G$ \\
{\tt NormalSubgroups(G);} & Returns all normal subgroups of $G$ \\
{\tt IsGroup(G);} & Checks if the input is a valid group \\
{\tt Isabelian(G);} & Checks if the input is an abelian group \\
{\tt Generators(G);} & Returns the generators of group $G$ \\
{\tt Center(G);} & Returns the center $Z(G)$ of group $G$ \\
{\tt AutomorphismGroup(G);} & Returns the automorphism group of $G$ \\
{\tt StructureDescription(G);} & Describing the structure of group $G$ \\
\bottomrule
\end{tabular}
\caption{Some {\tt GAP} built-in functions used in this research}\label{GAP-functions}
\end{table}

\newpage

\section{Discrete Symmetries of NHDM}

\subsection{General Remarks}\label{general-remarks}

In the NHDM, we introduce $N$ Higgs doublets $\phi_i$, $i = 1, \dots, N$, 
all possessing the same gauge quantum numbers, and
construct the Higgs Lagrangian, which includes the gauge-kinetic term for all the doublets and 
the renormalizable scalar self-interaction potential
\begin{equation}
V = m_{ij}^2 \fdf{i}{j} + \lambda_{ijkl}\fdf{i}{j}\fdf{k}{l}\,,\label{V-NHDM}
\end{equation}
where all indices go from 1 to $N$, as well as the Yukawa sector.
In this work, we will focus on the scalar potential.

It may happen that the Higgs Lagrangian is invariant under a set of global transformations
$\phi_i \mapsto U_{ij}\phi_j$, where $U \in U(N)$. We are considering $U(N)$ matrices, not the general linear $GL(N,\mathbb{C})$ matrices because we wish to keep the Higgs kinematic term invariant to avoid messing with gauge bosons. 
Such a transformation represents a flavor symmetry of a given Higgs Lagrangian;
the set of all its flavor symmetries forms the symmetry group $\tilde G \subseteq U(N)$. 

The group $U(N)$ contains the $U(1)$ subgroup of the common phase rotations of all doublets by the same phase shift. 
By construction, any Higgs Lagrangian is automatically invariant under this $U(1)$ group.
Since we want to study additional structural symmetries of the NHDM scalar sectors, 
we disregard such common phase rotations. 
Thus, we are interested not in $\tilde G \subseteq U(N)$ but in factor groups 
$G = \tilde G/U(1) \subseteq U(N)/U(1) \simeq PSU(N)$.

We draw the readers attention to the fact that the traditionally used $SU(N)$,
which is obtained by imposing $\det U = 1$,
does not eliminate the above common phase rotations completely.
Indeed, the group $SU(N)$ contains a non-trivial center $Z(SU(N))$,
that is, the $\Z_N$ subgroup of the phase rotations $\exp(2\pi i k /N)\cdot \id_N$, $k = 0, 1, \dots, N-1$. 
Transformations from the center act trivially on all Higgs Lagrangians
and do not offer any insight into the structural properties of the Higgs sector.
Therefore, we are led again to the factor group $SU(N)/Z(SU(N)) \simeq PSU(N)$.
A detailed discussion of these matters can be found in \cite{Ivanov:2011ae}.

When determining the symmetry group $G$ of a given multi-Higgs potential, one must identify its full symmetry content. 
In this way, one must avoid the situations in which accidental symmetries appear which are not included in $G$.
To give a simple example, one can construct a 2HDM model with the global symmetry $\Z_p$ with any $p > 2$.
However the scalar potential of this model will be automatically invariant under the accidental continuous global symmetry $U(1)$,
which of course contains any $\Z_p$ as a subgroup.
The total symmetry content of such a model is $U(1)$, not $\Z_p$; 
labeling this models as $\Z_p$-invariant 2HDM would be a misnomer.
Thus, whenever we say that an NHDM has the symmetry group $G$, 
we always make sure that it does not possess any accidental symmetry beyond this group. In other words, we need to make sure that $G$ is the ``maximal'' symmetry of the potential under question. 

Another crucial aspect about NHDM models equipped with discrete symmetries is that sometimes two Higgs potentials look differently, but their symmetry group is the same and one potential can be mapped to the other under a Higgs basis change $\phi\mapsto \phi^\prime = A\phi$. For example, when $N=4$:
\begin{equation}
\begin{pmatrix}
\phi_1 \\
\phi_2 \\
\phi_3 \\
\phi_4 \\
\end{pmatrix} \mapsto 
\begin{pmatrix}
\phi_1^\prime \\
\phi_2^\prime \\
\phi_3^\prime \\
\phi_4^\prime \\
\end{pmatrix} = 
A
\begin{pmatrix}
\phi_1 \\
\phi_2 \\
\phi_3 \\
\phi_4 \\
\end{pmatrix} = 
\begin{pmatrix}
a_{11} & a_{12} & a_{13} & a_{14} \\
a_{21} & a_{22} & a_{23} & a_{24} \\
a_{31} & a_{32} & a_{33} & a_{34} \\
a_{41} & a_{42} & a_{43} & a_{44} \\
\end{pmatrix}
\begin{pmatrix}
\phi_1 \\
\phi_2 \\
\phi_3 \\
\phi_4 \\
\end{pmatrix}
\end{equation}
where the matrix $A$ doesn't depend on spacetime coordinates $x^\mu$. In other words, $V_1(\phi)\neq V_2(\phi)$, but $V_1(A\phi) = V_2(\phi)$. In this case, the two models lead to the same physical contents even though the Higgs potential looks different mathematically. This is similar to the situation when we want to describe the circular motion in mechanics: one can use the equation $r = const.$ in polar coordinates, or $x^2 + y^2 = const.$ in Cartesian coordinates. Although the two equations look different mathematically, they all represent the same physical process of circular motion. 

Therefore, when classifying discrete symmetries of NHDM, we are actually classifying up to a freedom of Higgs basis change. In other words, we are grouping all the Higgs potentials to classes, and within each class, Higgs potentials are correlated by a basis change. In the subsequent texts, whenever we are presenting a Higgs potential with certain symmetry, we are actually presenting one representative potential from a class of potentials, and we pick the one representative potential that looks the most concise mathematically.  We find the mathematical concise representative potential by using the following trick to adjust redundant phase shifts of Higgs doublets:

Suppose that a NHDM potential is invariant under an abelian group of phase shifts generated by diagonal matrix $a$
and an exchange of two doublets such as $\phi_1 \leftrightarrow\phi_2$ accompanied by phase shifts.
This unitary transformation $b$ can be represented in the $(\phi_1, \phi_2)$ subspace by the $2\times2$ block
\begin{equation}
b = \mmatrix{0}{e^{i(\alpha+\beta)}}{e^{i(\alpha-\beta)}}{0} = e^{i\alpha}\mmatrix{0}{e^{i\beta}}{e^{-i\beta}}{0}\,.
\end{equation}
Let us perform a basis change by writing $\phi_2 = e^{-i\beta} \tilde\phi_2$. 
Then, within the space of doublets $\tilde \phi_1 \equiv \phi_1$ and $\tilde\phi_2$,
the transformation $a$ is unchanged while the transformation $b$ becomes 
$b = e^{i\alpha} \mmatrix{0}{1}{1}{0}$.

A special case of this trick is when $\alpha = \pi$. 
Bringing the minus sign inside the matrix,
we can interpret it as $\beta = \pi$ instead of $\alpha=\pi$. This allows us to flip the sign of one the two doublets
and arrive at $b = \mmatrix{0}{1}{1}{0}$.

The same trick can be applied to the longer permutation cycles. For example, 
order-3 permutation accompanied by arbitrary phase shifts can be transformed, 
upon redefinition of the doublets, to $c = e^{i\alpha}\times$ the cyclic permutation. In this way, we can find the most mathematically concise representative potential from a class of potentials correlated by Higgs basis change. 

However, because the basis change is represented by a N-by-N matrix and the actual symmetry of a model is also represented by a N-by-N matrix, sometimes the two concepts are easily confused.  We remind the reader that the freedom of Higgs basis change should not be confused with actual symmetries of certain NHDM potential. 

Another important fact is that a certain discrete group may have different ways to be imposed on the Higgs potential, and the Higgs potentials are physically different in the sense that they can't be correlated by a Higgs basis change. We will see this example when the symmetry groups are $\Z_6$ and $\Z_4$ for instance. 

It may also happen that the Higgs potential possesses generalized $CP$ symmetries, that is,
it remains invariant under the global transformations of the type $\phi_i(\vec r, t) \mapsto X_{ij} \phi_j^*(-\vec r, t)$,
where $X_{ij}$ is again a unitary matrix \cite{Ecker:1981wv,Ecker:1983hz,Ecker:1987qp,Grimus:1995zi,Branco:1999fs}.
In particular, if it happens that all coefficients of the Higgs potential are real in some basis,
the scalar sector obviously possesses the usual $CP$ symmetry, the one corresponding to $X = \id_N$.
However it is also possible to construct multi-doublet Higgs potentials which 
are invariant under a higher-order $CP$ symmetry; such a model is still explicitly $CP$ invariant
in spite of having complex free parameters in any basis.
This opportunity leads to a new class of the 3HDM based on CP4, the generalized $CP$ symmetry of order 4, 
which was clearly demonstrated in \cite{Ivanov:2015mwl,Haber:2018iwr}.
Higher order $CP$ symmetries in NHDM with $N > 4$ were identified in \cite{Ivanov:2018qni}.

It is well known that any abelian subgroup of $U(N)$ or $SU(N)$ can be represented, in a suitable basis, 
as pure phase rotations, this is because commutative matrices can be diagonalized simultaneously and diagonal matrices commute with each other. They also correspond to phase rotations upon the homomorphism $SU(N) \to SU(N)/Z(SU(N)) \simeq PSU(N)$.
However there exists some additional abelian groups in $PSU(N)$, which cannot be represented as phase
rotations because its full pre-image inside $SU(N)$ is non-abelian. Once factored by the center $Z(SU(N))$, 
it produces the abelian group in $PSU(N)$. So far, we have discovered such groups in 3HDM and 4HDM. In 3HDM, $\Z_3\times\Z_3\subset PSU(3)$ is the unique abelian subgroup in $PSU(3)$ that has this property; in 4HDM, we found $\Z_4\times\Z_4$, $\Z_4\times\Z_2\times\Z_2$, and $\Z_2\times\Z_2\times\Z_2\times\Z_2 \subset PSU(4)$ all have such properties. It can be shown that the pre-images in $SU(N)$ of such abelian groups in $PSU(N)$ are called Nilpotent group of class-2, see again \cite{Ivanov:2011ae}.

\subsection{Abelian Symmetries in NHDM}\label{abelian-Symmetries-in-NHDM}

Let's first consider possible abelian symmetries in NHDM. By definition, abelian symmetries is the invariance of the Higgs potential under transformations, which are realized as matrices, that commute among each other. Because diagonal matrices commute with each other, we get one naive case of abelian symmetry in NHDM: those which are generated by diagonal matrices, also known as rephasing transformations or phase rotations. 

We now identify the question we are interested in: consider an arbitrary interaction Lagrangian which contains several complex fields, such as the Higgs potential Eq.~\eqref{V-NHDM}.
It may happen that certain global phase rotations leave this Lagrangian invariant.
How to identify all such rephasing symmetries of a given Lagrangian?

There exists a completely general algorithmic procedure which solves this task.
It is based on the so-called Smith normal forms (SNF) and, for any input Lagrangian,
gives its rephasing symmetry group and the list of ``charges'' under the corresponding finite or continuous phase rotations. 
The detailed exposition of the technique can be found in \cite{Ivanov:2011ae}, with illustrations
for the 3HDM, 4HDM, and the general NHDM scalar sector.
A somewhat simplified explanation of the general strategy can be also found in section 2 of \cite{Ivanov:2013bka}
with examples from the NHDM Yukawa sector.
Here, we briefly recapitulate the method and describe our Python code {\tt 4HDM Toolbox} 
which makes use of the SNF technique to identify all realizations of cyclic symmetry groups in the 4HDM scalar sector.

To give an example, consider the 4HDM scalar sector and suppose the potential contains $k$ rephasing-sensitive terms
(not counting the complex conjugated ones).
Let us perform phase shifts of the four doublets as $\phi_j \mapsto e^{i\alpha_j}\phi_j$, 
with $\alpha_1, \dots, \alpha_4$ all independent.
Then, the first term of the potential picks up a phase rotation which can be generically written as 
$d_{11}\alpha_1 + d_{12}\alpha_2 + d_{13}\alpha_1 + d_{14}\alpha_4$, where the integer coefficients $d_{1j}$
indicate the power of $\phi_j$ in the first interaction term.
For example, the term $m_{14}^2 \fdf{1}{4}$ corresponds to $d_{1j} = (-1, 0, 0, 1)$,
while the second term $\lambda\fdf{1}{2}\fdf{4}{3}$ corresponds to $d_{2j} = (-1, 1, 1, -1)$.

If we have $k$ rephasing-sensitive terms, we obtain $k$ such rows of coefficients $d_{ij}$ and write them together
as a rectangular integer-valued matrix $D$. For example,
\begin{equation}
m_{14}^2\fdf{1}{4} + \lambda\fdf{1}{2}\fdf{4}{3} + \lambda'\fdf{2}{3}^2\quad \Rightarrow \quad D = 
\begin{pmatrix}
-1 & 0 & 0 & 1 \\
-1 & 1 & 1 & -1 \\
0 & -2 & 2 & 0 
\end{pmatrix}
\label{example-SNF-1}
\end{equation}
The complex conjugated terms differ by the overall minus sign; we drop them because they do not change the subsequent analysis.
Notice that the monomial $\fdf{1}{2}\fdf{4}{3}$ and monomial $\fdf{1}{3}\fdf{4}{2}$ are both represented as $(-1,1,1,-1)$ 
in the matrix, and we do not repeat the same row of coefficients. 

The Lagrangian is invariant under some phase rotations if and only if there exist non-trivial solutions $(\alpha_1, \dots, \alpha_N)$
to the following matrix equation:
\begin{equation}
d_{ij}\alpha_j = 2\pi n_i\,, \quad n_i \in \mathbb{Z}\,.\label{SNF}
\end{equation}
As is explained in \cite{Ivanov:2011ae}, there exists a set of simple transformation rules
which can bring the matrix $D$ to its diagonal form and, at the same time, preserves
the set of solutions.
The sequence of these elementary steps can be represented by multiplication of the left and right integer-valued matrices:
\begin{equation}
S\cdot D\cdot T = N\,,\label{example-SNF-2}
\end{equation}
where the diagonal matrix $N$ is called the Smith normal form.
In the above example,
\begin{equation}
S\cdot D\cdot T = 
\begin{pmatrix}
-1 & 1 & 0\\
-1 & 2 & 1 \\
-2 & 2 & 1 
\end{pmatrix}
\begin{pmatrix}
-1 & 0 & 0 & 1 \\
-1 & 1 & 1 & -1 \\
0 & -2 & 2 & 0 
\end{pmatrix}
\begin{pmatrix}
0 & -1 & 4 & 1 \\
1 & 0 & -1 & 1 \\
0 & 0 & 1 & 1 \\
0 & 0 & 0 & 1 \\
\end{pmatrix}
 = 
\begin{pmatrix}
1 & 0 & 0 & 0\\
0 & 1 & 0 & 0\\
0 & 0 & 4 & 0\\
\end{pmatrix}
 = N\,.
 \label{example-SNF-3}
\end{equation}
Once we know $N$ and its diagonal entries $d_i$, the system Eq.~\eqref{SNF} transforms into several uncoupled equations:
\begin{equation}
d_i\tilde \alpha_i = 2\pi \tilde n_i\,, \quad \tilde n_i \in \mathbb{Z}\,,\label{SNF-2}
\end{equation}
with unconstrained $\tilde \alpha_j$ for $j > k$ or for any situation with $d_i = 0$.
The solution to this system is elementary, and it yields the group $\Z_{d_1}\times\Z_{d_2}\times \dots\times \Z_{d_k} \times [U(1)]^{N-k}$.
In the 4HDM, we always have the final $U(1)$ (the common phase rotation), and we are interested
in the rephasing group factored by it.

For example, in the above example Eq.~\eqref{example-SNF-3}, we deduce that the potential contains the symmetry group $\Z_4$.
The absence of zeros on the main diagonal indicates that there is no accidental continuous symmetry.
In addition, the third column of the matrix $T$ stores the $\Z_4$ charges of the four doublets, 
which can be expressed as powers of $i$:
\begin{equation}
a = \diag(i^4,i^{-1},i^1, i^0) = \diag(1, -i, i, 1)\,.
\end{equation}

In the scalar sector of the 4HDM, there are 30 rephasing-sensitive monomials (counted up to conjugation), 
which include 6 quadratic terms and 24 quartic terms. 
Some of these terms transform under rephasing in the same way, such as $\fdf{1}{2}\fdf{3}{4}$ and $\fdf{1}{4}\fdf{3}{2}$.
Thus, we have in total 27 differently transforming monomials.

If we want to exhaust all possible situations with respect to rephasing symmetry transformations,
we would need to pick up all possible combinations of several monomials from the list and compute the corresponding SNF matrix.
This would be a tedious task to do by hand.
In \cite{Ivanov:2011ae}, bypassing such case-by-case checks, a theorem was proved which showed that any
finite abelian symmetry group $A$ with order $|A| \le 8$ can be realized as a rephasing symmetry of the 4HDM scalar sector.
However for specific realizations of each of these $A$, one would still need to verify various cases.

To facilitate this study, we wrote a Python code {\tt 4HDM Toolbox} available at \cite{TheCode} which does it automatically. For the special case -- cyclic symmetry in rephasing symmetry: 
the user defines which cyclic symmetry group $\Z_n$ should be searched for,
and the code iteratively checks all combinations of three distinct monomials and finds which cases yield
the desired group $\Z_n$. Then, for these cases, it computes the $\Z_n$ charges and, finally,
completes the potential by adding all terms which are invariant under this particular realization of the group $\Z_n$.
In this way, we could verify, for any $\Z_n$, that we do not miss any specific realization. For the other case of rephasing symmetry, whose group is the direct product of cyclic groups: slightly adjusted version of the code is used to carry out several analysis done in the cyclic symmetry case in parallel, accounting for the multi generator nature of such symmetry. In our analysis for the 4HDM, we also made sure that we didn't miss any cases for rephasing symmetries whose groups are direct product of cyclic groups: $\Z_2\times\Z_2$, $\Z_4\times\Z_2$, $\Z_2\times\Z_2\times\Z_2$. 

However, as is pointed out by the end of the last section \ref{general-remarks}, there is another type of 4HDM abelian symmetries, which differs from the naive ones generated by diagonal matrices. So far we have identified 3 such symmetries. Although further investigation is needed, we are still confident in claiming that, in this work, we have fully classified non-abelian symmetries of 4HDM constructed by group extensions by rephasing groups. 

\begin{table}[H]
\centering
\begin{tabular}[t]{cc}
\toprule
$A$ & $\Aut(A)$ \\
\midrule
$\Z_2$ & $\{e\}$ \\
$\Z_3$ & $\Z_2$ \\
$\Z_4$ & $\Z_2$ \\
$\Z_5$ & $\Z_4$ \\
$\Z_6$ & $\Z_2$ \\
$\Z_7$ & $\Z_6$ \\
$\Z_8$ & $\Z_2\times \Z_2$ \\
\bottomrule
\end{tabular}
\quad
\begin{tabular}[t]{cc}
\toprule
$A$ & $\Aut(A)$ \\
\midrule
$\Z_2 \times \Z_2$ & $S_3$ \\
$\Z_2 \times \Z_4$ & $D_4$ \\
$\Z_2 \times \Z_2 \times \Z_2$ & $GL(3,2)\simeq PSL(2,7)$ \\
\midrule
$\Z_4 \times \Z_4$ & $GL(2,\Z_4)\simeq{\tt SmallGroup}(96,195)$ \\
$\Z_4\times\Z_2\times\Z_2$ & ${\tt SmallGroup}(192,1493)$ \\
$\Z_2\times\Z_2\times\Z_2\times\Z_2$ & $A_8$ \\
\bottomrule
\end{tabular}
\caption{All abelian symmetry groups $A$ of 4HDM potentials, and their automorphism groups $\Aut(A)$}
\label{table-abelian}
\end{table}

\subsection{Building Non-abelian Models for NHDM}\label{non-abelian-Symmetries-in-NHDM}

The main focus of this work is to provide a full list of non-abelian symmetry groups for 4HDM. Before delving into detailed calculations, we need to make some general remarks on non-abelian NHDM, and then show how we construct non-abelian models. 

First and foremost, unlike that for abelian models, the full classification of non-abelian NHDM models is still not achieved, and we are still in the realm of case studies. This is due to the absence of an effective and powerful mathematical method that encapsulate them all. However absent such method is, through some endeavor, some methods (using group extensions by abelian groups) are developed and with such method, the full classification of 2HDM is done through \cite{Ivanov:2005hg,Ivanov:2006yq,Ferreira:2010yh,Ferreira:2023dke} while that for 3HDM is done through \cite{Ivanov:2011ae,Ivanov:2012ry,Ivanov:2012fp,Ivanov:2014doa,Fallbacher:2015rea,deMedeirosVarzielas:2019rrp,Darvishi:2021txa,Bree:2024edl}. We will present the current method at hand which is used to obtain the full list of non-abelian symmetry groups for 3HDM. 

When trying to tackle something unknown, it's better to start from something already known. In the case of studying non-abelian symmetry groups of 3HDM, since all abelian symmetry groups for 3HDM is known, it's better to start from the following full list of abelian symmetry groups:
\begin{equation}
A \mbox{ in 3HDM: } \quad \Z_2\;,\quad \Z_3\;,\quad \Z_4\;,\quad \Z_2\times\Z_2\;,\quad \Z_3\times\Z_3\label{3HDM-abelian}
\end{equation}
As is mentioned before, the $\Z_3\times\Z_3$ is generated not only by diagonal matrices, and it's pre-image in $PSU(3)$ is a non-abelian group $\Delta(27)$. All other groups in the list above are generated by diagonal matrices. 

We see that the orders of all these abelian groups contain only primes: 2 and 3.
Therefore, according to Cauchy's theorem, the order of any non-abelian finite group $G$ must contain only these two primes: $|G| = 2^a 3^b$.
Then, Burnside’s $p^aq^b$-theorem claims that, the group $G$ is solvable (Theorem 7.8 in \cite{isaacs2008finite}); 
for a physicist-friendly introduction to solvable groups, please refer to section \ref{group-theory-subgroups}, or see section~3 of \cite{Ivanov:2012fp}. 

A solvable group contains a normal abelian subgroup $A$ so that one can define the factor group $G/A$. This is easily seen in the derived series $G \simeq G^{(0)}\triangleright G^\prime \triangleright G^{(2)}\triangleright ... \triangleright G^{(n)}\triangleright G^{(n+1)}\simeq \{e\}$: when $G^{(n+1)} = \{e\}$, the group $G^{(n)}$ must be abelian. However, 
This information by itself is not sufficient to limit the order of $|G|$ and deduce the structure of the factor group $G/A$.
Luckily, it turns out that in the 3HDM one can prove a stronger statement:
any finite non-abelian group $G$ must contain a normal {\em maximal} abelian subgroup \cite{Ivanov:2012fp}, thus $G$ gas a normal series only having two nodes: $G\triangleright A \triangleright [A,A]\simeq \{e\}$ where $A$ is the maximal normal abelian subgroup of $G$. 
This additional piece of information represents the key step in the procedure because
existence of a normal maximal abelian subgroup $A$ implies that $G/A \subseteq \Aut(A)$,
the automorphism group of $A$. For proofs, see \cite{Ivanov:2012fp}. 
Thus, one arrives at a systematic procedure to classify all available groups $G$: 
\begin{itemize}
\item Take $A$ from the list Eq.~\eqref{3HDM-abelian}, compute its automorphism group $\Aut(A)$, 
and list all subgroups $K \subseteq \Aut(A)$.
\item
If $G/A \simeq K$, then $G$ can be constructed as an extension of $A$ by $K$.
\item
In general, there are two types of extensions. The so-called split extension, also known as semi-direct product
$A \rtimes K$, implies that $G$ contains, among its subgroups, not only $A$ but also $K$.
The non-split extension denoted as $A\,.\,K$ implies that $G$ does not contain $K$,
or, in other words, that $K$ is not closed under multiplication when embedded in $G$.
We will see explicit examples of this situation in the calculations.
\item
Even if one takes specific $A$ and specific $K$, the extension is not unique. 
One needs to construct all the cases explicitly.
\item
By checking all $A$'s, all $K$'s, and performing all possible extensions, one obtains the full list 
of finite non-abelian groups $G$ for the 3HDM scalar sector.
\end{itemize}
This procedure was performed in \cite{Ivanov:2012fp,Ivanov:2012ry} and 
produced the following full list of finite non-abelian groups $G$ 
available in the 3HDM which do not lead to accidental symmetries:
\begin{equation}
\mbox{$G$ in 3HDM:}\qquad S_3\,,\quad 
D_4\,,\quad 
A_4\,,\quad 
S_4\,,\quad 
\Delta(54)/\Z_3\,,\quad 
\Sigma(36)\,.
\label{3HDM-non-abelian}
\end{equation}
Trying to build a 3HDM scalar sector on any other group absent from the lists 
\eqref{3HDM-abelian} and Eq.~\eqref{3HDM-non-abelian}
will unavoidably lead to a continuous accidental symmetry.

Before trying to carry out the same method to 4HDM case, it is worth noting that in a most recent work by Christian Döring and Andreas Trautner \cite{Doring:2024kdg}, a new way of extending groups is introduced. Known results from 2HDM and 3HDM are reproduced. This new method, namely unorthodox group extensions, and the method we developed might mutually work together towards a full classification of 4HDM discrete symmetries. 

Now, we try to carry out the same analysis for 4HDM. This is partially done in our previous papers, \cite{Shao:2023oxt,Shao:2024ibu}. The old method we have at hand, group extension by abelian groups of NHDM, is called normal extension in \cite{Doring:2024kdg}, which have the weak point of not including unsolvable groups. From Burnside's $p^aq^b$ theorem, we know that all symmetry groups for 2HDM and 3HDM are solvable, since their order are $2^a3^b$. However, the abelian symmetry groups of 4HDM is in table \ref{table-abelian}. We can see the prime factors of non-abelian groups can be 2, 3, 5, and 7, which is more than two. So we can't claim that every finite symmetry groups for 4HDM is solvable, let alone that each group exists a maximal normal abelian group, therefore, by using this method, we may not find a full classification of non-abelian symmetry groups for 4HDM. Even so, it's necessary to use the method we used in 3HDM to get a list of non-abelian symmetry groups for 4HDM and see what we can do towards a full classification. 

It's worthy pointing out that when using the method for 3HDM to construct non-abelian symmetry groups for 4HDM, other technique difficulties arise aside from the non-solvability of symmetry groups. We outline some and our solution to each as follows:
\begin{itemize}
\item The uniqueness of the $PSU(4)$ representations of the generators of abelian groups in the table \ref{table-abelian} up to the freedom of Higgs basis change. This is also a difficulty in 3HDM, but the problem is far more complicated with one additional Higgs doublet involved. We developed a python code \cite{TheCode} based on the SNF technique to ensure we don't miss any cases. 

\item The complicated automorphism group of $A$. With a more complicated abelian symmetry in table \ref{table-abelian}, their automorphism groups are more complicated to study, with the order of the group and the number of subgroups typically exceeding a hundred, making the analysis hard to be done by hand. For example, $\Aut((\Z_2)^3)\simeq GL(3,2) \simeq SL(3,2) \simeq PSL(2,7)$ has the order $2^3\times 3\times 7 = 168$ with 179 subgroups. During our study, we proved a useful theorem to simplify our analysis and used {\tt GAP} to help us tackle the structure of $\Aut(A)$ more easily. We will give details of this theorem including it's proof and implications immediately afterwards.

\item The multi-generator nature of some cases we have to deal with. For $A$ equal to $\Z_2\times\Z_2$, $\Z_4\times\Z_2$, $\Z_2\times\Z_2\times\Z_2$, and the three extra cases $\Z_4\times\Z_4$, $\Z_4\times\Z_2\times\Z_2$, and $\Z_2\times\Z_2\times\Z_2\times\Z_2$, there are multiple generators of $A$ we need to consider, and the $\Aut(A)$ also have multiple generators. As is discussed in section \ref{section-split}, we see how much trouble this might cause by studying the simplest possible example: $A\simeq \Z_2\times\Z_2$ and $\Aut(A)\simeq S_3$. The analysis is almost impossible to be done without the help of {\tt GAP}, which we used throughout our study. 

\item The three extra cases: $A\simeq \Z_4\times\Z_4$, $\Z_4\times\Z_2\times\Z_2$, or $\Z_2\times\Z_2\times\Z_2\times\Z_2$. This three groups are technically generated by matrices that are not diagonal. In fact, some of the generators are called monomial matrices as mentioned before when we are defining the normalizer of a group $N_G(H)$. We will see more details in the calculation section. In general, matrices that are not diagonal is more difficult to study. We developed {\tt Python} code to tackle this difficulty. 

\end{itemize}

Now, we provide more details of the new theorem we proved to simplify the second difficulty mentioned above. 

In this paper, our main interest is in finite groups $G$ with an abelian normal subgroup $A$ such that $C_G(A) = A$, this is an alternative way to say the normal abelian subgroup is maximal. For such a group, the natural homomorphism $\pi: G \rightarrow \Aut(A)$ has kernel equal to $A$. We denote then $G = A\,.\,K$, where $K = \pi(G) \cong G/A$ is the action of $G$ on $A$.

The following result shows that it will suffice to consider the groups $K$ up to conjugacy in $\Aut(A)$. This is useful when analyzing larger groups. For instance, although $\Aut(\Z_2\times\Z_2\times\Z_2)\simeq GL(3,2)$ has 179 subgroups, there are only 13 nontrivial conjugacy classes. So, with the help of the theorem, we only need to study 13 cases when trying to build non-abelian groups by $\Z_2\times\Z_2\times\Z_2$ instead of 179 cases. 

\begin{theorem}
Let $A$ be a finite abelian group and $K_1$, $K_2$ be two subgroups of $\Aut(A)$ which are conjugate to each other, meaning there exists $q \in \Aut(A)$ such that $q^{-1}K_1 q = K_2$. 

Suppose $G_1 = A\,.\,K_1$ is a group extension such that $C_{G_1}(A) = A$, and $K_1$ is the action of $G_1$ on $A$. Then $G_1$ is isomorphic to a group of the form $G_2 = A\,.\,K_2$, where $C_{G_2}(A) = A$, and $K_2$ is the action of $G_2$ on $A$.
\end{theorem}\label{the-theorem}

\begin{proof}
We consider $A$ as an additive group $(A,+)$. For $K \leq \Aut(A)$, a \emph{$2$-cocycle} for $K$ is a map $f: K \times K \rightarrow A$ satisfying $f(x,1) = 0 = f(1,x)$ and $$f(xy,z) + f(x,y) = x(f(y,z)) + f(x,yz)$$ for all $x,y,z \in K$. Given a $2$-cocycle $f$, we can form the group $G_{f,K} = \{(a,x) : a \in A, x \in K \}$ with the group operation defined by $$(a,x) \cdot (a',x') = (a+x(a')+f(x,x'),xx')$$ for all $a,a' \in A$ and $x,x' \in K$. For the fact that $G_{f,K}$ is a group, see \cite[Theorem 7.30]{rotman2012introduction}. 

We can identify $A$ as a normal subgroup of $G_{f,K}$ via the isomorphism $a \mapsto (a,1)$. We have $$(a_0,x) \cdot (a,1) \cdot (a_0,x)^{-1} = (x(a),1)$$ for all $a,a_0 \in A$ and $x \in K$, therefore the action of $G_{f,K}$ on $A$ is equal to the group $K$. This also implies that $C_{G_{f,K}}(A) = A$. It follows from \cite[Theorem 7.30]{rotman2012introduction} that if $G = A\,.\,K$ is a group extension such that $C_G(A) = A$ and $K$ is the action of $G$ on $A$, then $G$ is isomorphic to $G_{f,K}$ for some $2$-cocycle $f$. 

Therefore, for the proof of the theorem we may assume that $G_1 = G_{f,K_1}$ for some $f$. Since $q^{-1}K_1 q = K_2$, we can define $f': K_2 \times K_2 \rightarrow A$ by $$f'(x,y) = q^{-1} \left(f(qxq^{-1}, qyq^{-1})\right)$$ for all $x,y \in K_2$. Using the fact that $f$ is a $2$-cocycle, a straightforward check shows that $f'$ is a $2$-cocycle for $K_2$. Now define a map $\tau: G_{f,K_1} \rightarrow G_{f',K_2}$ by $$\tau(a,x) = (q^{-1}(a), q^{-1} x q)$$ for all $a \in A$ and $x \in K_1$. It follows from the definitions that $\tau$ is an isomorphism, so by taking $G_2 = G_{f',K_2}$ we have $G_1 \cong G_2$, where $G_2$ has the required properties.\end{proof}

To make this formal proof more accessible to the physics community, we would like to point out that in the case of split extensions, also called the semidirect products $G = A\rtimes K$, the proof simplifies. In this case, the group $G$ contains 
not only the normal subgroup $A$ but also a subgroup $H$ isomorphic to $K$ which is a complement for $A$ in $G$, that is, 
$H \cap A = \{e\}$ and $G = AH$. Then one can define in a straightforward way the group operation 
not only on the set of $A$-cosets, but also on set of representative elements of these cosets. 
A simple way to obtain the split extension is to take
the trivial 2-cocycle $f(x,x') = 0$ for all $x,x' \in K$. 
In this case, the multiplication law is fully defined by the action
$x(a)$, that is, how $x \in K$ permutes the elements of $A$. The theorem then reduces to the calculation which relates the elements $g_1 \in G_1$
to the elements $g_2 \in G_2$. 

However when we build a non-split extension, the set of elements of $G$ of the form $(0,x)$ is not closed under the same group operation as $x \in K$. The 2-cocycle $f(x,x')$ is the construction that describes this failure to reproduce the group structure of $K$ inside $G$. Thus, to define the structure of the group $G$, it is not enough to specify $x(a)$, the action of elements of $K$ on $A$; we also need to define the 2-cocycle $f(x,x')$. As a result, in order to prove the isomorphism between $G_1$ and $G_2$, we need to demonstrate not only the relation between the actions $x_1(a)$, $x_1 \in K_1$, and $x_2(a)$, $x_2 \in K_2$, but also the unambiguous link between the corresponding 2-cocycles $f$ and $f'$. This is what the body of the proof does.

This theorem allows us to reduce the number of extensions we need to consider, especially when $\Aut(A)$ is large.
Namely, we need to list not all individual subgroups of $\Aut(A)$ but only 
all conjugacy classes of these subgroups. 
For each conjugacy class, we can select one representative subgroup and find all of its extensions of $A$.
The list of extensions of any other subgroup from the same conjugacy class will be the same.

There is, however, an important caveat which does not render the classification problem as easy as it may sound.
When we apply the above result to construction of the symmetry-based 4HDMs, we deal not with abstract groups
but with their four-dimensional representations, that is, subgroups of $PSU(4)$.
Even if $A$, $K_1$, and $K_2$ can be faithfully represented as groups of transformations from $PSU(4)$,
the transformation $q$ linking $K_1$ and $K_2$ and, consequently, the transformation $\tau$ linking $G_1$ and $G_2$
are not guaranteed to belong to $PSU(4)$. 
In our previous paper \cite{Shao:2023oxt}, we encountered examples of group transformations 
which do not fit the desired representation.
For example, the group $A = \Z_8 \in PSU(4)$ has the automorphism group $\Aut(\Z_8) = \Z_2\times \Z_2$.
We can take any non-trivial element $q$ from $\Aut(\Z_8)$ and define its action on the group $\Z_8$
in abstract group-theoretic terms. However, once we write $\Z_8$ as a subgroup of $PSU(4)$ and try to construct
$q \in PSU(4)$ satisfying the desired relations, we end up with a system of equations which does not have solutions. 
Thus, such $q$ does not fit $PSU(4)$.

If it happens that $\tau$, which maps $G_1 \to G_2$ by conjugation, can indeed be represented by a $PSU(4)$ transformation, 
then the situation simplifies.
Indeed, the invariance of $V_1(\phi)$ under $G_1$ means
that $V_1(g_1(\phi)) = V_1(\phi)$ for any $g_1 \in G_1$. 
The transformation $\tau$ now acts in the same space and defines a basis change: $\phi \mapsto \tau(\phi)$.
This is not a symmetry of the potential: $V_1(\tau(\phi)) \equiv V_2(\phi) \not = V_1(\phi)$.
However the potential $V_2(\phi)$ defined in this way is invariant under the group $G_2$.
Indeed, picking up $g_2 \in G_2$ and representing it as $\tau^{-1}g_1 \tau$ for some $g_1 \in G_1$, we obtain
\begin{equation}
  V_2\left(g_2(\phi)\right) = V_2\left(\tau^{-1}(g_1(\tau(\phi)))\right) = V_1\left(g_1(\tau(\phi))\right) =
  V_1\left(\tau(\phi)\right) = V_2(\phi)\,.
\end{equation}
Therefore, the potential $V_2(\phi)$ invariant under $G_2$ has the same symmetry content 
as $V_1(\phi)$ invariant under $G_1$. 
Since the two potentials, $V_1(\phi)$ and $V_2(\phi) =  V_1(\tau(\phi))$, are related by a mere basis change,
their physics consequences are identical. 

With this helpful result, we update our procedure for building 4HDM models based on extensions of the type $A\,.\,K$,
where $K \subseteq \Aut(A)$. We need to consider not the individual subgroups $K$ nor the entire conjugacy classes 
of $K$ inside $\Aut(A)$, but the conjugacy classes in which we only use transformations $\tau \in \Aut(A)$ 
expressible as $PSU(4)$ transformations. It is then sufficient to consider only one representative $K$ 
from each such conjugacy class and build all the group extensions available.

So far, during our studies for the more complicated 4HDM, we have developed tools to tackle technique and calculational details with the aid of a theorem and computers using {\tt GAP} and {\tt Python}, and completed the extension by rephasing groups, and it is time to take a step back to summarize what we have developed and to list difficulties we have encountered, aiming at informing and stimulating the innovation of follow-up researches. We will discuss more details in the discussion section after we carry out calculations. But before delving into calculations, one final piece of information is needed: we have been discussing groups, which should be realized as symmetry of 4HDM potentials. Next, we will discuss how to link the groups and 4HDM potentials.

\subsection{Connecting Groups to 4HDM Potential}\label{Connecting-groups-to-potential}

Since most of the abelian groups $A$ which we discuss in this work are represented, in a suitable basis, by pure phase rotations (except for $\Z_4\times\Z_4$, $\Z_4\times\Z_2\times\Z_2$, and $\Z_2\times\Z_2\times\Z_2\times\Z_2$, which requires special care),
let us break the $A$-symmetric scalar potential of the 4HDM into two part: the rephasing-insensitive part $V_0$, 
invariant under all phase rotations of individual doublets, 
and the rephasing-sensitive part $V(A)$ which stays invariant only
under the phase rotations which form the group $A$.
The rephasing-insensitive part $V_0$ can be written in the following form:
\begin{equation}
V_0 = \sum_{i=1}^4 \left[m_{ii}^2\fdf{i}{i} + \Lambda_{ii}\fdf{i}{i}^2 \right]+ 
\sum_{i < j} \left[\Lambda_{ij}\fdf{i}{i}\fdf{j}{j} + \tilde \Lambda_{ij}\fdf{i}{j}\fdf{j}{i}\right]\,.\label{V0-general}
\end{equation}
In total, this potential has 20 real free parameters.
The rephasing-sensitive part $V(A)$ depends on the group $A$ and will be given for every choice of $A$.

When we extend the group $A$ by its automorphisms, we will search for such transformations $b$ which
satisfy $b^{-1}A b = A$. Such transformations can be represented by certain permutations of the four doublets,
accompanied perhaps by additional phase factors.
The requirement of invariance under $b$ applies to $V_0$ as well as to $V(A)$.
Depending on which doublets are permuted under $b$, we will have one of the following constraints on $V_0$. 
\begin{itemize}
\item 
The permutation $\phi_1 \leftrightarrow \phi_2$ leads to the following constraints:
\begin{equation}
m_{11}^2 = m_{22}^2\,, \quad \Lambda_{11} = \Lambda_{22}\,, \quad \Lambda_{1k} = \Lambda_{2k}\,, 
\quad \tilde\Lambda_{1k} = \tilde\Lambda_{2k}\,, \quad k = 3,4.
\label{V0-2-1-1}
\end{equation} 
The potential $V_0$ then has 14 free parameters left.
\item 
The permutation $\phi_1 \leftrightarrow \phi_2$ and, simultaneously, $\phi_3 \leftrightarrow \phi_4$, leads to the following constraints:
\begin{eqnarray}
&& m_{11}^2 = m_{22}^2\,, \quad m_{33}^2 = m_{44}^2\,, \quad \Lambda_{11} = \Lambda_{22}\,, \quad \Lambda_{33} = \Lambda_{44}\,, \nonumber\\
&&\Lambda_{13} = \Lambda_{24}\,, \quad \Lambda_{14} = \Lambda_{23}\,, \quad 
\tilde\Lambda_{13} = \tilde\Lambda_{24}\,, \quad \tilde\Lambda_{14} = \tilde\Lambda_{23}\,.
\label{V0-2-2}
\end{eqnarray} 
The potential $V_0$ has only 12 free parameters left.
\item 
If we build a 4HDM with three Higgs doublets transforming as a triplet under some discrete group,
we encounter the cyclic permutation of three doublets $\phi_1 \mapsto \phi_2\mapsto \phi_3 \mapsto \phi_1$. 
It leads to the following constraints:
\begin{eqnarray}
&& m_{11}^2 = m_{22}^2 = m_{33}^2\,, \quad \Lambda_{11} = \Lambda_{22} = \Lambda_{33}\,, 
\quad \Lambda_{12} = \Lambda_{23} = \Lambda_{13}\,, 
\quad \tilde\Lambda_{12} = \tilde\Lambda_{23}  = \tilde\Lambda_{13}\,, \nonumber\\
&&
\Lambda_{14} = \Lambda_{24} = \Lambda_{34}\,, 
\quad \tilde\Lambda_{14} = \tilde\Lambda_{24}  = \tilde\Lambda_{34}\,.
\label{V0-3-1}
\end{eqnarray} 
In this case, the number of the free parameters is reduced to 8.
Notice that the potential $V_0$ then automatically becomes invariant under all permutations of the first three doublets
(group $S_3$), not only the cyclic ones.
\item 
Finally, if we encounter the cyclic permutation of all four doublets 
$\phi_1 \mapsto \phi_4\mapsto \phi_2 \mapsto \phi_3 \mapsto \phi_1$, we arrive at the following constraints:
\begin{eqnarray}
&& m_{11}^2 = m_{22}^2 = m_{33}^2 = m_{44}^2 \equiv m^2\,, \quad 
\Lambda_{11} = \Lambda_{22} = \Lambda_{33} = \Lambda_{44} \equiv \Lambda\,,
\nonumber\\
&& 
\Lambda_{13} = \Lambda_{14} = \Lambda_{23} = \Lambda_{24} \equiv \Lambda'\,, 
\quad 
\Lambda_{12} = \Lambda_{34} \equiv \Lambda''\,,\nonumber\\
&&
\tilde\Lambda_{13} = \tilde\Lambda_{14}  = \tilde\Lambda_{23}  = \tilde\Lambda_{24} \equiv \tilde\Lambda'\,,
\quad 
\tilde \Lambda_{12} = \tilde \Lambda_{34} \equiv \tilde\Lambda''\,.
\label{V0-4-0}
\end{eqnarray} 
In this case, the potential $V_0$ contains only 6 free parameters.
In addition to the cyclic permutations, it also becomes automatically invariant 
under the simultaneous exchange $\phi_1 \leftrightarrow \phi_4$ and $\phi_2 \leftrightarrow \phi_3$.
All these transformations form the group $D_4$, the symmetry group of the square
whose vertices are labeled by the four doublets.
\end{itemize}
This list of options is complete up to renaming of doublets.
Now, we provide the strategy we will use in the calculations:

\begin{itemize}
\item 
Pick up the abelian group $A \subset PSU(4)$ and write down its generators. 
It is convenient to represent them as $SU(4)$ rather than $PSU(4)$ transformations, 
but we need to keep in mind that all relations are defined modulo to
the center of $SU(4)$, which is the group $\Z_4$ generated by $i\cdot \id_4$.
\item 
It may happen that there are more than nonequivalent ways a given abelian group $A$ can be implemented in the 4HDM scalar sector
without leading to accidental continuous symmetries. For example, in our previous work \cite{Shao:2023oxt} we found
that $\Z_4$ can be implemented in three distinct ways, which cannot be linked by any basis change. 
Each implementation leads to the Higgs potential with different number of free parameters and, eventually, different
options for the non-abelian extensions.
In order to be sure that we do not miss any implementation, we rely on our own Python code {\tt 4HDM Toolbox} \cite{TheCode},
which was described in \cite{Shao:2023oxt} and is freely available at GitHub.
\item
List all the conjugacy classes of subgroups of $\Aut(A)$ using only such transformations which can be expressed as $PSU(4)$ transformations. 
For each class, take a representative subgroup with its generators,
which we generically write as $b$.
By defining how each generator $b$ acts in $A$ and by choosing whether the properties of $b$'s, now seen as the elements of the extension group,
reproduce their properties inside the parent group, the subgroup of $\Aut(A)$, construct all possible non-abelian extensions.
\item 
Using a specific implementation of the group $A$, which is defined by the expression of its generators $a_i$,
write the $b$ action on $a_i$ in the form of matrix equations. Solve this equations.
If a unitary solution for $b$ exists, we have constructed a desired extension.
\item 
Now turn to the Higgs potential invariant under $A$ and require that, in addition, it be invariant
under the generator $b$ just constructed. Check whether accidental symmetries appear.
If they do not, we find a viable 4HDM with the desired non-abelian symmetry group. Note that since $\Z_4\times\Z_4$, $\Z_4\times\Z_2\times\Z_2$, and $\Z_2\times\Z_2\times\Z_2\times\Z_2$ are not generated purely by rephasing transformations, their invariant potentials are not written as $V_0 + V(A)$; so extra care is needed. 
\end{itemize}

\newpage

\section{Calculating Extensions by $\Z_8$}\label{section-Z8}

\subsection{The $\Z_8$-invariant 4HDM}

We begin our analysis with the abelian group $\Z_8$ and its possible non-abelian extensions.
In this first example, we will expose the procedure step by step, 
and in the subsequent examples, we will rely on this sequence of steps.

The starting point is to write the 4HDM invariant under the symmetry group $\Z_8$. 
Following the discussion in Section~\ref{abelian-Symmetries-in-NHDM},
we represent the generator of this group by an $SU(4)$ phase rotation matrix $a$ which, in a suitable basis, 
has the following form:
\begin{equation}
a = \eta^{1/4}\, \cdot \diag(\eta,\, \eta^2,\, \eta^4,\, 1)\,, \quad \eta \equiv e^{i\pi/4} = \sqrt{i}\,, \quad \eta^8 = 1\,.
\label{Z8-a3}
\end{equation}
We draw the reader's attention to the fact that the choice of the $\Z_8$ charges used here is not arbitrary and is,
in fact, unique up to the doublets permutation and the possible $i = \eta^2$ factors.
These charges are fixed by the Smith normal form technique developed in \cite{Ivanov:2011ae} and summarized in Section~\ref{abelian-Symmetries-in-NHDM}. 
To double check that we do not miss other charge assignments, we used our Python code 
{\tt 4HDM Toolbox}, which checks all possible 
monomial combinations and identifies the rephasing symmetry group and its charge assignments, 
see details in Section~\ref{abelian-Symmetries-in-NHDM}. This code confirmed the uniqueness of the $\Z_8$ charge choice.

The Higgs potential invariant under so-defined $\Z_8$ contains, apart from the rephasing invariant terms $V_0$, 
the following rephasing-sensitive terms: 
\begin{equation}
V({\Z_8}) = \lambda_{1} \fdf{2}{1} \fdf{4}{1} + \lambda_{2} \fdf{3}{2} \fdf{4}{2} + \lambda_{3} \fdf{4}{3}^2 + h.c.
\label{VZ8}
\end{equation}
Here, the potential is a representative among all other $\Z_8$ invariant potentials up to the freedom of Higgs basis change. All the coefficients in $V(\Z_8)$ can, in principle, be complex. However upon a suitable rephasing of the four doublets, 
all $\lambda_i$ can be made real, without affecting the symmetry group $\Z_8$. 
We conclude that imposing $\Z_8$ on the scalar sector of the 4HDM automatically leads to explicit $CP$ conservation.
For generic values of the coefficients, this potential has no other symmetries.

\subsection{The automorphism group of $\Z_8$}

If $a$ generates $\Z_8$, then $a^3$, $a^5$, and $a^7 = a^{-1}$ can also play the role of its generator.
Therefore, the transformation $b$ which maps $a\mapsto a^3$ and the transformation $c$ which maps $a\mapsto a^{-1}$
upon conjugation
\begin{equation}
b^{-1}ab = a^3\,, \quad c^{-1}ac = a^{-1}
\end{equation}
leave the $\Z_8$ group unchanged and represent its automorphisms. 
One immediately checks that applying $b$ twice maps $a$ to $a^9 = a$. 
Therefore, $b^2$ is the trivial automorphism, and so is $c$.
Finally, applying $bc$ (or $cb$) maps $a\mapsto a^5$. Clearly, $(bc)^2$ is also the trivial automorphism.
Thus, we find that $\Aut(\Z_8) \simeq \Z_2\times \Z_2$ generated by $b$ and $c$.

\subsection{Attempts at extending $\Z_8$}

Let us consider the split extension (semidirect product) 
\begin{equation}
\Z_8 \rtimes \Z_2 = \langle a, b \,|\, a^8 = e, b^2 = e, b^{-1} a b = a^3\rangle\,.\label{extension-Z8Z2-1}
\end{equation}
We want to represent $b$ as a matrix in $SU(4)$ taking into account that $i^r$ factors can always accompany
the identity in matrix equations. 
Therefore, we need $b$ to solve the matrix equation
\begin{equation}
ab = ba^3\cdot i^r
\end{equation}
with any integer $r$. Noting that $i = \eta^2$ and writing this matrix equation explicitly, we get
\begin{equation}
\eta^{1/4}
\left(\!\! \begin{array}{cccc} 
\eta\,{\gray b_{11}} & \eta\,{\gray b_{12}} & \eta\,{\gray b_{13}} & \eta\,{\gray b_{14}}\\ 
\eta^2{\gray b_{21}} & \eta^2{\gray b_{22}} & \eta^2{\gray b_{23}} & \eta^2{\gray b_{24}}\\ 
\eta^4{\gray b_{31}} & \eta^4{\gray b_{32}} & \eta^4{\gray b_{33}} & \eta^4{\gray b_{34}}\\ 
{\gray b_{41}} & {\gray b_{42}} & {\gray b_{43}} & {\gray b_{44}}\\
\end{array}\!\!\right)
= \eta^{3/4} \eta^{2r}
\left(\!\! \begin{array}{cccc} 
\eta^3{\gray b_{11}} & \eta^6{\gray b_{12}} & \eta^4{\gray b_{13}} & {\gray b_{14}}\\ 
\eta^3{\gray b_{21}} & \eta^6{\gray b_{22}} & \eta^4{\gray b_{23}} & {\gray b_{24}}\\ 
\eta^3{\gray b_{31}} & \eta^6{\gray b_{32}} & \eta^4{\gray b_{33}} & {\gray b_{34}}\\ 
\eta^3{\gray b_{41}} & \eta^6{\gray b_{42}} & \eta^4{\gray b_{43}} & {\gray b_{44}}\\
\end{array}\!\!\right)\,.
\label{extension-Z8Z2-2}
\end{equation}
We need a non-trivial solution to this matrix equation on $b$, at least for some $r$.
Since the entries of $b$ are the same on both sides of the equality,
we compare the two matrices element by element and
either equate the rephasing factors or set the corresponding entry of $b$ to zero.
For example, the factors in front of $b_{11}$ are $\eta^{5/4}$ in the left hand side and $\eta^{3+2r+3/4}$ in the right hand side.
Since they cannot be made equal by any integer $r$, we must set $b_{11} = 0$.

A quick inspection reveals that all the elements must be set to zero because of the mismatch 
between the coefficients $\eta^{1/4}$ and $\eta^{3/4}$ which cannot be compensated by any integer power of $\eta$.
Therefore, the extension given by Eq.~\eqref{extension-Z8Z2-1}
cannot be realized within the 4HDM.

The second option is to extend $\Z_8$ to $\Z_8\rtimes \Z_2$ using $c$. 
This leads to the matrix equation $ac = ca^{-1}\cdot i^r$. Writing it in the same way,
we again notice the mismatch between $\eta^{1/4}$ and $\eta^{-1/4}$ which cannot be compensated by
any integer power of $i$. Thus, this extension does not fit the 4HDM.

Finally, we try to extend $\Z_8$ to $\Z_8\rtimes \Z_2$ using $d \equiv bc$:
\begin{equation}
\Z_8 \rtimes \Z_2 = \langle a, d \,|\, a^8 = e, d^2 = e, d^{-1} a d = a^5\rangle\,.\label{extension-Z8Z2-3}
\end{equation}
Now, the matrix equation $ad = da^5\cdot i^r$ does not run into the above problem because $\eta^{1/4}$
and $\eta^{5/4}$ do differ by a single power of $\eta$.
However, there is an insufficient number of non-zero elements of $b_{ij}$ for any value of $r$, 
making the matrix $d$ non-invertible. For example, for $r=0$, only $b_{14} \not = 0$, while all other elements must be set to zero.
Therefore, this choice does not represent a solution of $d^{-1} a d = a^5$.

We conclude that none of the possible (split) extensions of $\Z_8$ by its automorphisms 
can lead to a viable symmetry group in the 4HDM scalar sector.
Dropping the assumption $b^2=e$ 
and passing to non-split extensions, $b^2 \in \Z_8$, does not help:
the problems exposed above do not depend on the form of $b^2$.
The origin of the obstacle is the very particular way the $\Z_8$ group is embedded in $PSU(4)$.

\newpage

\section{Calculating Extensions by $\Z_7$}

\subsection{The $\Z_7$-invariant 4HDM}

As before, we begin by constructing the 4HDM potential based on the symmetry group $\Z_7$.
According to \cite{Ivanov:2011ae}, there is a basis in which its generator $a$ acts 
on the four Higgs doublets as
\begin{equation}
a = \diag(\eta,\, \eta^2,\, \eta^4,\, 1)\,, \quad \eta \equiv e^{2\pi i/7}\,, \quad \eta^7 = 1.
\label{Z7-a1}
\end{equation}
This choice is unique up to multiplication by $i$, doublet permutations, and automorphisms of the $\Z_7$ group.
The Higgs potential invariant under this symmetry is defined by the following rephasing-sensitive terms:
\begin{equation}
V({\Z_7}) = \lambda_{1} \fdf{2}{1} \fdf{4}{1} + \lambda_{2} \fdf{3}{2} \fdf{4}{2} + \lambda_{3} \fdf{1}{3} \fdf{4}{3} + h.c.
\end{equation}
It differs from $V({\Z_8})$ in Eq.~\eqref{VZ8} only by the last term.
As before, using the rephasing basis change freedom, one can make all three $\lambda_i$ real and positive,
which implies that $\Z_7$-invariant 4HDM is explicitly $CP$ conserving.

\subsection{The automorphism group of $\Z_7$}

The group $\Z_7$ is of prime order, and any of its non-unit elements can play the role of the generator.
Therefore, all the maps $a \mapsto a^q$, with $q = 1, \dots, 6$, are automorphisms of $\Z_7$
and form the group $\Aut(\Z_7) \simeq \Z_6$.
Since we are interested in extending $\Z_7$ not only by the full $\Aut(\Z_7)$ but also by its subgroups,
let us study separately the order-2 and the order-3 elements from $\Aut(\Z_7)$ defined by
\begin{equation}
b^{-1} a b = a^6 = a^{-1}\,, \quad 
c^{-1} a c = a^2\,. \label{Z7-bc}
\end{equation}
These two transformations commute, and their product, which maps $a\mapsto a^5$, is of order 6.

\subsection{Extension of $\Z_7$ by $\Z_2$}

Let us first construct the split extension
\begin{equation}
\Z_7 \rtimes \Z_2 = \langle a, b \,|\, a^7 = e, b^2 = e, b^{-1} a b = a^{-1}\rangle\,.\label{extension-Z7Z2-1}
\end{equation}
The relation between $a$ and $b$ leads to the matrix equation on $b$:
\begin{equation}
ab = ba^{-1}\cdot i^r
\end{equation}
with any integer $r$. This time, $i = \eta^{7/4}$, 
and the only chance to find non-trivial solutions to this equation is to set $r=0$.
Then, writing the matrix $b$ explicitly as in Eq.~\eqref{extension-Z8Z2-2}, we get
\begin{equation}
\left(\!\! \begin{array}{cccc} 
\eta\,{\gray b_{11}} & \eta\,{\gray b_{12}} & \eta\,{\gray b_{13}} & \eta\,{\gray b_{14}}\\ 
\eta^2{\gray b_{21}} & \eta^2{\gray b_{22}} & \eta^2{\gray b_{23}} & \eta^2{\gray b_{24}}\\ 
\eta^4{\gray b_{31}} & \eta^4{\gray b_{32}} & \eta^4{\gray b_{33}} & \eta^4{\gray b_{34}}\\ 
{\gray b_{41}} & {\gray b_{42}} & {\gray b_{43}} & {\gray b_{44}}\\
\end{array}\!\!\right)
= 
\left(\!\! \begin{array}{cccc} 
\eta^{-1}{\gray b_{11}} & \eta^{-2}{\gray b_{12}} & \eta^{-4}{\gray b_{13}} & {\gray b_{14}}\\ 
\eta^{-1}{\gray b_{21}} & \eta^{-2}{\gray b_{22}} & \eta^{-4}{\gray b_{23}} & {\gray b_{24}}\\ 
\eta^{-1}{\gray b_{31}} & \eta^{-2}{\gray b_{32}} & \eta^{-4}{\gray b_{33}} & {\gray b_{34}}\\ 
\eta^{-1}{\gray b_{41}} & \eta^{-2}{\gray b_{42}} & \eta^{-4}{\gray b_{43}} & {\gray b_{44}}\\
\end{array}\!\!\right)\,.
\end{equation}
For future convenience, let us recast this matrix equation in the form which brings in the spotlight the powers of $\eta$
in each entry: 
\begin{equation}
\left(\!\! \begin{array}{cccc} 
1 & 1 & 1 & 1\\ 
2 & 2 & 2 & 2\\ 
4 & 4 & 4 & 4\\ 
0 & 0 & 0 & 0\\ 
\end{array}\!\!\right)
\simeq 
\left(\!\! \begin{array}{cccc} 
-1 & -2 & -4 & 0\\ 
-1 & -2 & -4 & 0\\ 
-1 & -2 & -4 & 0\\ 
-1 & -2 & -4 & 0\\ 
\end{array}\!\!\right)\ \mod 7\,.
\label{extension-Z7Z2-3}
\end{equation}
We see that the only non-zero element is $b_{44}$, which means that there is no viable extension $\Z_7 \rtimes \Z_2$
in the 4HDM scalar sector.
As a consequence, the extension $\Z_7\rtimes\Z_6$ is also impossible within the 4HDM.

\subsection{Extension of $\Z_7$ by $\Z_3$}

Next, we try extending by $c$:
\begin{equation}
\Z_7 \rtimes \Z_3 = \langle a, c \,|\, a^7 = e, c^3 = e, c^{-1} a c = a^2\rangle\,.\label{extension-Z7Z3-1}
\end{equation}
This leads to $ac = ca^2$, where we already took into account that no $i^r$ can help find solutions. 
Tracking the powers of $\eta$ in this matrix equation equation, we obtain
\begin{equation}
\left(\!\! \begin{array}{cccc} 
1 & 1 & \underline{1} & 1\\ 
\underline{2} & 2 & 2 & 2\\ 
4 & \underline{4} & 4 & 4\\ 
0 & 0 & 0 & \underline{0}\\ 
\end{array}\!\!\right)
\simeq 
\left(\!\! \begin{array}{cccc} 
2 & 4 & \underline{1} & 0\\ 
\underline{2} & 4 & 1 & 0\\ 
2 & \underline{4} & 1 & 0\\ 
2 & 4 & 1 & \underline{0}\\ 
\end{array}\!\!\right)\ \mod 7\,.
\label{extension-Z7Z3-2}
\end{equation}
Now we find four elements of $c$ with matching powers of $\eta$, which are underlined in the above equation.
Therefore, the matrix $c$ has the following form:
\begin{equation}
c = \left(\!\! \begin{array}{cccc} 
0 & 0 & c_{13} & 0\\ 
c_{21} & 0 & 0 & 0\\ 
0 & c_{32} & 0 & 0\\ 
0 & 0 & 0 & c_{44}\\ 
\end{array}\!\!\right)\,. \label{extension-Z7Z3-3}
\end{equation}
In plain words, this automorphism is given by the cyclic permutation of the first three doublets
up to arbitrary phase rotations.
The phases of the non-zero elements cannot be constrained by the group-theoretic relations.
However, using the fact that we have already fixed $\lambda_{1,2,3}$ in $V(\Z_7)$ to be real and positive, 
we obtain the unique matrix $c$ by setting all its non-zero elements to be $1$.
Thus, we arrive at the 4HDM model with the symmetry group
\begin{equation}
\Z_7\rtimes \Z_3 \simeq T_7\,, \quad \mbox{generated by\ }
a = \left(\!\! \begin{array}{cccc} 
\eta & 0 & 0 & 0\\ 
0 & \eta^2 & 0 & 0\\ 
0 & 0 & \eta^4 & 0\\ 
0 & 0 & 0 & 1\\ 
\end{array}\!\!\right)\,,
\quad
c = \left(\!\! \begin{array}{cccc} 
0 & 0 & 1 & 0\\ 
1 & 0 & 0 & 0\\ 
0 & 1 & 0 & 0\\ 
0 & 0 & 0 & 1\\ 
\end{array}\!\!\right)\,,
\label{extension-Z7Z3-4}
\end{equation}
with the rephasing-sensitive potential containing only one independent coefficient:
\begin{equation}
V(T_7) = \lambda \left[\fdf{2}{1} \fdf{4}{1} + \fdf{3}{2} \fdf{4}{2} + \fdf{1}{3} \fdf{4}{3} + h.c.\right]\,.
\label{VT7-1}
\end{equation}
The rephasing-insensitive part of the potential $V_0$ must also be invariant under these permutations,
see Eq.~\eqref{V0-3-1}. 
The full Higgs potential contains only 9 real free parameters and may lead to intriguing phenomenology,
which we delegate to a future paper.

We also remark that the symmetry group $T_7$ has already received some attention in the context of neutrino masses and mixing 
model building, see e.g. \cite{Ishimori:2010au,Luhn:2007sy,Hagedorn:2008bc,Cao:2010mp,Vien:2014gza,Bonilla:2014xla}.
Here, we demonstrate that the group can also arise in the 4HDM scalar sector alone.

\subsection{Searching for non-split extensions of $\Z_7$}

Let us see if we can use $c$ to build a non-abelian non-split extension of $\Z_7$.
A non-split extension $G = \Z_7\,.\,\Z_3$ means that $\Z_7$ is normal in $G$
and that $G/\Z_7 \simeq \Z_3$. However we do not require $G$ to contain a copy of $\Z_3$.
This means that $c^3$ is no longer required to be $e$ but can in fact be equal to 
a non-unit element of $\Z_7$. 
The relation $c^{-1} a c = a^2$ remains intact, so that the above solution Eq.~\eqref{extension-Z7Z3-3} is valid in this case, too.
Now, when we calculate $c^3$, we obtain the diagonal matrix 
$c^3 = \diag(x,x,x,y)$, where $x = c_{13}c_{21}c_{32}$ and $y = c_{44}^3$.
The equality of the first three entries makes it clear that $c^3$
cannot be set equal to any of the non-trivial powers of $a$.

An alternative, and shorter, way is to notice that if $c^3$ is equal to any non-identity element of 
$\Z_7$, which is a valid generator of $\Z_7$, then all elements of $G = \Z_7\,.\,\Z_3$ 
can be presented as some power of $c$. This gives us the abelian group $\Z_{21}$.
But this group is absent in the list of viable $A$'s in the 4HDM.
In either way, we conclude that no non-split extension $\Z_7\,.\,\Z_3$ can be constructed for the 4HDM.

\newpage
\section{Calculating Extensions by $\Z_6$}

\subsection{The two options for the $\Z_6$ generators}

The same technique of \cite{Ivanov:2011ae} allows us to find the generator $a$ of the group $\Z_6$.
However, unlike the previous two cases, the group $\Z_6$ now admits several realizations inside the 4HDM scalar sector.
Using the code {\tt 4HDM Toolbox}, we found two in-equivalent choices of $\Z_6$ charges:
\begin{eqnarray}
\mbox{$\Z_6$ option 1:} && a_1 = \diag(\eta,\, \eta^2,\, \eta^3,\, 1)\,, \quad \eta \equiv e^{\pi i/3}\,, \quad \eta^6 = 1 \label{Z6-a1}\\
\mbox{$\Z_6$ option 2:} && a_2 = \eta^{-1/4}\cdot\diag(\eta, \eta^2, \eta^4, 1)\label{Z6-a2}
\end{eqnarray}
That these two generators lead to different $\Z_6$ invariant models can be understood by comparing these generators cubed.
Indeed, $a_1^3 = \diag(-1, 1, -1, 1)$, which has two 2-dimensional invariant subspaces, 
while $a_2^3 = \sqrt{-i}\cdot\diag(-1, 1, 1, 1)$, which has a 3-dimensional invariant subspace. 
We also checked that other choices of the $\Z_6$ generator, which could not be brought to these ones by basis changes,
unavoidably lead to continuous accidental symmetries and are, therefore, disregarded in this study.

In the following sections, we will extend $\Z_6$ generated by either of the two options. 
Because $\Aut(\Z_6)\simeq\Z_2$, we will construct the split extensions $\Z_6\rtimes\Z_2\simeq D_6$ and 
attempt at building the non-split extension $\Z_6\,.\,\Z_2\simeq Q_6$.

\subsection{Extending $\Z_6$: the first option}

The Higgs potential invariant under $a_1$ in Eq.~\eqref{Z6-a1} contains not three but five different terms:
\begin{equation}
V({\Z_6}) = \lambda_{1} \fdf{2}{1} \fdf{4}{1} + \lambda_{2} \fdf{1}{2} \fdf{3}{2} + \lambda_{3} \fdf{4}{3}^2
+  \lambda_{4} \fdf{1}{3} \fdf{2}{4}  + \lambda_{5} \fdf{1}{4} \fdf{2}{3} 
+ h.c.\label{V-Z6-1}
\end{equation}
All the coefficients can be complex. In contrast to the previous cases, 
it is in general impossible to set all of them real by rephasing.
This implies that the $\Z_6$-invariant 4HDM can come either in the explicitly $CP$ conserving 
or $CP$ violating versions.

Notice that in this case we encounter for the first time the monomials which involve all four doublets:
$\fdf{1}{3} \fdf{2}{4}$ and $\fdf{1}{4} \fdf{2}{3}$. These two terms transform in the same way under phase rotation basis changes.
Therefore, if the coefficients in front of them, $\lambda_{4}$ and $\lambda_{5}$,
have a relative phase, it can never be eliminated by a phase rotation. 
In short, the presence of such terms, by itself, already allows the model to be $CP$ violating.

The automorphism group $\Aut(\Z_6) \simeq \Z_2$ is generated by $b$ which maps $a \mapsto a^{-1}$,
where $a$ is understood as $a_1$ in Eq.~\eqref{Z6-a1}.
This relation leads to the matrix equation 
\begin{equation}
ab = ba^{-1}\cdot i^r\label{Z6-b}
\end{equation}
with integer $0\le r < 4$. Noting that $i = \eta^{3/2}$, we translate it into
comparison of powers of $\eta$:
\begin{equation}
\left(\!\! \begin{array}{cccc} 
1 & 1 & 1 & 1\\ 
2 & 2 & 2 & 2\\ 
3 & 3 & 3 & 3\\ 
0 & 0 & 0 & 0\\ 
\end{array}\!\!\right)
\simeq 
\left(\!\! \begin{array}{cccc} 
-1 & -2 & -3 & 0\\ 
-1 & -2 & -3 & 0\\ 
-1 & -2 & -3 & 0\\ 
-1 & -2 & -3 & 0\\ 
\end{array}\!\!\right)\ + \frac{3}{2}\,r \ \mod 6\,.
\label{extension-Z6Z2-1}
\end{equation}
A non-trivial solution exists only for $r = 2$ and leads to the following matrix $b$:
\begin{equation}
b = \left(\!\! \begin{array}{cccc} 
0 & b_{12} & 0 & 0\\ 
b_{21} & 0 & 0 & 0\\ 
0 & 0 & 0 & b_{34}\\ 
0 & 0 & b_{43} & 0\\ 
\end{array}\!\!\right)\,. \label{extension-Z6Z2-2}
\end{equation}
Since $b^2 = \diag(x, x, y, y)$, with $x = b_{12}b_{21}$ and $y = b_{34}b_{43}$,
it cannot be represented by any non-zero power of $a$. 
Thus, non-split extensions are excluded.

Since $b$ is unitary, it induces the simultaneous exchanges $\phi_1\leftrightarrow\phi_2$ and $\phi_3\leftrightarrow\phi_4$,
up to possible phase shifts.
Contrary to the previous cases, we have not fixed the phases of $\lambda_i$; 
therefore, we still have the rephasing basis change freedom
to set $b_{12} = b_{21} \equiv \exp(i\alpha_{12})$ and $b_{34} = b_{43} \equiv \exp(i\alpha_{34})$,
see the SNF trick we mentioned in Section~\ref{abelian-Symmetries-in-NHDM} where the details are spelled out.
However the requirements of $\det b = 1$ and $b^2 = i^q \mathbf{1}_4$ allow us to set $\alpha_{12}=\alpha_{34}=0$ without loss of generality.
Thus, we arrive at the following viable extension of $\Z_6$:
\begin{equation}
\Z_6\rtimes \Z_2 \simeq D_6 \simeq S_3 \times \Z_2\,, \quad \mbox{generated by\ }
a = \left(\!\! \begin{array}{cccc} 
\eta & 0 & 0 & 0\\ 
0 & \eta^2 & 0 & 0\\ 
0 & 0 & \eta^3 & 0\\ 
0 & 0 & 0 & 1\\ 
\end{array}\!\!\right)\,,
\quad
b = \left(\!\! \begin{array}{cccc} 
0 & 1 & 0 & 0\\ 
1 & 0 & 0 & 0\\ 
0 & 0 & 0 & 1\\ 
0 & 0 & 1 & 0\\ 
\end{array}\!\!\right)\,.
\label{extension-Z6Z2-3}
\end{equation}
We stress once again that, although we write these transformations as $SU(4)$ matrices,
they are meant to represent the corresponding $\Z_4$ cosets.
As a result, the direct calculation shows that $b^{-1}a b = \eta^3 a^{-1} = - a^{-1}$
which belongs to the same coset as $a^{-1}$.
Therefore, $b$ indeed maps the coset $a\Z_4$ to $a^{-1}\Z_4$.
Notice also that, with the same convention, $b$ commutes with $a^3$, which allowed us to represent 
the group as $S_3 \times \Z_2$.

In order for the rephasing-sensitive potential $V(\Z_6)$ to be invariant under $\Z_6 \rtimes \Z_2\simeq  D_6$, 
we must require, in this basis, that 
\begin{equation}
\lambda_1 = \lambda_2\,, \quad \lambda_3 = \lambda_3^*\,,
\label{VZ6Z2-1}
\end{equation}
while $\lambda_4$ and $\lambda_5$ are unconstrained because these terms are, individually, 
invariant under $b$.
In addition, we require the rephasing-insensitive part of the potential $V_0$ 
to satisfy the conditions Eq.~\eqref{V0-2-2}.

\subsection{Extending $\Z_6$: the second option}

The potential invariant under $\Z_6$ generated by $a_2$ in Eq.~\eqref{Z6-a2} also contains five terms:
\begin{equation}
V_2(\Z_6) = \lambda_1\fdf{1}{3}^2 + \lambda_2\fdf{1}{2}\fdf{1}{4} + \lambda_3\fdf{2}{3}\fdf{2}{4} + 
\lambda_4\fdf{3}{2}\fdf{3}{4} + \lambda_5\fdf{4}{2}\fdf{4}{3} + h.c.
\label{V-Z6-2}
\end{equation}
It differs from Eq.~\eqref{V-Z6-1} in that it does not contain terms involving all four doublets.
We look for solutions for the matrix Eq.~\eqref{Z6-b} with $a_2$. The matrix equation is:
\begin{equation}
-\frac{1}{4} + 
\left(\!\! \begin{array}{cccc} 
1 & 1 & 1 & 1\\ 
2 & 2 & 2 & 2\\ 
4 & 4 & 4 & 4\\ 
0 & 0 & 0 & 0\\ 
\end{array}\!\!\right)
\simeq 
\left(\!\! \begin{array}{cccc} 
-1 & -2 & -4 & 0\\ 
-1 & -2 & -4 & 0\\ 
-1 & -2 & -4 & 0\\ 
-1 & -2 & -4 & 0\\ 
\end{array}\!\!\right)\ + \frac{3}{2}\,r + \frac{1}{4} \ \mod 6\,.
\label{extension-Z6Z2-4}
\end{equation}
The non-trivial solution for $b$ exists for $r=1$ and leads to the following matrix:
\begin{equation}
b = \left(\!\! \begin{array}{cccc} 
b_{11} & 0 & 0 & 0\\ 
0 & 0 & 0 & b_{24}\\ 
0 & 0 & b_{33} & 0\\ 
0 & b_{42} & 0 & 0\\ 
\end{array}\!\!\right)\,. \label{extension-Z6Z2-5}
\end{equation}
This solution means that the $\Z_6\rtimes\Z_2$ invariant potential should remain unchanged under the exchange of $\phi_2$ and $\phi_4$ up to phase factors. 

We first consider the split extension, which requires that $b^2 = i^q\mathbf{1}_4$. 
Parametrizing the elements of $b$ by phase factors, determining them from the invariance of the potential Eq.~\eqref{V-Z6-1}, 
and imposing $\det b = 1$, $b^\dagger b = \mathbf{1}_4$, we obtain $b$ as
\begin{equation}
b = i^c \left(\!\! \begin{array}{cccc}
1 & 0 & 0 & 0\\ 
0 & 0 & 0 & 1\\ 
0 & 0 & \sigma & 0\\ 
0 & 1 & 0 & 0\\ 
\end{array}\!\!\right)\,, \label{extension-Z6Z2-6}
\end{equation}
where $c$ is half-integer for $\sigma = 1$ and integer for $\sigma = -1$. 
Therefore, the $D_6$ invariant potential is given by Eq.~\eqref{V-Z6-1} subject to the extra condition
\begin{equation}
\lambda_3 = \sigma \lambda_5\,.\label{extension-Z6Z2-7}
\end{equation}

We also tried to construct the non-split extension $Q_6$ by imposing $b^2 = a_2^3$. 
This could only be done by setting $\lambda_1 = \lambda_2 = 0$. But then the potential
acquires the accidental continuous symmetry given by the arbitrary phase rotations of the first doublet.
Thus, no non-split extensions are possible for $\Z_6$.

\newpage

\section{Calculating Extensions by $\Z_5$}

From \cite{Ivanov:2011ae}, we know that the generator of $\Z_5$ can be chosen in the following way:
\begin{equation}
a = \eta \cdot \diag(\eta,\, \eta^2,\, \eta^3,\, 1) = \diag(\eta^2,\, \eta^{-2},\, \eta^{-1},\, \eta)\,, \quad \eta \equiv e^{2 \pi i/5}\,, \quad \eta^5 = 1.
\label{Z5-a1}
\end{equation}
We verified with the code {\tt 4HDM Toolbox} that all $\Z_5$ invariant 4HDMs can be brought to this choice. 
The Higgs potential invariant under $\Z_5$ contains six terms:
\begin{equation}
\begin{aligned}
V({\Z_5}) & = \lambda_{1} \fdf{2}{1} \fdf{4}{1} + \lambda_{2} \fdf{3}{4}\fdf{2}{4}  + \lambda_{3} \fdf{1}{2} \fdf{3}{2} + \lambda_{4} \fdf{4}{3}\fdf{1}{3}  \\
& +  \lambda_{5} \fdf{1}{3} \fdf{2}{4}  + \lambda_{6} \fdf{4}{1} \fdf{3}{2} +  h.c.\\
\end{aligned}
\label{VZ5}
\end{equation}
The first line here contains the four terms which are linked by the cyclic permutation 
\begin{equation}
\phi_1 \mapsto \phi_4 \mapsto \phi_2 \mapsto \phi_3 \mapsto \phi_1\,.\label{Z5-b-0}
\end{equation}
The second line contains the two possible terms involving all four doublets; these two terms transform into one another upon the same cyclic permutation.
The usefulness of such grouping will become clear once we study the automorphisms of $\Z_5$.
All the coefficients $\lambda_i$ can be complex and, in general, cannot be simultaneously set real by a basis change. 
Thus, the $\Z_5$-invariant 4HDM can be either explicitly $CP$ conserving or $CP$ violating.

The automorphism group for $\Z_5$ is $\Aut(\Z_5) \simeq \Z_4$. The generator $b$ of this automorphism group sends $a$ to $a^2$ or to $a^3$;
the two choices are equivalent.
There is also another automorphism $c$, of order 2, which acts as $a \mapsto a^4 = a^{-1}$.
This automorphism generates the $\Z_2$ subgroup of $\Z_4$; we immediately recognize that $c = b^2$.
Thus, when constructing extensions of $\Z_5$, we can extend it by $\Z_4$ or by $\Z_2$.

\subsection{Extension $\Z_5 \rtimes \Z_4$}

We begin with the extension 
$$
\Z_5\rtimes\Z_4 = \langle a, b\, |\, a^5 = e, b^4=e, b^{-1}a b = a^2\rangle\,.\label{extension-Z5Z4-1}
$$
This group of order 20 is also known as $GA(1,5)$, the general affine group over the finite field $\mathbb{F}_5$,
and it has the {\tt GAP Id [20,3]}.
The relation $b^{-1}a b = a^2$ leads to the matrix equation for $b$:
\begin{equation}
ab = ba^{2}\cdot i^r\label{Z5-b-1}\,.
\end{equation}
Since $i = \eta^{5/4}$, we can rewrite this equation as comparison of the powers of $\eta$:
\begin{equation}
\left(\!\! \begin{array}{cccc} 
2 & 2 & 2 & 2\\ 
-2 & -2 & -2 & -2\\ 
-1 & -1 & -1 & -1\\ 
1 & 1 & 1 & 1\\ 
\end{array}\!\!\right)
\simeq 
\left(\!\! \begin{array}{cccc} 
-1 & 1 & -2 & 2\\ 
-1 & 1 & -2 & 2\\ 
-1 & 1 & -2 & 2\\ 
-1 & 1 & -2 & 2\\ 
\end{array}\!\!\right)\ + \frac{5}{4}\,r \ \mod 5\,.
\label{extension-Z5Z4-2}
\end{equation}
A solution for the matrix $b$ exists only for $r=0$, 
\begin{equation}
b = \left(\!\! \begin{array}{cccc} 
0 & 0 & 0 & b_{14}\\ 
0 & 0 & b_{23} & 0\\ 
b_{31} & 0 & 0 & 0\\ 
0 & b_{42} & 0 & 0\\ 
\end{array}\!\!\right)\,. \label{extension-Z5Z4-3}
\end{equation}
We find that $b$ realizes the cyclic permutation given in Eq.~\eqref{Z5-b-0}. 
Each non-zero entry here is a pure phase factor, and
since $b^4 = b_{14}b_{23}b_{31}b_{42}\, \mathbf{1}_4$, we conclude that non-split extensions are not possible. 
By performing an appropriate phase shift basis change, we can set all non-trivial entries 
to be equal to the same phase factor $\exp(i\alpha)$, see Section~\ref{abelian-Symmetries-in-NHDM}.
We take it out as a universe prefactor and determine $\alpha$ from the condition $\det b = 1$.
In this way we arrive at the extension
\begin{equation}
\Z_5\rtimes \Z_4 \simeq GA(1,5)\,, \quad \mbox{generated by\ }
a = \left(\!\! \begin{array}{cccc} 
\eta^2 & 0 & 0 & 0\\ 
0 & \eta^{-2} & 0 & 0\\ 
0 & 0 & \eta^{-1} & 0\\ 
0 & 0 & 0 & \eta\\ 
\end{array}\!\!\right)\,,
\quad
b = i^{1/2}\left(\!\! \begin{array}{cccc} 
0 & 0 & 0 & 1\\ 
0 & 0 & 1 & 0\\ 
1 & 0 & 0 & 0\\ 
0 & 1 & 0 & 0\\ 
\end{array}\!\!\right)\,.
\label{extension-Z5Z4-4}
\end{equation}
In order for the potential $V(\Z_5)$ to be invariant under this group, we must require
\begin{equation}
        \lambda_1  = \lambda_2 = \lambda_3 = \lambda_4\,,\quad \lambda_5  = \lambda_6\,.\label{extension-Z5Z4-5}
\end{equation}
These coefficients can still remain complex.
In addition, we require that the rephasing-insensitive part of the potential $V_0$ is invariant
under Eq.~\eqref{V0-4-0}. Notice that although the potential $V_0$ constrained by Eq.~\eqref{V0-4-0}
possesses accidental symmetries such as $\phi_1 \leftrightarrow \phi_4$, $\phi_2 \leftrightarrow \phi_3$,
this symmetry is not shared by the potential $V(\Z_5)$ even when constrained by Eq.~\eqref{extension-Z5Z4-5}.

As the full scalar potential of the $GA(1,5)$-symmetric 4HDM contains only 10 real free parameters,
one can expect many correlations among scalar properties. 
We delegate the phenomenological analysis of this model to a future paper.
We are not aware of any multi-Higgs study which uses the group $GA(1,5)$.

\subsection{Extension $\Z_5 \rtimes \Z_2$}

Let us now construct the second extension:
$$
\Z_5\rtimes\Z_2\simeq D_{5} = \langle a, c\, |\, a^5 = e, c^2=e, c^{-1}a c = a^{-1}\rangle\,.\label{Z5-c-1}
$$
Instead of relying on $b^2$, we write and solve the matrix equation on $c$ and arrive at the solution
\begin{equation}
c = \left(\!\! \begin{array}{cccc} 
0 & c_{12} & 0 & 0\\ 
c_{21} & 0 & 0 & 0\\ 
0 & 0 & 0 & c_{34}\\ 
0 & 0 & c_{43} & 0\\ 
\end{array}\!\!\right)\,, \label{extension-Z5Z2-1}
\end{equation}
which is of the same form as $b$ in Eq.~\eqref{extension-Z6Z2-2}. All the considerations given there apply to this case,
and we arrive at the 4HDM invariant under 
\begin{equation}
\Z_5\rtimes\Z_2\simeq D_{5}\,, \quad \mbox{generated by\ }
a = \left(\!\! \begin{array}{cccc} 
\eta^2 & 0 & 0 & 0\\ 
0 & \eta^{-2} & 0 & 0\\ 
0 & 0 & \eta^{-1} & 0\\ 
0 & 0 & 0 & \eta\\ 
\end{array}\!\!\right)\,,
\quad
c = \left(\!\! \begin{array}{cccc} 
0 & 1 & 0 & 0\\ 
1 & 0 & 0 & 0\\ 
0 & 0 & 0 & 1\\ 
0 & 0 & 1 & 0\\ 
\end{array}\!\!\right)\,. \label{extension-Z5Z2-2}
\end{equation}
This symmetry arises if the conditions Eq.~\eqref{V0-2-2} on $V_0$ are fulfilled and the parameters $\lambda$ in Eq.~\eqref{VZ5} satisfy
\begin{equation}
\lambda_1 = \lambda_3\,, \quad  
\lambda_2 = \lambda_4\,,\label{extension-Z5Z2-3}
\end{equation}
while $\lambda_5$ and $\lambda_6$ are unconstrained.
The rephasing-insensitive potential $V_0$ must satisfy the conditions Eq.~\eqref{V0-2-2}.

\newpage

\section{Calculating Extensions by $\Z_4$}\label{section-Z4}

\subsection{The three versions of $\Z_4$ in the 4HDM}

In the previous sections, we considered the groups $\Z_n$, $n = 5,6,7,8$, which were not available in the 3HDM scalar sector \cite{Ivanov:2011ae}.
The case of $\Z_4$ is different. It was already present in the 3HDM study, where it was the largest cyclic symmetry group 
realizable in the 3HDM scalar sector, with the generator $\diag(i, -i, 1)$. 
When embedding it in the 4HDM scalar sector, one could just take this generator and assume that the fourth doublet transforms under $\Z_4$
by multiplication of some power of $i$: $1$, $-1$, or $i$ (with the choice $-i$ being equivalent to $i$ upon $\phi_1\leftrightarrow \phi_2$). 
This construction leads us to three nonequivalent $\Z_4$ symmetries in the 4HDM, which we present in a slightly rearranged form:
\begin{eqnarray}
\mbox{$\Z_4$ option 1:} && a_1 = \sqrt{i}\cdot\diag(i,-1, -i, 1)\label{extension-Z4-a1}\\
\mbox{$\Z_4$ option 2:} && a_2 = \diag(i,-i, 1, 1)\label{extension-Z4-a2}\\
\mbox{$\Z_4$ option 3:} && a_3 = i^{3/4}\,\diag(i,i,-i, 1)\label{extension-Z4-a3}
\end{eqnarray}
The powers of $i$ are introduced to ensure that $\det a = 1$.
Option 1 can be called the ``fully represented $\Z_4$'' because the four doublets transform
as the four possible singlets of $\Z_4$.
Options 2 and 3 contain 2D invariant subspaces. 

Any other choice of the $\Z_4$ generator either can be reduced to one of those or will lead
to a continuous rephasing symmetry. For example, if one chooses
the generator $\sqrt{i}\cdot \diag(i,i,1,1)$ with two 2D invariant subspaces and tries to construct
a potential invariant under it, one will unavoidably obtain the accidental symmetry
$(e^{i\alpha}, e^{i\alpha}, e^{-i\alpha}, e^{-i\alpha})$.

The three options for the $\Z_4$ generator represent truly distinct cases, which cannot be linked by any basis change. 
Indeed, $a_1$ does not contain any 2D invariant subspace, while $a_3$ differs from $a_2$
by the fact that $a_3^2$ has a 3D invariant subspace.
This situation is reminiscent of two nonequivalent $U(1)$ groups which exist already in the 3HDM \cite{Ivanov:2011ae}.
With the aid of the code {\tt 4HDM Toolbox}, 
we verified that, in all the cases of the symmetry group $\Z_4$, we obtain the generator of the form
Eq.~\eqref{extension-Z4-a1}, Eq.~\eqref{extension-Z4-a2}, or Eq.~\eqref{extension-Z4-a3}, up to permutations and rephasing.

Below, we will extend all three versions of the $\Z_4$ group. Since $\Aut(\Z_4) \simeq \Z_2$, we can have
only two possibilities for non-abelian extensions: the split extension $\Z_4\rtimes \Z_2 \simeq D_4$ and the non-split extension
$\Z_4\,.\,\Z_2 \simeq Q_4$, see Section~\ref{section-split}. Thus, when constructing the generator $b$ of the $\Z_2$ group,
we will, in principle, need to check two options: $b^2 = e$ or $b^2 = a^2$, for each version of the $\Z_4$ group.

\subsection{Extending the fully represented $\Z_4$}\label{subsection-fully-rep-Z4}

The potential invariant under the generator $a_1$ in Eq.~\eqref{extension-Z4-a1} contains, apart from the rephasing invariant $V_0$,
the following terms:
\begin{equation}
\begin{aligned}
V_1(\Z_4) & = \lambda_1\fdf{1}{2}\fdf{1}{4} + \lambda_2\fdf{2}{1}\fdf{2}{3} + \lambda_3\fdf{1}{3}^2 \\ 
& + \lambda_4\fdf{4}{1}\fdf{4}{3} + \lambda_5\fdf{3}{2}\fdf{3}{4} + \lambda_6\fdf{2}{4}^2 \\
& + \lambda_7\fdf{1}{2}\fdf{4}{3} + \lambda_8\fdf{1}{3}\fdf{4}{2} + \lambda_9\fdf{1}{3}\fdf{2}{4} + \lambda_{10}\fdf{1}{4}\fdf{2}{3} + h.c.
\end{aligned}
\label{V1-Z4}
\end{equation}
These terms can be found by writing the $\Z_4$ charges of all possible quadratic monomials $\fdf{i}{j}$
and multiplying the terms with opposite charges to obtain the $\Z_4$-invariant terms.
As this potential has ten complex free parameters, it is in general impossible to set them simultaneously real
by any basis change; thus, the $\Z_4$ 4HDM model can be either explicitly $CP$ conserving or $CP$ violating. 

The automorphism group of $\Z_4$ is $\Z_2$, whose generator $b$ sends $a_1$ to $a_1^3$. 
Thus, we have matrix equation for the generator $b$ of $\Z_2$,
\begin{equation}
a_1b = ba_1^3\cdot i^r\,,
\end{equation}
which can be represented as the equation for the powers of $i$:
\begin{equation}\frac{1}{2} + 
\left(\!\! \begin{array}{cccc} 
1 & 1 & 1 & 1\\ 
2 & 2 & 2 & 2\\ 
3 & 3 & 3 & 3\\ 
0 & 0 & 0 & 0\\ 
\end{array}\!\!\right)
\simeq \frac{3}{2} + 
\left(\!\! \begin{array}{cccc} 
3 & 2 & 1 & 0\\ 
3 & 2 & 1 & 0\\ 
3 & 2 & 1 & 0\\ 
3 & 2 & 1 & 0\\ 
\end{array}\!\!\right)\ + \,r \ \mod 4\,.
\label{extension-Z4Z2-2}
\end{equation}
This set of matching conditions has one solution for each $r = 0,1,2,3$.
However solutions for $r = 0$ and $2$ are mapped onto each other by renaming the doublets,
and so are solutions for $r = 1$ and $3$. Thus, we have two non-equivalent classes of $b$,
which, using the rephasing basis change freedom, can be represented as
\begin{equation}
b = 
\begin{pmatrix}
0 & e^{i\alpha} & 0 & 0 \\
e^{i\alpha} & 0 & 0 & 0 \\
0 & 0 & 0 & e^{i\beta} \\
0 & 0 & e^{i\beta} & 0 
\end{pmatrix}
\,,\quad 
b' = 
\begin{pmatrix}
0 & 0 & e^{i\alpha} & 0 \\
0 & e^{i\gamma_2} & 0 & 0 \\
e^{i\alpha} & 0 & 0 & 0 \\
0 & 0 & 0 & e^{i\gamma_4} 
\end{pmatrix}
\label{extension-Z4Z2-bb}
\end{equation}
Let us first consider $b$. Squaring it leads to $b^2 = \diag(x,x,y,y)$, which cannot match any nontrivial power of $a_1$,
which forbids the non-split extension. 
Next, by requiring $\det b = 1$ and $b^2 = i^{q}\mathbf{1}_4$, and making use of rephasing basis change freedom,
we obtain the unique expression for $b$, up to the omnipresent powers of $i$:
\begin{equation}
b = 
\begin{pmatrix}
0 & 1 & 0 & 0 \\
1 & 0 & 0 & 0 \\
0 & 0 & 0 & 1 \\
0 & 0 & 1 & 0 
\end{pmatrix}\,.
\label{extension-Z4Z2-b2}
\end{equation}
Thus, we obtain the group $D_4$ generated by $a_1$ in Eq.~\eqref{extension-Z4-a1}
and by $b$ in Eq.~\eqref{extension-Z4Z2-b2}.
Requiring that the potential $V_1(\Z_4)$ be invariant under $b$ leads to the following relations:
\begin{equation}
\lambda_1 = \lambda_2\,,\quad \lambda_3 = \lambda_6\,,\quad \lambda_4 = \lambda_5\,,\quad \lambda_7,\lambda_8\in\mathbb{R}\,.
\label{extension-Z4Z2-b3p}
\end{equation}
The number of real parameters is reduced from 20 in Eq.~\eqref{V1-Z4} to 12.
The potential still contains complex parameters, so that the model allows for explicit $CP$ violation.

Next, we consider $b'$ in Eq.~\eqref{extension-Z4Z2-bb} and still aim to construct 
the split extension $\Z_4 \rtimes \Z_2 \simeq D_4$. Repeating the above analysis, we get 
the following generic expression for $b'$ which covers all sign choices:
\begin{equation}
b' = i^c
\begin{pmatrix}
0 & 0 & 1 & 0 \\
0 & \sigma & 0 & 0 \\
1 & 0 & 0 & 0 \\
0 & 0 & 0 & 1 
\end{pmatrix}\,.
\label{extension-Z4Z2-b3}
\end{equation}
Here, $\sigma = \pm 1$ and $c$ is integer for $\sigma = - 1$ and half-integer for $\sigma = +1$. 
Thus, the two transformations $a_1$ and $b'$ generate another version of the $D_4$-invariant 4HDM.
In order for $V_1(\Z_4)$ to be invariant under $b'$, its parameters must satisfy
\begin{equation}
\lambda_5 = \sigma \lambda_1\,, \quad 
\lambda_{10} = \sigma \lambda_7^*\,, \quad 
\lambda_9 = \sigma \lambda_8^*\,, \quad 
\lambda_3 \in \mathbb{R} \,.\label{extension-Z4Z2-b4}
\end{equation}
We are left with 13 real free parameters.

In addition, we can use $b'$ from Eq.~\eqref{extension-Z4Z2-bb} to construct the non-split extension
$\Z_4\,.\,\Z_2 \simeq Q_4$; this is the first example in which we encounter this possibility. 
Let us denote this version of $b'$ as $b''$.
Then, $(b'')^2 = \diag(x,y,x,z)$ and it can match $a_1^2 = \diag(-i, i, -i, i)$.
Combining this matching with $\det b'' = 1$, we obtain the following general expression for $b''$:
\begin{equation}
b'' = i^c
\begin{pmatrix}
0 & 0 & 1 & 0 \\
0 & -i\sigma & 0 & 0 \\
1 & 0 & 0 & 0 \\
0 & 0 & 0 & i 
\end{pmatrix}\,,
\label{extension-Z4Z2-b5}
\end{equation}
with the same convention for the power $c$.
Imposing $b''$ symmetry on the potential leads to the same type of restrictions as in Eq.~\eqref{extension-Z4Z2-b4}
and, in addition, eliminates the $\lambda_2$ and $\lambda_4$ terms. Thus, we are left with 9 real free parameters
in the rephasing sensitive part of the $Q_4$-invariant 4HDM.

\subsection{Extending $\Z_4$ option 2}

Next we turn to the $\Z_4$ symmetry group generated by $a_2$ defined in Eq.~\eqref{extension-Z4-a2}.
The rephasing-sensitive part of the potential invariant under this generator is
\begin{equation}
\begin{aligned}
V_2(\Z_4) & = m_{34}^2\fdf{3}{4} + \lambda_1\fdf{1}{2}^2 + \lambda_2\fdf{3}{4}^2
+ \lambda_3\fdf{3}{1}\fdf{3}{2} + \lambda_4\fdf{4}{1}\fdf{4}{2} \\
 & + \lambda_5\fdf{1}{3}\fdf{2}{4} + \lambda_6 \fdf{1}{4}\fdf{2}{3} + h.c.
\end{aligned}
\label{V2-Z4}
\end{equation}
This potential contains seven complex free parameters;
notice the presence of a new quadratic term. 

To extend this $\Z_4$ by a $\Z_2$ generated by a new transformation $c$, we follow 
the same strategy as before and obtain the following generic solution:
\begin{equation}
c = 
\begin{pmatrix}
0 & c_{12} & 0 & 0 \\
c_{21} & 0 & 0 & 0 \\
0 & 0 & c_{33} & c_{34} \\
0 & 0 & c_{43} & c_{44} \\
\end{pmatrix}\,.\label{extension-Z4-c1}
\end{equation}
The presence of the $2\times 2$ block in the $(\phi_3,\phi_4)$ subspace is a clear consequence of the fact that this subspace
is invariant under $\Z_4$.
However, $c^2$ must be diagonal: proportional either to $\mathbf{1}_4$ for the split extension $D_4$ 
or to $a_2^2$ for the non-split extension $Q_4$.
This leaves us with two possible shapes of this $2\times2$ block:
\begin{equation}
\mmatrix{c_{33}}{0}{0}{c_{44}}\quad \mbox{or}\quad \mmatrix{0}{c_{34}}{c_{43}}{0}\,.\label{extension-Z4-c2}
\end{equation}
Repeating the above analysis, we found that both options allow for a split and a non-split extension.
For example, the first option leads to the split extension $D_4$ which leaves $\phi_4$ invariant.
Therefore, this is exactly the same $D_4$ as was found in the 3HDM \cite{Ivanov:2012fp}.
For the non-split extension, we get
\begin{equation}
c = 
i^q\begin{pmatrix}
0 & i & 0 & 0 \\
i & 0 & 0 & 0 \\
0 & 0 & \sigma & 0 \\
0 & 0 & 0 & 1 \\
\end{pmatrix}\,,\label{extension-Z4-c3}
\end{equation}
where $\sigma = \pm 1$, with the corresponding value of $q$.
Imposing this symmetry on $V_{2}(\Z_4)$ eliminates the $\lambda_3$ and $\lambda_4$ terms, and the rephasing sensitive potential 
can then be compactly written as
\begin{equation}
V_2(Q_4) = \delta_{\sigma,1}m_{34}^2 \fdf{3}{4} + \lambda_1\fdf{1}{2}^2 + \lambda_2\fdf{3}{4}^2
+ \lambda_5\left[ \fdf{1}{3}\fdf{2}{4} -\sigma \fdf{1}{4}\fdf{2}{3}\right] + h.c.
\label{V2-Q4}
\end{equation}
with a real $\lambda_1$ and the quadratic term present only for $\sigma = +1$.
Notice the crucial role of $\phi_4$: although it is a $Q_4$ singlet, its presence allows us to build terms which
were absent in the 3HDM. As a result, we now have a realizable $Q_4$-invariant 4HDM without any accidental continuous symmetry,
the situation found in \cite{Ivanov:2012fp} impossible in the 3HDM.
The second option in Eq.~\eqref{extension-Z4-c2}, too, leads to viable split and non-split extensions.

\subsection{Extending $\Z_4$ option 3}

Finally, we consider the $\Z_4$ symmetry group generated by $a_3$ defined in Eq.~\eqref{extension-Z4-a3}.
The rephasing-sensitive part of the potential is
\begin{eqnarray}
V_3(\Z_4) &= & m_{12}^2\fdf{1}{2} + \lambda_1\fdf{1}{2}^2 + \lambda_2\fdf{1}{3}^2 + \lambda_3\fdf{2}{3}^2\nonumber\\
&& + \lambda_4\fdf{1}{4}\fdf{3}{4} + \lambda_5\fdf{2}{4}\fdf{3}{4} + \lambda_6\fdf{1}{3}\fdf{2}{3} + h.c.
\label{V3-Z4}
\end{eqnarray}
This potential contains seven complex free parameters, including one in the quadratic part. 

In order to extend this $\Z_4$, we need to solve $a_3b = ba_3^3\cdot i^r$ for $b$.
However, the prefactor $i^{3/4}$ prevents us from finding a non-trivial solution, in a way similar to our $\Z_8$ analysis.
Thus, we arrive at a peculiar $\Z_4$-invariant 4HDM which does not admit any extension.

To summarize our discussion of possible extensions of the symmetry group $\Z_4$ available in the 4HDM, we found that 
there are two non-equivalent $\Z_4$ symmetry groups in the 4HDM which can be extended both to $D_4 \simeq \Z_4\rtimes \Z_2$ 
and to $Q_4 \simeq \Z_4\,.\,\Z_2$ in a variety of non-equivalent way.
We explicitly constructed the generators of these groups and showed how the rephasing-sensitive part of the potential
is shaped by them.
We stress that the quaternion symmetry group $Q_4$ is a novel option for the 4HDM model building, 
which was unavailable within the 3HDM.
In addition, there exists a peculiar realization of $\Z_4$ in the 4HDM which does not admit any extension by $\Aut(\Z_4)$.

\newpage

\section{Calculating Extensions by $\Z_3$}\label{subsection-Z3}

Just as for the case of $\Z_4$, the symmetry group $\Z_3$ was also realizable in the 3HDM \cite{Ivanov:2011ae}.
Its generator in the 3HDM can be written as $\diag(1, \eta, \eta^{-1})$, where $\eta \equiv \exp(2\pi i/3)$, which is often denoted $\omega$.
Since $\eta^3 = 1$, this geerator can be written also as $\diag(1, \eta, \eta^2)$. This choice of $\Z_3$ charges is unique up to permutation;
if one tries to build a 3HDM based on, say, $\diag(1, 1, \eta)$, one would end up with a potential invariant under the continuous $U(1)$ symmetry.

Within the $\Z_3$ symmetric 4HDM, we again need to assign all three $\Z_3$ charges to the four Higgs doublets, 
and one of the charges will be used twice. As a result, we have only one possibility, up to permutations and the verall charge shifts.
The generator can be written as
\begin{equation}
a = \diag(\eta,\, \eta^2,\,1,\, 1)\,, \quad \eta \equiv e^{2 \pi i/3}\,, \quad \eta^3 = 1.
\label{Z3-a}
\end{equation}
The $\Z_3$-invariant 4HDM potential contains, in addition to $V_0$ given in Eq.~\eqref{V0-general}, the following terms:
\begin{eqnarray}
V_1(\Z_3) &= & m_{34}^2\fdf{3}{4} + \lambda_1\fdf{3}{4}^2 
+ \lambda_2\fdf{1}{2}\fdf{1}{3} + \lambda_3\fdf{2}{3}\fdf{2}{1} \\
&& + \lambda_4\fdf{1}{2}\fdf{1}{4} + \lambda_5\fdf{2}{4}\fdf{2}{1} 
+ \lambda_6\fdf{3}{1}\fdf{3}{2} \nonumber\\
&&+ \lambda_7\fdf{4}{1}\fdf{4}{2} 
+ \lambda_8\fdf{3}{1}\fdf{4}{2} + \lambda_9\fdf{4}{1}\fdf{3}{2}+ h.c.
\label{V1-Z3}
\end{eqnarray}
All of these ten coefficients can be complex.

The automorphism group of $\Z_3$ is $\Z_2$. 
Group-theoretically, we get only one non-abelian extension:
$\Z_3 \rtimes \Z_2 \simeq S_3$. The non-trivial automorphism is given by
\begin{equation}
b = 
i^q\begin{pmatrix}
0 & 1 & 0 & 0 \\
1 & 0 & 0 & 0 \\
0 & 0 & \sigma & 0 \\
0 & 0 & 0 & 1 \\
\end{pmatrix}\,,\label{extension-S3-b}
\end{equation}
where $\sigma = \pm 1$ and $q$ is half-integer for $\sigma = + 1$ and integer for $\sigma = -1$.
The version with $\sigma = -1$ can be called the fully-represented $S_3$ because in this case 
the irreducible representation decomposition of the four Higgs doublets is $2 + 1 + 1'$.

The two versions of the generator $b$ constrain the potential $V_1(Z_3)$ in Eq.~\eqref{V1-Z3}
in a slightly different manner.
In both cases, we have 
\begin{equation}
\lambda_2 = \sigma \lambda_3\,, \quad \lambda_4 = \lambda_5\,, \quad \lambda_8 = \sigma\lambda_9\,.\label{extension-S3-1}
\end{equation}
In addition, the fully represented $S_3$ model forbids the non-trivial quadratic term: $m_{34}^2 = 0$.
Also, in both cases, the rephasing-insensitive potential $V_0$ given in Eq.~\eqref{V0-general}
satisfies the conditions of the type of Eq.~\eqref{V0-2-1-1} but written for indices $3,4$ insteasd of $1,2$.

Let us finally mention that the symmetry group $\Z_2$ doees not have non-trivial automorphisms and cannot be extended.
We only remark that there exist two inequivalent implementations of $\Z_2$ in the 4HDM corresponding to the 
generators $\sqrt{i}\diag(1,1,1,-1)$ and $\diag(1,1,-1,-1)$. They lead to different constraits on the scalaar potential,
which are straightforward to write down.

\newpage

\section{Calculating Extensions by $\Z_2\times \Z_2$}\label{section-Z2Z2}

The symmetry group $\Z_2\times \Z_2 = \{e, a_1, a_2, a_1a_2 \} $ is easy to implement in a multi-Higgs model 
as its transformations can be defined as sign flips of certain doublets. 
Within the 3HDM, this construction is unique: the two generators of $\Z_2\times \Z_2$ can be always brought to the form
\begin{equation}
\mbox{3HDM:}\quad a_1 = 
\begin{pmatrix}
	-1 & 0 & 0 \\
	0 & -1 & 0 \\
	0 & 0 & 1 \\
\end{pmatrix}\,,\quad
a_2 = 
\begin{pmatrix}
	1 & 0 & 0 \\
	0 & -1 & 0 \\
	0 & 0 & -1 \\
\end{pmatrix}\,.\quad
\label{Z2Z2-1}
\end{equation}
Since the overall sign flip of all doublets has no effect on the model,
one can also view $a_1$ as the sign flip of $\phi_3$ alone and $a_2$ as the sign flip of $\phi_1$ alone.
This symmetry group was used in the famous Weinberg's model \cite{Weinberg:1976hu} which triggered 
an intense exploration of models with more than two scalar doublets.

With four Higgs doublets, we encounter two non-equivalent implementations of $\Z_2 \times \Z_2$,
which differ by the presence or absence of a 2D invariant subspace. 
We label these two implementations as
\begin{eqnarray}
\mbox{fully represented $\Z_2\times\Z_2$:} && a_1 = \diag(-1,-1, 1, 1)\,,\quad a_2 = \diag(1,-1, -1, 1)\,,\label{extension-Z2Z2-1}\\
\mbox{$\Z_2\times\Z_2$ with a 2D inv. subspace:} && 
a_1 = \sqrt{i}\cdot\diag(-1,1, 1, 1)\,,\ 
a_2 = \sqrt{i}\cdot\diag(1,-1, 1, 1)\,.\label{extension-Z2Z2-2}
\end{eqnarray}
We call the first option as the ``fully represented $\Z_2\times\Z_2$'' because the four doublets
transform as the four distinct singlets of $\Z_2\times\Z_2$:
\begin{equation}
\phi_1 \sim 1_{-+}\,, \quad
\phi_2 \sim 1_{--}\,, \quad
\phi_3 \sim 1_{+-}\,, \quad
\phi_4 \sim 1_{++}\,.
\end{equation}
Using the code {\tt 4HDM Toolbox} \cite{TheCode}, we verified by direct check of all combinations of three monomials
that any $\Z_2 \times \Z_2$ symmetry group in the 4HDM scalar sector can indeed be represented by one of the two above options.
Notice that if one removes the fourth doublet, then the two implementations of $\Z_2 \times \Z_2$ 
lead to the same group of the 3HDM, up to the overall phase change. 
Thus, it is the presence of the fourth doublet which distinguishes the two options, despite the fact  
that $\varphi_4$ transforms trivially under $\Z_2 \times \Z_2$. 

Turning now to the automorphism group $\Aut(\Z_2\times\Z_2) \simeq S_3$, where $S_3$ acts on the three non-trivial elements
of $\Z_2\times\Z_2$ by permutations, 
we notice that all three options for the group extension
\begin{equation}
(\Z_2\times\Z_2)\rtimes \Z_2 \simeq D_4\,, \quad
(\Z_2\times\Z_2)\rtimes \Z_3 \simeq A_4\,, \quad
(\Z_2\times\Z_2)\rtimes S_3 \simeq S_4
\end{equation}
were already available for the 3HDM model building \cite{Ivanov:2012fp}.
The extension by $\Z_2$ can be defined by an order-2 transformation $b$, 
which sends $a_1 \mapsto a_2$ and $a_2 \mapsto a_1$.
The extension by $\Z_3$ involves an order-3 transformation $c$, which generates
cyclic permutations of the three elements such as $a_1\mapsto a_2\mapsto a_1a_2 \mapsto a_1$.
The extension by $S_3$ involves both $b$ and $c$.

All three extensions can be readily exported to the 4HDM with the fully represented implementation of $\Z_2\times\Z_2$.
However the other implementation Eq.~\eqref{extension-Z2Z2-2} only admits the extension $(\Z_2\times\Z_2)\rtimes \Z_2 \simeq D_4$.
Trying to extend it by $\Z_3$ would lead to the equation $c^{-1}a_1 c = a_2$ and $c^{-1}a_2 c = a_1a_2$.
While the former equation can be solved, the latter one has no non-trivial solution because $a_2$ contains the prefactor $\sqrt{i}$
while $a_1a_2$ does not.
Thus, although this peculiar realization of the $\Z_2\times\Z_2$ symmetry leaves a 2D subspace invariant,
its non-abelian extensions can only be $D_4$, although different ones that in the fully represented case. 
Even if one considers non-split extensions by $\Z_2$,
one still arrives only at $D_4$ and not at the quaternion group $Q_4$ because $\Z_2\times\Z_2 \not \subset Q_4$. 

The scalar potential invariant under each implementation of the $\Z_2\times\Z_2$ and its extension
is rather lengthy but can be readily written out, as it borrows its structure from the 3HDM case.

\newpage

\section{Calculating Extensions by $\Z_4\times\Z_2$}\label{section-Z4Z2}

\subsection{The three implementations of $\Z_4\times\Z_2$ in the 4HDM}

The abelian group $\Z_4\times\Z_2$ was not available in the 3HDM but appears in the 4HDM, and the analysis of its extensions is much more involved.
Let us begin by describing the group itself, its automorphism group $\Aut(\Z_2\times\Z_4) \simeq D_4$, 
and the conjugacy classes of the subgroups of $D_4$.

The group $\Z_4\times\Z_2$ is generated by $a_1$ of order 4 and $a_2$ of order 2, which commute with each other.
This group contains, among its subgroups, both $\Z_4$ and $\Z_2\times\Z_2$.
We already know from \cite{Shao:2023oxt} that there are three distinct 4HDM implementations of $\Z_4$.
We also established in the previous section that there exist two nonequivalent implementations of $\Z_2\times\Z_2$.
Thus, it is natural to explore implementations of $\Z_4\times\Z_2$ by combining these choices.

It turns out, however, that not all combinations lead to viable models.
For example, let us take the fully represented $\Z_4$ and the fully represented $\Z_2\times\Z_2$.
This can be done by choosing the following generators of the $\Z_4\times\Z_2$:
\begin{equation}
a_1 = \sqrt{i}\cdot\diag(i, -1, -i, 1)\,, \quad a_2 = \diag(1, -1, -1, 1)\,.\label{extension-Z4Z2-invalid}
\end{equation}
One immediately sees that $a_1^2$ and $a_2$ are the pair which gives the fully generated $\Z_2\times\Z_2$.
Next, we take the corresponding $\Z_4$-invariant potential from \cite{Shao:2023oxt} 
and leave only those terms which remain invariant under $a_2$:
\begin{equation}
\begin{aligned}
	\tilde V & = \lambda_3\fdf{1}{3}^2 + \lambda_6\fdf{2}{4}^2 + \lambda_7\fdf{1}{2}\fdf{4}{3} + \lambda_8\fdf{1}{3}\fdf{4}{2} \\
	& + \lambda_9\fdf{1}{3}\fdf{2}{4} + \lambda_{10}\fdf{1}{4}\fdf{2}{3} + h.c.
\end{aligned}
\label{V1-Z4Z2-invalid}
\end{equation}
We remind the reader that, in addition to these terms, we full potential also includes 
the rephasing-insensitive part $V_0$ given in Eq.~\eqref{V0-general}.
However, the resulting potential $V_0 + \tilde V$ possesses an accidental $U(1)$ symmetry:
\begin{equation}
U(1)_{\rm acc.} = \diag\left(e^{i\alpha}, e^{-i\alpha}, e^{i\alpha}, e^{-i\alpha}\right)\,.\label{U1-accidental}
\end{equation}
Thus, this situation cannot be classified as a $\Z_4\times\Z_2$ symmetric model.

After studying all the combinations of the $\Z_4$ and $\Z_2\times\Z_2$ generators and verifying the results with the code
{\tt 4HDM Toolbox} \cite{TheCode}, we found that only three
in-equivalent options for the $\Z_4\times\Z_2$-symmetric 4HDM exist. We list below there potential and generators:
\begin{eqnarray}
\mbox{option 1:}&& V_1 = \lambda_1\fdf{1}{3}^2 + \lambda_2\fdf{2}{4}^2 + \lambda_3\fdf{1}{2}\fdf{3}{2} + \lambda_4\fdf{1}{4}\fdf{3}{4} + h.c. \nonumber\\
&& a_1^{(1)} = \sqrt{i}\cdot\diag(i,-1,-i,1)\,,\quad a_2^{(1)} = \sqrt{i}\cdot\diag(1,-1,1,1)\,, \label{extension-Z4Z2-a1}\\[2mm]
\mbox{option 2:}&& V_2 = \lambda_1\fdf{1}{2}^2 + \lambda_2\fdf{3}{4}^2 + \lambda_3\fdf{1}{3}\fdf{2}{4} + \lambda_4\fdf{1}{4}\fdf{2}{3}  + h.c.\nonumber\\
&& a_1^{(2)} = \diag(i,i,-1,1)\,,\quad a_2^{(2)} = \diag(-1,1,-1,1)\,,  \label{extension-Z4Z2-a2}\\[2mm]
\mbox{option 3:}&& V_3 = \lambda_1\fdf{1}{2}^2 + \lambda_2\fdf{1}{3}^2 + \lambda_3\fdf{2}{3}^2 + \lambda_4\fdf{1}{4}\fdf{3}{4}  + h.c.\nonumber\\
&& a_1^{(3)} = i^{3/4}\cdot\diag(i,i,-i,1)\,,\quad a_2^{(3)} = \diag(-1,1,-1,1)\,.  \label{extension-Z4Z2-a3}
\end{eqnarray}
Below we will build non-abelian extensions for each of these three options.

Let us stress here an important fact. It follows from the general analysis of rephasing transformations 
\cite{Ivanov:2011ae} that if an NHDM scalar potential contains less than $N-1$ rephasing-sensitive terms,
it acquires a continuous rephasing symmetry. Thus, if we want to avoid accidental continuous symmetries in the 4HDM scalar sector,
the potential must have at least three rephasing-sensitive terms.
Each of the three $\Z_4\times\Z_2$-symmetric options shown above contains four terms,
and no continuous symmetry is present.
However if a specific extension requires that any two of these coefficients vanish,
the potential will automatically acquire an accidental continuous rephasing symmetry. 

\subsection{The automorphism group of $\Z_4\times\Z_2$ and its conjugacy classes}

The automorphism group of $\Z_4\times\Z_2$ is 
\begin{equation}
\Aut(\Z_4\times\Z_2)\simeq D_4 = \langle b, c\,|\, b^4 = c^2 = e,\, c b c = b^{-1}\rangle\,.
\end{equation}
The two automorphisms $b$ and $c$, which generate $\Aut(\Z_4\times\Z_2)$, act on the group $\Z_4\times\Z_2$ in the following way:
\begin{eqnarray}
b: \ \left\{
\begin{aligned}
	&a_1 \mapsto a_1 a_2\\
	&a_2 \mapsto a_2a_1^2
\end{aligned}
\right.,
\qquad 
\tilde b \equiv b^2: \ \left\{
\begin{aligned}
	&a_1 \mapsto a_1^{-1}\\
	&a_2 \mapsto a_2
\end{aligned}
\right.,
\qquad 
c: \ \left\{
\begin{aligned}
	&a_1 \mapsto a_1\\
	&a_2 \mapsto a_2a_1^2
\end{aligned}
\right.
\label{extension-Z4Z2-bc}
\end{eqnarray}
Next, we need to list all subgroups of $D_4$, and also arrange them into conjugacy classes. 
We remind the reader that $D_4$ (the symmetry group of the square) contains only one 
subgroup $\Z_4$ (the rotations of the square) but five subgroups $\Z_2$ (reflections of the square).
These five $\Z_2$'s form three conjugacy classes:
two reflections parallel to the sides (generated by $c$ or $b^2 c$), 
two reflections along the diagonals (generated by $bc$ or $b^3 c$), 
and the unique point reflection $\tilde b$, which commutes with 
any symmetry of the square and, group-theoretically, corresponds to the center of $D_4$.
It is straightforward to check that there are also two $\Z_2 \times \Z_2$ subgroups, each forming its own conjugacy class.
Thus, we obtain eight proper non-trivial subgroups of $D_4$, which are explicitly listed in Table~\ref{table-subgroups-D4}.

\begin{table}[H]
\centering
\begin{tabular}[t]{cc|ccc}
  \toprule
  conjugacy classes & groups & option 1 & option 2 & option 3 \\
  \midrule
  $\Z_2$ & $\{e,\,b^2 \}$ & $\checkmark$ & $\checkmark$ & {\gray $\times$} \\[1mm]
  $\Z_2$ & $\{e,\,b^2c\}$ & $\checkmark$ & $\checkmark$ & {\gray $\times$} \\
       & $\{e,\,c\}$ & $\checkmark$ & $\checkmark$ & {\gray $\times$} \\[1mm]
  $\Z_2$ & $\{e,\,bc\}$ & {\gray $\times$} & $\checkmark$ & $\checkmark$ \\
         & $\{e,\,b^3c\}$ & {\gray $\times$} & $\checkmark$ & {\gray $\times$} \\[1mm]
  $\Z_4$ & $\{e,\,b,\,b^2,\,b^3\}$ & {\gray $\times$} & $\checkmark$ & {\gray $\times$} \\[1mm]
  $\Z_2 \times \Z_2$ & $\{e,\,b^2,\,c,\,b^2c \}$ & $\checkmark$ & $\checkmark$ & {\gray $\times$} \\[1mm]
  $\Z_2 \times \Z_2$ & $\{e,\,b^2,\,bc,\,b^3c \}$ & {\gray $\times$} & $\checkmark$ & {\gray $\times$} \\
  \bottomrule
\end{tabular}
\caption{The non-trivial proper subgroups of $D_4$ arranged by the conjugacy classes.
For each subgroup, the extensions are possible (marked with $\checkmark$)
or impossible ({\gray $\times$}) to construct, depending on which option we choose out of the three possible options 
for the $\Z_4\times \Z_2$ group, given in Eqs.~\eqref{extension-Z4Z2-a1}--\eqref{extension-Z4Z2-a3}.
}
\label{table-subgroups-D4}
\end{table}

As we will find below, not all subgroups can be used to build extensions. 
Moreover, the ability of a subgroup of $D_4$ to produce a non-abelian extension by $\Z_4\times \Z_2$ depends 
on the particular implementation of $\Z_4\times \Z_2$ in the 4HDM, given by Eqs.~\eqref{extension-Z4Z2-a1}--\eqref{extension-Z4Z2-a3}.
In what follows, we will first select this implementation (options 1, 2, or 3) and then go through 
the list of subgroups of $D_4$, trying each time to build an extension.
To help the reader navigate through the rather laborious procedure, we summarize in the same Table~\ref{table-subgroups-D4}
which subgroups for which option yield a non-trivial extension.

\subsection{Building $\Z_4\times\Z_2$ extensions: option 1}

\subsubsection{Three choices of $\Z_2$ extensions with no other extensions available}

We begin constructing extensions for the first implementation of the group $\Z_4\times \Z_2$, which is given in Eq.~\eqref{extension-Z4Z2-a1}.
For the reader's convenience, we repeat it here:
\begin{eqnarray}
	\mbox{option 1:}&& V_1 = \lambda_1\fdf{1}{3}^2 + \lambda_2\fdf{2}{4}^2 + \lambda_3\fdf{1}{2}\fdf{3}{2} + \lambda_4\fdf{1}{4}\fdf{3}{4} + h.c. \nonumber\\
	&& a_1 = \sqrt{i}\cdot\diag(i,-1,-i,1)\,,\quad a_2 = \sqrt{i}\cdot\diag(1,-1,1,1)\,, \label{extension-Z4Z2-a1-again}
\end{eqnarray}
The prefactors $\sqrt{i}$ in $a_1$ and $a_2$ make it clear that the equation of the form $p^{-1}a_1p = a_1a_2\cdot i^r$ does not have
any solution for $p$ for any integer $r$. In general, one must have either an even number of $a_i$'s or an odd number of $a_i$'s simultaneously on the left-hand and the right-hand sides of this equation. Looking at the definitions of the automorphisms $b$ and $c$ in Eq.~\eqref{extension-Z4Z2-bc},
we see that $c$ and even powers of $b$ can be used, while the automorphisms $b$, $b^3$, $b^3c$, $bc$ cannot be represented as $PSU(4)$ transformations. 
It is this reasoning which allows us to cross out half of the subgroups from Table~\ref{table-subgroups-D4}, option 1.

Another remark concerns the possible relation of the second and the third lines of this list.
These two $\Z_2$ subgroups, $\lr{b^2c}$ and $\lr{c}$, belong to one conjugacy class and are conjugate inside $D_4$.
That is, the equation $q^{-1} c q = b^2 c$ has a solution --- in fact, four solutions --- inside $D_4$: 
$q = b,\, b^3,\, bc,\, b^3c$. However these are precisely the elements of $D_4$ which cannot be represented by $PSU(4)$ transformations.
Therefore, we will not be able to use the short-cut argument given by the theorem we proved in Section~\ref{non-abelian-Symmetries-in-NHDM}
and will need to consider the second and third lines of Table~\ref{table-subgroups-D4} separately.

\subsubsection{Using the first $\Z_2$} 

Consider the first $\Z_2$ group from Table~\ref{table-subgroups-D4}, which is generated by $\tilde b \equiv b^2$.
The action of $\tilde b$ is defined in Eq.~\eqref{extension-Z4Z2-bc}. The corresponding matrix in $SU(4)$
must satisfy the following equations: 
\begin{equation}
\left\{
\begin{aligned}
	& \tilde b^{-1}a_1 \tilde b = a_1^3\cdot i^{r_1}\\
	& \tilde b^{-1}a_2 \tilde b = a_2\cdot i^{r_2}
\end{aligned}
\right.
\quad \Rightarrow \quad
\left\{
\begin{aligned}
	& a_1 \tilde b = \tilde b a_1^3 \cdot i^{r_1}\\
	& a_2 \tilde b = \tilde b a_2 \cdot i^{r_2}
\end{aligned}
\right.\,.\label{extension-Z4Z2-1-eq1}
\end{equation}
Here, $r_1$ and $r_2$ are arbitrary integers.
The presence of factors $i^{r_i}$ reflects the fact that,
although we work with matrices from $SU(4)$, all equalities are defined modulo to the center of $SU(4)$,
that is, modulo to powers of $i$, see details in \cite{Shao:2023oxt}.
Using the methods developed in \cite{Shao:2023oxt}, we solve this system of linear matrix equations
and find that $\tilde b$ must be of the form
\begin{equation}
\tilde b = 
\begin{pmatrix}
	0 & 0 & b_{13} & 0 \\
	0 & b_{22} & 0 & 0 \\
	b_{31} & 0 & 0 & 0 \\
	0 & 0 & 0 & b_{44} 
\end{pmatrix}\,,
\label{extension-Z4Z2-1-eq2}
\end{equation}
where all entries are pure phase factors. Their phases are not constrained by group theory
but can be determined from the phases of the complex coefficients of the corresponding potential $V_1$.

Knowing the action of $\tilde b$ on the generators $a_1$ and $a_2$ does not specify the extension 
$G = (\Z_4 \times \Z_2)\,.\,\Z_2$ uniquely.
We still need to define the value of $\tilde b^2$ inside $G$. One choice is to set $\tilde b^2 = e$, 
which results in a split extension,
also called the semi-direct product: $(\Z_4 \times \Z_2)\rtimes \Z_2 \simeq D_4 \times \Z_2$.
In this expression, the $D_4$ factor is generated by $a_1$ and $\tilde b$, while the last $\Z_2$ factor is the same 
$\Z_2$ subgroup generated by $a_2$.

Consider now the potential $V_1$ in Eq.~\eqref{extension-Z4Z2-a1-again}.
Upon a suitable rephasing, we can set the coefficient $\lambda_1$ real.
In this basis, we observe that 
\begin{equation}
\tilde b = \sqrt{i}
\begin{pmatrix}
	0 & 0 & 1 & 0 \\
	0 & -1 & 0 & 0 \\
	1 & 0 & 0 & 0 \\
	0 & 0 & 0 & -1 \\
\end{pmatrix}
\label{Z4xZ2_Option1_by_Z2_Z2xD4}
\end{equation}
becomes a symmetry of the potential without any additional requirement on the coefficients of $V_1$.
In other words, if we implement the group $\Z_4\times \Z_2$ using option 1, 
$V_1$ acquires an accidental {\em discrete} symmetry,
so that its total symmetry content is automatically enhanced to $D_4 \times \Z_2$.
Thus, the only condition for a $\Z_4\times\Z_2$ model, option 1, to become invariant under $D_4 \times \Z_2$
is that the rephasing-insensitive part $V_0$ is invariant under $\phi_1 \leftrightarrow\phi_3$.

Having obtained the generic form of $\tilde b$ in Eq.~\eqref{extension-Z4Z2-1-eq2}, 
we can also assume that $\tilde b^2 \not = e$ 
but instead lies inside $\Z_4 \times \Z_2$. 
Since the elements $(\tilde b^2)_{11} = (\tilde b^2)_{33}$, the only available options for $\tilde b^2$ are 
$a_2$, $a_1^2$, or $a_1^2a_2$. All three choices will result in non-split extensions.
Under the first choice $\tilde b^2 = a_2$, the generator $\tilde b$ takes, 
in a suitable real-$\lambda_1$ basis, the following form:
\begin{equation}
\tilde b = i^{1/4}
\begin{pmatrix}
	0 & 0 & 1 & 0 \\
	0 & i & 0 & 0 \\
	1 & 0 & 0 & 0 \\
	0 & 0 & 0 & 1 \\
\end{pmatrix}\,.
\label{Z4xZ2_Option1_by_Z2_Z2xQ4}
\end{equation}
Imposing this symmetry on $V_1$ flips the signs of the $\lambda_2$ and $\lambda_3$ terms. 
Thus, we are forced to set $\lambda_2 = \lambda_3 = 0$. But then we have too few rephasing sensitive terms,
and an accidental $U(1)$ symmetry emerges.
Similarly, under the second choice $\tilde b^2 = a_1^2$, we find that we are forced 
to set $\lambda_3 = \lambda_4 = 0$. 
The last choice $\tilde b^2 = a_1^2a_2$ leads to the vanishing $\lambda_2$ and $\lambda_4$. 
Thus, all attempts of construct a model based on a finite non-split extension of the first $\Z_2$ subgroup fail.

The bottom line is: extending $\Z_2 = \{e, \tilde b\}$ by $\Z_4 \times \Z_2$, implemented as in Eq.~\eqref{extension-Z4Z2-a1-again}, 
is only possible for the split extension $(\Z_4 \times \Z_2)\rtimes \Z_2 \simeq D_4 \times \Z_2$.
The extra generator is the permutation $\tilde b$ given by Eq.~\eqref{Z4xZ2_Option1_by_Z2_Z2xD4}.
The only condition for this symmetry group is that $V_0$ is invariant under $\phi_1 \leftrightarrow\phi_3$.
The $V_1$ part of the potential is automatically invariant under $\tilde b$.

\subsubsection{Using the second $\Z_2$}

Next, we pick $\Z_2\simeq \langle b^2c\rangle$ and denote $d = b^2c$. The corresponding equations are
\begin{equation}
	\left\{
	\begin{aligned}
		& d^{-1}a_1 d = a_1^3\cdot i^{r_1}\\
		& d^{-1}a_2 d = a_2a_1^2\cdot i^{r_2}
	\end{aligned}
	\right.
	\quad \Rightarrow \quad
	\left\{
	\begin{aligned}
		& a_1 d = d a_1^3 \cdot i^{r_1}\\
		& a_2 d = d a_2a_1^2 \cdot i^{r_2}
	\end{aligned}
	\right.\,,\label{extension-Z4Z2-1-eq2-1}
\end{equation}
which only has solutions for $r_1 = 1$, $r_2 = 1$:
\begin{equation}
	d = 
	\begin{pmatrix}
		d_{11} & 0 & 0 & 0 \\
		0 & 0 & 0 & d_{24} \\
		0 & 0 & d_{33} & 0 \\
		0 & d_{42} & 0 & 0 
	\end{pmatrix}\,.
	\label{extension-Z4Z2-1-eq2-2}
\end{equation}
In terms of the Higgs doublets, this transformation permutes $\phi_2\leftrightarrow \phi_4$. 
Since $d^2$ has the form $\diag(x, y, z, y)$, it can be either $e$ or 
a non-trivial element of $\Z_4\times\Z_2$, namely, $a_1^2$, $a_1a_2$, $a_1^3a_2$. 

We first consider the split extension by setting $d^2 = e$. The non-abelian group obtained in this way is labeled as
\begin{equation}
G = (\Z_4 \times \Z_2)\rtimes \Z_2 = {\tt SmallGroup(16,13)}\,\label{G1}
\end{equation}
where we indicated the {\tt GAP} id of this group of order 16. 
This group is also known as the Pauli group $G_1$ generated by the three Pauli matrices under multiplication.
In group theoretic terms, it can be also defined as the central product of $D_4$ and $\Z_4$, 
denoted by $\Z_4\circ D_4$. 
The transformation $d$, in a suitable basis, has the form
\begin{equation}
	d = \sqrt{i}
	\begin{pmatrix}
		-1 & 0 & 0 & 0 \\
		0 & 0 & 0 & 1 \\
		0 & 0 & -1 & 0 \\
		0 & 1 & 0 & 0 \\
	\end{pmatrix}\,.
	\label{extension-Z4Z2-eq2-3}
\end{equation}
We select a basis with real $\lambda_2$, so the invariance under Eq.~\eqref{extension-Z4Z2-eq2-3} only requires $\lambda_3 = \lambda_4$,
together with the constraints in $V_0$ obtained by imposing the invariance under $\phi_2\leftrightarrow\phi_4$. 

We can also try to build non-split extensions by solving matrix equation $d^2 = i^r\cdot a_1^2$. 
A solution exists for $r=-1$ 
and has the following form:
\begin{equation}
	d = 
	\begin{pmatrix}
		i & 0 & 0 & 0 \\
		0 & 0 & 0 & 1 \\
		0 & 0 & i & 0 \\
		0 & 1 & 0 & 0 \\
	\end{pmatrix}\,.
	\label{extension-Z4Z2-eq2-4}
\end{equation}
Invariance under this $d$ requires $\lambda_3 = -\lambda_4$ in the real $\lambda_2$ basis. 
This construction also leads to the same group $G_1$ as in Eq.~\eqref{G1}, 
but we arrived at it using non-split extension procedure.
This is not a coincidence: a potential with a real $\lambda_2$ and $\lambda_3 = -\lambda_4$ 
can be transformed into a potential with another real $\lambda_2$ and $\lambda_3 = \lambda_4$
by the basis change $\phi_{1,2,3} \mapsto \phi_{1,2,3}$, $\phi_4 \mapsto i \phi_4$.
Alternatively, we can note that a potential invariant under $d$ from Eq.~\eqref{extension-Z4Z2-eq2-4}
is also invariant under $d' = d a_2$, whose square is $(d')^2 = -i\cdot \id_4$.
Thus, the group $G = A\,.\,\Z_2$ with $\Z_2 = \lr{d}$ can also be represented as 
$G = A\rtimes\Z'_2$, where $\Z'_2 = \lr{d'}$.
In short, the attempt at a non-split extension leads us to the same result as the split extension
because the two models are related by a mere basis change and represent the same physical situation.

We also checked that the other non-split extension attempts, $d^2 = i^r\cdot a_1a_2$ and $d^2 = i^r\cdot a_1^3a_2$, 
lead to continuous symmetries and are not realizable as 4HDM discrete symmetries.

\subsubsection{Using the third $\Z_2$}

Next, let us take the $\Z_2$ group $\{e,c\}$ from the third line in Table~\ref{table-subgroups-D4}.
The action of $c$ is defined in Eq.~\eqref{extension-Z4Z2-bc}, and the corresponding equations are
\begin{equation}
\left\{
\begin{aligned}
	& c^{-1}a_1 c = a_1\cdot i^{r_1}\\
	& c^{-1}a_2 c = a_2a_1^2\cdot i^{r_2}
\end{aligned}
\right.
\quad \Rightarrow \quad
\left\{
\begin{aligned}
	& a_1 c = c a_1 \cdot i^{r_1}\\
	& a_2 c = ca_2a_1^2 \cdot i^{r_2}
\end{aligned}
\right.\,.\label{extension-Z4Z2-1-eq3}
\end{equation}
This system has solutions only for $r_1 = 2$, $r_2 = 1$, leading to
\begin{equation}
c =
\begin{pmatrix}
	0 & 0 & c_{13} & 0 \\
	0 & 0 & 0 & c_{24} \\
	c_{31} & 0 & 0 & 0 \\
	0 & c_{42} & 0 & 0 \\
\end{pmatrix}\,,
\label{extension-Z4Z2-1-eq4}
\end{equation}
which permutes $\phi_1\leftrightarrow \phi_3$ and $\phi_2\leftrightarrow\phi_4$ simultaneously.
The split extension, $c^2 = e$, leads to the same Pauli group $G_1$ mentioned above. 
In the basis, where $\lambda_1$ and $\lambda_2$ are real,
the transformation $c$ takes the form
\begin{equation}
c =
\begin{pmatrix}
	0 & 0 & 1 & 0 \\
	0 & 0 & 0 & 1 \\
	1 & 0 & 0 & 0 \\
	0 & 1 & 0 & 0 \\
\end{pmatrix}\,,
\label{extension-Z4Z2-1-eq5}
\end{equation}
The only condition we must impose on the potential $V_1$ in Eq.~\eqref{extension-Z4Z2-a1-again}
is $\lambda_3 = \lambda_4$, which must be accompanied with condition that the rephasing-insensitive part $V_0$ 
be invariant under the simultaneous change $\phi_1 \leftrightarrow \phi_3$ and $\phi_2 \leftrightarrow \phi_4$.
Note that this construction is nearly identical to the second $\Z_2$ extension; 
they differ only in $V_0$, not in $V_1$.

Attempts to build non-split extensions proceeds along the same lines as above. 
The only possibility is to set $c^2 = a_1^2$. In the real $\lambda_1$ and $\lambda_2$ basis, 
it leads to 
\begin{equation}
c = \sqrt{i}
\begin{pmatrix}
	0 & 0 & i & 0 \\
	0 & 0 & 0 & 1 \\
	i & 0 & 0 & 0 \\
	0 & 1 & 0 & 0 \\
\end{pmatrix}\,,
\label{extension-Z4Z2-1-eq6}
\end{equation}
Then, if $\lambda_3 = - \lambda_4$ and if, in addition, $V_0$ is invariant under 
$\phi_1 \leftrightarrow \phi_3$ together with $\phi_2 \leftrightarrow \phi_4$,
the full potential becomes invariant under this $c$.
The total symmetry group is again $G_1$; thus, we recover the split extension in disguise.

\subsubsection{Using $\Z_2 \times \Z_2$}

We have already established that $V_1$ is automatically invariant under $\tilde b$ in Eq.~\eqref{Z4xZ2_Option1_by_Z2_Z2xD4}.
Let us now continue with the above case symmetric under $c$ and combine it with $\tilde b$,
which brings us to the first $\Z_2 \times \Z_2$ group of Table~\ref{table-subgroups-D4}.
In order to achieve this symmetry, we need to impose an additional constraint on $V_0$.
As we saw, $c$ requires invariance of simultaneous permutation $\phi_1\leftrightarrow \phi_3$ and $\phi_2\leftrightarrow \phi_4$,
which is a less stringent constraint than invariance under $\phi_1\leftrightarrow \phi_3$ and $\phi_2\leftrightarrow \phi_4$, separately. 
Indeed, the $\fdf{1}{1}\fdf{2}{2} + \fdf{3}{3}\fdf{4}{4}$ in $V_0$ are invariant under $c$ but not under $\tilde{b}\in \Z_2\times\Z_2$. 

Let us now identify the symmetry group emerging in this case.
With $\tilde b \equiv b^2$ given in Eq.~\eqref{Z4xZ2_Option1_by_Z2_Z2xD4} and $c$ given in Eq.~\eqref{extension-Z4Z2-1-eq5},
we can verify that $\tilde b^2 = c^2 = e$ as well as $[\tilde b, c] = e$.
Thus, we get the split extension
\begin{equation}
	G = (\Z_4\times\Z_2)\rtimes (\Z_2 \times \Z_2) = {\tt SmallGroup(32,49)}\,,\label{extension-Z4Z2-1-extra}
\end{equation}
which is known as the extra-special group of order 32, plus-type, and is labeled as $\mathbf{2^{1+4}_+}$.

It is also possible to build such $c$ that $c^2=e$, but the commutator $[\tilde b,c] \not = e$ although it still lies inside $\Z_4 \times \Z_2$.
This allows us to consider non-split extensions of the form $ (\Z_4\times\Z_2)\,.\,(\Z_2 \times \Z_2)$. 
We checked all choices for $[\tilde b,c]$ and found that many lead to continuous symmetries.
For example, if $[\tilde b,c] = a_1a_2$, then $c$ requires $\lambda_1$ to be imaginary while $\tilde b$ requires
it to be real. Thus, we must set $\lambda_1 = 0$, but in this case the potential $V_1$ acquires the continuous symmetry 
of the form $\diag(e^{i\alpha}, 1, e^{-i\alpha}, 1)$.
The net result is that, by combining $\tilde b$ and any $c$ of the form of Eq.~\eqref{extension-Z4Z2-1-eq4}, 
we can only arrive at ${\tt SmallGroup(32,49)}$ through a split or non-split extension procedure. 
However since the group is the same, we end up only at the split extension of
$\Z_2\times \Z_2$ by $\Z_4\times\Z_2$.

This construction wraps up all the extension cases we have with the first option for $\Z_4\times\Z_2$.

\subsection{Building $\Z_4\times\Z_2$ extensions: option 2}\label{section-Z4Z4-option2}

Next, we consider the second way the $\Z_4\times \Z_2$ group can be implemented in the 4HDM. 
The potential and the generators $a_1$ and $a_2$ are given in Eqs.~\eqref{extension-Z4Z2-a2};
we repeat them here for the reader's convenience:
\begin{eqnarray}
\mbox{option 2:}&& V_2 = \lambda_1\fdf{1}{2}^2 + \lambda_2\fdf{3}{4}^2 + \lambda_3\fdf{1}{3}\fdf{2}{4} + \lambda_4\fdf{1}{4}\fdf{2}{3}  + h.c.\nonumber\\
&& a_1 = \diag(i,i,-1,1)\,,\quad a_2 = \diag(-1,1,-1,1)\,.  \label{extension-Z4Z2-a2-again}
\end{eqnarray}
Unlike in option 1, these generators do not carry the prefactors $\sqrt{i}$. 
As a result, all automorphisms of the $\Z_4\times\Z_2$ can be represented as $PSU(4)$ transformations.
This simplifies our task as we do not need to consider the subgroups in Table~\ref{table-subgroups-D4}
belonging to the same conjugacy class.

We will now describe the results giving fewer details than before because the methods are the same.
We start again with the first $\Z_2$ from Table~\ref{table-subgroups-D4}.
Following the similar steps, we can solve for the matrix $\tilde b$, which in this cases exchanges 
$\phi_1 \leftrightarrow \phi_2$ and $\phi_3 \leftrightarrow \phi_4$.
In the basis of real $\lambda_1$ and $\lambda_2$, this transformation is automatically a symmetry of $V_2$.
Thus, we only need to require $V_0$ to be invariant under these permutations, and in this way we again
obtain the non-abelian group $G = (\Z_4 \times \Z_2)\rtimes \Z_2 = D_4\times \Z_2$.

An attempt to build a non-split extension leads to the condition $\tilde b^2 = a_1^2$. 
But such a $\tilde b$ forces us to set $\lambda_3 = \lambda_4 = 0$, leading to an accidental continuous symmetry.

From the next conjugacy class we select the third $\Z_2$ subgroup $\{e, c\}$.
In the real $\lambda_1$ basis, the matrix $c$ only exchanges $\phi_1 \leftrightarrow \phi_2$. 
We need to require that $\lambda_3 = \lambda_4$ and to make sure that $V_0$ is invariant under $\phi_1 \leftrightarrow \phi_2$.
In this way, we again arrive at the the Pauli Group $G_1 = {\tt SmallGroup(16,13)}$ as in Eq.~\eqref{G1}.
Attempts at non-split extensions do not produce any new options.

Unlike for option 1, the third conjugacy class is now available, lines 4 and 5 of Table~\ref{table-subgroups-D4}.
We choose the subgroup $\{e, bc\}$ to construct extensions. 
The solution for the generator $d = bc$ corresponds, in a suitable basis, to the simultaneous exchange 
$\phi_1 \leftrightarrow \phi_4$ and $\phi_2 \leftrightarrow \phi_3$:
\begin{equation}
	d =  
	\begin{pmatrix}
		0 & 0 & 0 & 1 \\
		0 & 0 & 1 & 0 \\
		0 & 1 & 0 & 0 \\
		1 & 0 & 0 & 0 \\
	\end{pmatrix}\,.
	\label{extension-Z4Z2-2-eq0}
\end{equation}
The conditions for this symmetry to be present are $\lambda_1 = \lambda_2^*$ and
real $\lambda_3$ and $\lambda_4$, plus matching conditions on $V_0$. 
The symmetry group we obtain by combining $\Z_4\times \Z_2$ generated by $a_1$, $a_2$ and
$\Z_2$ generated by $d$ is also a semi-direct product $(\Z_4\times \Z_2)\rtimes \Z_2$ but a different one:
\begin{equation}
G = (\Z_4\times \Z_2)\rtimes \Z_2 = {\tt SmallGroup(16,3)}\,.
\end{equation} 
The non-split extension procedure with $d^2 = a_1^2a_2$ bring us again to this group,
which is thus a split extension in disguise.

If $V_0$ is invariant under 
$\phi_1 \leftrightarrow \phi_2$ and, independently, under $\phi_3 \leftrightarrow \phi_4$,
the full symmetry group is further enhanced,
through either a split or a non-split extension procedure.
We found that extension by line 7 of Table~\ref{table-subgroups-D4} leads to the same group {\tt SmallGroup(32,49)}
as in Eq.~\eqref{extension-Z4Z2-1-extra}, while extension by line 8 produces a new option, 
the group denoted as {\tt SmallGroup(32,27)}.

With option 2, we can also use the subgroup $\Z_4$ to build an extension, which was impossible for option 1.
The action of the generator $b$ defined in Eq.~\eqref{extension-Z4Z2-bc} tells us that the symmetry group 
we are going to construct is the following group of order 32:
\begin{equation}
	G = (\Z_4\times\Z_2)\rtimes\Z_4\simeq {\tt SmallGroup(32, 6)}\,,\label{extension-Z4Z2-2-eq1}
\end{equation}
which is also known as the faithful semi-direct product $(\Z_2)^3\rtimes\Z_4$.
The equations $b^{-1}a_1b =a_1 a_2 \cdot i^{r_1}$ and 
$b^{-1}a_2b = a_1^2 a_2 \cdot i^{r_2}$ have the following generic solution:
\begin{equation}
b = 
\begin{pmatrix}
	0 & 0 & b_{13} & 0 \\
	0 & 0 & 0 & b_{24} \\
	0 & b_{32} & 0 & 0 \\
	b_{41} & 0 & 0 & 0 \\
\end{pmatrix}\,,
\label{extension-Z4Z2-2-eq2}
\end{equation}
that is, the cyclic permutation $\phi_1 \mapsto \phi_3 \mapsto \phi_2 \mapsto \phi_4 \mapsto \phi_1$,
possibly corrected by the phase rotations.
Since $b^4$ is proportional to $\id_4$, the extension can only be split.
This transformation is the symmetry of $V_2$ if $\lambda_1 = \lambda_2$ are real
and $\lambda_3 = \lambda_4^*$. In addition, it strongly constrains the rephasing-insensitive part $V_0$
reducing it to
\begin{eqnarray}
V_0 &=& m^2 \left(\fdfn{1}{1} + \fdfn{2}{2} + \fdfn{3}{3} + \fdfn{4}{4}\right)
+ \Lambda \left(\fdfn{1}{1} + \fdfn{2}{2} + \fdfn{3}{3} + \fdfn{4}{4}\right)^2\nonumber\\
&&+\Lambda' \left(\fdfn{1}{1} + \fdfn{2}{2}\right)\left(\fdfn{3}{3} + \fdfn{4}{4}\right) 
+\Lambda'' \left[\fdf{1}{1}\fdf{2}{2} + \fdf{3}{3}\fdf{4}{4}\right]\nonumber\\
&& + \tilde\Lambda' \left(|\fdfn{1}{3}|^2 + |\fdfn{2}{3}|^2 + |\fdfn{1}{4}|^2 + |\fdfn{2}{4}|^2 \right)
+ \tilde\Lambda''\left(|\fdfn{1}{2}|^2 + |\fdfn{3}{4}|^2\right)\,.\label{extension-Z4Z2-2-eq3}
\end{eqnarray}

The final step in our classification of the extensions based on $\Z_4\times \Z_2$ is to use the full $D_4$.
For the split extensions, the resulting group is
\begin{equation}
G = (\Z_4\times\Z_2) \rtimes D_4 \simeq {\tt SmallGroup(64, 138)}\,.\label{extension-Z4Z2-2-D4}
\end{equation}
This is a group of order 64 also known as the unitriangular matrix group of degree 4 over the field $\mathbb{F}_2$, 
denoted as $UT(4,2)$. 
In order to obtain this symmetry group, we just need to impose invariance under $b$ and $c$.
This is achieved by using $V_0$ given in Eq.~\eqref{extension-Z4Z2-2-eq3} and $V_2$ of the form
\begin{equation}
	V_2 = \lambda_1\left[\fdf{1}{2}^2 + \fdf{3}{4}^2 + h.c.\right] + 
	\lambda_3\left[\fdf{1}{3}\fdf{2}{4} + \fdf{1}{4}\fdf{2}{3} + h.c.\right], \label{extension-Z4Z2-2-eq4}
\end{equation}
with all parameters real.
One can describe the full symmetry content of the potential $V_0 + V_2$ as invariance under $a_1$ and $a_2$ 
given in Eq.~\eqref{extension-Z4Z2-a2-again} as well as the following types of permutations: 
$\phi_1 \leftrightarrow \phi_2$, $\phi_3 \leftrightarrow \phi_4$, and the cyclic permutation
$\phi_1 \mapsto \phi_3 \mapsto \phi_2 \mapsto \phi_4 \mapsto \phi_1$.
This potential contains only eight free parameters and leads to remarkably constrained scalar sector of the model.

\subsection{Building $\Z_4\times\Z_2$ extensions: option 3}

Option 3 for implementing the group $\Z_4\times\Z_2$ in the 4HDM, given in Eq.~\eqref{extension-Z4Z2-a3},
is a special one. Due to the presence of the factor $i^{3/4}$ in the definition of $a_1$, 
most of actions defined in Eq.~\eqref{extension-Z4Z2-bc} do not admit solutions.
The only automorphism which can have solution if $d = bc$, which maps $a_1$ to $a_1a_2$ and keeps $a_2$ unchanged.
Using the same methods as before, we find that the generator $d$ swaps $\phi_1$ and $\phi_3$, 
possibly accompanied with phase factors.

If $d^2 = e$, we deal with the split extension $(\Z_4\times \Z_2)\rtimes \Z_2 = {\tt SmallGroup(16,3)}$,
just as we already encountered in option 2. In order to arrive at this symmetry group, we need to require that the coefficients of $V_3$ 
in Eq.~\eqref{extension-Z4Z2-a3}, in a suitable basis, satisfy $\lambda_1^* = \lambda_3$ and $\lambda_2$ is real.
As for the non-split extensions, the only choice which does not lead to continuous accidental symmetries is 
$d^2 = a_1^2 a_2$, which leads to the same symmetry group ${\tt SmallGroup(16,3)}$.

\newpage

\section{Calculating Extensions by $\Z_2\times\Z_2\times\Z_2$}\label{section-Z2Z2Z2}

\subsection{$(\Z_2)^3$ as a vector space and its automorphisms}

The group $\Z_2\times\Z_2\times\Z_2$, or $(\Z_2)^3$ for short, can be implemented within the 4HDM just via sign flips of individual doublets.
We have three, not four, $\Z_2$ factors just because flipping the signs of the first three doublets is equivalent to flipping the signs of fourth doublet.
The three generators of the group $(\Z_2)^3$ can be selected as
\begin{equation}
a_1 = \sqrt{i}\cdot\diag(-1,1,1,1),\quad a_2 = \sqrt{i}\cdot\diag(1,-1,1,1),\quad a_3 = \sqrt{i}\cdot\diag(1,1,-1,1)\,,\label{E8-eq1}
\end{equation}
where the factors $\sqrt{i}$ are required to guarantee $\det a_i = 1$.
The potential invariant under sign flips includes, in addition to the rephasing-insensitive part $V_0$, the following collection of quartic terms:
\begin{equation}
V_1 = \lambda_{12}\fdf{1}{2}^2 + \lambda_{13}\fdf{1}{3}^2 + \lambda_{23}\fdf{2}{3}^2 
+ \lambda_{14}\fdf{1}{4}^2 + \lambda_{24}\fdf{2}{4}^2 + \lambda_{34}\fdf{3}{4}^2 + h.c.\,,
\label{E8-V1}
\end{equation}
where all the coefficients can be complex.
Using the package {\tt 4HDM Toolbox} \cite{TheCode}, we verified that any implementation of the group $\Z_2\times\Z_2\times\Z_2$
in the 4HDM can indeed be represented, in a suitable basis, by Eq.~\eqref{E8-eq1}.

It is convenient to view the group $(\Z_2)^3$ as a three-dimensional vector space over the finite field with two elements $\mathbb{F}_2 = \{0, 1\}$.
In this construction, we view $\hat{a}_1$, $\hat{a}_2$, $\hat{a}_3$ as basis vectors, and think of elements of $(\Z_2)^3$ as their linear combinations
$\hat{a} = m_1 \hat{a}_1 + m_2 \hat{a}_2 + m_3 \hat{a}_3$, where the numbers $m_i \in \mathbb{F}_2$.
In the usual multiplicative notation, such an element $g = a_1^{m_1}a_2^{m_2}a_3^{m_3}$. It is straightforward to check that the axioms 
of a vector space are satisfied.

An automorphism of $(\Z_2)^3$ is a rule which maps each $g$ defined by $m_i$ to $g'$ defined by $m'$. Since the group laws must be satisfied,
we conclude that the map $m_i \mapsto m'_i$ is linear and, therefore, it can be represented as a $3\times 3$ matrix with elements in $\mathbb{F}_2$.
For example, the exchange of generators $a_1 \leftrightarrow a_3$, $a_2 \mapsto a_2$ is indeed an automorphism and is represented by the matrix
\begin{equation}
{\mathfrak b} = 
\begin{pmatrix}
	0 & 0 & 1\\
	0 & 1 & 0\\
	1 & 0 & 0
\end{pmatrix}\,.\label{E8-b-example1}
\end{equation}
Here and below, we use the fraktur letters such as ${\mathfrak b}$ to represent automorphisms as acting 
in the vector space $(\mathbb{F}_2)^3$, which are 
not to be confused with the matrices $b \in SU(4)$ acting on the four doublets.
In order to find $b$ from the known form of ${\mathfrak b}$, we apply the methods introduced before.
For example, if ${\mathfrak b}$ is given by Eq.~\eqref{E8-b-example1},
we need to solve the following system of equations
\begin{equation}
\left\{
\begin{aligned}
	& b^{-1}a_1b = a_3\cdot i^{r_1}\\
	& b^{-1}a_2b = a_2\cdot i^{r_2}\\
	& b^{-1}a_3b = a_1\cdot i^{r_3}\\
\end{aligned}
\right.\;,\quad 
\mbox{with the solution} \quad 
b = 
\begin{pmatrix}
	0 & 0 & b_{13} & 0\\
	0 & b_{22} & 0 & 0\\
	b_{31} & 0 & 0 & 0\\
	0 & 0 & 0 & b_{44}\\
\end{pmatrix}\ \mbox{for}\ r_i=0\,.\label{E8-b-example2}
\end{equation}
Since automorphisms are invertible, $\det {\mathfrak b} \not = 0$. Then, as we work over the field $\mathbb{F}_2$, 
this determinant can only be equal to one.
This is why the automorphism group of $(\Z_2)^3$, the collection of all matrices ${\mathfrak b}$ with $\det {\mathfrak b} = 1$,
can be written as $GL(3,2) = SL(3,2)$.  
The order of this group is easy to establish. The group $(\Z_2)^3$ contains seven non-trivial elements of order 2. 
When defining an automorphism $f$, we can map $a_1$ to any of these seven, then we map $a_2$ to any of the remaining six,
and finally map $a_3$ to any of the remaining elements barring $f(a_1)f(a_2)$. Thus, 
$|\Aut((\Z_2)^3)| = 7\times 6 \times 4 = 168$.

Next, we use the database {\ttfamily GAP} to study some properties of $\Aut((\Z_2)^3)$ and its subgroups.
The group itself is labeled as {\ttfamily SmallGroup(168,42)} and has 179 subgroups.
Fortunately, many of these subgroups are conjugate to each other. Thanks to the Theorem proved in Section~\ref{abelian-Symmetries-in-NHDM},
we only need to classify the conjugacy classes of these subgroups and then consider only one example in each class.
Using {\ttfamily GAP}, we found that this group has 13 conjugacy classes of the non-trivial proper subgroups, which we list in Table~\ref{table-subgroup_GL32}.
\begin{table}[H]
\centering
\begin{tabular}[t]{c|cccccccccc}
	\toprule
	\makecell[c]{Representative\\Subgroups} & $\Z_2$ & $\Z_3$ & $\Z_4$ & $\Z_2\times\Z_2$ & $S_3$ & {\color{gray}$\Z_7$} & $D_4$ & $A_4$ & {\color{gray}$\Z_3\rtimes\Z_7$} & $S_4$ \\
	\midrule
	\makecell[c]{Number of \\ conjugacy classes} & 1 & 1 & 1 & 2 & 1 & {\color{gray}1} & 1 & 2 & {\color{gray}1} & 2 \\
	\bottomrule
\end{tabular}
\caption{The conjugacy classes of all non-trivial proper subgroups of $GL(3,2)$. The classes of subgroups which contain $\Z_7$ are not available in the 4HDM (see main text)
	and are shown in gray.}
\label{table-subgroup_GL32}
\end{table}

It turns out that not all automorphisms of the abstract group $(\Z_2)^3$ can be defined when $(\Z_2)^3$ is implemented 
as the symmetry group of the 4HDM as in Eq.~\eqref{E8-eq1}. Let us consider, for example, the automorphism
$a_1 \mapsto a_3$, $a_2 \mapsto a_1$, $a_3 \mapsto a_2a_3$. 
Its matrix $\mathfrak{f}$ and the system of equations for $f$ are
\begin{equation}
\mathfrak{f} = 
\begin{pmatrix}
	0 & 0 & 1\\
	1 & 0 & 0\\
	0 & 1 & 1
\end{pmatrix},\;
\mbox{defining}\;
\left\{
\begin{aligned}
	& f^{-1}a_1f = a_3\\
	& f^{-1}a_2f = a_1\\
	& f^{-1}a_3f = a_2a_3\,.
\end{aligned}
\right.\label{E8-f-fail}
\end{equation}
The map $\mathfrak{f}$ is well defined and has order 7, which can be verified by direct multiplication.
However the system of equations has no solutions for $f$. The obstacle is the last equation: 
even with the freedom of multiplication by integer powers of $i$, the equation $a_3f=fa_2a_3\cdot i^r$ cannot produce invertible matrices $f$.
The root of the problem is that $a_2a_3$ in the right-hand side does not possess the $\sqrt{i}$ factor to match this factor from $a_3$ in the left-hand side.
We conclude that extensions of $\Z_7$ --- and in fact of all groups which contain $\Z_7$ as a subgroup --- are impossible in the 4HDM.  

The lesson we draw from the above example is that, when constructing matrices $\mathfrak{b}$, we can only use four rows,
$(1,1,1)$, $(1,0,0)$, $(0,1,0)$, and $(0,0,1)$, and we must pick up three different ones, in any order. 
In this way, we can construct automorphisms of orders 2, 3 and 4.

Having done this exercise, we found two distinct families of transformations of order 2.
The three transformations
\begin{equation}
	{\mathfrak b}_1' = 
\begin{pmatrix}
	1 & 1 & 1\\
	0 & 0 & 1\\
	0 & 1 & 0
\end{pmatrix}\,,
\quad
	{\mathfrak b}_2' = 
\begin{pmatrix}
	0 & 0 & 1\\
	1 & 1 & 1\\
	1 & 0 & 0
\end{pmatrix}\,,
\quad
	{\mathfrak b}_3' = 
\begin{pmatrix}
	0 & 1 & 0\\
	1 & 0 & 0\\
	1 & 1 & 1
\end{pmatrix}
\label{E8-b'-examples}
\end{equation}
form the first family and share the property that they can be written as squares of order-4 transformations from $\Aut(A)$.
The second family contains six transformations of order 2 such as ${\mathfrak b}$ in Eq.~\eqref{E8-b-example1} 
and products of the type $({\mathfrak b}_1')^{-1}{\mathfrak b}{\mathfrak b}_1'$;
the members of this family cannot be written as a square of any order-4 transformation.

It turns out that all the transformations within each family are linked by some transformations $q \in \Aut(A)$
which can be represented by $\tau \in PSU(4)$. Therefore, these two families are exactly the two conjugacy classes
of physically equivalent models which we described in Section~\ref{non-abelian-Symmetries-in-NHDM}.
To build the full list of $\Z_2$-based extensions, it suffices to consider one representative transformation
from each family.

As for the transformations of order 3 and order 4, we found that all transformations of the same order can be linked 
by a $PSU(4)$ transformations. Thus, we need to consider only one representative $\Z_3$ and $\Z_4$ group.

\subsection{Building extensions}

Let us begin with extending $\Z_2$. Table~\ref{table-subgroup_GL32} tells us that all $\Z_2$ subgroups are conjugate to each other inside $GL(3,2)$,
but the above discussion suggests that we need to separately consider two representative groups,
which are not linked by any $PSU(4)$ transformation.

For the first example, we can select the automorphism ${\mathfrak b}$ in Eq.~\eqref{E8-b-example1}, 
which leads to $b$ of the form given in Eq.~\eqref{E8-b-example2}.
Squaring $b$, we get the diagonal matrix of the type $b^2=\diag(x,y,x,z)$, which can be proportional to $\mathbf{1}_4,\;  a_2,\;  a_1a_3,\;  a_1a_2a_3$.
The choice $b^2 = \mathbf{1}_4$ leads to the split extension $\Z_2\times D_4$. 
In a suitable basis, $b$ corresponds to the exchange $\phi_1 \leftrightarrow \phi_3$
and a sign flip, for example, of $\phi_2$. In order for $V_1$ in Eq.~\eqref{E8-V1} to be invariant under this transformation,
we require, in the real $\lambda_{13}$  basis, that $\lambda_{12} = \lambda_{23}^*$ and $\lambda_{14} = \lambda_{34}$.
Also, the rephasing-insensitive part of the potential, $V_0$, must be invariant under the exchange $\phi_1 \leftrightarrow \phi_3$.

For a non-split extension, we can select $b$ as in Eq.~\eqref{Z4xZ2_Option1_by_Z2_Z2xQ4}, so that $b^2 = a_2$.
In the real $\lambda_{13}$ basis, $V_1$ acquires this symmetry if $\lambda_{12} = -\lambda_{23}^*$, $\lambda_{14} = \lambda_{34}$,
and in addition $\lambda_{24} = 0$. The conditions for the other two non-split extensions can be immediately constructed.
In all three cases of non-split extensions, the total symmetry group is {\tt SmallGroup(16,3)}, which we already encountered
in Section~\ref{section-Z4Z4-option2} when extending $\Z_2$ by $\Z_4 \times \Z_2$.

The second $\Z_2$ example is ${\mathfrak b}_2'$ in Eq.~\eqref{E8-b'-examples}, which corresponds to 
the simultaneous transformation $\phi_1 \leftrightarrow \phi_3$ and $\phi_2 \leftrightarrow \phi_4$.
Clearly, it is equivalent to the cyclic permutation $\phi_1 \mapsto \phi_2 \mapsto \phi_3 \mapsto \phi_4 \mapsto \phi_1$ applied twice.
In this case, we can only have a split extension, and the total symmetry group is $\Z_2\times D_4$.

Table~\ref{table-subgroup_GL32} indicates, and the above discussion confirms, 
that we need to consider only one example of the $\Z_3$ subgroup.
A suitable order-3 automorphism is 
\begin{equation}
{\mathfrak c} = 
\begin{pmatrix}
	0 & 1 & 0\\
	0 & 0 & 1\\
	1 & 0 & 0
\end{pmatrix}
\end{equation}
which, in a suitable basis, leads to 
\begin{equation}
c = 
\begin{pmatrix}
	0 & 1 & 0 & 0\\
	0 & 0 & 1 & 0\\
	1 & 0 & 0 & 0\\
	0 & 0 & 0 & 1\\
\end{pmatrix}\,.\label{E8-c-1}
\end{equation}
This $c$ leads to the split extension $(\Z_2)^3\rtimes\Z_3\simeq A_4 \times \Z_2$, which reduces the potential
$V_1$ to 
\begin{equation}
V_1 = \lambda\left[\fdf{1}{2}^2 + \fdf{2}{3}^2 + \fdf{3}{1}^2\right]
+ \lambda'\left[\fdf{1}{4}^2 + \fdf{2}{4}^2 + \fdf{3}{4}^2\right] + h.c.\,.
\label{E8-V1-Z3}
\end{equation}
Similarly, extension of $\Z_4$ can be constructed with the aid of 
\begin{equation}
{\mathfrak d} = 
\begin{pmatrix}
	0 & 0 & 1\\
	1 & 1 & 1\\
	0 & 1 & 0
\end{pmatrix}\,,
\quad\mbox{leading to} \quad
d = 
\begin{pmatrix}
	0 & 0 & 1 & 0\\
	0 & 0 & 0 & 1\\
	0 & 1 & 0 & 0\\
	1 & 0 & 0 & 0\\
\end{pmatrix}\,,\label{E8-d-1}
\end{equation}
that is, the same cyclic permutation $\phi_1 \mapsto \phi_3 \mapsto \phi_2 \mapsto \phi_4 \mapsto \phi_1$ as we encountered 
in Eq.~\eqref{extension-Z4Z2-2-eq2} leading to the same group $(\Z_2)^3\rtimes\Z_4 \simeq {\tt SmallGroup(32, 6)}$.
The potential $V_1$ simplifies then to 
\begin{equation}
V_1 = \lambda\left[\fdf{1}{2}^2 + \fdf{3}{4}^2\right]
+ \lambda'\left[\fdf{1}{3}^2 + \fdf{3}{2}^2 + \fdf{2}{4}^2 + \fdf{4}{1}^2\right] + h.c.\,,
\label{E8-V1-Z4}
\end{equation}
with real $\lambda$,
while the rephasing-insensitive part $V_0$ takes the form as in Eq.~\eqref{extension-Z4Z2-2-eq3}.

Continuing with the subgroups in Table~\ref{table-subgroup_GL32}, we deal next with the group $\Z_2\times\Z_2$, 
which corresponds to two distinct conjugacy classes in $GL(3,2)$.
With the aid of {\ttfamily GAP}, we select these two pairs of generators: 
\begin{equation}
\Z_2\times\Z_2 \subset GL(3,2),\,
\mbox{option 1:}\qquad 
\mathfrak{b}_1 = 
\begin{pmatrix}
	0 & 1 & 0\\
	1 & 0 & 0\\
	0 & 0 & 1
\end{pmatrix}\;,\quad 
\mathfrak{b}_3' 
\,,\label{E8-Z2Z2-1}
\end{equation}
\begin{equation}
\Z_2\times\Z_2 \subset GL(3,2),\,
\mbox{option 2:}\qquad 
\mathfrak{b}_1'\,, \quad 
\mathfrak{b}_3'\,.\label{E8-Z2Z2-2}
\end{equation}
As usual, in each case, we have options for split vs. non-split extensions, similar to the cases we have found earlier.
Skipping technical details, we only provide the final result:
the first option leads only to $(\Z_2)^3\rtimes(\Z_2\times\Z_2) \simeq {\tt SmallGroup(32,49)}$,
constructed either as split or non-split extension,
while the second option produces the groups {\tt SmallGroup(32,27)} (split extension) and {\tt SmallGroup(32,34)} (non-split extension).
The constraints on the potential can also be established using the methods we have already used before.
We only stress here that all of these groups are different from group $\Sigma(32) = {\tt SmallGroup(32,11)}$ which is sometimes used
in bSM model building.

Moving on, we select a representative $S_3 \subset GL(3,2)$ generated by $b$ in Eq.~\eqref{E8-b-example2}
and $c$ in Eq.~\eqref{E8-c-1}. We checked that only split extension is possible, leading to total symmetry group 
$(\Z_2)^3\rtimes S_3\simeq S_4\times \Z_2$. In essence, this is the same symmetry group $S_4$ acting on the first three doublets,
which we had already in the 3HDM, times the $\Z_2$ group of independent sign flip of $\phi_4$.
The potential $V_1$ in this case is the same as for the $(\Z_2)^3\rtimes \Z_3\simeq A_4\times \Z_2$
and was given in Eq.~\eqref{E8-V1-Z3}. The only extra condition now is that $\lambda$ in Eq.~\eqref{E8-V1-Z3} must be real.

A representative subgroup $D_4 \subset GL(3,2)$ can be generated by the same $\mathfrak{d}$ as in Eq.~\eqref{E8-d-1} and $\mathfrak{b}_1$
in Eq.~\eqref{E8-Z2Z2-1} as they satisfy $\mathfrak{b}_1^{-1}\mathfrak{d}\mathfrak{b}_1 = \mathfrak{d}^{-1}$.
The resulting group is $G = (\Z_2)^3 \rtimes D_4 \simeq UT(4,2) \simeq {\tt SmallGroup(64, 138)}$, the same group
as in Eq.~\eqref{extension-Z4Z2-2-D4}. The potential $V_0$ takes the form Eq.~\eqref{extension-Z4Z2-2-eq3} while the $V_1$ part 
is the same as in Eq.~\eqref{E8-V1-Z4} but now both $\lambda$ and $\lambda'$ being real.

The next subgroup to consider is $A_4 \subset GL(3,2)$. Table~\ref{table-subgroup_GL32} indicates two distinct conjugacy classes
for $A_4$. However it turns out that the same constraint which forbade the $\Z_7$ subgroup
forbids also one of the $A_4$ conjugacy classes.
The remaining one can be generated by the familiar $\mathfrak{b}_3'$ in Eq.~\eqref{E8-b'-examples} and $\mathfrak{c}$ in Eq.~\eqref{E8-c-1}
because they satisfy the relations defining the $A_4$ group:
$(\mathfrak{b}_3')^{3} = \mathfrak{c}^2 = (\mathfrak{cb'}_3)^3 = \mathfrak{e}$.
In terms of $b$ and $c$, we arrive at the defining presentation of $A_4$ as a group
of positive-signature permutations of four doublets.

The total symmetry group obtained through this construction is $(\Z_3)^3\rtimes A_4 \simeq {\tt SmallGroup(92,70)}$ of order 92.
Imposing invariance under the $A_4$ group of permutations and sign flips of individual doublets 
dramatically constrains the potential, with 
\begin{eqnarray}
V_0 &=& m^2 \left(\fdfn{1}{1} + \fdfn{2}{2} + \fdfn{3}{3} + \fdfn{4}{4}\right)
+ \Lambda \left(\fdfn{1}{1} + \fdfn{2}{2} + \fdfn{3}{3} + \fdfn{4}{4}\right)^2\nonumber\\
&&+\Lambda' \left[(\fdfn{1}{1})^2 + (\fdfn{2}{2})^2 + (\fdfn{3}{3})^2 + (\fdfn{4}{4})^2\right]\nonumber\\
&& + \Lambda'' \left(|\fdfn{1}{3}|^2 + |\fdfn{2}{3}|^2 + |\fdfn{1}{4}|^2 + |\fdfn{2}{4}|^2 
+ |\fdfn{1}{2}|^2 + |\fdfn{3}{4}|^2\right)\label{E8-A4-V0}
\end{eqnarray}
and 
\begin{equation}
V_1 = \lambda\left[\fdf{1}{2}^2 + \fdf{2}{3}^2 + \fdf{3}{1}^2 + \fdf{1}{4}^2 + \fdf{2}{4}^2 + \fdf{3}{4}^2 + h.c.\right]\,.
\label{E8-A4-V1}
\end{equation}
However, upon a quick inspection, it becomes clear that this potential is invariant under {\em all} permutations of the four doublets,
not only the positive-signature ones. Therefore, the group $(\Z_3)^3\rtimes A_4$ we just constructed is not realizable in the 4HDM because
it automatically leads a additional discrete symmetries. The total symmetry content of this potential is
\begin{equation}
G = (\Z_2)^3\rtimes S_4 \simeq {\tt SmallGroup(192,955)}\,.\label{E8-S4}
\end{equation}
It is remarkable that the 4HDM scalar sector, with so many symmetries and so few free parameters, 
does not possess an accidental continuous symmetry.

\newpage

\section{Attempting to Extend by $\Z_4\times\Z_4$, $\Z_4\times\Z_2\times\Z_2$, and $(\Z_2)^4$}

As is mentioned by the end of Section \ref{general-remarks}, we found out three special abelian symmetry of 4HDM: $\Z_4\times\Z_4$, $\Z_4\times\Z_2\times\Z_2$, and $(\Z_2)^4$. 
We discovered the three groups from previous experience in studying 3HDM discrete symmetry and from studying group extensions for 4HDM: in 3HDM, we have the so-called ``extra-special'' group $(\Z_3\times\Z_3)\rtimes \Z_3$, and the pre-image of the abelian group $\Z_3\times\Z_3$ in $SU(3)$ is not abelian, and, therefore, we expect that in 4HDM, we have a $\Z_4\times\Z_4$ symmetry as well, which turns out to be true; in the extension $(\Z_4\times\Z_2)\rtimes D_4\simeq UT(4,2)$ where the symmetry is maximal, we have $(\Z_2)^4$ as a subgroup of $UT(4,2)$. This means that the $UT(4,2)$ symmetric potential, which has no continuous symmetry, is also symmetric under $(\Z_2)^4$ as a subgroup of $UT(4,2)$. This means that when we carefully relax the constraints of the $UT(4,2)$ symmetric potential by ignoring the constraints from generators in $UT(4,2)$ but not in $(\Z_2)^4$, we can obtain the $(\Z_2)^4$ symmetric potential, proving the existence of $(\Z_2)^4$ as 4HDM abelian symmetry. The abelian group $\Z_4\times\Z_2\times\Z_2$ actually comes from the study of $\Z_4\times\Z_4$ extensions in a very similar way to how we discovered $(\Z_2)^4$. We will discuss $\Z_4\times\Z_2\times\Z_2$ later during our attempts to build extensions based on $\Z_4\times\Z_4$. 

It's worthy pointing out that the three groups can not be generated only by diagonal matrices as is proven in \cite{Ivanov:2011ae}: the abelian symmetry group of the 4HDM generated by phase transformations must have an order satisfying $|G|\leq 8$. Typically, non-diagonal matrices do not commute with each other or with diagonal matrices. However, the generators of \(\Z_4 \times \Z_4\), \(\Z_4 \times \Z_2 \times \Z_2\), and \(\Z_2 \times \Z_2 \times \Z_2 \times \Z_2\) exhibit special behavior: \( [a_k, a_l] = i^{n_{kl}} \id_4 \); although these generators are non-commutative in $SU(4)$, they commute in $PSU(4)$ since the identity of $PSU(4)$ is $i^r\mathbf{1}_4$. Thus, \(\Z_4 \times \Z_4\), \(\Z_4 \times \Z_2 \times \Z_2\), and \(\Z_2 \times \Z_2 \times \Z_2 \times \Z_2\) actually describes the group structures in $PSU(4)$, with their pre-image groups in $SU(4)$ being non-abelian. In fact, \cite{Ivanov:2011ae} also proves that these pre-image groups in $SU(4)$ are nilpotent groups of class-2.

A natural question arise: are there any other abelian subgroups of $PSU(4)$, whose pre-images in $SU(4)$ are non-abelian, that can also be realized as 4HDM symmetry just as the three groups discovered? This question still remains open, and, therefore, the Table \ref{table-abelian} may not be exhaustive. This, together with the fact that the Burnside's$-p^aq^b$ theorem is inapplicable to proving the completeness of the classification of all discrete symmetries in the 4HDM, indicates that we cannot ensure a complete classification of 4HDM discrete symmetries based solely on extensions of abelian groups. These points illustrate the challenges involved in the full classification of 4HDM discrete symmetries. 

With that being said, we are still confident in claiming that a sub-case of classification is complete under the method of group extensions by rephasing groups, and the follow-up steps towards a full classification is clear: try extending by the three groups $\Z_4\times\Z_4$, $\Z_4\times\Z_2\times\Z_2$ and $(\Z_2)^4$, hopefully discovering new symmetries, which might provide us insights into a method that exhaust all 4HDM symmetries. 

We start from studying $\Z_4\times\Z_4$ extensions, whose generators are as follows:
\begin{equation}
a_1 = \sqrt{i}\left(\begin{array}{cccc} 
1 & 0 & 0 & 0\\ 
0 & i & 0 & 0\\ 
0 & 0 & -1 & 0\\ 
0 & 0 & 0 & -i\\ 
\end{array}\right)\,,\quad
a_2 = \sqrt{i}\left(\begin{array}{cccc} 
0 & 1 & 0 & 0\\ 
0 & 0 & 1 & 0\\ 
0 & 0 & 0 & 1\\ 
1 & 0 & 0 & 0\\ 
\end{array}\right)\,.\label{extension-Z4xZ4-generators}
\end{equation}
we can immediately see that $[a_1, a_2] = -i\mathbf{1}_4$, which is the identity in $PSU(4)$ but not in $SU(4)$. 

Next, we need to figure out the automorphism group of $\Z_4\times\Z_4$. Based on our discussion of $\Aut(\Z_2 \times \Z_2 \times \Z_2)$, we can conjecture that
$$
\Aut(\overbrace{\Z_p \times \Z_p \times \dots \times \Z_p}^{n}) \simeq GL(n, p),
$$
where $p$ is a prime number. This implies that the automorphism group of the abelian group $(\Z_p)^n$ is equivalent to the general linear group $GL(n, p)$, defined on an $n$-dimensional vector space over the finite field $\mathbb{F}_p$. This conjecture is correct, and its proof can be constructed in a manner similar to the proof for $\Aut(\Z_2 \times \Z_2 \times \Z_2) \simeq GL(3, 2)$. A key point here is that $p$ is a prime, as only in this case can the abelian group $\Z_p$ be promoted to the finite field $\mathbb{F}_p$, and since vector spaces are defined over fields, we obtain the above conclusion.

What happens when $p$ is a general natural number $m$? In this case, $\Z_m$ no longer has a finite field structure. Consequently, the array $(k_1, \dots, k_n)$, which represents an element $ g = \prod_{i=1}^n a_i^{k_i} $ of $(\Z_m)^n$, where $k_i \in \Z_m$, no longer forms a vector. Although $(k_1, \dots, k_n)$ is no longer a vector, the set of all invertible transformations on the space of $(k_1, \dots, k_n)$ can also be named as $\Aut((\Z_m)^n)$, even though this space is not a vector space. In mathematical terms, the space of $(k_1, \dots, k_n)$ where $k_i\in \Z_m$ is called a module, and $GL(n, \Z_m)$ is then defined as the automorphism group of the module. Therefore, we have $\Aut((\Z_m)^n) \simeq GL(n, \Z_m)$, meaning the automorphism group of the abelian group $(\Z_m)^n$ is isomorphic to the module automorphism group $GL(n, \Z_m)$. Hence, we have $\Aut(\Z_4 \times \Z_4) \simeq GL(2, \Z_4)$. It's worthy noting that $GL(2,\Z_4)$ should not be confused with the group $GL(2,4)$. The number 4 in the notation $GL(2,4)$ means the finite field of 4 elements, which can be obtained through field extension $\Z_2[x]/\langle x^2+1\rangle\simeq \mathbb{F}_4$. The field $\mathbb{F}_4$ is not $\Z_4$, which is not even a field. For a beginner friendly introduction to field extensions in algebra, see Chapter 20, 21, 22 of {\it Contemporary Abstract Algebra} \cite{gallian2021contemporary}.

Although the above definition is somewhat uncommon, the group $GL(2,\Z_4)$ is easy to handle in practical terms: we use $2 \times 2$ matrices to represent an automorphism of $\Z_4 \times \Z_4$, where all elements of the matrix are taken from $\Z_4 = \{0, 1, 2, 3\}$. The operations among the numbers $0, 1, 2, 3$ should be understood as modular arithmetic under modulo 4 addition and multiplication; for instance, $2 \times 3 = 2$ (mod) 4.

Then, we need to study the subgroup structure of $GL(2,\Z_4)$. With the aid of {\tt GAP}, we arrive at the table \ref{table-subgroup_GL24_1}. 

\begin{table}[H]
\normalsize
\centering
\begin{tabular}[t]{cc}
    \toprule
    Groups & N \\
    \midrule
    $\Z_2$ & 6 \\
    $\Z_3$ & 1 \\
    $\Z_4$ & 3 \\
    $\Z_2\times\Z_2$ & 13 \\
    $\Z_6$ & 2 \\
    $S_3$ & 1 \\
    $(\Z_2)^3$ & 6 \\
    $\Z_4\times\Z_2$ & 3 \\
    $D_4$ & 6 \\
    $A_4$ & 1 \\
    $Dic_{12}$ & 1 \\
    $\Z_6\times\Z_2$ & 1 \\
    \bottomrule
\end{tabular}
\quad 
\begin{tabular}[t]{cc}
    \toprule
    Groups & N \\
    \midrule
    $D_{6}$ & 1 \\
    $(\Z_2)^4$ & 1 \\
    $D_4\times\Z_2$ & 3 \\
    {\tt SmallGroup(16,3)} & 3 \\
    $A_4\times\Z_2$ & 2 \\
    $S_4$ & 1 \\
    {\tt SmallGroup(24,8)} & 1 \\
    {\tt SmallGroup(32,27)} & 1 \\
    $A_4\times\Z_2\times\Z_2$ & 1 \\
    $S_4\times\Z_2$ & 1 \\
    $SL(2,\Z_4)$ & 1 \\
    $GL(2,\Z_4)$ & 1 \\
    \bottomrule
\end{tabular}
\caption{Subgroup structure of $GL(2,\Z_4)$; listed all proper subgroups up to conjugacy classes; the number $N$ means the number of distinct conjugacy classes for each representative subgroup.}
\label{table-subgroup_GL24_1}
\end{table}

We see 24 proper subgroups up to conjugacy class, which is more than twice as large as that for $GL(3,2)$; what's more, the number $N$'s of each distinct conjugacy classes are also larger: for instance, there are 6 conjugacy classes for $\Z_2$ subgroup, and within each conjugacy class among the 6, there are numerous mutually conjugate $\Z_2$ subgroups. But luckily, with the aid of the theorem~\ref{the-theorem}, we only need to study the 6 cases without delving into mutually conjugate subgroups within each class. 

Before start doing group extensions, we need to give the $\Z_4\times\Z_4$ symmetrical 4HDM potential, which is straight forward to obtain from the option 1 $\Z_4$ symmetric potential. This is because the option 1 $\Z_4$ generator Eq.~\eqref{extension-Z4-a1} is equal to $a_1$ in Eq.~\eqref{extension-Z4xZ4-generators}. Therefore, the $\Z_4\times\Z_4$ symmetric potential $V(\Z_4\times\Z_4)$ is obtained by imposing the $a_2$ invariance on $V_0 + V_1(\Z_4)$ where the rephasing sensitive part $V_1(\Z_4)$ is given in Eq.~\eqref{V1-Z4}. 

Note that the rephasing insensitive part $V_0$ is not contained in $V(\Z_4\times\Z_4)$ because $a_2$ is not a rephasing transformation. For the convenience of applying efficient computer-aided calculations, we rearrange $V(\Z_4\times\Z_4)$ in the following way such that the coefficients $r_i$ is adjusted to real. 
$$
V(\Z_4\times\Z_4) = (r_9+ir_{10})W_9 + (r_9-ir_{10})W_{10} + \sum_{k=0}^{8}r_k W_k
$$
\begin{equation}
\begin{aligned}
  & W_0 = (\phi_1^\dagger\phi_1) + (\phi_2^\dagger\phi_2) + (\phi_3^\dagger\phi_3) + (\phi_4^\dagger\phi_4)\\
  & W_1 = W_0^2\\
  & W_2 = (\phi_1^\dagger\phi_1)(\phi_1^\dagger\phi_1) + (\phi_1^\dagger\phi_1)(\phi_2^\dagger\phi_2) + (\phi_1^\dagger\phi_1)(\phi_3^\dagger\phi_3) + (\phi_1^\dagger\phi_1)(\phi_4^\dagger\phi_4) \\
  & + (\phi_1^\dagger\phi_2)(\phi_2^\dagger\phi_1) + (\phi_1^\dagger\phi_3)(\phi_3^\dagger\phi_1) + (\phi_1^\dagger\phi_4)(\phi_4^\dagger\phi_1) + (\phi_2^\dagger\phi_2)(\phi_2^\dagger\phi_2)\\ 
  & + (\phi_2^\dagger\phi_2)(\phi_3^\dagger\phi_3) + (\phi_2^\dagger\phi_2)(\phi_4^\dagger\phi_4) + (\phi_2^\dagger\phi_3)(\phi_3^\dagger\phi_2) + (\phi_2^\dagger\phi_4)(\phi_4^\dagger\phi_2)\\ 
  & + (\phi_3^\dagger\phi_3)(\phi_3^\dagger\phi_3) + (\phi_3^\dagger\phi_3)(\phi_4^\dagger\phi_4) + (\phi_3^\dagger\phi_4)(\phi_4^\dagger\phi_3) + (\phi_4^\dagger\phi_4)(\phi_4^\dagger\phi_4)\\
  & W_3 = (\phi_1^\dagger\phi_1)(\phi_3^\dagger\phi_3) + (\phi_2^\dagger\phi_2)(\phi_4^\dagger\phi_4)\\
  & W_4 = (\phi_1^\dagger\phi_3)(\phi_3^\dagger\phi_1) + (\phi_2^\dagger\phi_4)(\phi_4^\dagger\phi_2)\\
  & W_5 = (\phi_1^\dagger\phi_2)(\phi_4^\dagger\phi_3) + (\phi_1^\dagger\phi_4)(\phi_2^\dagger\phi_3) + (\phi_2^\dagger\phi_1)(\phi_3^\dagger\phi_4) + (\phi_3^\dagger\phi_2)(\phi_4^\dagger\phi_1)\\
  & W_6 = (\phi_1^\dagger\phi_3)(\phi_2^\dagger\phi_4) + (\phi_1^\dagger\phi_3)(\phi_4^\dagger\phi_2) + (\phi_2^\dagger\phi_4)(\phi_3^\dagger\phi_1) + (\phi_3^\dagger\phi_1)(\phi_4^\dagger\phi_2)\\
  & W_7 = (\phi_1^\dagger\phi_2)(\phi_2^\dagger\phi_1) + (\phi_1^\dagger\phi_4)(\phi_4^\dagger\phi_1) + (\phi_2^\dagger\phi_3)(\phi_3^\dagger\phi_2) + (\phi_3^\dagger\phi_4)(\phi_4^\dagger\phi_3)\\
  & W_8 = (\phi_1^\dagger\phi_3)^2 + (\phi_2^\dagger\phi_4)^2 + (\phi_3^\dagger\phi_1)^2 + (\phi_4^\dagger\phi_2)^2\\
  & W_9 = (\phi_1^\dagger\phi_2)(\phi_1^\dagger\phi_4) + (\phi_2^\dagger\phi_1)(\phi_2^\dagger\phi_3) + (\phi_3^\dagger\phi_2)(\phi_3^\dagger\phi_4) + (\phi_4^\dagger\phi_1)(\phi_4^\dagger\phi_3)\\ 
  & W_{10} = (\phi_1^\dagger\phi_2)(\phi_3^\dagger\phi_2) + (\phi_1^\dagger\phi_4)(\phi_3^\dagger\phi_4) + (\phi_2^\dagger\phi_1)(\phi_4^\dagger\phi_1) + (\phi_2^\dagger\phi_3)(\phi_4^\dagger\phi_3)\\\end{aligned}
\label{W_coeff}
\end{equation}
Among them, $r_0W_0 + r_1W_1 + r_2W_2$ is invariant under $SU(4)$, which plays a row of $V_0$ in the study of $\Z_4\times\Z_4$, $\Z_4\times\Z_2\times\Z_2$, and $(\Z_2)^4$. Terms that transform into linear combinations of each other are $W_k$'s where $k=3,4,5,6,7,8,9,10$. If further symmetries are imposed, there will be relations among coefficients $r_3\sim r_{10}$. 

Next, let's start building extensions. As a nice example, we consider the $\Z_4$ generated by the following matrix:
\begin{equation}
\mathfrak{b} = 
\begin{pmatrix}
0 & 1 \\
3 & 0
\end{pmatrix}\label{extension-Z4xZ4-1}
\end{equation}
which defines a set of matrix equations as shown in Eq.~\eqref{extension-Z4xZ4-2}:
\begin{equation}
\left\{
\begin{aligned}
& b^{-1}a_1b = a_2 \\
& b^{-1}a_2b = a_1^3 
\end{aligned}
\right.\label{extension-Z4xZ4-2}
\end{equation}

Now we encountered another difficulty in addition to the more complicated subgroup structure of $\Z_4\times\Z_4$: the matrix equations are harder to solve. This is because the $a_2$ matrix is a monomial matrix which is not diagonal, therefore, when written term-by-term (Eq.~\eqref{extension-Z4xZ4-3} and Eq.~\eqref{extension-Z4xZ4-4}), we can't solve the equation easily by simply comparing elements on each $i,j$ spot in the matrix. See Eq.~\eqref{extension-Z8Z2-2} for compassion. We see that in Eq.~\eqref{extension-Z8Z2-2}, the elements in the matrix equation satisfy $b_{ij}\propto b_{ij}$ while elements in Eq.~\eqref{extension-Z4xZ4-3} and Eq.~\eqref{extension-Z4xZ4-4}, which take a generic form $b_{ij}\propto b_{kl}$, don't have such nice property. 
\begin{equation}
b^{-1}a_1b = a_2\quad \Rightarrow \quad 
\begin{pmatrix}
b_{11} & b_{12} & b_{13} & b_{14} \\
ib_{21} & ib_{22} & ib_{23} & ib_{24} \\
-b_{31} & -b_{32} & -b_{33} & -b_{34} \\
-ib_{41} & -ib_{42} & -ib_{43} & -ib_{44} \\
\end{pmatrix} = 
\begin{pmatrix}
b_{14} & b_{11} & b_{12} & b_{13} \\
b_{24} & b_{21} & b_{22} & b_{23} \\
b_{34} & b_{31} & b_{32} & b_{33} \\
b_{44} & b_{41} & b_{42} & b_{43} \\
\end{pmatrix} \label{extension-Z4xZ4-3}
\end{equation}
\begin{equation}
b^{-1}a_2b = a_1^3\quad \Rightarrow \quad 
\begin{pmatrix}
b_{21} & b_{22} & b_{23} & b_{24} \\
b_{31} & b_{32} & b_{33} & b_{34} \\
b_{41} & b_{42} & b_{43} & b_{44} \\
b_{11} & b_{12} & b_{13} & b_{14} \\
\end{pmatrix} = i
\begin{pmatrix}
b_{11} & -ib_{12} & -b_{13} & ib_{14} \\
b_{21} & -ib_{22} & -b_{23} & ib_{24} \\
b_{31} & -ib_{32} & -b_{33} & ib_{34} \\
b_{41} & -ib_{42} & -b_{43} & ib_{44} \\
\end{pmatrix} \label{extension-Z4xZ4-4}
\end{equation}

What's more, there are in total 61 sets of matrix equations from Table~\ref{table-subgroup_GL24_1} we need to solve, all of which are of the general form $b_{ij}\propto b_{kl}$. 
Therefore, we need a new method to massively solve the matrix equations. 

We developed a computer code in {\tt Python} enabling us to solve such matrix equations, based on the following observations: given equations Eq.~\eqref{extension-Z4xZ4-3} and Eq.~\eqref{extension-Z4xZ4-4}, it is not difficult to see that, when combined, they form a large linear system of equations expressed as $M\tilde{b} = 0$. Here, $\tilde{b}$ is the column vector obtained by reshaping the $4 \times 4$ matrix $b$ into a $16 \times 1$ vector:
$$
\tilde{b} = (b_{11} , b_{12} , b_{13} , b_{14}, b_{21} , b_{22} , b_{23} , b_{24} ,b_{31} , b_{32} , b_{33} , b_{34} , b_{41} , b_{42} , b_{43} , b_{44})^T
$$
and the matrix $M$ is a $32\times 16$ matrix. This is because we have 2 matrix equations each of which contributes to 16 rows of the matrix $M$. In general, if we have $n$ matrix equations, the size of $M$ is $16n\times 16$.

To solve the set of matrix equations is equivalent to determine the basis vectors of the null space of the $16n \times 16$ matrix $M$. Once the null space basis vectors are identified, we apply an inverse transformation to reshape the $16 \times 1$ null space basis vectors back into $4 \times 4$ matrices. The null space of a matrix can be readily computed using computer assistance, giving us an efficient solution to large systems of matrix equations. 

It is worth emphasizing that this method is broadly applicable for solving any system of matrix equations. In this particular case, the shape of matrix $M$ is $16n \times 16$ because $n$ matrix equations have been combined into the system. Typically, the dimension of the null space of the coefficient matrix for an over-determined system of linear equations is very small. However, in this scenario, we obtain a null space with a surprisingly high dimension of 4, which means the matrix equations have 4 solutions. Although the 4 solutions are linearly independent as 16 dimensional vectors, they are related by Higgs basis change as $PSU(4)$ matrices, meaning that we only need to list one solution, hopefully the most mathematically concise one, which is shown in Eq.~\eqref{extension-Z4xZ4-5}.
\begin{equation}
b = \frac{i^{-1/4}}{2}\cdot
\begin{pmatrix}
    i & i & i & i\\
    i & 1 & -i & -1\\
    i & -i & i & -i\\
    i & -1 & -i & 1
\end{pmatrix}
\label{extension-Z4xZ4-5}
\end{equation}

The follow-up step is to impose the $b$ invariance on the potential $V(\Z_4\times\Z_4)$ to obtain the $(\Z_4\times\Z_4)\rtimes\Z_4$ symmetric potential. It's worth noting that we will encounter this very same problem in later extensions by subgroups shown in Table~\ref{table-subgroup_GL24_1}. Therefore, it would be ideal if we have a code that can output the desired relations among $r_i$'s given the input of $b$ from solving the matrix equations. This is exactly what we did. See our code \cite{TheCode}.  The relation among coefficients after imposing the invariance under $b$ in Eq.~\eqref{extension-Z4xZ4-5} is:
\begin{equation}
\left\{
\begin{aligned}
  & r_{10} = 0\\
  & r_3 = r_6 + 2r_8 - 2r_9 \\
  & r_4 = -r_6 - 2r_9 \\
  & r_5 = -r_6 - r_7 - 2r_8
\end{aligned}
\right.\label{extension-Z4xZ4-5-relation}
\end{equation}

The final step in studying $\Z_4$ extension is to identify the name of the group after extension. By using {\tt GAP}, we can easily get ${\tt SmallGroup(64,34)}\simeq (\Z_4\times\Z_4)\rtimes\Z_4$. However, it turns out that this group is not the maximal symmetry of the potential with constraints Eq.~\eqref{extension-Z4xZ4-5-relation}. This is because the potential $V(\Z_4\times\Z_4)$ is automatically symmetric under another transformation absent in $\Z_4\times\Z_4$: the one generated by $\mathfrak{w}$, which happens to generate the center of $GL(2,\Z_4)$.
$$
\mathfrak{w} = \begin{pmatrix} 3 & 0 \\ 0 & 3 \\ \end{pmatrix}\;,\quad Z(GL(2,\Z_4)) \simeq \langle \mathfrak{w}\rangle \simeq \Z_2
$$
This is easy to verify using the method we just developed. So the actual symmetry for the potential $V(\Z_4\times\Z_4)$ is {\tt SmallGroup(32,34)}, also known as the Generalized Dihedral Group for $\Z_4\times\Z_4$. The term ``generalized'' indicates the way this group is constructed. In compassion: the generalized dihedral group for $\Z_4$ is the ordinary dihedral group $D_4$. 

Therefore, the full symmetry content of the potential with constraints given in Eq.~\eqref{extension-Z4xZ4-5-relation} is $(\Z_4\times\Z_4)\rtimes(\Z_4\times\Z_2)\simeq {\tt SmallGroup(128,856)}$ instead of the smaller group ${\tt SmallGroup(64,34)}$.  

This is where we have discovered that $\Z_4\times\Z_2\times\Z_2$ is also a possible abelian symmetry of 4HDM potential: because the group {\tt SmallGroup(128,856)} has $\Z_4\times\Z_2\times\Z_2$ as subgroup.

We have also encountered this situation in extending by $\Z_2\simeq \langle \mathfrak{b}^\prime\rangle$. 
\begin{equation}
\mathfrak{b}^\prime = 
\begin{pmatrix}
1 & 0 \\
2 & 1
\end{pmatrix}
\;,\quad 
\tilde{\mathfrak{b}} = 
\begin{pmatrix}
1 & 2 \\
2 & 1
\end{pmatrix}
\end{equation}
The actual symmetry content we obtained from $\langle\mathfrak{b}^\prime\rangle$ extension is $(\Z_4\times\Z_4)\rtimes(\Z_2\times\Z_2)\simeq{\tt SmallGroup(64,216)}$, which also has $\Z_4\times\Z_2\times\Z_2$ as abelian subgroup. 

After demonstrating a few examples, it's time to apply the method we developed for every case in Table~\ref{table-subgroup_GL24_1}. In the six $\Z_2$ cases, three of whose matrix equations have no solutions, and one is the automatically satisfied $\Z_2\simeq \langle \mathfrak{w}\rangle$, so there are only two distinct $\Z_2$ symmetries, $\langle\mathfrak{b}^\prime\rangle$ and $\langle\tilde{\mathfrak{b}}\rangle$, each of which actually lead to $(\Z_4\times\Z_4)\rtimes (\Z_2\times\langle\mathfrak{w}\rangle)$ symmetrical 4HDM potentials. The constraints of the coefficients from $(\Z_4\times\Z_4)\rtimes \langle\mathfrak{b}^\prime, \mathfrak{w}\rangle$ are $r_9 = r_{10} = 0$, and the constraints of the coefficients from $(\Z_4\times\Z_4)\rtimes \langle\tilde{\mathfrak{b}},\mathfrak{w}\rangle$ are $r_9 = 0, r_3 + r_4 = 2r_8$.  

That three $\Z_2$'s have no solutions to matrix equations placed a strong constraints on other groups in Table~\ref{table-subgroup_GL24_1}. For instance, only 3 among the 13 different choices of $\Z_2\times\Z_2$ is possible. This can be seen by using the {\tt GAP} code we provided by the end of Section~\ref{GAP-intro}, which still requires a cumbersome case-by-case evaluation. Actually, there is another alternative for constructing $\Z_2\times\Z_2$ extensions, which can also be applied to other constructions: this method starts from the potential instead of from subgroups in Table~\ref{table-subgroup_GL24_1}.   

As we go further down Table~\ref{table-subgroup_GL24_1}, it's easy to see that while $GL(2,\Z_4)$ has rich subgroups, the potential $V(\Z_4\times\Z_4)$ already has a large symmetry {\tt SmallGroup(32,34)}$\simeq (\Z_4\times\Z_4)\rtimes\langle\mathfrak{w}\rangle$, leaving limited space for further symmetries. Therefore, when the subgroups are getting complicated but the potential is getting constraint, it's easier to see results starting from the potential instead of from extending each subgroups in Table~\ref{table-subgroup_GL24_1}. 

So, knowing that the feasible $\Z_2\times\Z_2$ extensions only comes from the combination of $\mathfrak{b}^\prime, \tilde{\mathfrak{b}}$, and $\mathfrak{w}$, and their constraints on the potential:
\begin{itemize}
\item $\mathfrak{w}$: no constraints on $r_i$ parameters
\item $\mathfrak{b}^\prime$: $r_9 = 0$, $r_{10} = 0$
\item $\tilde{\mathfrak{b}}$: $r_9 = 0$, $r_3 +r_4 = 2r_8$
\end{itemize}
we quickly get the following possible combinations:
\begin{itemize}
\item $V$ with $r_9 = r_{10} = 0$: symmetric under $(\Z_4\times\Z_4)\rtimes\langle\mathfrak{w}, \mathfrak{b}^\prime\rangle\simeq$ {\tt SmallGroup(64,216)}
\item $V$ with $r_9 = 0, r_3+r_4 = 2r_8$: symmetric under $(\Z_4\times\Z_4)\rtimes\langle\mathfrak{w}, \tilde{\mathfrak{b}}\rangle\simeq$ {\tt SmallGroup(64,216)}
\item $V$ with $r_9 = r_{10} = 0, r_3 + r_4 = 2r_8$: \\symmetric under $(\Z_4\times\Z_4)\rtimes \langle\mathfrak{w}, \mathfrak{b}^\prime, \tilde{\mathfrak{b}}\rangle\simeq (\Z_4\times\Z_4)\rtimes(\Z_2)^3\simeq$ {\tt SmallGroup(128,2264)}
\end{itemize}

We see that when we put the constraints of $\mathfrak{b}^\prime$ and $\tilde{\mathfrak{b}}$ together, the actual symmetry of the potential is {\tt SmallGroup(128,2264)} $\simeq (\Z_4\times\Z_4)\rtimes(\Z_2)^3$  instead of $(\Z_4\times\Z_4)\rtimes(\Z_2\times\Z_2)$. 

This shows the power of the new perspective compared with examining all groups in Table~\ref{table-subgroup_GL24_1}: we simply need to try extend by cyclic groups first, and see if they have solutions to the matrix equations, and compare their constraints on $r_i$ coefficients. Hopefully many options will be eliminated due to non-solvable matrix equations or continuous symmetries resulting from too stringent coefficients relations such as $r_3 = \dots =r_{10} = 0$. Then, we will try to combine feasible generators together like we did in analyzing $\Z_2\times\Z_2$ extensions to form larger subgroups of $GL(2,\Z_4)$ on the Table~\ref{table-subgroup_GL24_1}. In this way, we will easily arrive at extensions from larger subgroups without discussing case-by-case the complicated generator-relations in group extensions. 

With this method, we obtained that 2 among the three $\Z_4$'s and the single $\Z_3$ are possible extensions. Guided by these feasible generators whose population is greatly reduced, we can eliminate many subgroups in Table~\ref{table-subgroup_GL24_1} and obtain a list of symmetries from $\Z_4\times\Z_4$ extensions. 

Although it seems the calculations have been smooth so far, there are still difficulties in the analysis, making it hard to continue the classification. For one thing, given the method we introduced, the classification is still complicated and time consuming, therefore, new insights into the structure of 4HDM discrete symmetry is still wanted. In addition, it remains uncertain whether we have other $\Z_4\times\Z_4$, $\Z_4\times\Z_2\times\Z_2$, or $(\Z_2)^4$ potentials that are physically in-equivalent to what we have discovered. What's more, the automorphism group of $\Z_4\times\Z_2\times\Z_2$ is {\tt SmallGroup(192,1493)}, to which the linear representation theory doesn't apply; and the automorphism group of $(\Z_2)^4$ is $A_8$, which is a group of a huge order: 20160. Although $A_8\simeq SL(4,2)$, which could be studied using linear representation theory just as we did in $\Z_4\times\Z_4$ and $(\Z_3)^2$, the difficulty in analyzing $SL(4,2)$ is that some of it's subgroups can't be inputs of many of the very useful {\tt GAP} built-in functions as is shown in Table~\ref{GAP-functions}, making the analyze difficult. The above difficulties are not the most challenging task: the theorem we proved in Section~\ref{non-abelian-Symmetries-in-NHDM} claims that two extensions are related by Higgs basis change, therefore physically equivalent, only if the $q\in \Aut(A)$ which $q^{-1}K_1 q = K_2$ can be represented by $PSU(4)$ matrices.  In the case of $\Z_4\times\Z_4$, we are still able to check whether the $q$ automorphisms can be represented by $PSU(4)$ matrices without endeavoring much effort, since the order of $GL(2,\Z_4)$ is only 96. We can imagine how tedious it becomes when it comes to {\tt SmallGroup(192,1493)}, which can't be studied using linear representation easily, and $SL(4,2)$, some of whose subgroups are not available as {\tt GAP} function inputs.

\newpage

\section{Conclusions}

Multi-Higgs-doublet models, first proposed half a century ago, have become a popular framework for building models beyond the Standard Model. Equipping such models with global symmetry groups proved to be particularly helpful, as each symmetry group leads to its characteristic phenomenology. As a result, full classification of symmetry options available in the $N$-Higgs-doublet models for each $N$ emerges as a useful and rewarding task. For $N=2$ and 3, this task has already been solved, and models built on these symmetries are now being explored. For $N=4$, despite nearly a hundred of publications which make use of four Higgs doublets, this task has not been attempted, and researchers were guided by trial and error.

In this thesis, we presented the first systematic attack on this task within the 4HDM. Our main goal is to classify non-abelian finite global symmetry groups that can be imposed on the scalar sector of the 4HDM without leading to accidental continuous symmetries. We used the group extension formalism developed and successfully applied to the 3HDM in \cite{Ivanov:2012fp} and adapted it to the 4HDM case. Our procedure includes the following steps.
\begin{itemize}
\item
We start with the exhaustive list of finite abelian symmetry groups $A$, which have already been established for the 4HDM in \cite{Ivanov:2011ae}.
\item
For each $A$, we compute its automorphism group $\Aut(A)$, then list all its subgroups $K \subset \Aut(A)$, and then build all distinct non-abelian group extensions of $K$ by $A$, including not only semidirect products but also non-split extensions.
\item
For each construction, we build the Higgs potential and check whether a continuous accidental group emerges.
\end{itemize}
When following this strategy, we found three main classes of the abelian groups $A$:
\begin{itemize}
\item
cyclic groups $\Z_n$, $n = 2, \dots, 8$;
\item
products of cyclic groups: $\Z_2\times\Z_2$, $\Z_2\times\Z_4$, $\Z_2\times\Z_2\times\Z_2$;
\item
three additional groups, namely $\Z_4\times\Z_4$, $\Z_2\times\Z_2\times\Z_4$, $(\Z_2)^4$, that are abelian subgroups of $PSU(4)$ but whose pre-images inside $SU(4)$ are not abelian but are nilpotent groups of class 2.
\end{itemize}
In this thesis, we completely solved the classification task for non-abelian extension emerging from the first two classes of abelian groups $A$. The resulting list of non-abelian groups is presented in the summary Table~\ref{table-non-abelian}.

Since this extended version of the thesis is aimed at the physics audience, we tried to keep the exposition as pedagogical as possible. We started with a detailed introduction to the finite group theory concepts that are relevant to our analysis and to the helpful computational tools offered by the {\tt GAP} programming language. We then provided an overview of the about strategy for symmetry classification, with an illustration for the 3HDM, and the guidelines for the 4HDM. We also stated and proved here a useful theorem, which helps us significantly reduce the amount of case by case checks.

With these preparatory material, we then embarked on the construction of group extensions for each $A$ from the first two classes of the abelian groups. We showed all important technical steps, the nuts and bolts of building symmetry-based NHDMs in a constructive way, without any guessing. We carefully discussed the situations when the same group $A$ can have several nonequivalent realizations within the 4HDM, with different implications for the non-abelian group construction. We showed numerous examples when the same $A$ and $K$ lead to several distinct extensions, including the rather exotic nonsplit extensions, which were absent in the 3HDM. In short, we equipped the reader not only with the results given in Table~\ref{table-non-abelian}, but also with powerful methods for symmetry group construction and analysis.

Despite all these results, we still have not yet finished full classification of all possible finite non-abelian groups in the 4HDM scalar sector. However, we outlined the steps which need to be done to complete it, and we expect that few additional finite groups will arise in addition to our Table~\ref{table-non-abelian}. In this way, one thesis paved the way for full classification of all symmetry-based phenomenological situations which are possible with four Higgs doublets.

\begin{table}[H]
\centering
\renewcommand\arraystretch{1.2}
\begin{tabular}[t]{cccc}
    \toprule
    rephasing group $A$ & group extensions & extended group $G$ & order $|G|$ \\
    \midrule
    $\Z_2$ & --- & --- & --- \\[1mm]
    $\Z_3$ & $\Z_3\rtimes \Z_2$ & $S_3$ & 6 \\[1mm]
    $\Z_4$ & $\Z_4\rtimes \Z_2$ & $D_4$ & 8 \\ 
           & $\Z_4\,.\, \Z_2$ & $Q_4$ & 8  \\[1mm]
    $\Z_5$ & $\Z_5\rtimes \Z_4$ & $GA(1,5)$ & 20  \\
           & $\Z_5\rtimes \Z_2$ & $D_5$ & 10 \\[1mm]
    $\Z_6$ & $\Z_6\rtimes \Z_2$ & $D_6$ & 12 \\[1mm]
    $\Z_7$ & $\Z_7\rtimes \Z_3$ & $T_7$ & 21 \\[1mm]
    $\Z_8$ & --- & --- & ---  \\
    \midrule
    \multirow{3}{*}{$\Z_2\times\Z_2$} & $(\Z_2\times\Z_2)\rtimes\Z_2$ & $D_4$ & 8 \\[1mm]
    & $(\Z_2\times\Z_2)\rtimes\Z_3$ & $A_4$ & 12  \\[1mm]
    & $(\Z_2\times\Z_2)\rtimes S_3$ & $S_4$ & 24 \\[1mm]
    \midrule
    \multirow{5}{*}{$\Z_4\times\Z_2$} 
    & $(\Z_4\times\Z_2)\rtimes\Z_2$ & $\Z_2\times D_4$ & 16 \\[1mm]
    & $(\Z_4\times\Z_2)\rtimes\Z_2$ & {\tt SmallGroup(16,3)} & 16 \\[1mm]
    & $(\Z_4\times\Z_2)\rtimes\Z_2$ & {\tt SmallGroup(16,13)} & 16 \\[1mm]
    & $(\Z_4\times\Z_2)\rtimes\Z_4$ & {\tt SmallGroup(32,6)} & 32 \\[1mm]
    & $(\Z_4\times\Z_2)\rtimes (\Z_2\times\Z_2)$ & $\mathbf{2^5_+}$ & 32 \\[1mm]
    & $(\Z_4\times\Z_2)\rtimes (\Z_2\times\Z_2)$ & {\tt SmallGroup(32,27)} & 32 \\[1mm]
    & $(\Z_4\times\Z_2)\rtimes D_4$ & $UT(4,2)$ & 64 \\[1mm]
    \midrule
    \multirow{6}{*}{$(\Z_2)^3$} & $(\Z_2)^3\rtimes\Z_2$ & $\Z_2\times D_4$ & 16 \\[1mm]
    & $(\Z_2)^3\,.\,\Z_2$ & {\tt SmallGroup(16,3)} & 16 \\[1mm]
    & $(\Z_2)^3\rtimes\Z_3$ & $\Z_2\times A_4$ & 24 \\[1mm]
    & $(\Z_2)^3\rtimes\Z_4$ &{\tt SmallGroup(32,6)} & 32 \\[1mm]
    & $(\Z_2)^3\rtimes(\Z_2\times\Z_2)$ & $\mathbf{2^5_+}$ & 32 \\[1mm]
    & $(\Z_2)^3\rtimes(\Z_2\times\Z_2)$ & {\tt SmallGroup(32,27)} & 32 \\[1mm]
    & $(\Z_2)^3\,.\,(\Z_2\times\Z_2)$ & {\tt SmallGroup(32,34)} & 32 \\[1mm]
    & $(\Z_2)^3\rtimes S_3$ & $\Z_2\times S_4$ & 48 \\[1mm]
    & $(\Z_2)^3\rtimes D_4$ & $UT(4,2)$ & 64 \\[1mm]
    & $(\Z_2)^3\rtimes S_4$ & {\tt SmallGroup(192,955)} & 192 \\[1mm]
    \bottomrule
\end{tabular}
\caption{List of 4HDM non-abelian discrete symmetries coming from group extensions by rephasing groups. }
\label{table-non-abelian}
\end{table}


\newpage

\begin{thebibliography}{99}

\bibitem{Born:1926uzf}
M.~Born, W.~Heisenberg and P.~Jordan,
Z. Phys. \textbf{35}, no.8-9, 557-615 (1926)
doi:10.1007/BF01379806

\bibitem{CMS:2012qbp}
S.~Chatrchyan \textit{et al.} [CMS],
Phys. Lett. B \textbf{716}, 30-61 (2012)
doi:10.1016/j.physletb.2012.08.021
[arXiv:1207.7235 [hep-ex]].

\bibitem{ATLAS:2012yve}
G.~Aad \textit{et al.} [ATLAS],
Phys. Lett. B \textbf{716}, 1-29 (2012)
doi:10.1016/j.physletb.2012.08.020
[arXiv:1207.7214 [hep-ex]].

\bibitem{Srednicki:2007qs}
M.~Srednicki,
Cambridge University Press, 2007,
ISBN 978-0-521-86449-7, 978-0-511-26720-8
doi:10.1017/CBO9780511813917

\bibitem{Logan:2014jla}
H.~E.~Logan,
``TASI 2013 lectures on Higgs physics within and beyond the Standard Model,''
[arXiv:1406.1786 [hep-ph]].

\bibitem{Langacker:2010zza}
P.~Langacker,
``The standard model and beyond'', ISBN: 978-142-007-906-7

\bibitem{Clowe:2006eq}
D.~Clowe, M.~Bradac, A.~H.~Gonzalez, M.~Markevitch, S.~W.~Randall, C.~Jones and D.~Zaritsky,
Astrophys. J. Lett. \textbf{648}, L109-L113 (2006)
doi:10.1086/508162
[arXiv:astro-ph/0608407 [astro-ph]].

\bibitem{Rubakov:2017xzr}
V.~A.~Rubakov and D.~S.~Gorbunov,
World Scientific, 2017,
ISBN 978-981-320-987-9, 978-981-320-988-6, 978-981-322-005-8
doi:10.1142/10447

\bibitem{SupernovaSearchTeam:2004lze}
A.~G.~Riess \textit{et al.} [Supernova Search Team],
Astrophys. J. \textbf{607}, 665-687 (2004)
doi:10.1086/383612
[arXiv:astro-ph/0402512 [astro-ph]].

\bibitem{2309.03870}
A.~Crivellin and B.~Mellado,
Nature Rev. Phys. \textbf{6}, no.5, 294-309 (2024)
doi:10.1038/s42254-024-00703-6
[arXiv:2309.03870 [hep-ph]].

\bibitem{Lee:1973iz}
T.~D.~Lee,
Phys. Rev. D \textbf{8}, 1226-1239 (1973)
doi:10.1103/PhysRevD.8.1226

\bibitem{Weinberg:1976hu}
S.~Weinberg,
Phys. Rev. Lett. \textbf{37}, 657 (1976)
doi:10.1103/PhysRevLett.37.657

\bibitem{Bjorken:1977vt}
J.~D.~Bjorken and S.~Weinberg,
Phys. Rev. Lett. \textbf{38}, 622 (1977)
doi:10.1103/PhysRevLett.38.622

\bibitem{Wyler:1979fe}
D.~Wyler,
Phys. Rev. D \textbf{19}, 3369 (1979)
doi:10.1103/PhysRevD.19.3369

\bibitem{Leurer:1992wg}
M.~Leurer, Y.~Nir and N.~Seiberg,
Nucl. Phys. B \textbf{398}, 319-342 (1993)
doi:10.1016/0550-3213(93)90112-3
[arXiv:hep-ph/9212278 [hep-ph]].

\bibitem{GonzalezFelipe:2014mcf}
R.~Gonz\'alez Felipe, I.~P.~Ivanov, C.~C.~Nishi, H.~Ser\^odio and J.~P.~Silva,
Eur. Phys. J. C \textbf{74}, no.7, 2953 (2014)
doi:10.1140/epjc/s10052-014-2953-9
[arXiv:1401.5807 [hep-ph]].

\bibitem{Cree:2011uy}
G.~Cree and H.~E.~Logan,
Phys. Rev. D \textbf{84}, 055021 (2011)
doi:10.1103/PhysRevD.84.055021
[arXiv:1106.4039 [hep-ph]].

\bibitem{Dutta:2018yos}
B.~Dutta and Y.~Mimura,
Phys. Lett. B \textbf{790}, 589-594 (2019)
doi:10.1016/j.physletb.2019.01.065
[arXiv:1810.08413 [hep-ph]].

\bibitem{Arroyo-Urena:2019lzv}
M.~A.~Arroyo-Ure\~na, J.~L.~Diaz-Cruz, B.~O.~Larios-L\'opez and M.~A.~P.~de Le\'on,
Chin. Phys. C \textbf{45}, no.2, 023118 (2021)
doi:10.1088/1674-1137/abcfae
[arXiv:1901.01304 [hep-ph]].

\bibitem{Drees:1988fc}
M.~Drees,
Int. J. Mod. Phys. A \textbf{4}, 3635 (1989)
doi:10.1142/S0217751X89001448

\bibitem{Griest:1989ew}
K.~Griest and M.~Sher,
Phys. Rev. Lett. \textbf{64}, 135 (1990)
doi:10.1103/PhysRevLett.64.135

\bibitem{Griest:1990vh}
K.~Griest and M.~Sher,
Phys. Rev. D \textbf{42}, 3834-3849 (1990)
doi:10.1103/PhysRevD.42.3834

\bibitem{Nelson:1993vc}
A.~E.~Nelson and L.~Randall,
Phys. Lett. B \textbf{316}, 516-520 (1993)
doi:10.1016/0370-2693(93)91037-N
[arXiv:hep-ph/9308277 [hep-ph]].

\bibitem{Krasnikov:1993qd}
N.~Krasnikov, G.~Kreyerhoff and R.~Rodenberg,
Nuovo Cim. A \textbf{107}, 589-596 (1994)
doi:10.1007/BF02768793

\bibitem{Masip:1995sm}
M.~Masip and A.~Rasin,
Phys. Rev. D \textbf{52}, R3768-R3772 (1995)
doi:10.1103/PhysRevD.52.R3768
[arXiv:hep-ph/9506471 [hep-ph]].

\bibitem{Aranda:2000zf}
A.~Aranda and M.~Sher,
Phys. Rev. D \textbf{62}, 092002 (2000)
doi:10.1103/PhysRevD.62.092002
[arXiv:hep-ph/0005113 [hep-ph]].

\bibitem{Marshall:2010qi}
G.~Marshall and M.~Sher,
Phys. Rev. D \textbf{83}, 015005 (2011)
doi:10.1103/PhysRevD.83.015005
[arXiv:1011.3016 [hep-ph]].

\bibitem{Kawase:2011az}
H.~Kawase,
JHEP \textbf{12}, 094 (2011)
doi:10.1007/JHEP12(2011)094
[arXiv:1110.3861 [hep-ph]].

\bibitem{Clark:2011cv}
T.~E.~Clark, S.~T.~Love and T.~ter Veldhuis,
Phys. Rev. D \textbf{85}, 015014 (2012)
doi:10.1103/PhysRevD.85.015014
[arXiv:1107.3116 [hep-ph]].

\bibitem{Yagyu:2012qp}
K.~Yagyu,
 Doctorate Thesis, ``Studies on Extended Higgs Sectors as a Probe of New Physics Beyond the Standard Model,''
[arXiv:1204.0424 [hep-ph]].

\bibitem{Grossman:2014xia}
Y.~Grossman and C.~Peset,
JHEP \textbf{04}, 033 (2014)
doi:10.1007/JHEP04(2014)033
[arXiv:1401.1818 [hep-ph]].

\bibitem{rotman2012introduction}
Rotman, Joseph J
An introduction to the theory of groups
Vol. 148. Springer Science \& Business Media, 2012
, ISBN: 978-038-794-285-8

\bibitem{massey2019basic}
Massey, William S. 
A basic course in algebraic topology. 
Vol. 127. Springer, 2019,
ISBN: 978-038-797-430-9

\bibitem{isaacs2008finite}
Isaacs, I Martin. 
Finite group theory
Vol. 92. American Mathematical Soc, 2008,
ISBN: 978-082-184-344-4

\bibitem{Rodejohann:2019izm}
W.~Rodejohann and U.~Salda\~na-Salazar,
JHEP \textbf{07}, 036 (2019)
doi:10.1007/JHEP07(2019)036
[arXiv:1903.00983 [hep-ph]].

\bibitem{Goncalves:2023ydf}
B.~L.~Gon\c{c}alves, M.~Knauss and M.~Sher,
Phys. Rev. D \textbf{107}, no.9, 095001 (2023)
doi:10.1103/PhysRevD.107.095001
[arXiv:2301.08641 [hep-ph]].

\bibitem{Porto:2007ed}
R.~A.~Porto and A.~Zee,
Phys. Lett. B \textbf{666}, 491-495 (2008)
doi:10.1016/j.physletb.2008.08.001
[arXiv:0712.0448 [hep-ph]].

\bibitem{Porto:2008hb}
R.~A.~Porto and A.~Zee,
Phys. Rev. D \textbf{79}, 013003 (2009)
doi:10.1103/PhysRevD.79.013003
[arXiv:0807.0612 [hep-ph]].

\bibitem{CarcamoHernandez:2021osw}
A.~E.~C\'arcamo Hern\'andez, I.~de Medeiros Varzielas, M.~L.~L\'opez-Ib\'a\~nez and A.~Melis,
JHEP \textbf{05}, 215 (2021)
doi:10.1007/JHEP05(2021)215
[arXiv:2102.05658 [hep-ph]].

\bibitem{Vien:2020aif}
V.~V.~Vien,
Nucl. Phys. B \textbf{956}, 115015 (2020)
doi:10.1016/j.nuclphysb.2020.115015

\bibitem{Deshpande:1977rw}
N.~G.~Deshpande and E.~Ma,
Phys. Rev. D \textbf{18}, 2574 (1978)
doi:10.1103/PhysRevD.18.2574

\bibitem{Meloni:2011cc}
D.~Meloni, S.~Morisi and E.~Peinado,
Phys. Lett. B \textbf{703}, 281-287 (2011)
doi:10.1016/j.physletb.2011.07.084
[arXiv:1104.0178 [hep-ph]].

\bibitem{Lavoura:2011ry}
L.~Lavoura,
J. Phys. G \textbf{39}, 025202 (2012)
doi:10.1088/0954-3899/39/2/025202
[arXiv:1109.6854 [hep-ph]].

\bibitem{Meloni:2010sk}
D.~Meloni, S.~Morisi and E.~Peinado,
Phys. Lett. B \textbf{697}, 339-342 (2011)
doi:10.1016/j.physletb.2011.02.019
[arXiv:1011.1371 [hep-ph]].

\bibitem{Boucenna:2011tj}
M.~S.~Boucenna, M.~Hirsch, S.~Morisi, E.~Peinado, M.~Taoso and J.~W.~F.~Valle,
JHEP \textbf{05}, 037 (2011)
doi:10.1007/JHEP05(2011)037
[arXiv:1101.2874 [hep-ph]].

\bibitem{deAdelhartToorop:2011ad}
R.~de Adelhart Toorop, F.~Bazzocchi and S.~Morisi,
Nucl. Phys. B \textbf{856}, 670-681 (2012)
doi:10.1016/j.nuclphysb.2011.11.020
[arXiv:1104.5676 [hep-ph]].

\bibitem{Bonilla:2023pna}
C.~Bonilla, J.~Herms, O.~Medina and E.~Peinado,
JHEP \textbf{06}, 078 (2023)
doi:10.1007/JHEP06(2023)078
[arXiv:2301.10811 [hep-ph]].

\bibitem{Ivanov:2012hc}
I.~P.~Ivanov and V.~Keus,
Phys. Rev. D \textbf{86}, 016004 (2012)
doi:10.1103/PhysRevD.86.016004
[arXiv:1203.3426 [hep-ph]].

\bibitem{GonzalezFelipe:2013yhh}
R.~Gonzalez Felipe, H.~Serodio and J.~P.~Silva,
Phys. Rev. D \textbf{88}, no.1, 015015 (2013)
doi:10.1103/PhysRevD.88.015015
[arXiv:1304.3468 [hep-ph]].

\bibitem{GonzalezFelipe:2013xok}
R.~Gonz\'alez Felipe, H.~Ser\^odio and J.~P.~Silva,
Phys. Rev. D \textbf{87}, no.5, 055010 (2013)
doi:10.1103/PhysRevD.87.055010
[arXiv:1302.0861 [hep-ph]].

\bibitem{Ma:2001dn}
E.~Ma and G.~Rajasekaran,
Phys. Rev. D \textbf{64}, 113012 (2001)
doi:10.1103/PhysRevD.64.113012
[arXiv:hep-ph/0106291 [hep-ph]].

\bibitem{He:2006dk}
X.~G.~He, Y.~Y.~Keum and R.~R.~Volkas,
JHEP \textbf{04}, 039 (2006)
doi:10.1088/1126-6708/2006/04/039
[arXiv:hep-ph/0601001 [hep-ph]].

\bibitem{Grimus:2008tt}
W.~Grimus and L.~Lavoura,
JHEP \textbf{09}, 106 (2008)
doi:10.1088/1126-6708/2008/09/106
[arXiv:0809.0226 [hep-ph]].

\bibitem{Grimus:2008nf}
W.~Grimus and L.~Lavoura,
Phys. Lett. B \textbf{671}, 456-461 (2009)
doi:10.1016/j.physletb.2008.12.041
[arXiv:0810.4516 [hep-ph]].

\bibitem{Grimus:2008vg}
W.~Grimus and L.~Lavoura,
JHEP \textbf{04}, 013 (2009)
doi:10.1088/1126-6708/2009/04/013
[arXiv:0811.4766 [hep-ph]].

\bibitem{Grimus:2009sq}
W.~Grimus and L.~Lavoura,
Phys. Lett. B \textbf{687}, 188-193 (2010)
doi:10.1016/j.physletb.2010.03.025
[arXiv:0912.4361 [hep-ph]].

\bibitem{Grimus:2009pg}
W.~Grimus, L.~Lavoura and P.~O.~Ludl,
J. Phys. G \textbf{36}, 115007 (2009)
doi:10.1088/0954-3899/36/11/115007
[arXiv:0906.2689 [hep-ph]].

\bibitem{Ferreira:2011hw}
P.~M.~Ferreira and L.~Lavoura,
``Seesaw Neutrino Masses from an $A_4$ Model with Two Equal Vacuum expectation values,''
[arXiv:1111.5859 [hep-ph]].

\bibitem{Park:2011zt}
N.~W.~Park, K.~H.~Nam and K.~Siyeon,
Phys. Rev. D \textbf{83}, 056013 (2011)
doi:10.1103/PhysRevD.83.056013
[arXiv:1101.4134 [hep-ph]].

\bibitem{BenTov:2012tg}
Y.~BenTov, X.~G.~He and A.~Zee,
JHEP \textbf{12}, 093 (2012)
doi:10.1007/JHEP12(2012)093
[arXiv:1208.1062 [hep-ph]].

\bibitem{Ivanov:2012fp}
I.~P.~Ivanov and E.~Vdovin,
Eur. Phys. J. C \textbf{73}, no.2, 2309 (2013)
doi:10.1140/epjc/s10052-013-2309-x
[arXiv:1210.6553 [hep-ph]].

\bibitem{Ivanov:2011ae}
I.~P.~Ivanov, V.~Keus and E.~Vdovin,
J. Phys. A \textbf{45}, 215201 (2012)
doi:10.1088/1751-8113/45/21/215201
[arXiv:1112.1660 [math-ph]].

\bibitem{TheCode}
Jiazhen Shao, The 4HDM Toolbox,
available at {\tt  https://github.com/JiazhenShao/4HDM-Toolbox.git}

\bibitem{GAP}
GAP - Groups, Algorithms, Programming
A System for Computational Discrete Algebra: 
 {\tt https://www.gap-system.org/}

\bibitem{Ecker:1981wv}
G.~Ecker, W.~Grimus and W.~Konetschny,
Nucl. Phys. B \textbf{191}, 465-492 (1981)
doi:10.1016/0550-3213(81)90309-6

\bibitem{Ecker:1983hz}
G.~Ecker, W.~Grimus and H.~Neufeld,
Nucl. Phys. B \textbf{247}, 70-82 (1984)
doi:10.1016/0550-3213(84)90373-0

\bibitem{Ecker:1987qp}
G.~Ecker, W.~Grimus and H.~Neufeld,
J. Phys. A \textbf{20}, L807 (1987)
doi:10.1088/0305-4470/20/12/010

\bibitem{Grimus:1995zi}
W.~Grimus and M.~N.~Rebelo,
Phys. Rept. \textbf{281}, 239-308 (1997)
doi:10.1016/S0370-1573(96)00030-0
[arXiv:hep-ph/9506272 [hep-ph]].

\bibitem{Branco:1999fs}
G.~C.~Branco, L.~Lavoura and J.~P.~Silva,
Int. Ser. Monogr. Phys. \textbf{103}, 1-536 (1999)

\bibitem{Ivanov:2015mwl}
I.~P.~Ivanov and J.~P.~Silva,
Phys. Rev. D \textbf{93}, no.9, 095014 (2016)
doi:10.1103/PhysRevD.93.095014
[arXiv:1512.09276 [hep-ph]].

\bibitem{Haber:2018iwr}
H.~E.~Haber, O.~M.~Ogreid, P.~Osland and M.~N.~Rebelo,
JHEP \textbf{01}, 042 (2019)
doi:10.1007/JHEP01(2019)042
[arXiv:1808.08629 [hep-ph]].

\bibitem{Ivanov:2018qni}
I.~P.~Ivanov and M.~Laletin,
Phys. Rev. D \textbf{98}, no.1, 015021 (2018)
doi:10.1103/PhysRevD.98.015021
[arXiv:1804.03083 [hep-ph]].

\bibitem{Ivanov:2013bka}
I.~P.~Ivanov and C.~C.~Nishi,
JHEP \textbf{11}, 069 (2013)
doi:10.1007/JHEP11(2013)069
[arXiv:1309.3682 [hep-ph]].

\bibitem{Ivanov:2005hg}
I.~P.~Ivanov,
Phys. Lett. B \textbf{632}, 360-365 (2006)
doi:10.1016/j.physletb.2005.10.015
[arXiv:hep-ph/0507132 [hep-ph]].

\bibitem{Ivanov:2006yq}
I.~P.~Ivanov,
Phys. Rev. D \textbf{75}, 035001 (2007)
[erratum: Phys. Rev. D \textbf{76}, 039902 (2007)]
doi:10.1103/PhysRevD.75.035001
[arXiv:hep-ph/0609018 [hep-ph]].

\bibitem{Ferreira:2010yh}
P.~M.~Ferreira, H.~E.~Haber, M.~Maniatis, O.~Nachtmann and J.~P.~Silva,
Int. J. Mod. Phys. A \textbf{26}, 769-808 (2011)
doi:10.1142/S0217751X11051494
[arXiv:1010.0935 [hep-ph]].

\bibitem{Ferreira:2023dke}
P.~M.~Ferreira, B.~Grzadkowski, O.~M.~Ogreid and P.~Osland,
Eur. Phys. J. C \textbf{84}, no.3, 234 (2024)
doi:10.1140/epjc/s10052-024-12561-8
[arXiv:2306.02410 [hep-ph]].

\bibitem{Ivanov:2012ry}
I.~P.~Ivanov and E.~Vdovin,
Phys. Rev. D \textbf{86}, 095030 (2012)
doi:10.1103/PhysRevD.86.095030
[arXiv:1206.7108 [hep-ph]].

\bibitem{Ivanov:2014doa}
I.~P.~Ivanov and C.~C.~Nishi,
JHEP \textbf{01}, 021 (2015)
doi:10.1007/JHEP01(2015)021
[arXiv:1410.6139 [hep-ph]].

\bibitem{Fallbacher:2015rea}
M.~Fallbacher and A.~Trautner,
Nucl. Phys. B \textbf{894}, 136-160 (2015)
doi:10.1016/j.nuclphysb.2015.03.003
[arXiv:1502.01829 [hep-ph]].

\bibitem{deMedeirosVarzielas:2019rrp}
I.~de Medeiros Varzielas and I.~P.~Ivanov,
Phys. Rev. D \textbf{100}, no.1, 015008 (2019)
doi:10.1103/PhysRevD.100.015008
[arXiv:1903.11110 [hep-ph]].

\bibitem{Darvishi:2021txa}
N.~Darvishi, M.~R.~Masouminia and A.~Pilaftsis,
Phys. Rev. D \textbf{104}, no.11, 115017 (2021)
doi:10.1103/PhysRevD.104.115017
[arXiv:2106.03159 [hep-ph]].

\bibitem{Bree:2024edl}
I.~Bree, D.~D.~Correia and J.~P.~Silva,
Phys. Rev. D \textbf{110}, no.3, 035028 (2024)
doi:10.1103/PhysRevD.110.035028
[arXiv:2407.09615 [hep-ph]].

\bibitem{Doring:2024kdg}
C.~D\"oring and A.~Trautner,
``Symmetries from outer automorphisms and unorthodox group extensions,''
[arXiv:2410.11052 [hep-ph]].

\bibitem{Shao:2023oxt}
J.~Shao and I.~P.~Ivanov,
JHEP \textbf{10}, 070 (2023)
doi:10.1007/JHEP10(2023)070
[arXiv:2305.05207 [hep-ph]].

\bibitem{Shao:2024ibu}
J.~Shao, I.~P.~Ivanov and M.~Korhonen,
J. Phys. A \textbf{57}, no.38, 385401 (2024)
doi:10.1088/1751-8121/ad7340
[arXiv:2404.10349 [hep-ph]].

\bibitem{Ishimori:2010au}
H.~Ishimori, T.~Kobayashi, H.~Ohki, Y.~Shimizu, H.~Okada and M.~Tanimoto,
Prog. Theor. Phys. Suppl. \textbf{183}, 1-163 (2010)
doi:10.1143/PTPS.183.1
[arXiv:1003.3552 [hep-th]].

\bibitem{Luhn:2007sy}
C.~Luhn, S.~Nasri and P.~Ramond,
Phys. Lett. B \textbf{652}, 27-33 (2007)
doi:10.1016/j.physletb.2007.06.059
[arXiv:0706.2341 [hep-ph]].

\bibitem{Hagedorn:2008bc}
C.~Hagedorn, M.~A.~Schmidt and A.~Y.~Smirnov,
Phys. Rev. D \textbf{79}, 036002 (2009)
doi:10.1103/PhysRevD.79.036002
[arXiv:0811.2955 [hep-ph]].

\bibitem{Cao:2010mp}
Q.~H.~Cao, S.~Khalil, E.~Ma and H.~Okada,
Phys. Rev. Lett. \textbf{106}, 131801 (2011)
doi:10.1103/PhysRevLett.106.131801
[arXiv:1009.5415 [hep-ph]].

\bibitem{Vien:2014gza}
V.~V.~Vien and H.~N.~Long,
JHEP \textbf{04}, 133 (2014)
doi:10.1007/JHEP04(2014)133
[arXiv:1402.1256 [hep-ph]].

\bibitem{Bonilla:2014xla}
C.~Bonilla, S.~Morisi, E.~Peinado and J.~W.~F.~Valle,
Phys. Lett. B \textbf{742}, 99-106 (2015)
doi:10.1016/j.physletb.2015.01.017
[arXiv:1411.4883 [hep-ph]].

\bibitem{gallian2021contemporary}
Gallian,~Joseph,
Contemporary abstract algebra, 
2021, Chapman and Hall/CRC,
ISBN: 978-130-565-796-0, 978-130-588-785-5

\end{thebibliography}
\end{document}